\pdfoutput=1
\documentclass[a4paper,fleqn]{cas-sc}
\usepackage[authoryear]{natbib}
\setcitestyle{authoryear,open=[,close=],aysep={ }}

\usepackage{pdflscape}

\usepackage[T1]{fontenc}
\usepackage{newtxtext}
\usepackage{float}

\usepackage{booktabs}
\usepackage{tabularx}
\usepackage{multirow}
\usepackage{siunitx}
\usepackage{colortbl,xcolor}

\DeclareRobustCommand{\VAN}[3]{#2}
\let\VANthebibliography\thebibliography
\def\thebibliography{\DeclareRobustCommand{\VAN}[3]{##3}\VANthebibliography}

\usepackage{graphicx}	
\usepackage{amsmath}	
\usepackage{amssymb}	
\usepackage{caption,subcaption}
\usepackage[english]{babel}
\usepackage[flushleft]{threeparttable}
\usepackage{xspace}
\usepackage{hyperref}

\newcommand{\chandra}{\textsl{\textsc{Chandra }}}

\newcommand\blfootnote[1]{%
  \begingroup
  \renewcommand\thefootnote{}\footnote{#1}%
  \addtocounter{footnote}{-1}%
  \endgroup
}

\newcommand{\cv}{C\,\textsc{v}\xspace}
\newcommand{\cvi}{C\,\textsc{vi}\xspace}
\newcommand{\nvi}{N\,\textsc{vi}\xspace}
\newcommand{\nvii}{N\,\textsc{vii}\xspace}
\newcommand{\ovii}{O\,\textsc{vii}\xspace}
\newcommand{\oviii}{O\,\textsc{viii}\xspace}
\newcommand{\neix}{Ne\,\textsc{ix}\xspace}
\newcommand{\nex}{Ne\,\textsc{x}\xspace}
\newcommand{\mgxi}{Mg\,\textsc{xi}\xspace}
\newcommand{\mgxii}{Mg\,\textsc{xii}\xspace}
\newcommand{\alxii}{Al\,\textsc{xii}\xspace}
\newcommand{\alxiii}{Al\,\textsc{xiii}\xspace}
\newcommand{\sixiii}{Si\,\textsc{xiii}\xspace}
\newcommand{\sixiv}{Si\,\textsc{xiv}\xspace}

\begin{document}

\let\WriteBookmarks\relax
\def\floatpagepagefraction{1}
\def\textpagefraction{.001}

\shorttitle{NICER cometary X-ray Survey}

\shortauthors{T. K. Deskins et~al.}

\title [mode = title]{Bayesian Modeling of NICER Cometary X-ray Spectra: A Legacy Survey of Solar-Wind Charge Exchange}         




%
\author[1]{T. K. Deskins}[type=editor,
                        role=Researcher,
                        orcid=0000-0001-6644-6484]
\credit{Conceptualization, Methodology, Software, Validation, Formal analysis, Data Curation, Writing - Original Draft, Writing - Review \& Editing, Visualization, Funding acquisition}

\author[1]{Dennis Bodewits}[type=editor, role=PI]
\credit{Conceptualization, Methodology, Supervision, Project administration, Writing - Review \& Editing, Funding acquisition}

\author[1]{Steven Bromley}[type=editor]
\credit{Writing - Review \& Editing}

\author[2]{Konrad Dennerl}[type=editor]
\credit{Writing - Review \& Editing}

\author[3]{Damian J. Christian}[type=editor]
\credit{Writing - Review \& Editing}

\author[4]{Du\v{s}an Odstr\v{c}il
}[type=editor]
\credit{Resources}

\affiliation[1]{organization={Department of Physics, Auburn University},
    addressline={Edmund C. Leach Science Center}, 
    city={Auburn},
    postcode={36832}, 
    state={AL},
    country={USA}}
    
\affiliation[2]{organization={Max Planck Institute for Extraterrestrial Physics},
    addressline={Gießenbachstraße 1}, 
    city={Garching},
    postcode={85748}, 
    state={Bavaria},
    country={Germany}}

\affiliation[3]{organization={Eureka Scientific},
    addressline={2452 Delmer Street Suite 100}, 
    city={Oakland},
    postcode={94602}, 
    state={CA},
    country={USA}}

\affiliation[4]{organization={Department of Physics \& Astronomy, George Mason University},
    addressline={Planetary Hall, 4400 University Drive, MSN 3F3},
    city={Fairfax},
    postcode={22030},
    state={VA},
    country={USA}}
    
\cortext[cor1]{Corresponding author. Email: \url{tkd0009@auburn.edu}}

\date{}

\begin{abstract}
We present a uniform, epoch-resolved analysis of soft X-ray observations of eight comets obtained with NICER, using Bayesian statistics to identify charge-exchange line components, measure relative ion fluxes, and infer nominal solar-wind freeze-in temperatures.
The sample exhibits recurring spectral morphologies that fall into distinct families: carbon-dominated, intermediate, and nitrogen-/oxygen-dominated.
Epoch-resolved flux ratios yield a robust separation between diagnostics:
carbon-derived freeze-in temperatures cluster near $T_{\mathrm{freeze}}(C)\approx 1.4-1.7$\,MK, while nitrogen- and oxygen-derived diagnostics are systematically higher, typically $T_{\mathrm{freeze}}(N,O)\approx 2.0-2.3$\,MK.
Short-timescale variability in inferred freeze-in conditions is common, indicating that instantaneous solar-wind charge-state fluctuations, rather than large changes in coma composition, dominate the spectral differences.
We discuss instrumental and modeling limitations, demonstrate how our Bayesian fitting method mitigates degeneracies via physically motivated priors and Bayesian model selection, and recommend laboratory measurements and coordinated high-resolution X-ray observations to refine charge-exchange diagnostics and validate low-resolution inferences.
\end{abstract}
\begin{highlights}
\item Epoch-resolved Bayesian Color-Model fits to NICER spectra of eight comets reveal three recurring spectral families (carbon-dominated, intermediate, nitrogen/oxygen-dominated) that trace upstream solar-wind charge-state conditions.

\item Carbon-based freeze-in temperatures cluster near $1.4$–$1.7$\,MK, systematically cooler than nitrogen/oxygen diagnostics at $2.0$–$2.3$\,MK; day-to-week variability argues for epoch-by-epoch analysis.

\item The Bayesian Color Model captures subtle neutral-dependent (H$_2$O vs.\ CO$_2$/CO) spectral shifts in low-resolution NICER data, while residual degeneracies highlight the need for improved charge-exchange cross sections and future high-resolution SWCX spectroscopy.
\end{highlights}


\maketitle
 
\section{Introduction}

\begin{figure}
    \centering
    \begin{subfigure}[b]{0.49\textwidth}
        \centering
        \includegraphics[width=\linewidth,trim={0.4cm .2cm 0.5cm 1.25cm},clip]{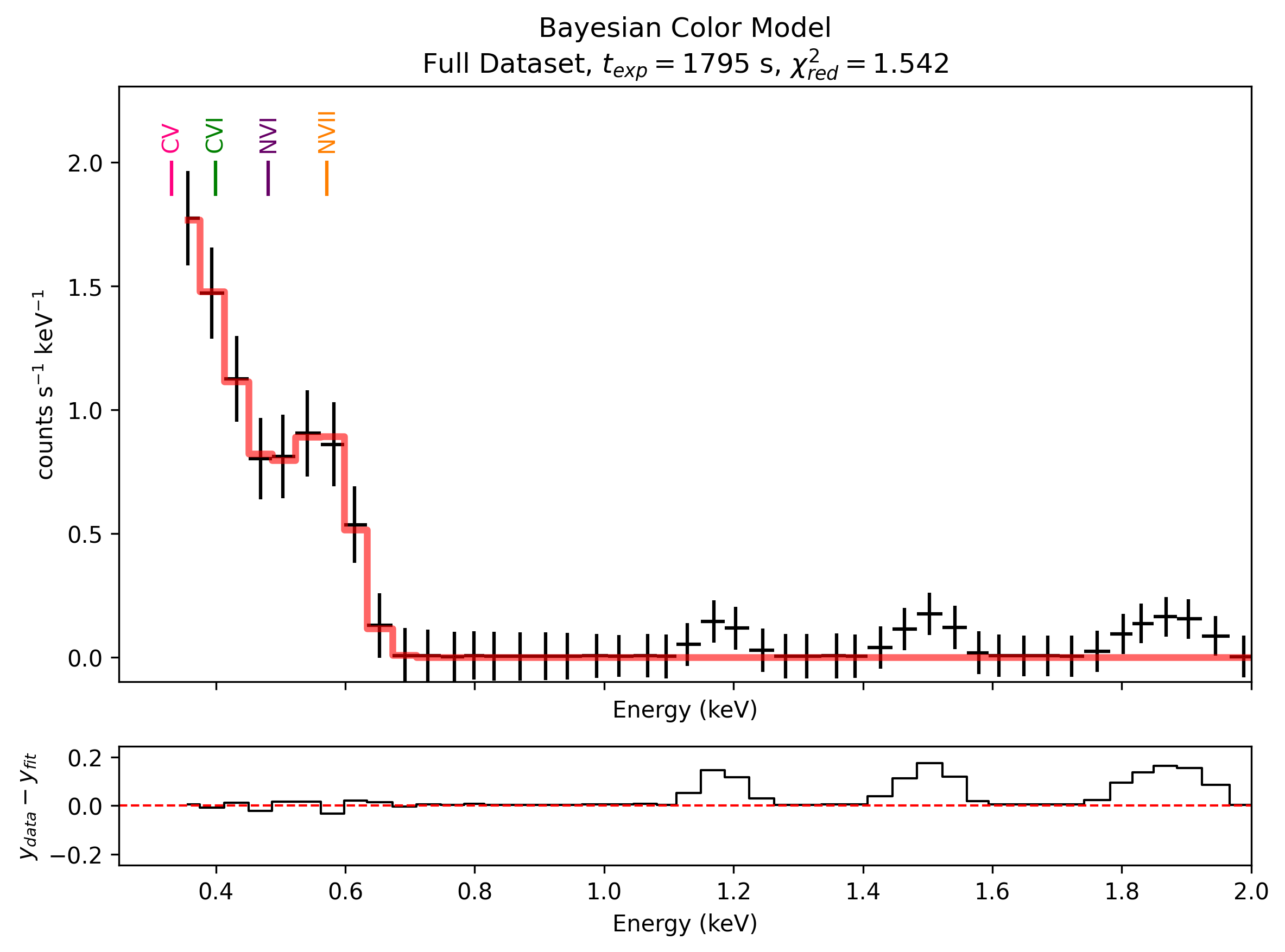}
        \caption{12P/Pons-Brooks.}
        \label{fig:spec_12P}
    \end{subfigure}\hfill
    \begin{subfigure}[b]{0.49\textwidth}
        \centering
        \includegraphics[width=\linewidth,trim={0.4cm .2cm 0.5cm 1.25cm},clip]{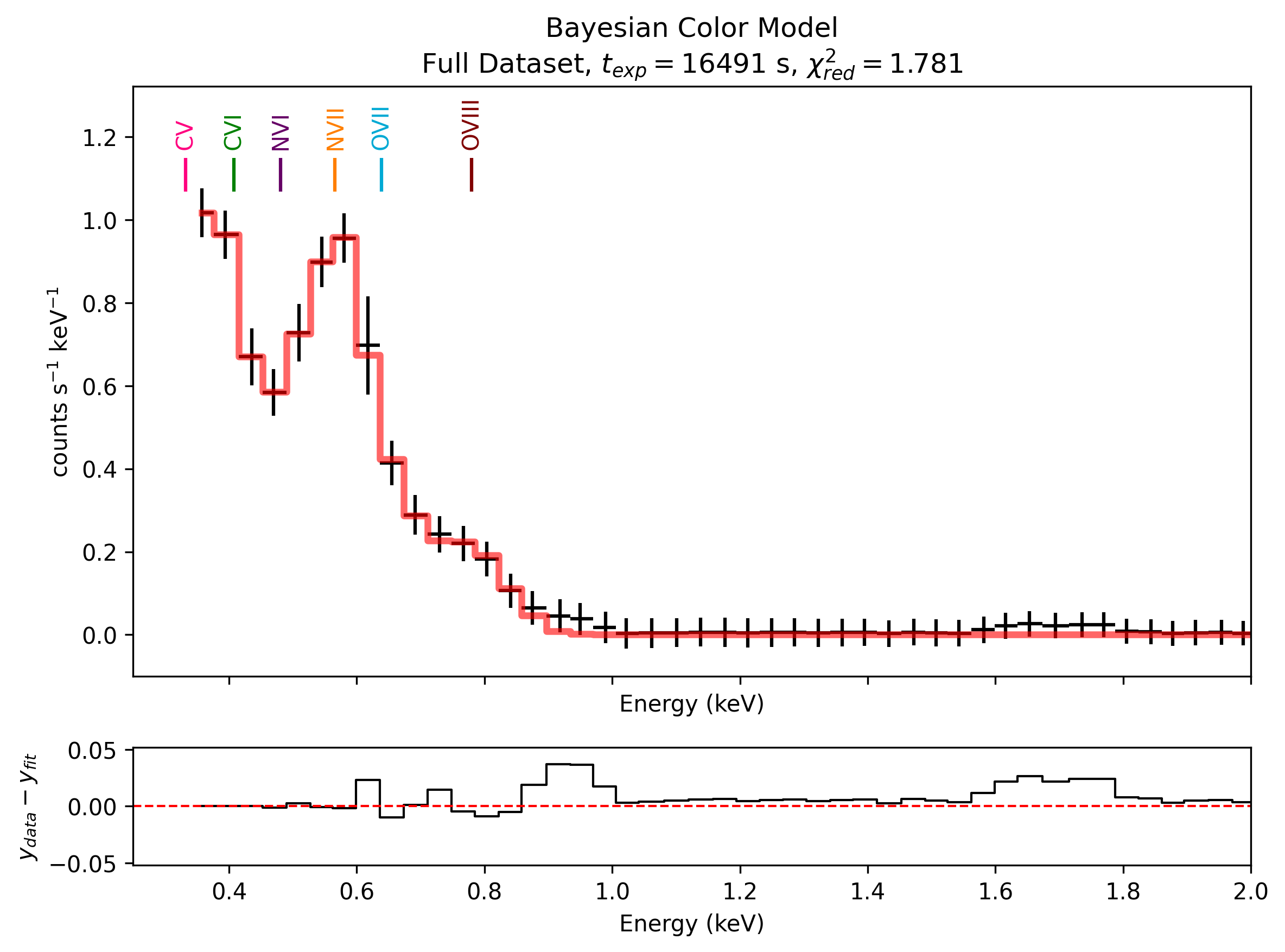}
        \caption{C/2017 K2 (PANSTARRS).}
        \label{fig:spec_K2}
    \end{subfigure}

    \caption{X-ray spectrum from the interaction between the solar wind and the atmospheres of 12P/Pons-Brooks (left) and C/2017 K2 (PANSTARRS) (right), each dataset fitted with the Bayesian Color Model (BCM). The BCM utilizes Bayesian principles to quantitatively determine optimal parameters for fitting the low-resolution data. By iterating through peak combinations associated with \cv, \cvi, \nvi, \nvii, \ovii, \oviii, \neix, \nex, \mgxi, \mgxii, \alxii, \alxiii, \sixiii, and \sixiv, the model selects the combination with the best fit and minimal complexity, as indicated by the minimum Bayesian Information Criterion (BIC). As the BCM’s BIC penalty function excluded the higher-energy Ne, Mg, Al, and Si lines, the fit is limited to data below 1\,keV. Note that the BCM found it best to exclude O from consideration in the fit of 12P/Pons-Brooks, likely due to the degeneracy between N and O.}
    \label{fig:spec_both}
\end{figure}

Comets are primordial remnants of the early Solar System, composed predominantly of dust and volatile ices.  As they approach the Sun, volatile sublimation generates an extensive, neutral coma, typically consisting of mainly H$_2$O, CO$_2$, and/or CO, and interacts directly with the solar wind. 
This interaction gives rise to plasma boundaries such as bow shocks and contact surfaces and drives charge‐exchange (CX) reactions between highly charged solar‐wind ions and cold cometary neutrals \citep{Cravens2002}. 
Solar events such as flares, shocks, and coronal mass ejections (CME) can cause variability in solar‐wind conditions, which can alter the coma environment on timescales of hours, leaving temporal and compositional signatures in the resulting X‑ray emission \citep{Cravens1987, Goetz2022}, which can be revealed in the spectra created from data collected by X-ray telescopes, an example of which may be seen in Figure~\ref{fig:spec_both}.

Charge exchange (CX) emission occurs anywhere an ionized plasma encounters a cold, neutral gas \citep{Dennerl2010}.
This emission has been observed in planetary atmospheres, the geocorona, supernova remnants, and star-forming regions.
Comets emit X-rays as highly charged ions in the solar wind undergo CX reactions with the cold (50\,K) cometary neutral gases \citep{Cravens2002}.
Our understanding of X-rays from comets evolved over the years through numerous scenarios. Among them were scattering/fluorescence of solar X-rays \citep{krasnopolsky1997nature}, thermal bremsstrahlung associated with collisions of solar-wind electrons with cometary neutral gas or dust \citep{owens1998evidence}, electron/proton K- and L-shell ionization \citep{krasnopolsky1997nature}, and Rayleigh-scattering of solar X-rays by attogram dust particles \citep{owens1998evidence} and charge exchange between highly ionized solar-wind minor ions and cometary neutral species \citep{cravens1997comet}. A thorough comparative study by \citet{krasnopolsky1997nature} demonstrated that none of these mechanisms except for the solar-wind charge-exchange emission model could account for more than 5\% of the observed luminosities.
The resulting solar wind charge exchange (SWCX) X-ray spectra mostly consist of Ly and K-series of H- and He-like ions. Detection of forbidden lines of He-like ions is an obvious marker of CX because electron capture in CX preferentially populates triplet states, whose slow radiative decay produces the forbidden transition, while ordinary electron-impact excitation typically favors the resonance line \citep{kharchenko2000spectra}. SWCX emission is useful as a diagnostic of properties of both the solar wind and the cometary environment. For example, the oxygen lines are commonly used for temperature and density diagnostics of thermal emission from diffuse plasmas \citep{Snowden2004}.
Recent in-situ measurements from NASA's Parker Solar Probe \citep{fox2016solar} and ESA's Solar Orbiter \citep{muller2020solar}, together with ongoing advances in three-dimensional MHD solar-wind models (e.g., ENLIL; \citep{odstrcil2003modeling}), have driven sustained interest in the detailed physics of the solar wind.


Since the ROSAT discovery of cometary X‑rays \citep{lisse1996discovery, dennerl1997x}, higher‐resolution instruments such as Chandra HETG, XMM‑Newton RGS have revealed individual emission lines and demonstrated the need for more detailed CX models \citep{lisse2001charge, Beiersdorfer2003p596, Bodewits2007}.
However, the diffuse nature of comets and the limited photon yields in observations hinder the capabilities of grating spectrometers.
With relatively small effective areas, on the order of 59\,cm$^2$ at 1\,keV for Chandra HETG, the resulting spectra often suffer from line blending and low signal‐to‐noise \citep{canizares2000high, dennerl2003xray}. 
Moderate‐resolution instruments such as NICER (E/$\Delta$E$\approx6-80$) and XMM‑Newton EPIC thus remain essential for surveying cometary SWCX across multiple targets \citep{Gendreau2016_nicerdesign, brinkman1996reflection}.

A growing body of work has also examined the role of non–CX mechanisms at higher X-ray energies. In particular, \citet{snios2018presence} demonstrated that while charge exchange overwhelmingly dominates cometary emission below 1\,keV, coherent scattering of solar X-rays by dust and ice grains can contribute significantly between 1–2\,keV during periods of elevated solar activity. Their combined CX+scattering models reproduced Chandra spectra of comets ISON and Ikeya–Zhang at energies where anomalously high charge states would otherwise be required.

XRISM and other next-generation high-resolution spectrometers will not only increase signal fidelity but will resolve individual SWCX lines, allowing direct measurement of triplet ratios and single-line fluxes that distinguish CX from thermal emission, constrain ion charge-state distributions and freeze-in temperatures, and empirically test state-selective capture cross sections \citep{xrism2020science}. This will also provide high-quality spectra needed to set physically motivated BCM priors and to validate abundance and temperature inferences derived from lower-resolution instruments, thereby allowing for reanalysis of archival data with greater confidence.
Despite the promise of next-generation instruments, those of the current generation are far from obsolete, as XRISM currently has no access to low energies due to its gate-valve issues \citep{ishisaki2022status}. 
As a result, Chandra and XMM-Newton provide, and will continue to provide, the most valuable low-energy data to the scientific community.
\begin{figure}
    \centering
    \includegraphics[width=0.6\linewidth,trim={0.0cm 0.0cm 0.0cm 0.5cm},clip]{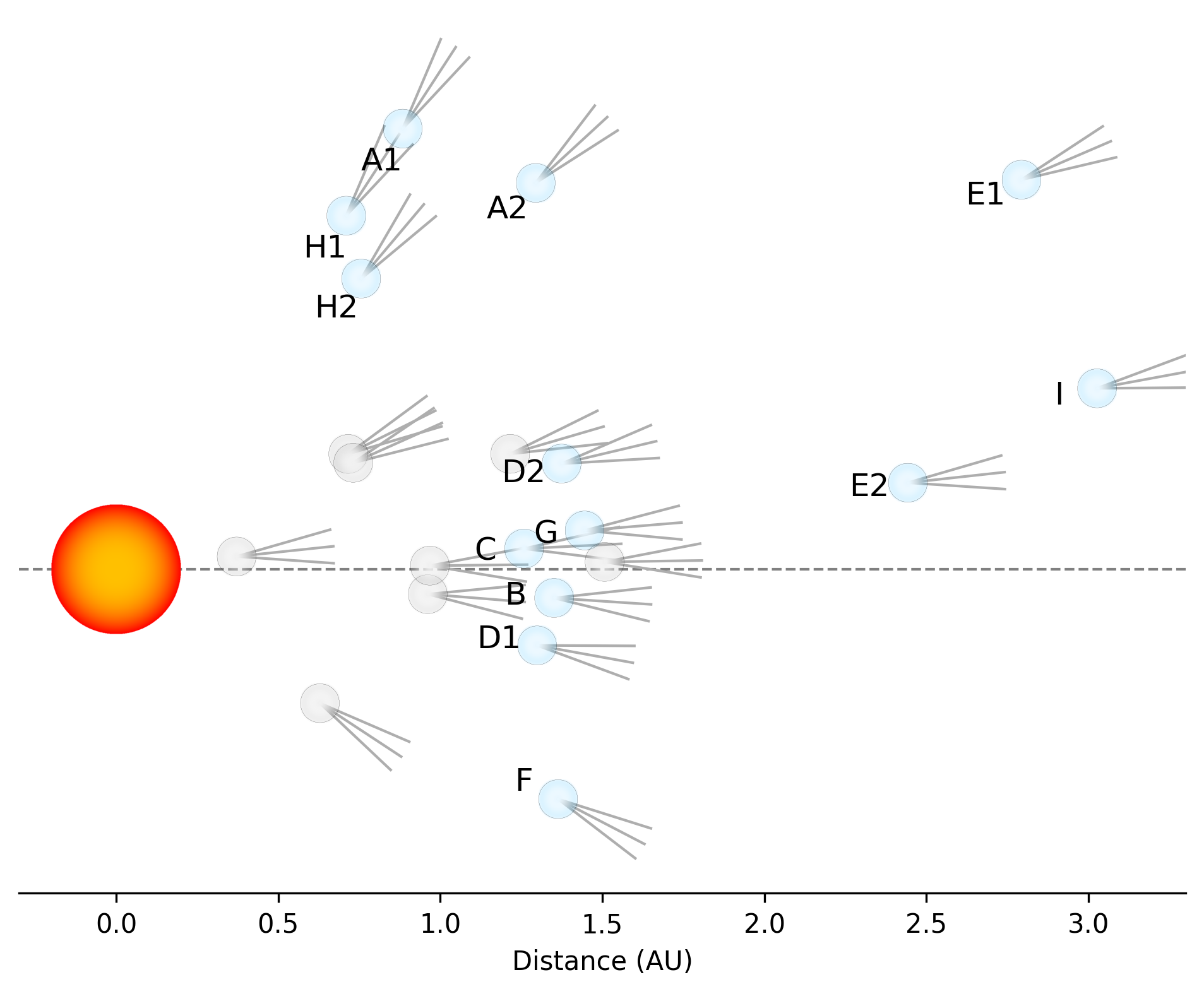}
    \caption{Heliocentric latitude projection of observed comet positions on their respective observation dates. The Sun is at the origin and the dashed grey line marks the ecliptic plane. Each comet is plotted at $x=r\cos\beta$, $y=r\sin\beta$ (with $r$ the heliocentric distance in AU and $\beta$ the heliocentric ecliptic latitude). Letters mark observations and correspond to:(A1) C/2017 T2: 2020-04-24;
    (A2) C/2017 T2: 2020-06-26;
    (B) 88P/Howell: 2020-10-01;
    (C) 67P/Churyumov–Gerasimenko: 2021-11-30;
    (D1) 19P/Borrelly: 2022-01-17;
    (D2) 19P/Borrelly: 2022-03-19;
    (E1) C/2017 K2: 2022-05-31;
    (E2) C/2017 K2: 2022-08-07;
    (F) C/2022 E3 (ZTF): 2023-03-21;
    (G) 62P/Tsuchinshan 1: 2024-02-23;
    (H1) 12P/Pons–Brooks: 2024-02-23;
    (H2) 12P/Pons–Brooks: 2024-03-04;
    (I) P/2010 H2 (Vales): 2025-04-02. 
    Observations labeled “1” and “2” indicate initial and final positions for targets that moved significantly during our observing plan; single-letter labels denote targets with negligible positional change. Comets colored gray and without accompanying letter labels are from the Chandra survey described in \citet{Bodewits2007} and include C/1999 S4 (LINEAR), C/1999 T1 (McNaught–Hartley), C/2000 WM1 (LINEAR), 153P/2002 (Ikeya–Zhang), 2P/Encke, C/2001 Q4 (NEAT), 9P/Tempel 1, and 73P/Schwassmann–Wachmann 3 fragment B. Axes are in AU.}

    \label{fig:sun_with_comets}
\end{figure}
A systematic \chandra\ survey of eight comets showed that cometary X-ray spectra can be linked to solar-wind freeze-in temperatures even with low-resolution solar-wind data \citep{Bodewits2007}, but the sample was small and largely confined to snap-shots at low heliolatitudes \ref{fig:sun_with_comets}.
Traditional forward-modeling approaches therefore struggle to disentangle contributions from different ion species for comets with low X-ray count rates, particularly when theoretical CX cross sections lack extensive laboratory validation at relevant solar-wind speeds \citep{Bodewits2007, Bromley2022}.
The NICER archive advances this state of the art. 
With sensitivity in the $0.2-2$\,keV band, a large effective area, and rapid response, NICER extends comet detections to fainter targets, larger heliocentric distances, and higher latitudes.
This unique dataset more than doubles the \chandra\ sample and provides the first opportunity to probe charge exchange systematically across diverse heliospheric environments and deliver a legacy archive for heliophysics and astrophysics.
Similarly, XMM-Newton EPIC studies of CMEs interacting with Earth’s exosphere have shown that complex background subtraction and time-dependent ion fluxes can bias inferred line strengths \citep{carter2010high, ishi2019suzaku}. 
These challenges underscore the need for a fitting framework that can incorporate prior knowledge of expected line-ratio distributions and background variability when analyzing low- to moderate-resolution data from current-generation instruments.

\begin{figure}
    \centering
    \includegraphics[width=0.7\linewidth]{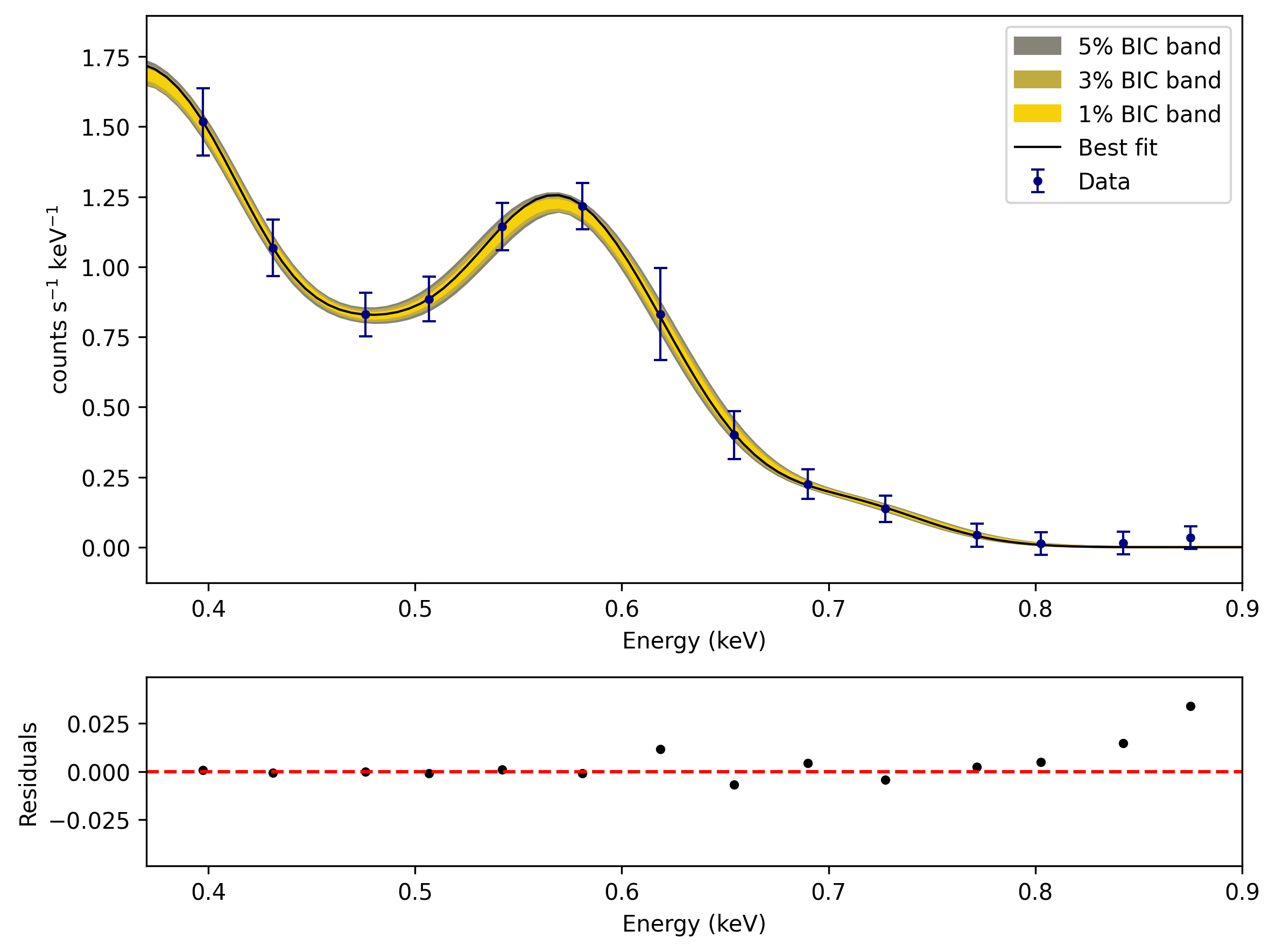}
    \caption{Top: X-ray spectrum for the full 62P/Tsuchinshan dataset from the solar wind's interaction with the atmosphere of comet 62P/Tsuchinshan, fitted with the Bayesian Color Model (BCM), which uses the Bayesian Information Criterion (BIC) as a model-selection tool to balance model complexity and goodness of fit. We only consider energies between 370\,eV and 900\,eV because no lines are fitted above 900\,eV. Colored sensitivity bands, calculated through a Monte Carlo perturbation analysis, show the degeneracy of the fit. Percent difference is an objective measurement of comparing the BICs of the many degenerate solutions; its calculation is discussed herein. Bottom: residual plot depicting the difference between the data, represented by dots, and the fit, represented by a dashed red line. \textbf{A quantitative assessment of goodness-of-fit is provided by the BIC comparison rather than visual inspection of residual scatter alone.}}
    \label{fig:deg_62P}
\end{figure}

To address these limitations, we have developed the Bayesian Color Model (BCM), which imposes physically motivated priors on line ratios and background levels, reduces degeneracy in multi‐component fits, and yields robust estimates of ion abundances and freeze‑in temperatures, even when applied to low‐resolution data \citep{deskins2025modeling}. 
Figure~\ref{fig:spec_12P} shows how the BCM fits the full dataset of comet 12P/Pons-Brooks, and Figure~\ref{fig:spec_K2} shows how it fits the dataset of comet C/2017 K2 (PANSTARRS). These two spectra were chosen to represent a sample of our dataset, as the spectra for the other comets presented in this survey lie between these two spectra. The other spectra are presented in Appendix~\ref{app:spectra}. 
A visual representation of the degenerate fits inherent to low-resolution data is depicted in Figure~\ref{fig:deg_62P}. 

Recent advances in theoretical cross‑section calculations and laboratory benchmarking have improved the fidelity of CX models;
nonetheless, parameter degeneracies remain a major obstacle. 
Progress is also constrained by gaps in direct, long‑term in‑situ measurement of solar‑wind ion composition: 
many legacy instruments are aging or partially disabled, and until very recently, new dedicated composition monitors have been rare. 
However, a number of missions under development or nearing deployment promise improvements in our ability to measure ion charge states and composition continuously \citep{koutroumpa2025empirical}.
\citet{Koutroumpa2024} provided compound cross sections for heavy solar‐wind ions interacting with H and He.
\citet{Bromley2022} employ a time‑dependent lattice method to compute cross‑sections for bare Ne collisions; experimental measurements remain pending.
Yet when applied to low‐resolution cometary spectra, model fits can yield non‐physical ion abundances without strong priors.
Bayesian hierarchical methods have shown promise in astrophysical spectroscopy more broadly—for example, in resolving crowded line complexes in galaxy cluster plasmas \citep{Carter2022earth, Stansby2019diagnosing}.
The BCM extends these techniques to SWCX by defining color parameters that link line intensities through physically motivated priors, reducing the likelihood of false correlations and improving the accuracy of freeze‑in temperature estimates.

To demonstrate the generality of BCM beyond a single comet, we apply it to a multi‐comet dataset spanning diverse heliocentric distances and solar‐wind regimes.
Our sample includes both short-period and long-period comets. 
Short-period objects include the Jupiter-family comets 19P/Borrelly, 62P/Tsuchinshan, 67P/Churyumov–Gerasimenko, and 88P/Howell, as well as the Halley-type 12P/Pons–Brooks, while the long-period targets are C/2017 K2 (PANSTARRS), C/2017 T2 (PanSTARRS), and C/2022 E3 (ZTF); all were observed with NICER.
Comet P/2010 H2 (Vales) is not analyzed as our signal-to-noise ratio was too low for meaningful interpretation.
We leverage in situ Rosetta measurements \citep{Taylor2017, Combi2020} and remote‐sensing light curves \citep{Womack2021visual} to anchor our priors on gas production rates and neutral composition, ensuring that our results are reasonable and do not wildly contradict the measurements conducted by other instruments.
By comparing the BCM‐derived ion fractions and freeze‑in temperatures across this heterogeneous sample of comets, we can assess the model’s robustness, identify any systematic biases, and explore how variations in neutral abundances, solar‐wind speed, and instrument response shape the inferred SWCX spectra.

\textbf{The BCM framework is also naturally extendable to interactions between CME plasma and neutral gas. CMEs often exhibit elevated freeze-in temperatures and enhanced abundances of high charge states relative to ambient slow or fast solar wind. 
In such cases, the charge-exchange emission mechanism remains unchanged, but the upstream charge-state distribution shifts toward higher ionization stages, leading to strengthened high-energy lines and modified line ratios. 
Within the BCM, this scenario is accommodated by broadening the prior space in freeze-in temperature and allowing for increased ion flux and potential temporal variability during the observation window.}

In this work, we extend the BCM framework beyond a single case study of comet 67P to a sample of nine comets observed with NICER, eight of which yield spectra suitable for SWCX analysis. We use this multi-comet, multi-epoch dataset to probe the robustness of the BCM and to explore how cometary spectra map onto solar-wind charge-state conditions.
Specifically, we aim to (1) demonstrate the efficacy of the BCM in extracting reliable ion composition and freeze-in temperatures from low-resolution SWCX spectra so these quantities can be compared across targets and epochs; 
(2) quantify how variations in solar-wind state produce systematic, time-dependent spectral signatures in different comae, thereby characterizing the cometary response function to impulsive and steady wind perturbations; and (3) evaluate the robustness of the BCM against conventional fitting methods and use the inferred timing and ion-flux histories to test and constrain solar-wind propagation predictions.
Achieving these goals will (1) provide empirical constraints on the time scales and compositional sensitivity of cometary SWCX, (2) produce a small set of well-characterized, multi-epoch SWCX spectra that can serve as data for priors used in archival low-resolution studies and for future high-resolution missions, and (3) offer an independent, remote sensing test of solar-wind propagation and CME impact forecasts that complements in-situ measurements. 


\section{Target Comet Sample}
In this section, we summarize the key properties and observational highlights of each comet in our survey. Each comet in the sample was chosen for its brightness, heliocentric distance, and heliocentric latitude: selecting a bright comet, often the brightest of each year, increases the likelihood of a high signal-to-noise ratio. We endeavored to select comets of varying heliocentric latitudes to sample different solar winds. Each of the following summaries provides the orbital context, photometric behavior, and gas production characteristics that inform our interpretation.
Figure~\ref{fig:sun_with_comets} places the nine targets in their heliocentric-latitude and distance context on the dates of observation; 
each point marks a comet and is labeled with the letter used throughout this paper. 
Compared with the benchmark \chandra\ survey \citep{Bodewits2007}, which consisted mainly of low-heliolatitude snap-shots, our NICER sample more than doubles the number of objects and samples a substantially larger range of heliocentric latitudes (Fig.~\ref{fig:sun_with_comets}) while covering similar or larger heliocentric distances.
Since the SWCX signal depends on local ion flux and composition, this broader latitudinal coverage, and the different solar-wind streamlines it samples, provides a critical test for how CX line ratios and inferred freeze-in temperatures vary with heliocentric latitude and coronal source region.
The positions in Figure~\ref{fig:sun_with_comets} therefore supply essential context for interpreting the spectral and brightness variations discussed below. 
Brief summaries and observational highlights for each labeled target follow.

\subsection{C/2017 T2}
C/2017 T2 (PanSTARRS) was observed by our program in April–June 2020, bracketing its perihelion passage on 2020 May 4 ($q=1.61$\,AU) \citep{bahyl2022comet}.
The comet reached visual brightness of order V $\approx8$ near perihelion and showed well-developed inner-coma structure and jets in ground-based imaging \citep{bahyl2022comet}. 
Time-series photometry of the nucleus yields a rotation period of about 5.676\,h \citep{bahyl2022comet}. 
SOHO/SWAN Lyman-alpha monitoring provides a measured water-production history for this object;
SWAN derived water production rates for C/2017 T2 that are of order $10^{28}\,\mathrm{molecules\,s^{-1}}$ near perihelion (peak values $\approx1\times10^{28}\,\mathrm{molecules\,s^{-1}}$), establishing that the coma was substantially active at $r_h\approx1.6$\,AU \citep{combi2021water}.
The comet’s relatively high northern ecliptic latitude (inclination $\approx57^\circ$) placed it favorably for our observing geometry.
The combination of (1) multiple epochs during the 2020 apparition, (2) documented inner-coma jets and concentric structures, and (3) SWAN-measured water production makes C/2017 T2 a useful comparison object in our sample for linking NICER SWCX diagnostics to contemporaneous photometric and gas-production behavior \citep{bahyl2022comet,combi2021water}.

\subsection{88P/Howell}
88P/Howell is a short-period comet that reached perihelion on 2020 Sep 26 and was observed by our program through the 2020 apparition (Sep–Nov 2020).
The comet remained close to the ecliptic plane and reached modest visual brightness, with a peak near $V\approx8-9$, with numerous amateur and survey images documenting a compact, green coma indicative of strong C$_2$ gas emission \citep{Yoshida2020_88P}.
Infrared spectroscopy from the AKARI survey places 88P among comets with relatively elevated CO$_2$ relative to H$_2$O, of order a few–$\approx$25\% in the AKARI sample, implying a non-negligible role for CO$_2$ in driving activity and gas-rich coma morphology \citep{ootsubo2012akari}.
As 88P is nearby, regularly returning, and relatively accessible, it has been highlighted as a potential sample-return target in mission studies \citep{sandford2017corsair}. 88P has not been the subject of extensive detailed compositional or high-resolution plasma studies, so our NICER observations contribute useful remote-SWCX constraints on a broadly accessible, CO$_2$-bearing Jupiter-family comet.

\subsection{67P/Churyumov–Gerasimenko}
67P is a short-period, Jupiter-family comet and the primary target of ESA’s Rosetta mission, which accompanied the nucleus through an extended 2014–2016 campaign and provided the definitive in-situ context for remote studies \citep{Taylor2017}.
Rosetta and complementary ground-based monitoring reveal spatially heterogeneous outgassing that make 67P less intrinsically bright than the largest long-period comets but richly structured in time and space \citep{Combi2020,jehin2022trappist,Noonan23,rosenbush2025comprehensive}. 
In situ plasma measurements documented efficient pickup of cometary ions and variable plasma sources and loss processes that strongly modulate the ion flux near the comet \citep{Nilsson2015,Heritier2018,Goetz2022}. 
The discovery of far-UV auroral emission further underscores the complex coupling between neutrals and electrodynamics at 67P \citep{Galand2020}.
These combined neutral and plasma constraints make 67P an ideal testing ground for how in-situ ion/neutral conditions map onto integrated SWCX X-ray diagnostics and for validating the NICER-based modeling presented in \citet{deskins2025modeling}.
Near perihelion the comet’s gas production is dominated by water \citep{combi2020surface}.
Detailed Rosetta measurements show that CO$_2$ typically makes up only a few percent of the total gas, or roughly 4–8\% averaged over the mission \citep{rubin2023volatiles}, while CO and O$_2$ are present at the percent level.
These fractions vary with season, latitude, and heliocentric distance and carry per-species uncertainties of order $\approx$20\% \citep{combi2020surface,lauter2020gas}.

\subsection{19P/Borrelly}
19P/Borrelly, the target of NASA’s Deep Space 1 flyby \citep{boice2000deep}, is an elongated ($8\times4\times4$\,km) Jupiter-family comet.  Its 2022 perihelion occurred on 2022 February 1 ($q=1.36$\,AU).  Narrowband photometry from TRAPPIST‐North and South \citep{jehin2022trappist} on 2022-01-17 ($r_h=1.32$\,AU, $\Delta=1.22$\,AU, $\Delta T=-7$\,d) yields $Q({\mathrm OH})=(2.76\pm0.41)\times10^{28}\,\mathrm{s^{-1}}$ and $A(0)f\rho_{\rm BC}=1847\pm35$\,cm.  Historical apparitions show a gradual post‐perihelion decline in water production \citep{Combi2020}.

\subsection{C/2017 K2 (PANSTARRS)}
K2 is a long‐period comet with perihelion at $q=1.79$\,AU on Dec. 19, 2022 \citep{hmiddouch2025variations}. 
C/2017 K2 is often described as dynamically new, and improved elements by Nakano (NK~5335) $1/a_{\rm origin}=+5.0\times10^{-5}\,\mathrm{AU^{-1}}$ support this, as this meets the classical criterion ($1/a_{\rm origin}<1\times10^{-4}\,\mathrm{AU^{-1}}$) \citep{Nakano2024_NK5335}.
Its highly inclined orbit (northern latitudes) and early activity detection at $\sim24$\,AU imply hypervolatile‐driven coma \citep{yang2021discovery}. 
Peak brightness reached $V\approx8$ near perihelion \citep{hmiddouch2025variations}. 
Our NICER campaign (May 2022–January 2023) spans $r_h=3.23$–$1.87$\,AU and includes multiple epochs beyond $2.5$\,AU, when activity was still plausibly dominated by species more volatile than H$_2$O.
Early observations (May–June 2022) likely predate significant H$_2$O production; $Q({\rm H_2O})$ rose from $\sim1.7\times10^{28}$ to $2.6\times10^{28}\,\mathrm{s^{-1}}$ by June, reaching $1.3\times10^{29}\,\mathrm{s^{-1}}$ in January 2023. 
Ground‐based O\,\textsc{I}\xspace green/red ratio measurements suggest CO/CO$_2$ contributions declined rapidly as water sublimation commenced \citep{cambianica2023co2,combi2025water}.
The combination of multiple epochs, including those outside the water–ice line, makes K2 a key case for probing spectral differences between volatile drivers \citep{Kuntz2019}.

\subsection{C/2022 E3 (ZTF)}
C/2022 E3 (ZTF) became a naked-eye object in early 2023, peaking at $V\approx4.5$ when at 0.28 AU from Earth.  Its retrograde, high‐inclination orbit ($i=109^\circ$) confined it to southern skies.  Perihelion occurred on 2023 January 12 ($q=1.11$\,AU).  Remote photometry shows an irregular activity profile: peak water production one month before perihelion followed by a factor-of-4 decline by $T_p+35$\,d ($Q\simeq3.5\times10^{28}\,\mathrm{s^{-1}}$; \citep{combi2025water}.  A rapid periodic modulation ($P\approx0.35$\,d) in $Q({\rm H_2O})$—varying between $5.5$–$6.5\times10^{28}\,\mathrm{s^{-1}}$—was attributed to nucleus rotation \citep{manzini2023rotation}.

\subsection{62P/Tsuchinshan}
62P/Tsuchinshan is a short-period ($P=6.37$\,yr) Jupiter-family comet with perihelion on 2023 December 25 ($q=1.38$\,AU). It peaked at $V\approx8$ at perihelion and was observed at $V\approx8$ in January 2024 \citep{Yoshida2024_62P}.
TRAPPIST observations on 2024-01-11 ($r_h=1.28$\,AU, $\Delta=0.51$\,AU, $\Delta T=+17$\,d) report $Q({\rm OH})=(9.06\pm0.98)\times10^{27}\,\mathrm{s^{-1}}$, $A(0)f\rho_{\rm RC}=311\pm7$\,cm, and $A(0)f\rho_{\rm BC}=255\pm18$\,cm \citep{jehin2024atel}.

\subsection{12P/Pons–Brooks}
12P/Pons–Brooks is a Halley-type comet with a 71.2-year orbital period and perihelion distance $q=0.78$\,AU.  Its orbit is highly inclined ($i=74^\circ$), and at perihelion in 2024 April it reached visual magnitude $V\approx3$.  During our NICER observations (2024 February–March), the comet was   $V\approx7$–8  but retained a distinct “horned” coma morphology \citep{jewitt2025multiple}.
Vander Donckt (under review) reports multiple photometric outbursts, including brightening events of 0.63 mag on 2024-02-03 and 0.88 mag on 2024-02-29.
OH-derived water production was measured at $(1.54\pm0.43)\times10^{29},\mathrm{s^{-1}}$ on 2024-02-25; other groups report increases to $\approx5\times10^{29},\mathrm{s^{-1}}$ by late March--April, with some aperture-dependent estimates approaching $1\times10^{30},\mathrm{s^{-1}}$ near perihelion \citep{combi2024water,li2025pre}.

\subsection{P/2010 H2 (Vales)}
P/2010 H2 (Vales) was initially classified as a quasi-Hilda asteroid until a $>7.5$\,mag outburst in 2010 revealed cometary activity \citep{jewitt2020outbursting}.
Its perihelion at $q=3.11$\,AU is too distant for strong H$_2$O sublimation, so CO and CO$_2$ are the more likely drivers of any present outgassing.
Although SWCX X-rays are produced whenever the ionized solar wind encounters neutral gas, regardless of the volatile composition \citep{cabot2023x}, the species and production rate determine the X-ray brightness and spectral morphology.
The absence of recent positive optical detections implies very weak or dormant behavior, so any X-ray signal would probe extremely low levels of activity.
Since NICER is the most sensitive soft-X-ray instrument currently operating, these observations provide a stringent test of our background-subtraction procedure and set practical limits on the detectability of tenuous, CO/CO$_2$-dominated cometary activity.
\section{Observations}
\subsection{NICER Instrument}

We used NICER's X-ray Timing Instrument (XTI), which pairs 56 X-ray concentrators with silicon-drift detectors, because its large soft X-ray collecting area ($\sim$1900\,cm$^2$ at 1.5\,keV), sub-microsecond timing, and ACIS/XRT-like spectral performance make it well suited to follow a moving target and track spectral evolution. For this study we restrict the analysis to 0.37 to 2.0\,keV: below 0.37\,keV the signal-to-noise ratio falls rapidly due to instrument sensitivity, detector background, and atmospheric absorption, while extending above 1\,keV is retained only to search for suggested Mg/Si/Al features despite ambiguity in their origin \citep{Gendreau2016_nicerdesign}.

Features in the 1 to 2\,keV band are treated as non-robust because line structure there is broad and easily confused with fluorescence from dust or scattered solar X-rays; 
when such features appear in our spectra they are inconsistent in energy and are not attributed to charge exchange. 
Spectral resolution and the noise model follow the instrument characterization and calculations given previously; 
full details of the XTI performance, data reduction, and the energy-selection rationale are provided in \citep{deskins2025modeling}.

Unfortunately, on June 24, 2025, NASA announced that NICER suspended science operations after performance degradation in one of its tracking motors began affecting observations on June 17 \citep{KazmierczakReddy2025NICERStatus}.
Engineers are currently investigating the root cause and potential remedies to restore the telescope’s pointing capability.
This reminder of the vulnerability of real-time in‑flight SWCX monitoring highlights the need for analysis methods that can leverage archival low‑resolution data when observations are interrupted. 

\subsection{Observing Campaign}
NICER observed nine comets over multiple epochs between 2020 and 2025, as summarized in Table~\ref{tab:obs_epochs_fullwidth}. Each epoch comprises one or more consecutive observations with similar viewing geometry, allowing exposures within the same epoch to be stacked to improve signal-to-noise ratio (SNR) in the 0.37–0.9,keV band after background subtraction. For each comet, the observing campaign was designed to capture solar-wind charge-exchange (SWCX) emission under favorable conditions—often near perihelion or during close approaches to Earth—while spanning a range of heliocentric distances, geocentric distances, and solar wind latitudes. Across all targets, NICER’s scheduling flexibility enabled coverage before, during, and after key orbital events, producing datasets with on-target exposures from a few kiloseconds to several tens of kiloseconds per epoch. 

We designed our observation campaign in order to reliably remove background X-ray emission. NICER did not track each comet's non-sidereal motion; we instead used ephemerides obtained from JPL Horizons to determine the comet's position at a specific time, denoted as $t_c$. For each observation session, we captured blank-sky images, each lasting around 600 seconds, centered around the comet's anticipated location. These observations were conducted 48 hours before and after the designated time $t_c$, ensuring that the ISS experiences similar geomagnetic effects at each time of observation.
Table~\ref{tab:obs_epochs_fullwidth} shows observations as grouped epochs and Appendix~\ref{app:obs-logs} contains detailed observation logs for each comet, showing observation times and parameters of interest. 

The soft X-ray background near Earth is dominated by SWCX between solar-wind ions and neutrals in the heliosphere, geocorona, and interplanetary medium, producing emission lines that overlap those from comets \citep{Kuntz2019}.
This confusion is worst near solar minimum, when reduced photoionization and radiation pressure raise neutral densities and the wind is dominated by slower, low-charge-state ions that enhance the same \ovii and \oviii lines seen in cometary spectra \citep{Carter2022earth};
consequently, standard `blank-sky' spectra do not reliably separate cometary charge-exchange from background \citep{Christian2010}. 

In practice, these terrestrial SWCX features are highly variable: 
NICER routinely detects rapid, direction-dependent K$\alpha$ emission from \ovii ($\approx574$\,eV), \oviii ($\approx654$\,eV), and Ne\,\textsc{ix} ($\approx898$\,eV) that can change on timescales of minutes and with pointing direction \citep{heasarc2024scorpeon}.
In addition, K$\alpha$ emission from neutral oxygen ($\approx533$\,eV) has been observed and is plausibly produced when solar X-rays ionize, upwelling neutral O in cusp-like regions \citep{heasarc2024scorpeon}; 
however, this emission can appear over much broader sky regions and is not predictable in a straightforward operational way. 
As noted by \citet{heasarc2024scorpeon},  due to the variable behavior of both SWCX and neutral-oxygen emission, these components must be modeled explicitly rather than treated as fixed background: 
practical NICER fitting frameworks therefore include separate normalization parameters for neutral-O and SWCX lines that default to zero and must be allowed to vary by the analyst. 

The soft X-ray background also varies with the ISS geomagnetic location, space weather, and solar-light loading \citep{Walsh2014}, and comets are moving targets, so accurate subtraction requires dedicated, contemporaneous off-source observations. 
In the case of NICER, low-energy interpretation is further complicated by features in the effective area, uncertainties below $\approx0.4$\,keV, and a variable low-energy excess tied to optical loading and the Sun/ISS geometry \citep{heasarc_nicer_responses_2024}.
These effects combine to make absolute low-energy fluxes sensitive to the chosen background and response treatment.
Further discussion of these effects and our background strategy is provided in \citep{deskins2025modeling}.

\begin{table*}
\centering
\begin{threeparttable}
\caption{NICER observing log for the comets analyzed in this work. For each epoch, we report the UTC date range, the number of NICER observation sets (N Obs., each set contains pre-background, on-comet, and post-background pointings), the total summed on-comet exposure time  (Total $t_{\rm exp}$, in seconds), the mean heliocentric distance (Mean $r_{\rm h}$), the mean heliocentric latitude (Mean $\beta_{\rm h}$, in degrees), mean geocentric distance (Mean $\Delta$), and a representative integrated V-band magnitude (from Aerith / Seiichi Yoshida; observer-reported). Background pointings are excluded; values are rounded to the precision shown.}
\label{tab:obs_epochs_fullwidth}
\scriptsize
\setlength{\tabcolsep}{5pt}
\renewcommand{\arraystretch}{1.00} 
\begin{tabularx}{\textwidth}{@{}l c c *{5}{>{\centering\arraybackslash}p{1.66cm}} c@{}}
\toprule
Comet & Epoch & Date range (UTC) & N Obs. & $t_{\rm exp}$ (s) & $r_{\rm h}$ (au) & $\beta_{\rm h}$ (deg) & $\Delta$ (au) & V (mag) \\
\midrule

\multirow{1}{*}{T2}
  & 1 & 2020-04-23 -- 2020-06-28 & 5 & 9685 & 1.678 & 49.374 & 1.707 & 9.5  \\
\midrule

\multirow{3}{*}{88P}
  & 1 & 2020-09-10 -- 2020-09-12 & 2 & 3876 & 1.364 & -3.030 & 1.360 & 9.0 \\
  & 2 & 2020-09-25 -- 2020-09-29 & 3 & 3488 & 1.363 & -3.641 & 1.363 & 9.1 \\
  & 3 & 2020-10-27 -- 2020-10-30 & 3 & 5481 & 1.364 & -4.321 & 1.361 & 9.3 \\
\midrule

\multirow{6}{*}{67P}
  & 1 & 2021-11-10 -- 2021-11-11 & 2 & 2120 & 1.216 & 1.942 & 0.4182 & 9.5 \\
  & 2 & 2021-11-12 -- 2021-11-12 & 1 & 759  & 1.218 & 2.024 & 0.4182 & 9.5 \\
  & 3 & 2021-11-13 -- 2021-11-15 & 3 & 4846 & 1.221 & 2.131 & 0.4184 & 9.5 \\
  & 4 & 2021-11-16 -- 2021-11-17 & 2 & 5277 & 1.226 & 2.261 & 0.4189 & 9.5 \\
  & 5 & 2021-11-18 -- 2021-11-21 & 4 & 8658 & 1.231 & 2.409 & 0.4198 & 9.4 \\
  & 6 & 2021-12-16 -- 2021-12-17 & 4 & 8395 & 1.333 & 3.409 & 0.4454 & 9.2 \\
\midrule

\multirow{3}{*}{19P}
  & 1 & 2022-01-17 -- 2022-01-22 & 6  & 3583  & 1.315 & -9.297 & 1.223 & 9.8  \\
  & 2 & 2022-02-13 -- 2022-02-17 & 5  & 6235  & 1.318 & 1.460  & 1.338 & 9.2  \\
  & 3 & 2022-03-17 -- 2022-03-19 & 3  & 3608  & 1.410 & 13.008 & 1.563 & 9.4  \\
\midrule

\multirow{3}{*}{K2}
  & 1 & 2022-05-31 -- 2022-06-05 & 5  & 8837  & 3.012 & 22.772 & 2.160 & 8.6  \\
  & 2 & 2022-07-29 -- 2022-07-31 & 3  & 3416  & 2.517 & 8.641  & 1.857 & 8.2  \\
  & 3 & 2022-08-03 -- 2022-08-07 & 3  & 4238  & 2.463 & 6.843  & 1.900 & 8.2  \\
\midrule

\multirow{1}{*}{E3}
  & 1 & 2023-03-21 -- 2023-03-21 & 10 & 16304 & 1.540 & -27.498 & 1.535 & 9.5  \\
\midrule

\multirow{2}{*}{62P}
  & 1 & 2024-02-18 -- 2024-02-18 & 4  & 1010  & 1.420 & 4.735  & 0.510 & 7.8  \\
  & 2 & 2024-02-27 -- 2024-02-27 & 4  & 5359  & 1.480 & 4.690  & 0.530 & 7.8  \\
\midrule

\multirow{2}{*}{12P} 
  & 1 & 2024-02-23 -- 2024-02-23 & 5  & 1124  & 1.290 & 56.958 & 1.730 & 7.0  \\
  & 2 & 2024-03-02 -- 2024-03-02 & 3  & 671   & 1.200 & 51.434 & 1.680 & 6.9  \\
\midrule

\multirow{1}{*}{H2}
  & 1 & 2025-04-02 -- 2025-04-03 & 3  & 2192  & 3.080 & 10.450 & 2.110 & 20.5 \\
\bottomrule
\end{tabularx}

\end{threeparttable}
\end{table*}

\subsection{Supporting observations}
\begin{figure}
    \centering
    \includegraphics[width=0.85\linewidth]{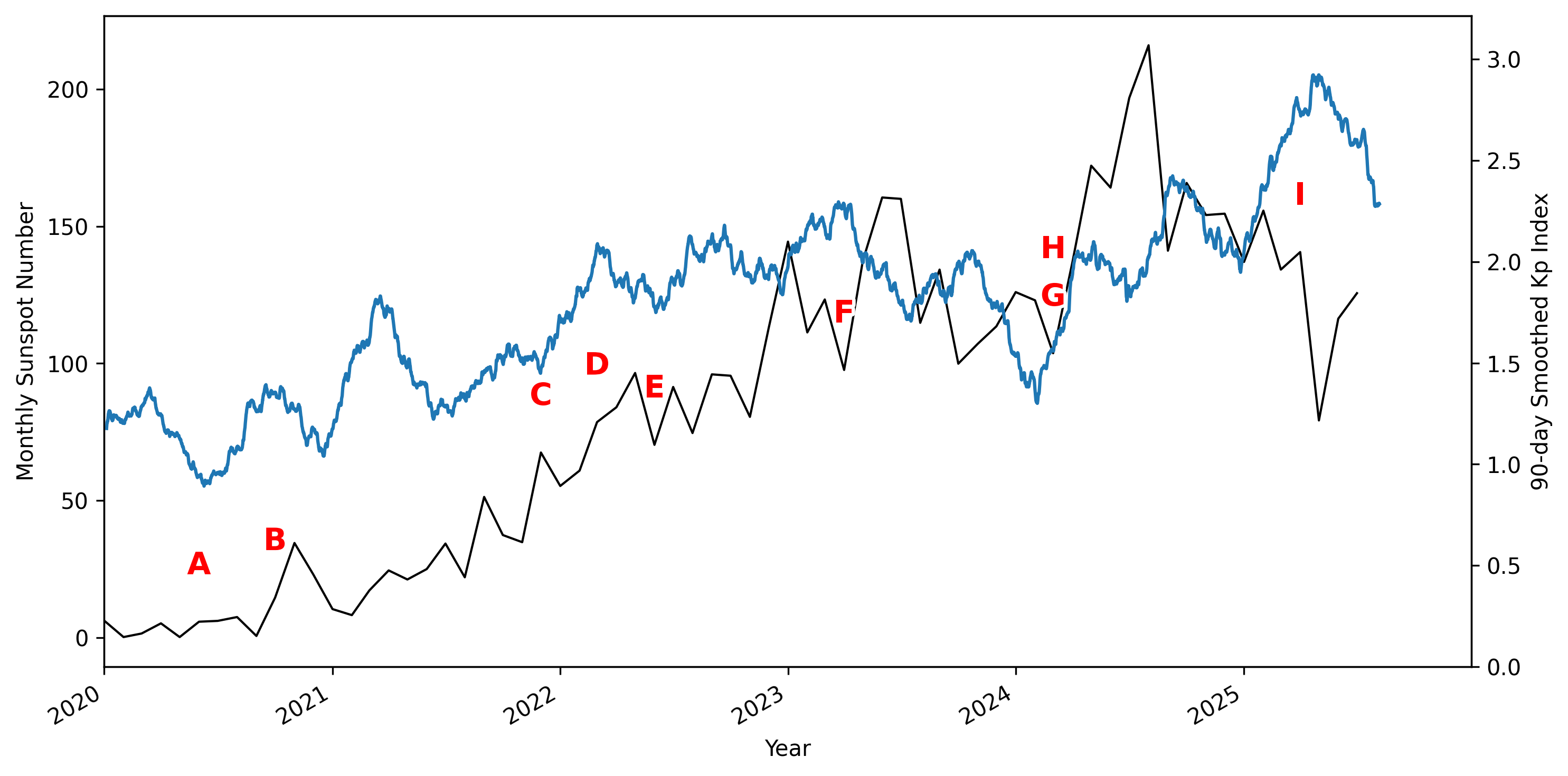}
    \caption{Monthly sunspot number (black) between 2020-01-01 and 2025-12-31 with observation annotations (red letters) marking representative NICER on-comet observation dates: 
    (A) C/2017 T2 (PANSTARRS): 2020-05-24;
    (B) 88P/Howell: 2020-10-01;
    (C) 67P/Churyumov–Gerasimenko: 2021-11-30;
    (D) 19P/Borrelly: 2022-02-15;
    (E) C/2017 K2 (PANSTARRS): 2022-06-15;
    (F) C/2022 E3 (ZTF): 2023-03-21;
    (G) 62P/Tsuchinshan: 2024-02-23;
    (H) 12P/Pons–Brooks: 2024-02-28;
    (I) P/2010 H2 (Vales): 2025-04-02.
    The right-hand axis shows the geomagnetic Kp index (blue), plotted as a 90-day centered rolling mean (each smoothed point is the mean over $\pm$45 days).}
    \label{fig:sunspot_kp}
\end{figure}
We reviewed solar-wind data to explore the influence of solar-wind conditions, including potential coronal mass ejections (CMEs) and solar flares, on our observations. 
We also considered several solar parameters at the times of observation, including the time of the solar cycle, the number of sun spots, and the Kp index. Table~\ref{tab:comet_sunspots_kp} allows for a direct comparison of the solar properties at each comet's time of observation and Figure~\ref{fig:sunspot_kp} shows the variation of these parameters over several years.

\begin{table}
    \centering
    \caption{Solar Activity and Geomagnetic Impact at the time of cometary observations.}
    \label{tab:comet_sunspots_kp}
    \small
    \setlength{\tabcolsep}{28pt}
    \renewcommand{\arraystretch}{1.00} 
    \begin{tabular}{@{} l c c @{}}
        \toprule
        Comet & Approximate Monthly Sunspot Number & Approximate Kp Index \\
        \midrule
        C/2017 T2             & 10  & 1.0 \\
        88P/Howell            & 25  & 1.4 \\
        67P/Churyumov--Gerasimenko & 60  & 1.5 \\
        19P/Borrelly          & 75  & 2.1 \\
        C/2017 K2 (PANSTARRS) & 75  & 2.0 \\
        C/2022 E3 (ZTF)       & 100 & 2.2 \\
        62P/Tsuchinshan       & 100 & 1.5 \\
        12P/Pons--Brooks      & 100 & 1.5 \\
        P/2010 H2 (Vales)     & 150 & 2.7 \\
        \bottomrule
    \end{tabular}
\end{table}

To verify the absence of significant solar flares during our NICER observing windows, we examined the Geostationary Operational Environmental Satellites (GOES) \citep{menzel1994introducing} soft X-ray flux in the 1.6–12.4 keV band (0.1–0.8 nm), as plotted in the figures in Appendix~\ref{app:GOES}. 
These figures show the continuous solar data from GOES, containing both raw 1-second counts and smoothed rolling averages to distinguish transient flares from baseline emission.
The data were obtained through the use of Fido, a unified data search-and-retrieval tool which is part of \texttt{SunPy},a solar-data-analysis environment for Python . These data validate the absence of strong flare-driven enhancements in the cometary X-ray spectra over all six epochs.
However, comets are one of the few probes that provide in-situ measurements of the SW at multiple distances from the Sun.  
Cometary X-ray spectra varies greatly for differing solar wind conditions, and provide a valuable diagnostic \citep{krasnopolsky2006x, Bodewits2007,krasnopolsky2015cxo}.
For example, the \ovii $\approx$570\,eV emission is much stronger than the \cvi $\approx$370\,eV emission for the slow solar wind ($\approx$300\,km/s), but these features are nearly equal for the fast ($\approx$700\,km/s) solar wind \citep{Bodewits2007}.

SolarMonitor$^1$\blfootnote{1 https://www.solarmonitor.org} \citep{gallagher2002active} and the NOAA Space Weather Prediction Center$^2$\blfootnote{2 https://www.swpc.noaa.gov} supplied the Solar Region Summaries, Solar Event Lists, GOES 5-min X-rays (see the figures contained in Appendix.~\ref{app:GOES}), proton and electron data. We verified that there were no significant solar flares of class M or higher during our observations, when the comet was facing the same side of the Sun as Earth.
\begin{figure*}
    \centering
    \includegraphics[width=0.517\linewidth,trim={0.2cm .8cm 1cm 1.5cm},clip]{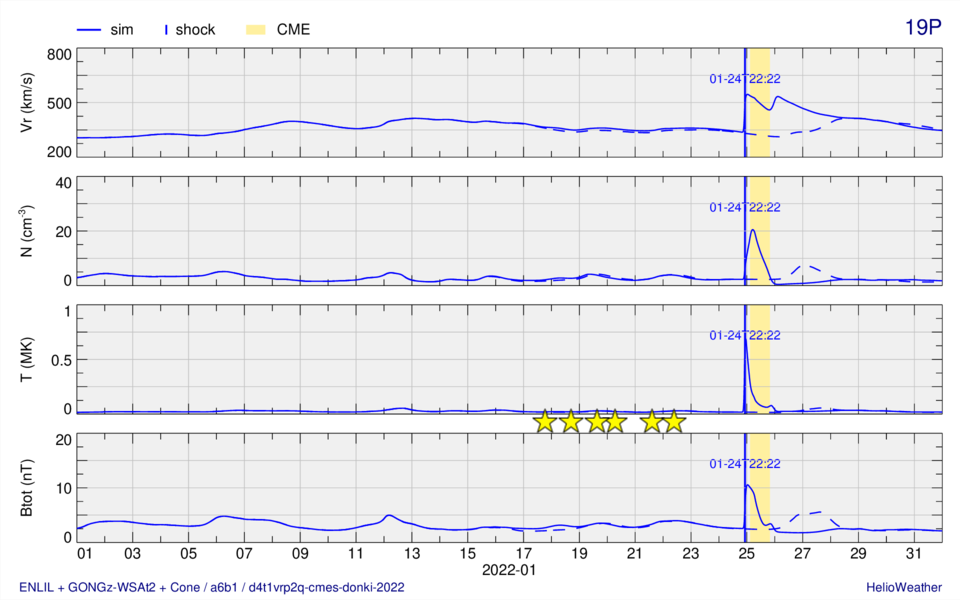}
    \hfill%
    \includegraphics[width=0.473\linewidth,trim={2.7cm .8cm 1.3cm 1.5cm},clip]{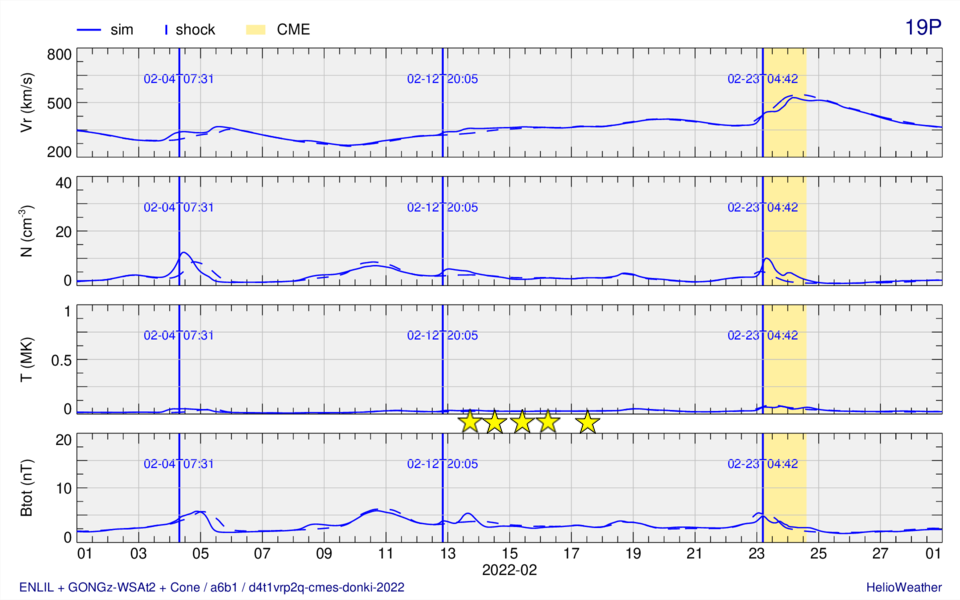}
    \caption{Simulated solar-wind parameters from the ENLIL model with GONG boundary conditions, evaluated at the position of Comet 19P/Borrelly in January (left) and February (right) 2022. The horizontal panels display the velocity ($V_r$), density ($N$), temperature ($T$), and magnetic field intensity ($|B|$) of the solar wind. K Shocks are marked by vertical blue lines with annotated times, while coronal mass ejections are highlighted in yellow. Times accompanying vertical blue lines denote shocks. Variations in the solar-wind parameters are observed in both months, coninciding with shock/CME events and IMF polarity changes that suggest dynamic solar wind conditions. Figure provided by Du\v san Odstr\v cil.}
    \label{fig:CMEsMonths}
\end{figure*}

Figure~\ref{fig:CMEsMonths} shows a simulation that relies on the ENLIL model with boundary conditions fixed from observations to simulate solar-wind parameters at the position of Comet 19P/Borrelly during January and February, 2022.
The ENLIL model is based on a magnetohydrodynamic (MHD) description of the solar wind and is used to predict solar-wind dynamics. 
The model is primarily driven by input from the Wang-Sheeley-Arge (WSA) model, which derives the radial magnetic field and flow velocity using observations of the Sun’s photospheric magnetic field and other coronal structures. 
Specifically, the WSA model uses data from Global Oscillation Network Group (GONG) observations, which inform the model about the Sun’s magnetic field conditions. 

The ENLIL model is useful for predicting real-time solar-wind disturbances, but its proper use hinges upon the user's understanding of its limitations:
a single, corotating map drives the model's simulation, which assumes steady-state conditions for the solar wind. This can fail to consider small-scale, transient events that may occur closer to the Sun or in different regions of the solar wind.
Additionally, the model depends heavily on the accuracy of the initial magnetic field and solar-wind velocity conditions derived from WSA and GONG observations. 
In cases where these boundary conditions are inaccurate or incomplete, the predictions that rely on them may not completely capture the dynamics of solar-wind interactions with cometary atmospheres;
moreover, CMEs vary in type, causing the interpretation of spectra for CME interactions to be a complicated endeavor. 
Another limitation is the relatively coarse spatial and temporal resolution of the model, which may not show finer details of solar wind structures, such as the complex evolution of CME shocks or stream interaction regions. 
It is possible to overlay real-time data from spacecraft like ACE and STEREO on the ENLIL model's predictions to validate or refine its accuracy, but these observations are usually limited to regions near Earth. 
Taking this approach for objects distant from Earth as a means of estimation is an acceptable practice, but confidence in the estimate decreases with increasing distance.

\section{Data Preparation}
After collecting data from observations, we must prepare them for analysis.

\subsection{\textit{NICER}-recommended Tasks}

Raw FITS products were ingested and an observation log with comet ephemerides was produced, then we processed the data with the standard NICER pipeline, as recommended by \citet{heasarc_nicerl2}$^\dagger$\blfootnote{$\dagger$ www.heasarc.gsfc.nasa.gov/docs/nicer/analysis\_threads/nicerl2/}. 
We ran \texttt{nicerl2} with default settings recommended in the NICER Data Analysis Threads to apply up-to-date calibration, screening, and filtering; this yields a cleaned event list and an updated filter file. 
Per-detector count-rate plots produced from the \texttt{nicerl2} output were inspected to identify outliers or faulty detectors \citep{Prigozhin2016_nicerperformance}, and the cleaned event lists were passed to \texttt{nicerl3-spect} to generate the spectral products used in our analysis (spectrum, background estimate, ARF, RMF).

NICERDAS is distributed as part of HEASoft \citep{nicerdas}; for reproducibility we used HEASoft v6.31, NICERDAS v10a, and the NICER calibration database version \texttt{xti20221001}.

\subsection{Background Subtraction}

\begin{figure}
    \centering
	\includegraphics[width=0.6\columnwidth,trim={1cm 0 1.2cm 1.2cm},clip]{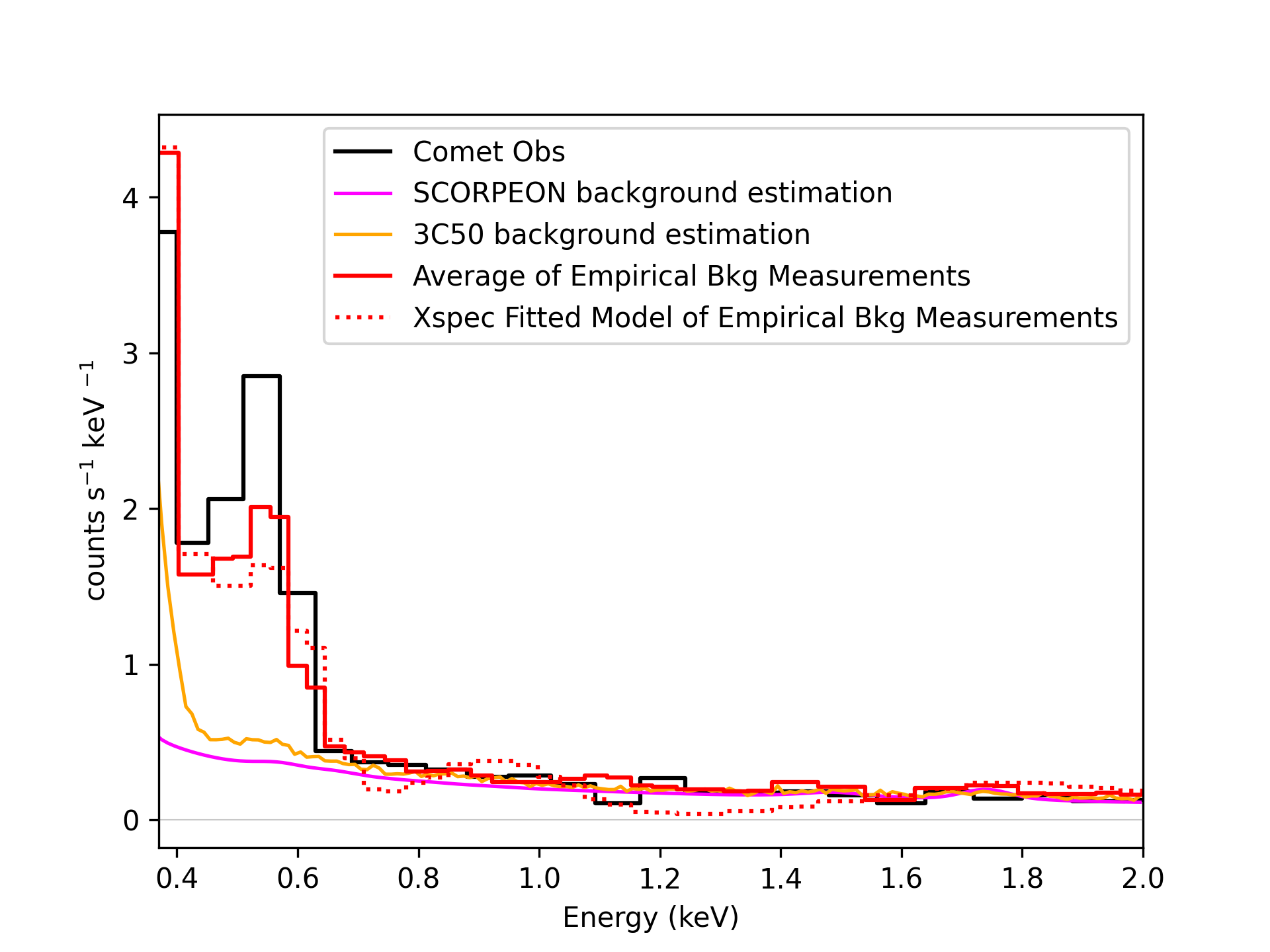}
    \caption{An X-ray spectrum shown between 0.37-2.00 keV is shown from one representative observation. The y-axis shows the normalized count rate, and the x-axis shows energy. The on-target observation is depicted in black, while the weighted average of pre- and post-background observations is presented in solid red. The dotted red line is the Xspec-fitted model of our empirical background measurements, while the pink and yellow lines represent the SCORPEON and 3C50 background estimates.}
    \label{fig:bg_comparison_SCORPEON}
\end{figure}

\begin{figure}
    \centering
    \includegraphics[width=0.6\linewidth,trim={0 0cm 0 0cm},clip]{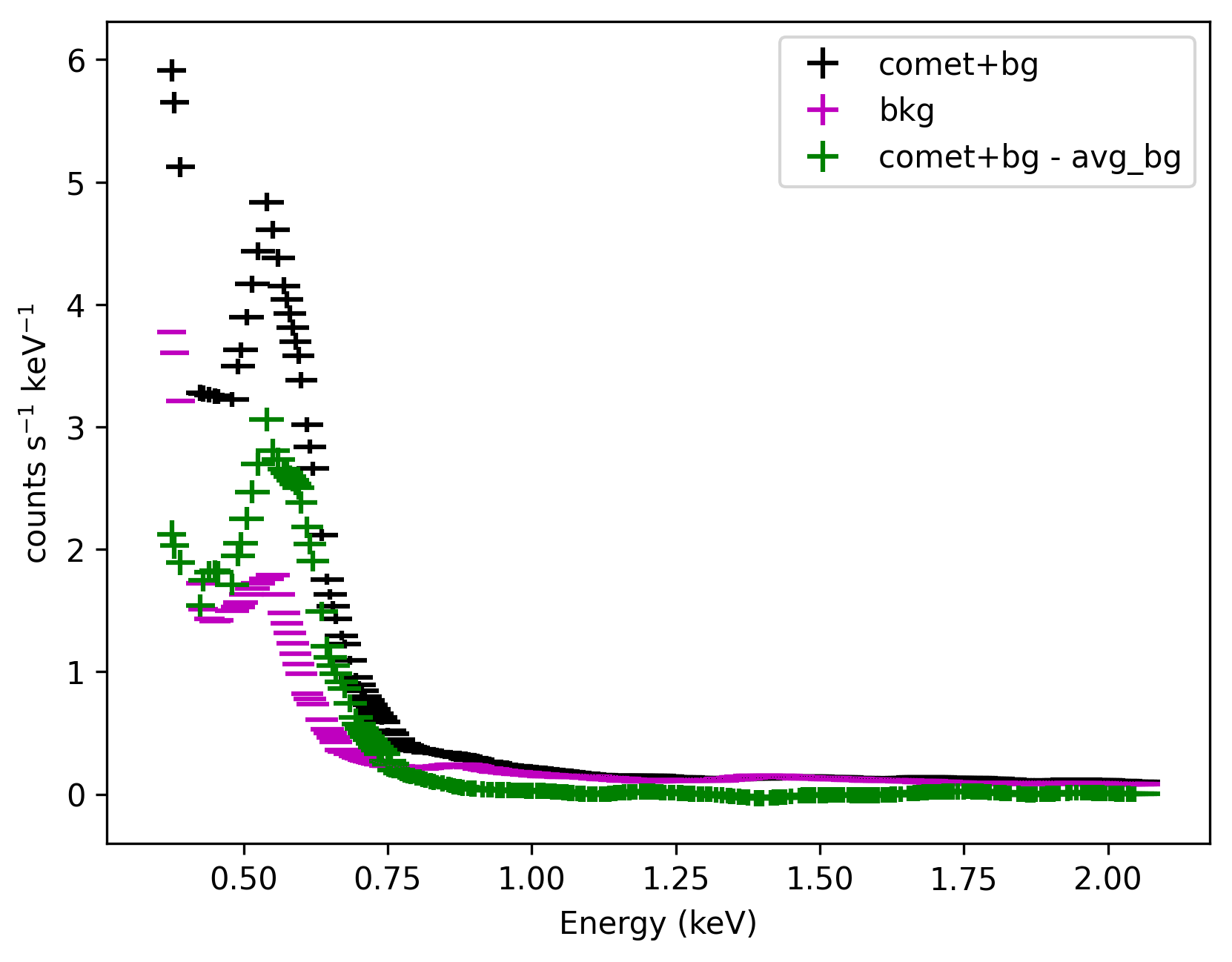}
    \caption{X-ray spectrum for the full dataset for C/2022 E3 (ZTF). The black data series is the spectrum before background subtraction; the purple data series is the average of the pre-comet and post-comet background spectra, which  dominates  above 1\,keV; the green data series is the spectrum attributable to the comet after background subtraction. The average background flux for the pre-comet and post-comet background observations is 0.65 counts per second per keV over the 0.37--2-keV energy range.}
    \label{fig:background_subtraction}
\end{figure} 

Background fluctuations arise from detector-intrinsic effects (gain drift, readout and thermal noise) and external particle events, so screening and filtering are required to remove spurious counts; NICER's detector resolution in the 0.37 to 2.0 keV band is reported between about 50 eV and 100 eV (FWHM). Cosmic rays and high-energy particles likewise produce transient signals that are mitigated by standard reduction steps \citep{instrumentfabbiano2006populations,Prigozhin2016_nicerperformance}. The diffuse cosmic X-ray background and Galactic contributions, including the halo and supernova remnants, vary across sight lines and can change background levels by tens of percent, and local near-Earth sources such as geomagnetic effects, radiation-belt proximity, auroral emission, and geocoronal SWCX add additional, temporally variable components \citep{ray2019discovery,moretti2009new,kaaret2020disk,halotoft2002x,spacecraftlu2019review,a2024ground,d2024magnetosheath,auroralpetrinec1999statistical,scorpeon_background}. NICER has detected neutral-oxygen K$\alpha$ emission near 533 eV that can appear unpredictably with spacecraft pointing and geomagnetic conditions, so theoretical background estimates tuned for higher-energy targets do not reliably reproduce our empirical, soft-band measurements; we compared our data to the SCORPEON and 3C50 model predictions and found significant mismatches \citep{Remillard2022_empirical}.

Because of these complexities we directly measure background for each pointing and accept the observing-cost tradeoff: roughly 25 percent of our total exposure is dedicated to off-source background measurements. 
Pre- and post-pointing backgrounds are generally quite consistent, despite being separated by several days. This suggests that solar-wind-driven changes occurring on timescales of hours, as suggested by \citet{Bonamente2021}, are small enough to allow for low-resolution spectroscopy. 
In practice we subtract the background from each observation before stacking, propagate observation-wise uncertainties, and form exposure-weighted averages when combining observations into epochs. 
We separated our observations into epochs based on solar-wind data, grouping observations that were near to each other in time. 
If two on-target observations were conducted more than 106,000 seconds ($\approx$1.2 days) apart from one another, then we did not consider them to belong to the same epoch. 
Further details of these background tests, comparisons to models, and our epoching procedure are given in \citep{deskins2025modeling}.

We obtained two 600 s exposures of the same field 48 hours before and 48 hours after each comet pointing, timed to match the International Space Station observing geometry, which repeats roughly every two days. These off-source pointings sample short-term variations in the soft X-ray background while preserving comparable observing conditions for the comet and background exposures. 
The two background spectra were combined with an exposure-time-weighted average and that average spectrum was subtracted from the comet spectra; an example is shown in Figure~\ref{fig:background_subtraction}. In practice nearly all the paired backgrounds show little variation beyond statistical noise, so the weighted-average subtraction provides an effective background correction. 
Our direct sampling of the background shows a difference compared to the background estimates from the NICER SCORPEON and 3C50 models$^\star$\blfootnote{$\star$ www.heasarc.gsfc.nasa.gov/docs/nicer/analysis\_threads/scorpeon-overview/}, as shown in Figure~\ref{fig:bg_comparison_SCORPEON}.

\section{Uncertainties}

Instrumental calibration and detector effects drive most systematic uncertainty. 
NICER exhibits a strong noise feature near 130\,eV that affects spectra up to about 200\,eV, and readout/trigger-efficiency effects and quantum-efficiency variations at roughly 0.4, 0.6, and 1.1\,keV introduce additional uncertainty \citep{Prigozhin2016_nicerperformance}. 
To mitigate these issues we adopt a conservative low-energy cutoff of 250\,eV and restrict analysis to 0.37--2.0\,keV; 
standard processing reduces but does not eliminate residual calibration-related errors.

Large line fluxes near 300\,eV, which we associate with \cv, are difficult to model because of entrance-window absorption and the trigger-efficiency behavior of the analog readout, so we treat that region with caution. 
Typical comet count rates in our data are about 0.8 counts s$^{-1}$ and the error introduced by background subtraction is on the order of 0.1 counts s$^{-1}$. 
Further details on these choices, noise characterization, and their impact on our analysis are provided in \citep{deskins2025modeling}. 

\section{Model selection and choice of the BCM}

Low-resolution charge-exchange spectra contain many closely spaced lines in our observable range of 0.37 and 1.0\,keV, but even more below 0.3\,keV \citep{Bodewits2007,Koutroumpa2024,sasseen2006search}. 
NICER-like detectors have an energy resolution of order 50\,eV in the soft band \citep{Prigozhin2016_nicerperformance}.
At this resolution multiple ionic features are blended and fit parameters become highly degenerate; for example, \nvii and \ovii are strongly anti-correlated \citep{deskins2025modeling}.
Consequently, model selection must trade physical fidelity against overfitting: unconstrained models with many free Gaussians fit noise, while overly rigid models miss real spectral structure.
We explored a hierarchy of models ranging from the nearly free-line ``Free Model'' to physics-constrained charge-state and species-based models (FCSM, FSSTM, FSMTM) and a BIC-regularized variant (FSMTBM), each differing in the number of free parameters and the degree to which line ratios or freeze-in physics are enforced \citep{deskins2025modeling}.

From that comparison we adopted the Bayesian Color Model (BCM) for the present work for three reasons. 
First, the BCM encodes prior physical knowledge by associating C, N, and O charge states with potential peaks within a specified energy range, searching the combinatorial space of peak selections, so it captures expected CX features without permitting arbitrary, noise-driven lines.
Second, the BCM explicitly minimizes the Bayesian Information Criterion (BIC), which enforces a quantitative balance between goodness-of-fit and model complexity and therefore reduces overfitting in low-SNR, low-resolution spectra \citep{Bonamente2021,Bodewits2007}. 
Third, in our previous analysis of the 67P dataset the BCM provided a compact, reproducible fit (six dominant peaks) that both highlighted the \ovii complex and produced stable epoch-to-epoch comparisons when the same peak set was fit to individual epochs. 
Practically, the BCM initializes Gaussian means from weighted-average line energies, constrains their allowed ranges based on known line lists, and sets widths using the instrument resolution; it then selects the peak combination with minimum BIC, yielding a physically interpretable fit with far fewer effective degrees of freedom than the unconstrained Free Model.

The BCM provided the optimal balance between interpretability and statistical simplicity in the prior study and enabled direct comparison of peak positions and amplitudes across epochs. For these reasons, it is adopted here as the primary fitting method. Full algorithmic details, peak-energy ranges, line-ratio inputs, and validation tests against the alternative models are given in \citep{deskins2025modeling}.


The BCM was developed and validated specifically to handle low-resolution SWCX spectra and to make model-selection conservative and reproducible \citep{deskins2025modeling}. 
When the BCM is applied to the larger ensemble in this work, and considering the associated BIC-driven complexity penalty, physically motivated priors, and Monte--Carlo sensitivity bands, it yields internally consistent line lists and uncertainties across datasets.
The larger sample improves the practical power of the method because repeated detection of the same features in independent observations reduces the odds that a given line identification is a simple result of degeneracy; 
conversely, when a potential line is only slightly supported by the data for a single object, the BCM’s BIC penalty will tend to exclude it. 
For example, in 12P, the data in the oxygen region are comparatively weak and the instrument response and blending make the \ovii/\oviii\ complex difficult to constrain;
in numerous Monte--Carlo trials the fits that included oxygen fell into higher-BIC sensitivity bands, so the nitrogen-only solution is the data-driven outcome rather than a failure of the algorithm. 
Instrumental sensitivity and the NICER/XTI low-energy calibration and resolution further reduce the leverage on oxygen lines in low-count regimes \citep{Gendreau2016_nicerdesign}.

We primarily validated the BCM with simulated Xspec datasets designed to mimic realistic SWCX spectra.
Tests included a compact six-ion case (C, N, O species) and a richer 14-ion case (adding Ne, Mg, Al, Si), and in both cases the BCM recovered the input parameters more reliably than the alternative models we tested.
Important practical limitations were also identified: under low spectral resolution the \ovii triplet near 0.56\,keV can be partially absorbed into neighboring features, and the BIC penalty makes inclusion of weak, high-energy ions (Mg/Al/Si above $\sim$0.9\,keV) unfavorable unless those lines are strongly detected.
For these reasons BCM fits in both the original study and the present work are restricted to the band where C, N and O dominate ($\sim$0.37–-1.0\,keV).
For implementation details, Monte-Carlo band construction, simulated-data tests, and a fuller discussion of limitations and caveats, the reader is referred to \citet{deskins2025modeling}.

Figure~\ref{fig:spec_12p_full} shows the spectrum of 12P, indicating a nitrogen-only detection, which illustrates a limitation tied to data quality and spectral degeneracy, not a systematic bias in BCM. 
To perform further validation of the BCM and to check this explicitly we inspected around the expected \ovii\ energy and found no persistent, statistically significant excess, and performed a test by adding a \ovii/\oviii\ complex of plausible amplitude to the 12P spectrum; only substantial additions above the data’s noise floor produced a change in BIC favorable to models including oxygen. 
These cross-checks confirm that BCM’s conservative model selection is functioning as intended: it avoids adding poorly constrained oxygen components that the data do not require, while still permitting their inclusion when the signal is robust.
The practical implication for 12P is therefore one of limited sensitivity rather than methodological failure; resolving the ambiguity would require higher spectral resolution or deeper exposures that raise the oxygen-band signal above the present detection threshold.

\section{Results}

We fitted SWCX lines in the 0.37--2.0\,keV band for nine comets considered in this study: C/2017 T2 (PANSTARRS) (T2), 88P/Howell (88P), 67P/Churyumov–Gerasimenko (67P), 19P/Borrelly (19P), C/2017 K2 (PANSTARRS) (K2), C/2022 E3 (ZTF) (E3), 62P/Tsuchinshan 1 (62P), 12P/Pons–Brooks (12P), and P/2010 H2 (Vales) (H2).
Results are presented here for T2, 88P, 67P, 19P, K2, E3, 62P, and 12P, but not for H2, as the signal-to-noise ratio was too low for meaningful interpretation.
For each comet we used the BCM to identify the best-fitting ion combination and to fit spectra. We also used flux ratios to infer nominal freeze-in temperatures.
The possible ions included in the best-fit combination are \cv, \cvi, \nvi, \nvii, \ovii, \oviii, \neix, \nex, \mgxi, \mgxii, \alxii, \alxiii, \sixiii, and \sixiv. The most common best-fit combination of ions was found to be \cv, \cvi, \nvi, \nvii, \ovii, and \oviii, and this was the case in 19P, 62P, E3, and K2. 12P was singular in that its best-fit combination lacked \ovii and \oviii, including only \cv, \cvi, \nvi, and \nvii. 

\subsection{Spectral Characteristics}

\begin{table*}
\centering
\caption{Gaussian parameters for the BCM fits to the full dataset of each comet. Centroids and sigmas are in keV. The amplitudes are fit normalizations and are in arbitrary units. \textbf{Measured centroids represent response-folded blended complexes in low-resolution SWCX spectra.} Only the ions included in each comet's best-fit model are listed.}
\label{tab:gauss_params_full}
\footnotesize
\setlength{\tabcolsep}{8pt}
\begin{tabular}{@{} l l *{3}{>{\centering\arraybackslash}p{3.3cm}} @{}}
\toprule
Comet & Ion & Mean (keV) & Sigma (keV) & Amplitude \\
\midrule

\multirow{6}{*}{C/2017 T2 (PANSTARRS)}
  & C V   & 0.332 & 0.045 & 0.038 \\
  & C VI  & 0.404 & 0.034 & 0.013 \\
  & N VI  & 0.481 & 0.032 & 0.011 \\
  & N VII & 0.538 & 0.035 & 0.016 \\
  & O VII & 0.639 & 0.049 & 0.019 \\
  & O VIII& 0.787 & 0.070 & 0.022 \\
\midrule

\multirow{6}{*}{88P / Howell}
  & C V   & 0.332 & 0.045 & 0.103 \\
  & C VI  & 0.396 & 0.045 & 0.063 \\
  & N VI  & 0.481 & 0.059 & 0.078 \\
  & N VII & 0.567 & 0.045 & 0.067 \\
  & O VII & 0.639 & 0.064 & 0.040 \\
  & O VIII& 0.787 & 0.070 & 0.007 \\
\midrule

\multirow{6}{*}{67P / Churyumov--Gerasimenko}
  & C V   & 0.320 & 0.046 & 0.154 \\
  & C VI  & 0.409 & 0.031 & 0.046 \\
  & N VI  & 0.481 & 0.055 & 0.044 \\
  & N VII & 0.580 & 0.055 & 0.098 \\
  & O VII & 0.639 & 0.020 & 0.001 \\
  & O VIII& 0.775 & 0.055 & 0.012 \\
\midrule

\multirow{6}{*}{19P / Borrelly}
  & C V   & 0.332 & 0.039 & 0.116 \\
  & C VI  & 0.405 & 0.032 & 0.060 \\
  & N VI  & 0.481 & 0.040 & 0.072 \\
  & N VII & 0.560 & 0.037 & 0.096 \\
  & O VII & 0.639 & 0.040 & 0.060 \\
  & O VIII& 0.755 & 0.040 & 0.017 \\
\midrule

\multirow{6}{*}{C/2017 K2 (PANSTARRS)}
  & C V   & 0.332 & 0.024 & 0.104 \\
  & C VI  & 0.407 & 0.020 & 0.053 \\
  & N VI  & 0.481 & 0.038 & 0.053 \\
  & N VII & 0.566 & 0.040 & 0.079 \\
  & O VII & 0.639 & 0.055 & 0.049 \\
  & O VIII& 0.780 & 0.055 & 0.028 \\
\midrule

\multirow{6}{*}{C/2022 E3 (ZTF)}
  & C V   & 0.306 & 0.023 & 0.072 \\
  & C VI  & 0.393 & 0.020 & 0.095 \\
  & N VI  & 0.453 & 0.045 & 0.171 \\
  & N VII & 0.550 & 0.043 & 0.278 \\
  & O VII & 0.639 & 0.045 & 0.138 \\
  & O VIII& 0.765 & 0.045 & 0.026 \\
\midrule

\multirow{6}{*}{62P / Tsuchinshan 1}
  & C V   & 0.332 & 0.040 & 0.126 \\
  & C VI  & 0.393 & 0.037 & 0.102 \\
  & N VI  & 0.474 & 0.040 & 0.062 \\
  & N VII & 0.559 & 0.040 & 0.097 \\
  & O VII & 0.612 & 0.040 & 0.050 \\
  & O VIII& 0.706 & 0.040 & 0.016 \\
\midrule

\multirow{4}{*}{12P / Pons--Brooks} 
  & C V   & 0.332 & 0.028 & 0.128 \\
  & C VI  & 0.400 & 0.033 & 0.104 \\
  & N VI  & 0.481 & 0.040 & 0.068 \\
  & N VII & 0.572 & 0.040 & 0.090 \\

\bottomrule
\end{tabular}
\end{table*}

The BCM fits each comet's spectrum with a set of Gaussian components, each corresponding to a particular ion. The possible ions considered in the modeling are \cv, \cvi, \nvi, \nvii, \ovii, \oviii, \neix, \nex, \mgxi, \mgxii, \alxii, \alxiii, \sixiii, and \sixiv. 
For each comet we report only the ions selected in that comet's best-fit combination and the Gaussian parameters of those components.
Table~\ref{tab:gauss_params_full} lists the numerical values of the fitted Gaussian parameters for the full dataset of each comet.
Spectra for each comet and their epochs are displayed in Appendix~\ref{app:spectra}.

C/2017 T2 (PANSTARRS) is fitted with \cv, \cvi, \nvi, \nvii, \ovii, and \oviii, each at relatively low amplitude. The \cv component is the largest contributor in this fit, and \cvi, \nvi, \nvii, \ovii, and \oviii\ all have amplitudes roughly half that or lower. The fitted widths span the typical range for the sample, with the \oviii component showing the largest width in the fit.

88P/Howell is fitted with \cv, \cvi, \nvi, \nvii, \ovii, and \oviii. Both carbon lines contribute substantially to the low-energy portion of the fit, while nitrogen components contribute appreciably in the mid-energy region. \ovii is detected at moderate amplitude and \oviii is small, so oxygen contributes at a lower level in the 88P fit.

67P/Churyumov--Gerasimenko is fitted with \cv, \cvi, \nvi, \nvii, \ovii, and \oviii. The \cv component has the largest amplitude in the 67P fit with a moderate width, while the higher-energy carbon and nitrogen components are present at lower amplitudes and broader widths. The \ovii component in the 67P fit has a very small fitted amplitude despite a narrow width, and \oviii is small, so the fitted oxygen components in 67P are minimal.

19P/Borrelly is fitted with \cv, \cvi, \nvi, \nvii, \ovii, and \oviii. The \cv component for 19P has a larger fitted width than in several other comets, and both \ovii and \oviii are present at measurable amplitudes. The resulting model contains low- and mid-energy carbon and nitrogen features together with detectable oxygen emission at higher energies.

C/2017 K2 (PANSTARRS) is fitted with \cv, \cvi, \nvi, \nvii, \ovii, and \oviii. The fitted \cv and \cvi components have relatively small widths compared with many other spectra in the sample, and the amplitudes for the higher-energy nitrogen and oxygen lines are measurable but modest; its morphology resembles the intermediate shapes seen in 19P and 62P but with smaller amplitudes in the Gaussians attributed to nitrogen and oxygen. The combination of narrower low-energy carbon components and non-zero nitrogen and oxygen components yields a multi-component fit spanning the low- to higher-energy range. 

C/2022 E3 (ZTF) is fitted with \cv, \cvi, \nvi, \nvii, \ovii, and \oviii. The carbon components in E3 are among the narrowest in the sample, and the fitted nitrogen amplitudes are large, yielding a fit with substantial mid- to high-energy contributions from nitrogen lines as tabulated in Table~\ref{tab:gauss_params_full}.

62P/Tsuchinshan 1 is fitted with \cv, \cvi, \nvi, \nvii, \ovii, and \oviii. The fit shows pronounced carbon components at low energies with weaker but non-zero oxygen components at higher energies. The relative amplitudes and widths differ from those in 19P, producing a spectrum in which the carbon features are prominent while oxygen is still present at measurable levels.

12P/Pons--Brooks is fitted with \cv, \cvi, \nvi, and \nvii. The low-energy spectrum is dominated by the pair of carbon ions, which are relatively strong and comparatively narrow in the fit, while the nitrogen lines contribute a broader, higher-energy tail; no oxygen components were selected in the best-fit combination for the full 12P dataset. 

Across the sample, the fitted centroids fall close to the expected charge-exchange line energies, while the fitted widths and amplitudes reveal several distinct, reproducible patterns.
Many comets, for example 19P and 62P, and the nitrogen components of 12P, show components with $\sigma\approx0.04$\,keV across multiple ions, producing moderately broad profiles in both the low- and mid-energy ranges.
By contrast, K2 and E3 exhibit substantially narrower carbon components (K2: \cv\ $\sigma = 0.024$\,keV, \cvi $\sigma = 0.020$\,keV; E3: \cv\ $\sigma = 0.023$\,keV, \cvi $\sigma = 0.020$\,keV), which produce comparatively sharp low-energy peaks in those fits. 
At the high-energy end, T2 and 88P have notably larger widths for the oxygen \oviii component, with $\sigma = 0.070$\,keV in both cases, giving those fits broader high-energy components than seen for many other comets.
Amplitude patterns are also heterogeneous: E3 stands out with very large nitrogen amplitudes and a relatively large \ovii amplitude, whereas 67P is the most carbon-dominated fit in the table and shows only minimal oxygen.
More generally, carbon and nitrogen amplitudes span a wide range across the sample while most oxygen amplitudes remain comparatively small, with the exceptions noted above.

The fitted Gaussian widths reported here primarily reflect the instrumental resolution and the degree of line blending; they do not reflect any intrinsic broadening of the CX features. 
The relative amplitudes of the low‐energy carbon and nitrogen features and high‐energy oxygen features in contrast do provide a measure of the spectral hardness, which is commonly quantified through the use of the hardness ratio (HR).
HR is the relative strength of emission above 0.55\,keV versus below roughly 0.55\,keV and it traces the balance between lower‐charge carbon/nitrogen ions and higher‐charge oxygen ions in the solar wind.  
A harder spectrum, corresponding to stronger \ovii and \oviii relative to C and N lines, therefore indicates a higher freeze‐in temperature of the solar wind source plasma, while a softer spectrum reflects a cooler, less ionized wind.

See Table~\ref{tab:gauss_params_full} for the full numerical values.

\subsection{Emission above 1\,keV}
Although cometary SWCX has been demonstrated to yield emission features above 1\,keV in previous studies \citep{ewing2013emission, carter2010high}, our model does not conclusively detect any such emission in the present dataset. 
Across all nine comets, weak residuals appear sporadically between 1.0 and 1.3\,keV, but none exceed the local noise level or meet the BIC threshold for inclusion in the optimal BCM solution.
For each spectrum, regardless of stacking, the addition of high-energy components (Ne, Mg, Al, or Si) failed to improve the BIC score, indicating that these lines are not statistically required by the data.
We therefore restrict our quantitative analysis to the 0.37--1.0\,keV range, where the signal-to-noise ratio is sufficient for reliable fitting.

\subsection{Short-term variability of C/2022 E3 (ZTF)}
\citet{manzini2023rotation} suggests that E3 has a measured nucleus rotation period of approximately 8.5 hours from coma morphology observations they conducted.
We examined whether the count rates measured in individual NICER observations of E3 show a periodic modulation associated with its nucleus rotation, so we plotted the count rate for each observation, spaced by about 1.4 hours over a 14-hour span.
No consistent periodic structure was evident: peaks and troughs in count rate do not repeat at the rotation period, and the amplitude of variation remains within statistical uncertainty for most epochs.  
We also compared spectral widths and amplitudes for low- vs mid-energy ions across observations but found no convincing correlation between these spectral shape parameters and phase in the rotation period.

\subsection{Relative Line Intensities}

\begin{table}
\centering
\caption{Relative intensities of six ions (\cv, \cvi, \nvi, \nvii, \ovii, and \oviii) across eight comets' full datasets, found using the Bayesian Color Model (BCM). Intensities are normalized to the \cvi feature.}
\label{tab:intensities_all_comets}
\footnotesize
\setlength{\tabcolsep}{3pt}
\renewcommand{\arraystretch}{1.05}
\begin{tabular}{@{} l c *{8}{>{\centering\arraybackslash}p{1.5cm}} @{}}
\toprule
&&\multicolumn{8}{c}{Intensity (normalized to \cvi feature)}\\\midrule
Ion & E.Range (eV) & T2 & 88P & 67P & 19P & K2 & E3 & 62P & 12P \\
\midrule
\cv    & 304--332 & 4.20 & 1.60 & 3.09 & 1.94 & 2.05 & 1.83 & 1.23 & 1.24 \\
\cvi   & 393--436 & 1.00 & 1.00 & 1.00 & 1.00 & 1.00 & 1.00 & 1.00 & 1.00 \\
\nvi   & 436--481 & 1.55 & 1.22 & 1.40 & 1.21 & 1.26 & 1.83 & 0.61 & 0.66 \\
\nvii  & 538--600 & 1.45 & 1.05 & 1.55 & 1.60 & 0.97 & 1.61 & 0.95 & 0.87 \\
\ovii  & 577--639 & 1.96 & 0.62 & 0.64 & 1.00 & 1.17 & 1.83 & 0.49 & -- \\
\oviii & 703--787 & 2.26 & 0.11 & 0.30 & 0.28 & 0.62 & 0.38 & 0.15 & -- \\
\bottomrule
\end{tabular}
\end{table}

The absolute flux of the ions discussed in this study and considered within the BCM, specifically \cv, \cvi, \nvi, \nvii, \ovii, and \oviii, largely depends on solar wind, but also varies based on the activity of the comet.
We do not consider the absolute flux in this study chiefly because we do not have much confidence in an interpretation of absolute flux, given that we draw the line for the low-energy cutoff.
A more reliable interpretation can be obtained from the relative intensity of the spectral lines, normalized to the \cvi feature, which is why we instead use those in this study.
Table~\ref{tab:intensities_all_comets} lists the relative intensities, normalized to \cvi, for eight comets considered in this survey.
Several clear patterns emerge.
C/2022 E3 (E3) stands out in the nitrogen diagnostics, exhibiting the largest relative \nvi and \nvii intensities in the sample (\nvi=1.83, \nvii=1.61).
C/2017 T2 (T2) displays the strongest oxygen signal (\ovii=1.96, \oviii=2.26) and a large \cv.
At the opposite extreme, 12P shows negligible oxygen in the best-fit models, producing spectra that are strongly carbon-dominated.
The remainder of the sample occupies intermediate morphologies: 19P and 62P have measurable oxygen but moderate \ovii and \oviii (both show \nvii and \cv$\approx1.60$ and $1.23$, respectively), while 88P and C/2017 T2 (T2) show intermediate nitrogen but different oxygen behavior: notably, T2 has \oviii$\approx2.26$ and \ovii$\approx1.96$ (\oviii/\ovii$\ge1$), a pattern that may indicate a relatively harder ion population or reflect fitting/systematic differences in that epoch.
C/2017 K2 (K2) occupies an intermediate position with moderate nitrogen and oxygen (\cvi$\approx1.00$, \nvi$\approx1.26$, \nvii$\approx0.97$, \ovii$\approx1.17$).
Since these intensities are normalized to \cvi, they primarily trace changes in the solar-wind charge-state distribution and the cometary response rather than absolute emission measures.
Small differences in fitted centers, instrumental resolution, and possible unresolved blends can affect individual ratios, especially for \ovii/\oviii, so detailed physical interpretation of single-ion anomalies, for example T2's high \oviii/\ovii, should be made with caution and, where possible, compared against contemporaneous solar-wind diagnostics.

\subsection{Epoch-Resolved Freeze‑In Temperatures}

\begin{table*}
\centering
\caption{Epoch-resolved flux ratios and inferred freeze-in temperatures for the comets considered in this work. The flux ratios are expressed for the indicated ion pairs; $T_{\rm freeze}$ values are given in MK and are inferred from the listed ratios using the ionization-balance model described in the text. Values are computed from BCM fits to each epoch's combined observations.}
\label{tab:flux_ratios_epochs}
\scriptsize
\setlength{\tabcolsep}{3pt}
\begin{tabular}{@{} l c *{6}{>{\centering\arraybackslash}p{1.85cm}} @{}}
\toprule
Comet & Epoch & \cvi\ / \cv & $T_{\rm freeze}$(C)\,(MK) & \nvii\ / \nvi & $T_{\rm freeze}$(N)\,(MK) & \oviii\ / \ovii & $T_{\rm freeze}$(O)\,(MK) \\
\midrule

\multirow{1}{*}{C/2017 T2 (PanSTARRS)}
  & 1 & 0.69 & $1.27\pm0.14$ & 1.11 & $2.03\pm0.23$ & 1.21 & $2.69\pm0.25$ \\
\midrule

\multirow{3}{*}{88P/Howell}
  & 1 & 0.13 & $0.97\pm0.10$ & 0.78 & $1.91\pm0.22$ & 0.23 & $2.12\pm0.20$ \\
  & 2 & 1.35 & $1.42\pm0.15$ & 0.35 & $1.66\pm0.19$ & 2.56 & $2.99\pm0.28$ \\
  & 3 & 2.15 & $1.53\pm0.16$ & 1.95 & $2.24\pm0.25$ & 0.06 & $1.76\pm0.17$ \\
\midrule

\multirow{6}{*}{67P/Churyumov--Gerasimenko}
  & 1 & 0.62 & $1.25\pm0.13$ & 2.84 & $2.39\pm0.27$ & 0.16 & $2.02\pm0.19$ \\
  & 2 & 1.71 & $1.48\pm0.16$ & 2.00 & $2.25\pm0.25$ & 0.69 & $2.48\pm0.23$ \\
  & 3 & 2.10 & $1.53\pm0.16$ & 1.04 & $2.01\pm0.23$ & 0.24 & $2.14\pm0.20$ \\
  & 4 & 0.85 & $1.32\pm0.14$ & 1.49 & $2.14\pm0.24$ & 0.42 & $2.31\pm0.22$ \\
  & 5 & 0.55 & $1.22\pm0.13$ & 0.74 & $1.90\pm0.21$ & 0.68 & $2.47\pm0.23$ \\
  & 6 & 0.27 & $1.09\pm0.12$ & 2.14 & $2.28\pm0.26$ & 0.82 & $2.54\pm0.24$ \\
\midrule

\multirow{3}{*}{19P/Borrelly}
  & 1 & 0.77 & $1.29\pm0.14$ & 1.60 & $2.17\pm0.25$ & 0.43 & $2.32\pm0.22$ \\
  & 2 & 1.45 & $1.44\pm0.15$ & 1.98 & $2.25\pm0.25$ & 0.28 & $2.18\pm0.20$ \\
  & 3 & 0.58 & $1.24\pm0.13$ & 0.61 & $1.84\pm0.21$ & 0.44 & $2.32\pm0.22$ \\
\midrule

\multirow{3}{*}{C/2017 K2 (PANSTARRS)}
  & 1 & 0.74 & $1.29\pm0.14$ & 0.80 & $1.92\pm0.22$ & 0.85 & $2.56\pm0.24$ \\
  & 2 & 0.86 & $1.32\pm0.14$ & 15.00 & $3.18\pm0.36$ & 14.13 & $3.81\pm0.36$ \\
  & 3 & 1.83 & $1.49\pm0.16$ & 2.04 & $2.26\pm0.26$ & 0.27 & $2.16\pm0.20$ \\
\midrule

\multirow{1}{*}{C/2022 E3 (ZTF)}
  & 1 & 3.79 & $1.68\pm0.18$ & 1.94 & $2.24\pm0.25$ & 0.20 & $2.08\pm0.20$ \\
\midrule

\multirow{2}{*}{62P/Tsuchinshan}
  & 1 & 1.47 & $1.44\pm0.15$ & 1.04 & $2.01\pm0.23$ & 1.18 & $2.68\pm0.25$ \\
  & 2 & 2.11 & $1.53\pm0.16$ & 2.17 & $2.28\pm0.26$ & 0.37 & $2.27\pm0.21$ \\
\midrule

\multirow{2}{*}{12P/Pons--Brooks}
  & 1 & 1.73 & $1.48\pm0.16$ & 44.85 & $3.84\pm0.43$ & \textemdash & \textemdash \\
  & 2 & 2.35 & $1.55\pm0.17$ & 1.86 & $2.22\pm0.25$ & \textemdash & \textemdash \\

\bottomrule
\end{tabular}
\end{table*}

\begin{figure}
    \centering
    \includegraphics[width=0.6
    \linewidth,trim={0cm 0cm 0cm 0cm},clip]{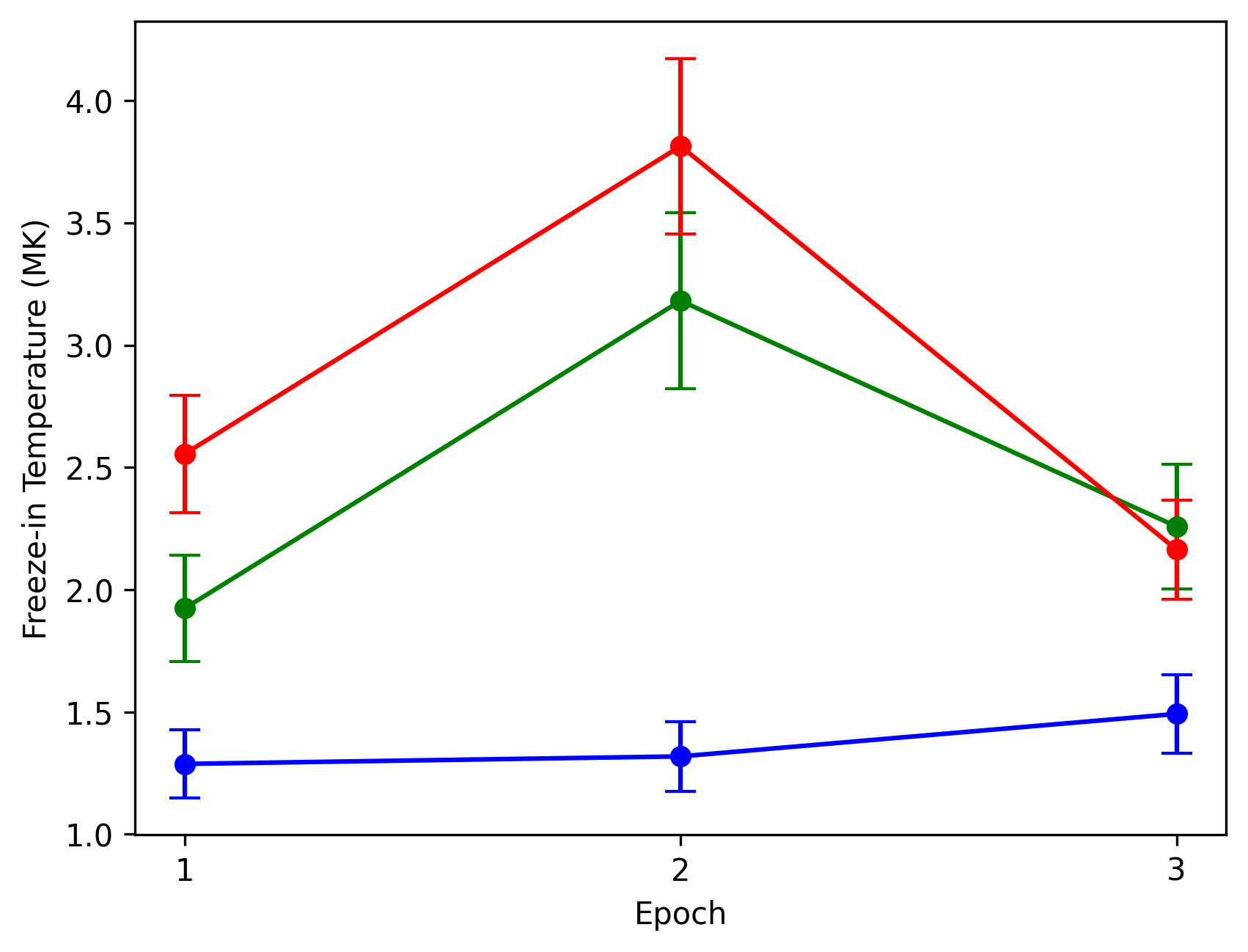}
    \caption{Freeze-in temperatures for comet C/2017 K2 (PANSTARRS) inferred from flux ratios, plotted versus observing epoch. Blue shows $T_{\rm freeze}$ derived from \cvi\ / \cv, green is from \nvii\ / \nvi, and red is from \oviii\ / \ovii; vertical bars are $1\sigma$ uncertainties propagated from the line-fit errors.}
    \label{fig:K2_freezeintemps_inpaper}
\end{figure}
\begin{figure}
    \centering
    \includegraphics[width=0.6
    \linewidth,trim={0cm 0cm 0cm 0cm},clip]{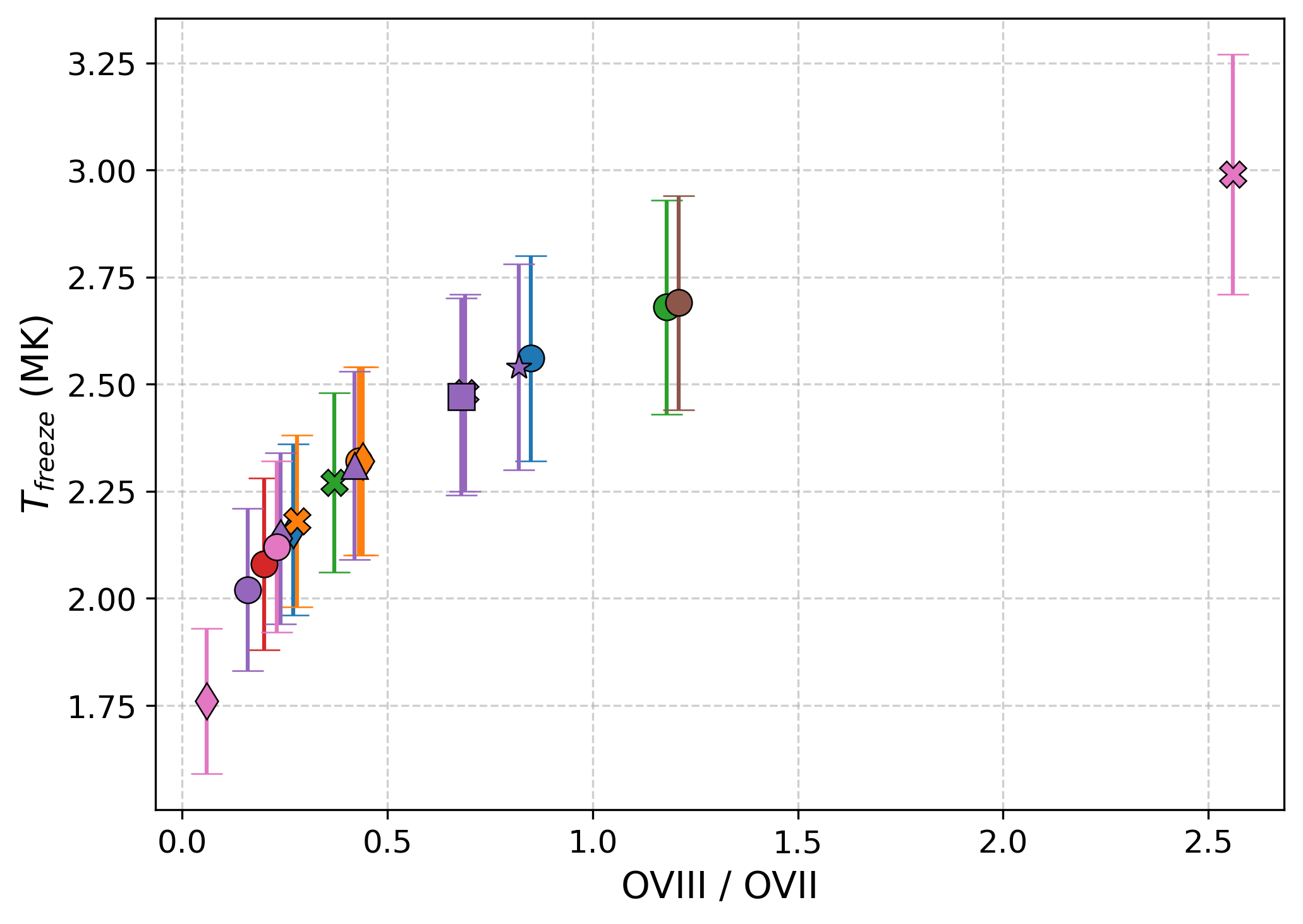}
    \caption{Freeze-in temperature inferred from the \oviii\ / \ovii flux ratio plotted against the measured oxygen flux ratio for each comet and observing epoch. Vertical error bars show 1-$\sigma$ uncertainties propagated from the line-fit errors. Colors identify comets: T2 (brown), 88P (magenta), 67P (purple), 19P (orange), K2 (blue), E3 (red), and 62P (green). Marker shapes denote observing epochs: circle = Epoch 1, X = Epoch 2, diamond = Epoch 3, triangle up = Epoch 4, square = Epoch 5, star = Epoch 6. Numerical values are shown in Table~\ref{tab:flux_ratios_epochs}. Note that Epoch 2 of K2 is a clear outlier, with a large O flux ratio and inferred $T_{\rm freeze}$, and is omitted from this figure. This epoch is treated in more depth in the Discussion section.}
    \label{fig:Tfreeze_v_O}
\end{figure}

Each ion pair, such as \cv and \cvi, encodes a freeze-in temperature through the flux ratio of the ion pair \citep{mazzotta1998ionization,bryans2009new}. 
Figure~\ref{fig:K2_freezeintemps_inpaper} shows the evolution of the freeze-in temperature for K2, one of the comets considered in this work, and Appendix~\ref{app:freezeintemps} shows the freeze-in temperature variations over epochs for all five comets. 
Figure~\ref{fig:Tfreeze_v_O} shows the relationship between the freeze-in temperature and the oxygen flux ratio for each epoch of each comet.
Table~\ref{tab:flux_ratios_epochs} shows our epoch-resolved analysis of each comet's flux ratios and inferred freeze-in temperatures.
The analysis reveals measurable temporal changes in the inferred freeze-in conditions for several comets. 

T2 is represented by a single epoch in our dataset and yields comparatively high oxygen ionization. 
The epoch diagnostics place the T2 observations among the more highly ionized SW in our sample for the oxygen diagnostic, while the carbon-derived temperature remains similar to other comets in the survey. 
The single-epoch coverage limits statements about short-term variability but the measured ratios indicate a relatively oxygen-rich charge-state distribution incident on the coma during the observation.

88P exhibits significant epoch-to-epoch variability in its freeze-in diagnostics.
The full-dataset diagnostics are $T_{\rm freeze}({\rm C}) = 1.49$\,MK, $T_{\rm freeze}({\rm N}) = 2.01$\,MK and $T_{\rm freeze}({\rm O}) = 2.05$\,MK.
The three epochs show contrasting conditions: Epoch 1 is relatively soft in carbon ($0.97$\,MK) with moderate nitrogen and oxygen ($1.91$\,MK and $2.12$\,MK, respectively); Epoch 2 shows a hardening in the oxygen diagnostic ($T_{\rm freeze}({\rm O}) = 2.99$\,MK, \oviii / \ovii $=2.56$) while carbon and nitrogen remain moderately low ($1.42$\,MK and $1.66$\,MK); Epoch 3 returns to a harder carbon/nitrogen state but a softer oxygen diagnostic.
88P demonstrates that the oxygen-based diagnostic can fluctuate by more than a factor of two between epochs, producing a full-dataset average that masks substantial short-term changes. 

67P shows a mixture of relatively cool carbon diagnostics and systematically higher nitrogen- and oxygen-derived diagnostics. 
The epoch-resolved oxygen diagnostics cluster near $2.0$--$2.6$\,MK. 
67P displays modest epoch-to-epoch variation but consistent behavior: carbon-derived temperatures remain lowest, nitrogen-derived values are intermediate, and oxygen-derived diagnostics are highest.

19P shows slight epoch-to-epoch variation, but a consistent separation between the carbon- and nitrogen-based diagnostics. The full-dataset diagnostics are $T_{\rm freeze}({\rm C}) = 1.44$\,MK and $T_{\rm freeze}({\rm N}) = 2.16$\,MK, and epoch-level carbon-derived temperatures span approximately $1.24$--$1.44$\,MK (typical uncertainties $0.13$--$0.15$\,MK) while nitrogen-derived temperatures span approximately $1.84$--$2.25$\,MK (typical uncertainties $0.21$--$0.25$\,MK). Where oxygen is detected for 19P, oxygen-based temperatures cluster around $2.1$--$2.3$\,MK. The second epoch exhibits the hardest nitrogen signature and the third epoch is noticeably fainter in overall line amplitude.

K2 exhibits substantial epoch-to-epoch variation. The diagnostics for the full dataset are $T_{\rm freeze}({\rm C}) = 1.44$\,MK, $T_{\rm freeze}({\rm N}) = 2.20$\,MK and $T_{\rm freeze}({\rm O}) = 2.43$\,MK. Figure~\ref{fig:K2_freezeintemps_inpaper} shows the epoch-to-epoch variation in the inferred freeze-in temperature.
Notably the freeze-in temperature inferred from the carbon flux ratio remains roughly the same across epochs, while the freeze-in temperature inferred from the nitrogen and oxygen ratios rises in epoch 2 and falls in epoch 3.
Across the three K2 epochs (3.2--1.9\,AU), the spectral morphology evolves from relatively strong carbon and nitrogen features toward a modest increase in oxygen-line contribution.  
These variations occur over the same heliocentric distance range where independent optical and infrared observations indicate a transition in the coma composition from CO$_2$/CO-dominated to H$_2$O-dominated sublimation \citep{combi2025water,cambianica2023co2,woodward2025jwst}.  
Interpretation of this compositional transition and its impact on the observed X-ray spectra is provided in the discussion.

E3 is the most highly ionized object in the sample in the carbon diagnostic. The full dataset is the lone epoch considered in this study, given that our observations of E3 did not extend more than one day. The flux ratios yield $T_{\rm freeze}({\rm C}) = 1.68 \pm 0.18$\,MK, $T_{\rm freeze}({\rm N}) = 2.24 \pm 0.25$\,MK and $T_{\rm freeze}({\rm O}) = 2.08 \pm 0.20$\,MK, and E3 shows the largest relative nitrogen line intensity of the sample. The higher carbon-derived temperature and the presence of emission attributed to nitrogen are consistent with encounter of a more highly ionized solar-wind during the observations. Our limited observations with NICER did not indicate a clear periodicity in the total X-ray flux.

62P has full-dataset values of $T_{\rm freeze}({\rm C}) = 1.55$\,MK, $T_{\rm freeze}({\rm N}) = 2.22$\,MK and $T_{\rm freeze}({\rm O}) = 2.23$\,MK. Epoch 1 is somewhat softer in nitrogen and harder in oxygen, with $T_{\rm freeze}({\rm C}) = 1.44 \pm 0.15$\,MK, $T_{\rm freeze}({\rm N}) = 2.01 \pm 0.23$\,MK and $T_{\rm freeze}({\rm O}) = 2.68 \pm 0.25$\,MK, while Epoch 2 shows higher nitrogen ionization ($T_{\rm freeze}({\rm N}) = 2.28 \pm 0.26$,MK) and a lower oxygen-derived temperature ($T_{\rm freeze}({\rm O}) = 2.27 \pm 0.21$\,MK). 

12P is characterized by relatively cool carbon-derived freeze-in temperatures in the full fit, with \cvi\ /\cv giving $1.55$\,MK and \nvii\ / \nvi giving $2.16$\,MK. The epoch-resolved values show a clear short-timescale hardening: Epoch 1 yields $T_{\rm freeze}({\rm C}) = 1.24 \pm 0.13$\,MK and $T_{\rm freeze}({\rm N}) = 2.08 \pm 0.24$\,MK, while Epoch 2 rises to $T_{\rm freeze}({\rm C}) = 1.61 \pm 0.17$\,MK and $T_{\rm freeze}({\rm N}) = 2.25 \pm 0.26$\,MK. The full-dataset numbers therefore fall between these epoch extremes, consistent with a change in the charge-state distribution incident on the coma between the two observing intervals (see Table~\ref{tab:flux_ratios_epochs}).

Taken together, the epoch-resolved results show a coherent trend across the sample: carbon-derived freeze-in temperatures cluster near $T_{\rm freeze}({\rm C})\approx1.4$--$1.7$\,MK, while nitrogen- and oxygen-derived diagnostics are systematically higher, typically $T_{\rm freeze}({\rm N,O})\approx2.0$--$2.3$\,MK. 
Notable departures from this behavior highlight the role of short-term solar-wind variability: 
E3 and T2 both show relatively elevated oxygen ionization, with E3 in the carbon diagnostic and with T2 in the oxygen diagnostic; 
67P, with six epochs, displays epoch-to-epoch variability; 
88P exhibits dramatic short-term variation across epochs, with its calculated $T_{\rm freeze}({\rm O})$ changing by more than a factor of two between epochs. 
K2 likewise shows strong variation in its second epoch that stands out as an outlier in the sample.

\subsection{Stacked Spectra}
To provide a single, empirical view of the spectral morphologies sampled by our survey, three groups have been created: 
(i) a stack of all available observations from the nine comets, known as the ``full'' stack,
(ii) a stack of the five epochs with the highest inferred freeze-in temperatures ($T_f$), and 
(iii) a stack of the five epochs with the lowest inferred freeze-in temperatures.
The BCM fits for these stacks were performed identically to the single-comet/epoch fits described above.

\begin{table}
\centering
\caption{Relative intensities of six ions (\cv, \cvi, \nvi, \nvii, \ovii, and \oviii) for the three stacked datasets. Intensities are normalized to the \cvi\ feature. Each ion's energy range shows the possible values for the Gaussian representing the ion.}
\label{tab:intensities_stacks}
\footnotesize
\setlength{\tabcolsep}{35pt}
\renewcommand{\arraystretch}{1.05}
\begin{tabularx}{\linewidth}{@{} l X c c c @{}} 
\toprule
&&\multicolumn{3}{c}{Intensity (normalized to \cvi)}\\
\midrule
Ion & E.Range (eV) & Full stack & High-$T_{f}$ & Low-$T_{f}$ \\
\midrule
\cv    & 304--332 & 2.57 & 2.82 & 0.86 \\
\cvi   & 393--436 & 1.00 & 1.00 & 1.00 \\
\nvi   & 436--481 & 1.44 & 1.34 & 0.75 \\
\nvii  & 538--600 & 1.65 & 1.26 & 1.05 \\
\ovii  & 577--639 & 1.00 & 0.82 & 0.65 \\
\oviii & 703--787 & 0.21 & 0.50 & 0.12 \\
\bottomrule
\end{tabularx}
\end{table}

\begin{table*}
\centering
\caption{Flux ratios and inferred freeze-in temperatures for the three stacked datasets. Flux ratios for each ion pair are given with the more highly ionized ion in the numerator; e.g. $C_{FR}$ represents \cvi\ / \cv. $T_{f}$ values are given in MK and use the same ionization-balance mapping as Table~\ref{tab:flux_ratios_epochs}.}
\label{tab:flux_ratios_stacks}
\setlength{\tabcolsep}{20pt}
\begin{tabularx}{\linewidth}{@{} l c c c c c c @{}}
\toprule
Dataset & $C_{FR}$ & $T_{f}$(C)\,(MK) & $N_{FR}$ & $T_{f}$(N)\,(MK) & $O_{FR}$ & $T_{f}$(O)\,(MK) \\
\midrule

Full
  & 1.13 & $1.38 \pm 0.15$ & 1.36 & $2.11 \pm 0.24$ & 0.22 & $2.11 \pm 0.20$ \\
\midrule

High $T_{f}$
  & 1.03 & $1.36 \pm 0.15$ & 1.11 & $2.03 \pm 0.23$ & 0.63 & $2.45 \pm 0.23$ \\
\midrule

Low $T_{f}$
  & 3.38 & $1.65 \pm 0.18$ & 1.66 & $2.18 \pm 0.25$ & 0.20 & $2.08 \pm 0.20$ \\

\bottomrule
\end{tabularx}
\end{table*}

Table~\ref{tab:intensities_stacks} reports the relative intensities of six ions across the three stacks, normalized to the \cvi\ feature. 
Table~\ref{tab:flux_ratios_stacks} gives the flux ratios for the three diagnostic ion pairs, along with the corresponding freeze-in temperatures ($T_{f}$) in MK. 

\begin{figure}
    \centering
    \includegraphics[width=0.6\textwidth,trim={0.35cm 0.25cm 0.2cm 1.35cm},clip]{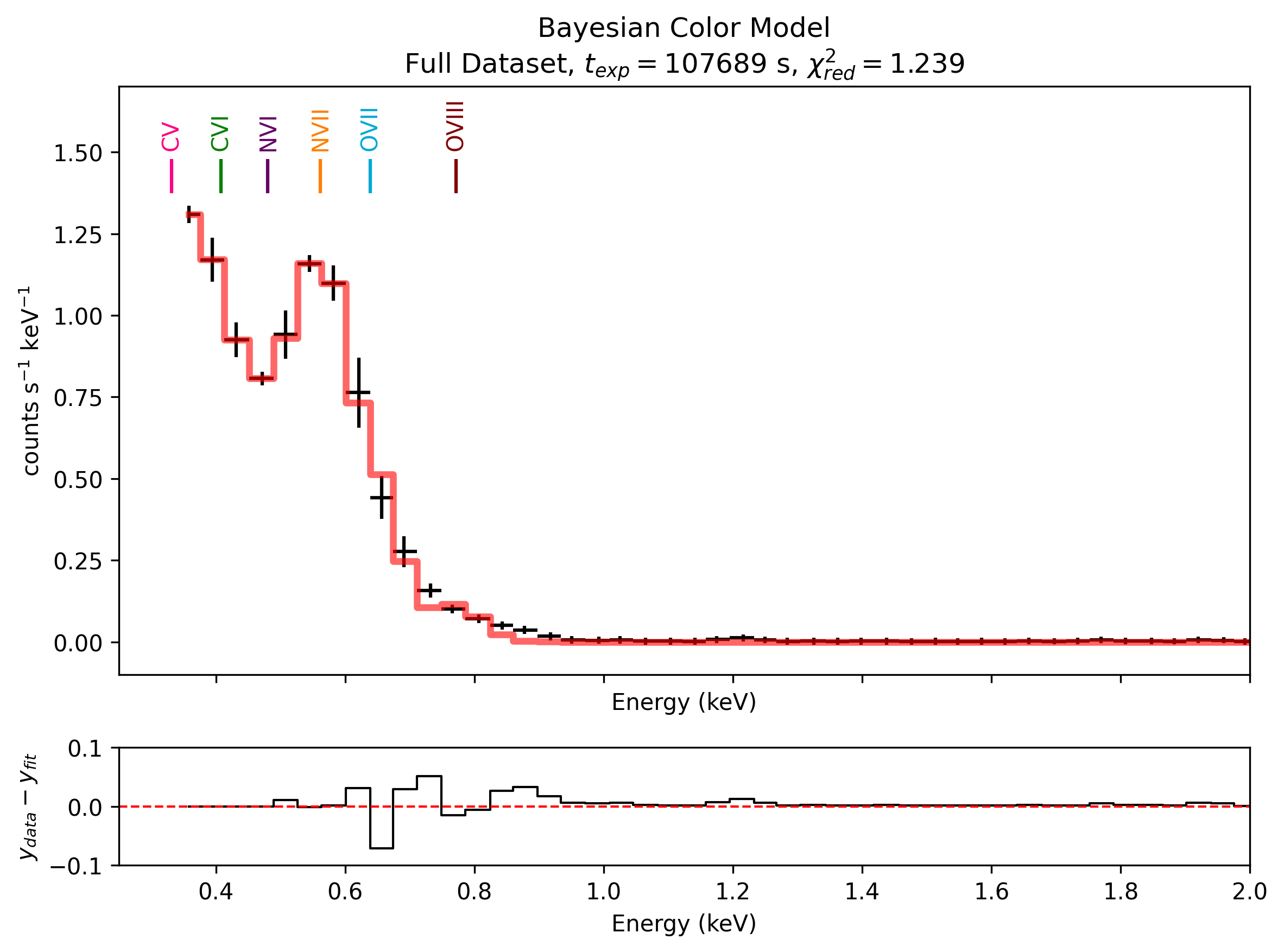}
    \caption{X-ray spectrum from the interaction between the solar wind and the atmospheres of all nine comets in the study, fitted with the Bayesian Color Model (BCM).}
    \label{fig:spectrum_fullstack}
\end{figure}

\begin{figure}
    \centering
    \begin{subfigure}[b]{0.49\textwidth}
        \centering
        \includegraphics[width=\linewidth,trim={0.4cm .2cm 0.5cm 1.25cm},clip]{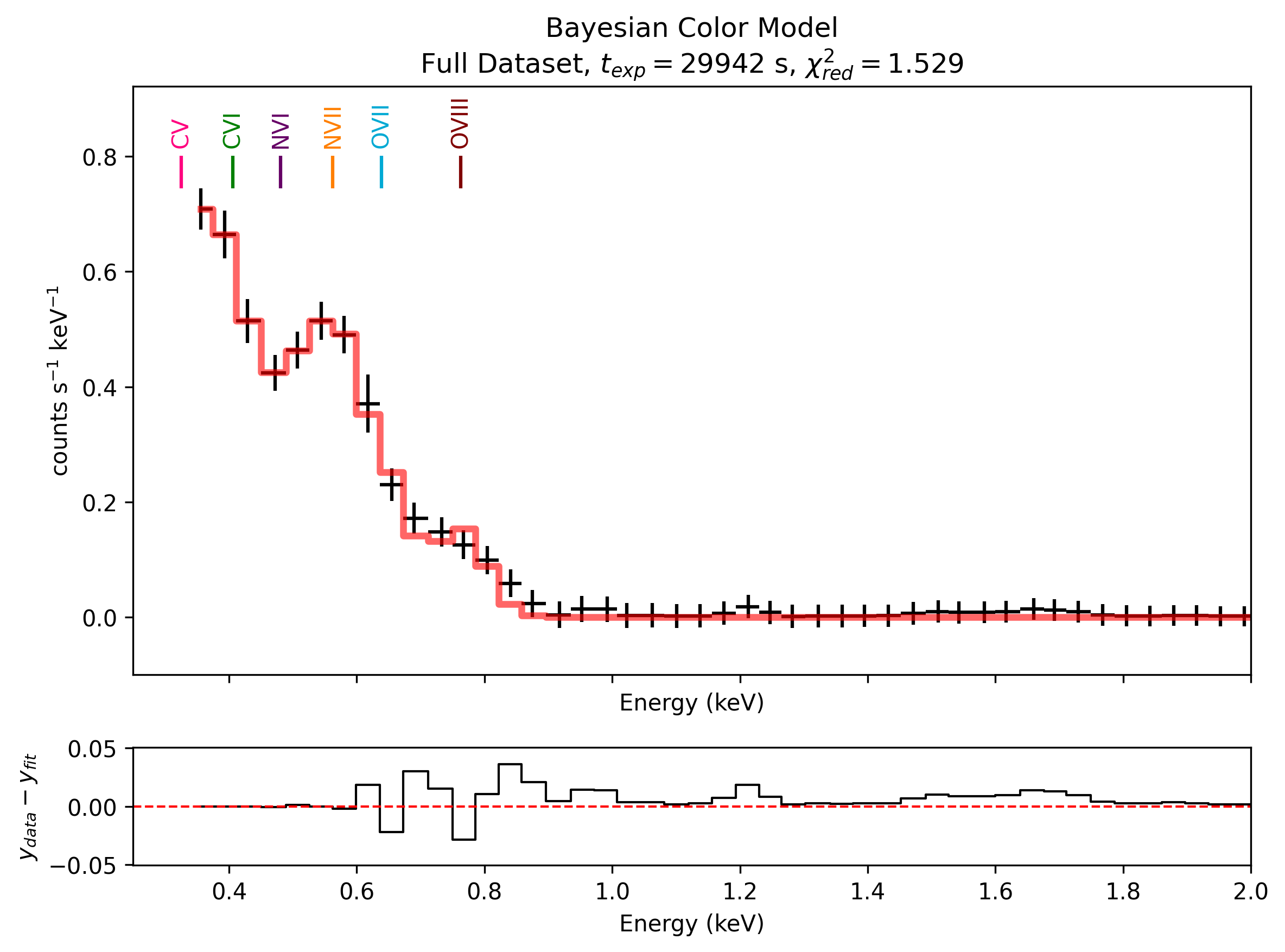}
        \caption{High-$T_{f}$ stack.}
        \label{fig:spec_hightf_stack}
    \end{subfigure}\hfill
    \begin{subfigure}[b]{0.49\textwidth}
        \centering
        \includegraphics[width=\linewidth,trim={0.4cm .2cm 0.5cm 1.25cm},clip]{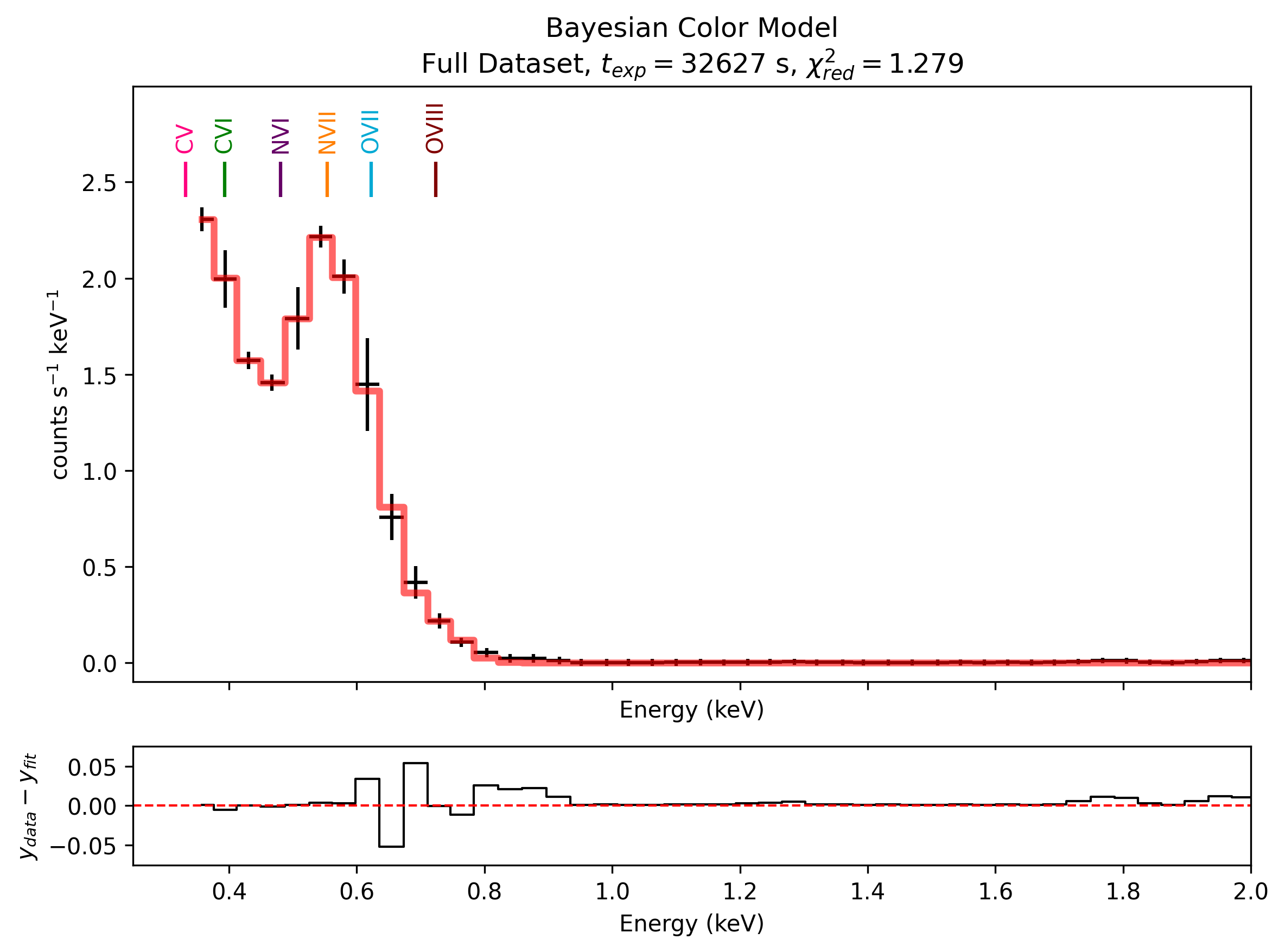}
        \caption{Low-$T_{f}$ stack.}
        \label{fig:spec_lowtf_stack}
    \end{subfigure}

    \caption{X-ray spectra from the interaction between the solar wind and the atmospheres of comets, separated by freeze-in temperature ($T_f$). Left: spectrum for the high $T_f$ stack. Right: spectrum for the low $T_f$ stack. Both spectra are fitted with the Bayesian Color Model (BCM).}
    \label{fig:spectrum_HighLowTf_stack}
\end{figure}

As shown in Figure~\ref{fig:spectrum_fullstack}, the spectrum of the full stack set of observations seems to show slight features at $\sim$1.20\,keV, $\sim$1.78\,keV, and $\sim$1.95\,keV.
Figure~\ref{fig:spectrum_HighLowTf_stack} shows the spectrum for the stack of the five epochs with the highest inferred freeze-in temperatures (left) and the spectrum for the stack of the five epochs with the lowest inferred freeze-in temperatures (right).

The classification of epochs into high-$T_{f}$ and low-$T_{f}$ groups is based specifically on the oxygen flux ratio, $O_{FR} =$ \oviii\ / \ovii. 
This choice is motivated by instrumental and modeling considerations. 
The carbon flux ratio, $C_{FR} =$ \cvi\ / \cv, is affected by greater uncertainty because the NICER XTI instrument has larger calibration errors at energies below 370\,eV, where the \cv\ feature lies. 
The nitrogen flux ratio, $N_{FR} =$ \nvii\ / \nvi, is also less reliable because of degeneracy from both lower and higher energies: counts at the low-energy side can be attributed to \cvi, while counts at the high-energy side can be attributed to \ovii. 
It is therefore most robust to define the classification on the basis of oxygen, where the features are more isolated. 

The five epochs comprising the high-$T_{f}$ group are: 88P epoch~2, T2 epoch~1, 62P epoch~1, K2 epoch~1, and 67P epoch~6. 
The five epochs comprising the low-$T_{f}$ group are: 88P epoch~3, 67P epoch~1, E3 epoch~1, 88P epoch~1, and 67P epoch~3. 

Comparisons among the three stacks show several trends. 
For carbon, the flux ratios ($C_{FR}$) range from $1.03$ in the high-$T_{f}$ stack to $3.38$ in the low-$T_{f}$ stack, with the full stack intermediate at $1.13$. 
For nitrogen, the ratios ($N_{FR}$) are closer in value across the stacks, ranging from $1.11$ (high $T_{f}$) to $1.66$ (low $T_{f}$), with the full stack at $1.36$. 
For oxygen, the flux ratio of which defines the grouping, the ratios ($O_{FR}$) span from $0.20$ in the low-$T_{f}$ stack to $0.63$ in the high-$T_{f}$ stack, with the full stack at $0.22$. 
The corresponding $T_{f}$(O) values are $2.08 \pm 0.20$\,MK for the low-$T_{f}$ stack, $2.45 \pm 0.23$\,MK for the high-$T_{f}$ stack, and $2.11 \pm 0.20$\,MK for the full stack.

In addition to the freeze-in temperatures inferred from the oxygen ratio, those inferred from the carbon and nitrogen ratios are also given in Table~\ref{tab:flux_ratios_stacks}. 
The $T_{f}$(C) values range from $1.36 \pm 0.15$\,MK (high-$T_{f}$ stack) to $1.65 \pm 0.18$\,MK (low-$T_{f}$ stack), while $T_{f}$(N) values range from $2.03 \pm 0.23$\,MK (high-$T_{f}$) to $2.18 \pm 0.25$\,MK (low-$T_{f}$), with the full stack at $2.11 \pm 0.24$\,MK. 

\subsection{Neutral–species comparison}

\begin{figure}
    \centering
    \begin{subfigure}[b]{0.49\textwidth}
        \centering
        \includegraphics[width=\linewidth,trim={0 0 0 0},clip]{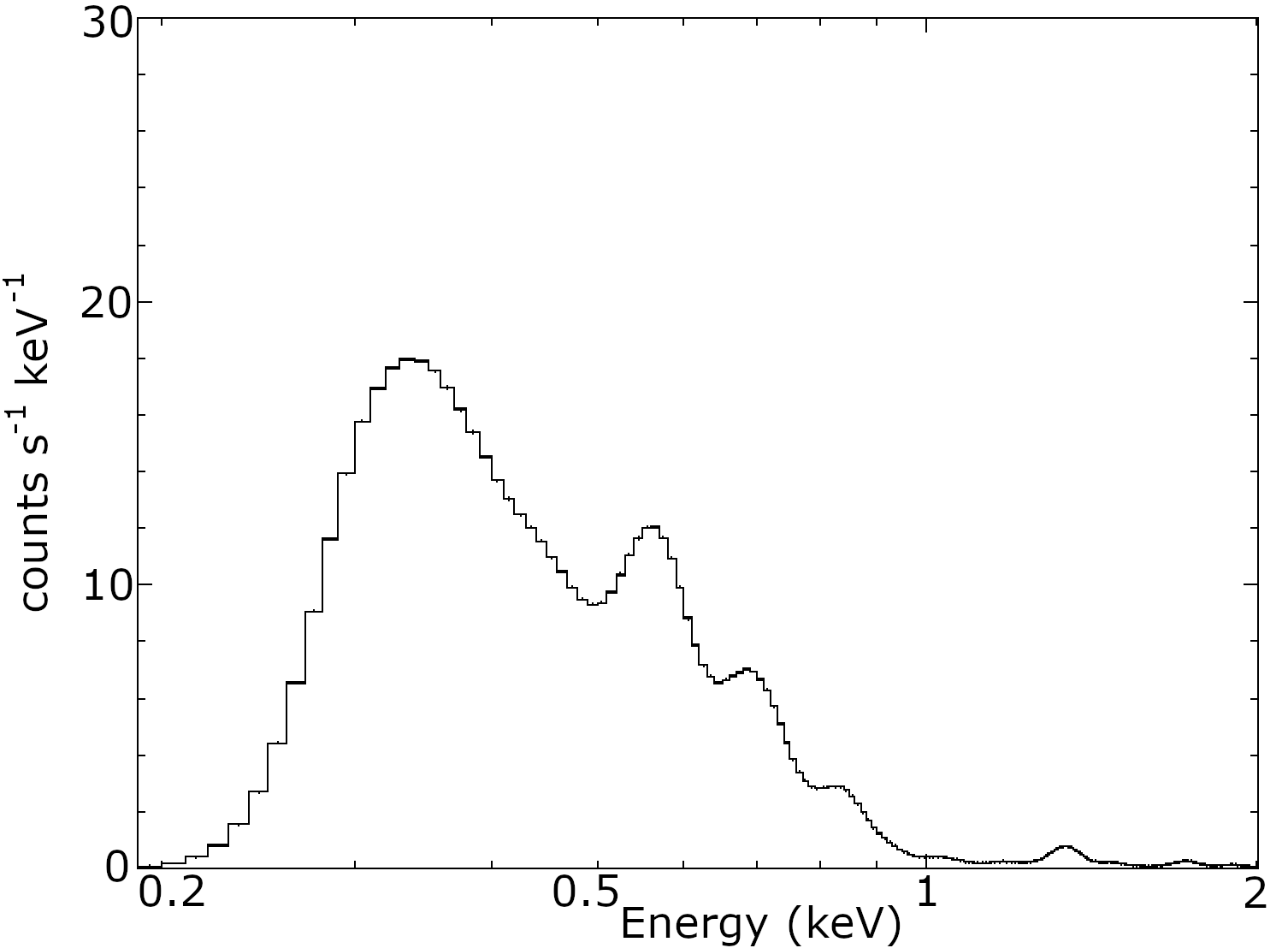}
        \caption{Simulated NICER Spectrum with H$_2$O as the target.}
        \label{fig:simspec_h2o}
    \end{subfigure}\hfill
    \begin{subfigure}[b]{0.49\textwidth}
        \centering
        \includegraphics[width=\linewidth,trim={0 0 0 0},clip]{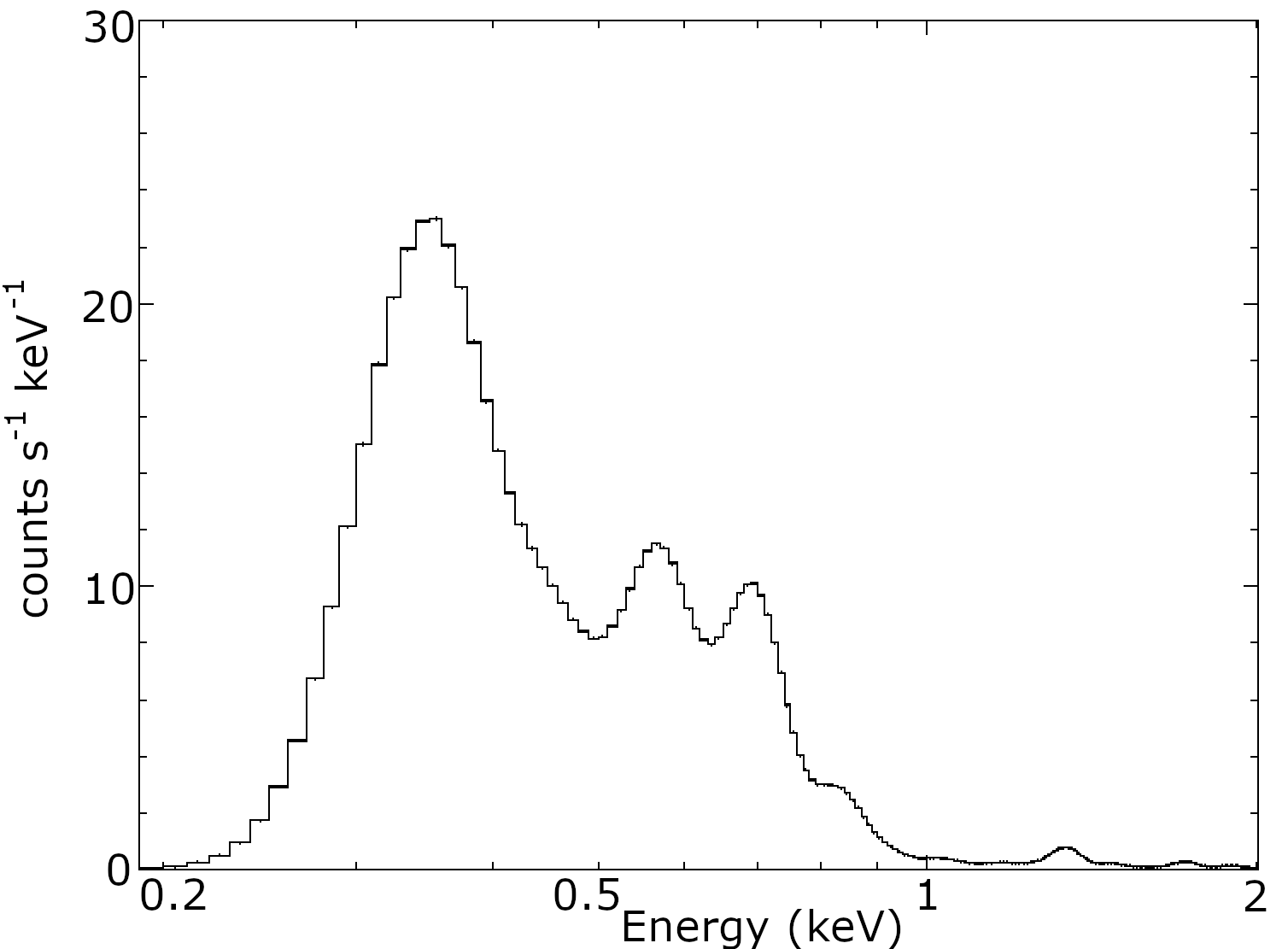}
        \caption{CO$_2$}
        \label{fig:simspec_co2}
    \end{subfigure}
    \caption{Simulated NICER XTI spectra for cometary charge-exchange interactions, representing a H$_2$O-dominated coma (left) and a CO$_2$-dominated coma (right). 
    The simulations were performed using the \textit{fakeit} command in \textit{XSPEC} with NICER response files from its calibration database. 
    Distinct spectral features arise from different neutral targets, highlighting the compositional dependence of cometary X-ray emission.}
    \label{fig:simulated_spectra}
\end{figure} 

In this subsection we present results for three comets, K2, 88P, and 19P, each fit under two assumptions for the dominant neutral species present in the coma after sublimation: CO$_{2}$+CO and H$_{2}$O. 
We consider CO$_{2}$ and CO together because there is little difference in the X-ray spectra resulting from solar-wind charge exchange with each as a neutral target \citep{Mullen2017}.
To apply the correct line ratios in our model, we must know the dominant neutral species in the coma. 
\textbf{We use the normalized line ratio for CO$_2$ and H$_2$O from the Kronos charge‑exchange database \citep{cumbee2017charge} to convert measured charge-state ratios to freeze-in temperature. 
In the Kronos framework, the upstream solar-wind charge-state distribution is assumed to follow coronal ionization equilibrium (CIE), in which electron-impact ionization is balanced by radiative and dielectronic recombination for a Maxwellian electron distribution.
These CIE-derived ion fractions are then used as inputs to detailed charge-exchange cascade calculations that predict emission line strengths as a function of freeze-in temperature and neutral target species. 
The BCM interpolates within these precomputed Kronos charge-exchange line-ratio grids to infer the effective freeze-in temperature corresponding to the measured spectral ratios.}
Under collision conditions of 987.3\,eV\,u$^{-1}$, corresponding to a collision velocity of 434.8\,km\,s$^{-1}$, this velocity is quite reasonable for the solar wind: typical ambient solar‑wind speeds near 1\,AU are $\sim$400\,km\,s$^{-1}$, with occasional faster streams up to $\sim$700\,km\,s$^{-1}$ \citep{tokumaru2010solar}.
As Figure~\ref{fig:simulated_spectra} shows, whether the coma is CO$_2$-/CO-dominated or H$_2$O-dominated affects the spectral morphology, and this subsection aims to show spectral differences based on the choice of the dominant neutral species.

The contrasting morphologies of the CO$_2$/CO- and H$_2$O-target simulations ultimately reflect differences in their state-selective charge-exchange cross sections and available electronic levels.
In the Kronos calculations, collisions with CO$_2$ preferentially populate higher-$n$ and higher-$\ell$ states and a larger fraction of singlet cascades, which enhances higher-series lines and produces a slightly harder overall spectrum, whereas H$_2$O tends to feed the triplet manifolds of C, N, and O, strengthening the He-like triplets relative to the high-$n$ lines \citep{greenwood2001experimental}. 
These trends agree with laboratory measurements of SWCX on simple versus more complex molecules and imply that, in principle, the relative strength of He-like triplets compared to higher-series lines can diagnose the dominant neutral species in the coma for instruments with sufficient resolution and signal-to-noise. 
In the NICER data, however, these neutral-dependent signatures are subtle and remain below the threshold for a firm compositional determination.

For each case, we provide relative ion intensities normalized to the \cvi\ feature, and flux ratios with the corresponding freeze-in temperatures $T_{f}$.
We note that the freeze-in temperature classification is based on the oxygen flux ratio (\oviii\ / \ovii), since the carbon ratio is more uncertain due to the high NICER XTI instrument error at $E<370$\,eV, and the nitrogen ratio is unreliable because of degeneracy: on the low-energy side counts may be attributed to \cvi, while on the high-energy side they may be attributed to \ovii.  
Therefore, the \oviii\ / \ovii-based $T_{f}$ provides the most robust classification.  
The high-$T_{f}$ and low-$T_{f}$ groups correspond to the five epochs with the highest and lowest O-based $T_{f}$, respectively.

\subsubsection{K2}

\begin{table}
\centering
\caption{Relative intensities of six ions for comet K2 (full dataset), under CO$_2$+CO and H$_2$O cases. Intensities are normalized to the \cvi\ feature.}
\label{tab:intensities_k2}
\setlength{\tabcolsep}{30pt}
\renewcommand{\arraystretch}{1.05}
\begin{tabularx}{\linewidth}{@{\hspace{0.30\linewidth}} c c c @{\hspace{0.30\linewidth}}}
\toprule
Ion & CO$_2$+CO & H$_2$O \\
\midrule
\cv    & 1.30 & 1.12 \\
\cvi   & 1.00 & 1.00 \\
\nvi   & 1.54 & 1.29 \\
\nvii  & 1.77 & 1.57 \\
\ovii  & 1.00 & 1.11 \\
\oviii & 0.36 & 0.32 \\
\bottomrule
\end{tabularx}
\end{table}

\begin{table}
\centering
\caption{Flux ratios and freeze-in temperatures for comet K2 (full dataset), under CO$_2$+CO and H$_2$O cases. $T_{f}$ values are given in MK.}
\label{tab:flux_ratios_k2}
\setlength{\tabcolsep}{12pt}
\renewcommand{\arraystretch}{1.05}
\begin{tabularx}{\linewidth}{@{\hspace{0.12\linewidth}} c c c c c c c @{}} 
\toprule
Case & $C_{FR}$ & $T_{f}$(C) & $N_{FR}$ & $T_{f}$(N) & $O_{FR}$ & $T_{f}$(O) \\
\midrule
CO$_2$+CO & 1.38 & $1.41 \pm 0.16$ & 1.15 & $2.05 \pm 0.23$ & 0.36 & $2.23 \pm 0.21$ \\
H$_2$O    & 1.21 & $1.39 \pm 0.16$ & 1.22 & $2.07 \pm 0.24$ & 0.29 & $2.19 \pm 0.21$ \\
\bottomrule
\end{tabularx}
\end{table}

Table~\ref{tab:intensities_k2} shows that for K2, the relative intensity of \cv\ is higher in the CO$_2$+CO case (1.30) compared to H$_2$O (1.12), while \nvi\ and \nvii\ show similar behavior with values of 1.54 and 1.77 (CO$_2$+CO) versus 1.29 and 1.57 (H$_2$O).  
The O-based flux ratio, as seen in Table~\ref{tab:flux_ratios_k2}, decreases from 0.36 (CO$_2$+CO) to 0.29 (H$_2$O), corresponding to $T_{f}$ values of $2.23 \pm 0.21$\,MK and $2.19 \pm 0.21$\,MK, respectively.  

Across the three epochs of K2, as seen in Table~\ref{tab:intensities_k2_epochs} and Table~\ref{tab:flux_ratios_k2_epochs}, both the relative intensities and flux ratios vary. In the CO$_2$+CO case, \cv\ increases from 1.20 (epoch 1) to 2.72 (epoch 3), while in the H$_2$O case \cv\ rises from 1.15 to 2.40 across the same epochs.  
The O-based flux ratio in CO$_2$+CO changes from 0.44 (epoch 1; $T_{f}=2.32 \pm 0.22$\,MK) to 0.27 (epoch 2; $T_{f}=2.17 \pm 0.20$\,MK) and 3.95 (epoch 3; $T_{f}=3.18 \pm 0.30$\,MK).  
For H$_2$O the corresponding sequence is 0.41 (epoch 1; $T_{f}=2.29 \pm 0.22$\,MK), 0.25 (epoch 2; $T_{f}=2.14 \pm 0.20$\,MK), and 3.51 (epoch 3; $T_{f}=3.15 \pm 0.30$\,MK).  
Thus, both cases show similar progression, with H$_2$O consistently yielding slightly lower \cv\ and O-based ratios compared to CO$_2$+CO.

\begin{table*}
\centering
\caption{Relative ion intensities for comet K2, resolved by epoch, under CO$_2$+CO and H$_2$O assumptions. Intensities are normalized to the \cvi\ feature.}
\label{tab:intensities_k2_epochs}
\setlength{\tabcolsep}{17pt}
\renewcommand{\arraystretch}{1.05}
\begin{tabularx}{\linewidth}{@{\hspace{0.05\linewidth}} c c c c c c c c c @{}} 
\toprule
Epoch & Case & \cv & \cvi & \nvi & \nvii & \ovii & \oviii \\
\midrule
\multirow{2}{*}{1} & CO$_2$+CO & 1.20 & 1.00 & 1.37 & 2.19 & 0.86 & 0.36 \\
                  & H$_2$O    & 1.15 & 1.00 & 1.32 & 2.12 & 0.90 & 0.34 \\
\midrule
\multirow{2}{*}{2} & CO$_2$+CO & 2.02 & 1.00 & 1.38 & 2.40 & 1.04 & 0.26 \\
                  & H$_2$O    & 1.84 & 1.00 & 1.33 & 2.18 & 0.97 & 0.22 \\
\midrule
\multirow{2}{*}{3} & CO$_2$+CO & 2.72 & 1.00 & 1.32 & 0.98 & 0.14 & 0.54 \\
                  & H$_2$O    & 2.40 & 1.00 & 1.27 & 0.90 & 0.16 & 0.48 \\
\bottomrule
\end{tabularx}
\end{table*}

\begin{table*}
\centering
\caption{Flux ratios and freeze-in temperatures for comet K2, resolved by epoch, under CO$_2$+CO and H$_2$O assumptions. $T_{f}$ values are given in MK.}
\label{tab:flux_ratios_k2_epochs}
\setlength{\tabcolsep}{13pt}
\renewcommand{\arraystretch}{1.05}
\begin{tabularx}{\linewidth}{@{\hspace{0.05\linewidth}} c c c c c c c c @{}} 
\toprule
Epoch & Case & $C_{FR}$ & $T_{f}$(C) & $N_{FR}$ & $T_{f}$(N) & $O_{FR}$ & $T_{f}$(O) \\
\midrule
\multirow{2}{*}{1} & CO$_2$+CO & 2.42 & $1.56 \pm 0.17$ & 1.91 & $2.23 \pm 0.25$ & 0.44 & $2.32 \pm 0.22$ \\
                  & H$_2$O    & 2.10 & $1.52 \pm 0.16$ & 1.85 & $2.21 \pm 0.25$ & 0.41 & $2.29 \pm 0.22$ \\
\midrule
\multirow{2}{*}{2} & CO$_2$+CO & 1.43 & $1.43 \pm 0.15$ & 2.07 & $2.26 \pm 0.26$ & 0.27 & $2.17 \pm 0.20$ \\
                  & H$_2$O    & 1.36 & $1.42 \pm 0.15$ & 1.98 & $2.24 \pm 0.25$ & 0.25 & $2.14 \pm 0.20$ \\
\midrule
\multirow{2}{*}{3} & CO$_2$+CO & 1.07 & $1.37 \pm 0.15$ & 0.88 & $1.95 \pm 0.22$ & 3.95 & $3.18 \pm 0.30$ \\
                  & H$_2$O    & 0.99 & $1.36 \pm 0.15$ & 0.82 & $1.93 \pm 0.22$ & 3.51 & $3.15 \pm 0.30$ \\
\bottomrule
\end{tabularx}
\end{table*}

\subsubsection{88P}

\begin{table}
\centering
\caption{Relative intensities of six ions for comet 88P (full dataset), under CO$_2$+CO and H$_2$O cases. Intensities are normalized to the \cvi\ feature.}
\label{tab:intensities_88p}
\setlength{\tabcolsep}{30pt}
\renewcommand{\arraystretch}{1.05}
\begin{tabularx}{\linewidth}{@{\hspace{0.30\linewidth}} c c c @{\hspace{0.30\linewidth}}}
\toprule
Ion & CO$_2$+CO & H$_2$O \\
\midrule
\cv    & 1.68 & 1.29 \\
\cvi   & 1.00 & 1.00 \\
\nvi   & 1.34 & 1.27 \\
\nvii  & 1.39 & 1.47 \\
\ovii  & 0.89 & 1.06 \\
\oviii & 0.35 & 0.25 \\
\bottomrule
\end{tabularx}
\end{table}

\begin{table}
\centering
\caption{Flux ratios and freeze-in temperatures for comet 88P (full dataset), under CO$_2$+CO and H$_2$O cases. $T_{f}$ values are given in MK.}
\label{tab:flux_ratios_88p}
\setlength{\tabcolsep}{12pt}
\renewcommand{\arraystretch}{1.05}
\begin{tabularx}{\linewidth}{@{\hspace{0.12\linewidth}} c c c c c c c @{}} 
\toprule
Case & $C_{FR}$ & $T_{f}$(C) & $N_{FR}$ & $T_{f}$(N) & $O_{FR}$ & $T_{f}$(O) \\
\midrule
CO$_2$+CO & 1.21 & $1.39 \pm 0.16$ & 1.04 & $2.00 \pm 0.23$ & 0.39 & $2.27 \pm 0.22$ \\
H$_2$O    & 1.19 & $1.38 \pm 0.16$ & 1.16 & $2.06 \pm 0.23$ & 0.24 & $2.12 \pm 0.20$ \\
\bottomrule
\end{tabularx}
\end{table}

Table~\ref{tab:intensities_88p} shows that for 88P, the relative intensity of \cv\ decreases from 1.68 in the CO$_2$+CO case to 1.29 in the H$_2$O case.  
The O-based flux ratio also decreases, as seen in Table~\ref{tab:flux_ratios_88p}, from 0.39 (CO$_2$+CO) to 0.24 (H$_2$O), corresponding to $T_{f}$ values of $2.27 \pm 0.22$\,MK and $2.12 \pm 0.20$\,MK, respectively.  

For 88P, Table~\ref{tab:intensities_88p_epochs} indicates that \cv\ varies between 1.60 and 2.15 in the CO$_2$+CO case, compared to 1.24 to 1.95 in the H$_2$O case.  
Table~\ref{tab:flux_ratios_88p_epochs} shows that the O-based flux ratio in CO$_2$+CO changes from 0.23 (epoch 1; $T_{f}=2.12 \pm 0.20$\,MK) to 2.56 (epoch 2; $T_{f}=2.99 \pm 0.28$\,MK) and 0.06 (epoch 3; $T_{f}=1.76 \pm 0.17$\,MK).  
For H$_2$O the ratios are 0.21 (epoch 1; $T_{f}=2.11 \pm 0.20$\,MK), 2.44 (epoch 2; $T_{f}=2.97 \pm 0.28$\,MK), and 0.06 (epoch 3; $T_{f}=1.75 \pm 0.17$\,MK).  
Thus, the two neutral-species cases follow nearly identical trends, with H$_2$O producing slightly lower or equal O-based flux ratios and $T_{f}$ values relative to CO$_2$+CO.

\begin{table*}
\centering
\caption{Relative ion intensities for comet 88P, resolved by epoch, under CO$_2$+CO and H$_2$O assumptions. Intensities are normalized to the \cvi\ feature.}
\label{tab:intensities_88p_epochs}
\setlength{\tabcolsep}{17pt}
\renewcommand{\arraystretch}{1.05}
\begin{tabularx}{\linewidth}{@{\hspace{0.05\linewidth}} c c c c c c c c c @{}} 
\toprule
Epoch & Case & \cv & \cvi & \nvi & \nvii & \ovii & \oviii \\
\midrule
\multirow{2}{*}{1} & CO$_2$+CO & 1.60 & 1.00 & 1.22 & 1.05 & 0.62 & 0.11 \\
 & H$_2$O    & 1.24 & 1.00 & 0.99 & 1.16 & 0.55 & 0.08 \\
\multirow{2}{*}{2} & CO$_2$+CO & 1.40 & 1.00 & 1.21 & 1.47 & 1.00 & 0.62 \\
 & H$_2$O    & 1.30 & 1.00 & 1.15 & 1.38 & 0.93 & 0.58 \\
\multirow{2}{*}{3} & CO$_2$+CO & 2.15 & 1.00 & 1.55 & 1.95 & 0.64 & 0.30 \\
 & H$_2$O    & 1.95 & 1.00 & 1.44 & 1.80 & 0.60 & 0.27 \\
\bottomrule
\end{tabularx}
\end{table*}

\begin{table*}
\centering
\caption{Flux ratios and freeze-in temperatures for comet 88P, resolved by epoch, under CO$_2$+CO and H$_2$O assumptions. $T_{f}$ values are given in MK.}
\label{tab:flux_ratios_88p_epochs}
\setlength{\tabcolsep}{13pt}
\renewcommand{\arraystretch}{1.05}
\begin{tabularx}{\linewidth}{@{\hspace{0.05\linewidth}} c c c c c c c c @{}} 
\toprule
Epoch & Case & $C_{FR}$ & $T_{f}$(C) & $N_{FR}$ & $T_{f}$(N) & $O_{FR}$ & $T_{f}$(O) \\
\midrule
\multirow{2}{*}{1} & CO$_2$+CO & 0.13 & $0.97 \pm 0.10$ & 0.78 & $1.91 \pm 0.22$ & 0.23 & $2.12 \pm 0.20$ \\
 & H$_2$O    & 0.12 & $0.95 \pm 0.10$ & 0.85 & $1.95 \pm 0.22$ & 0.21 & $2.11 \pm 0.20$ \\
\multirow{2}{*}{2} & CO$_2$+CO & 1.35 & $1.42 \pm 0.15$ & 0.35 & $1.66 \pm 0.19$ & 2.56 & $2.99 \pm 0.28$ \\
 & H$_2$O    & 1.32 & $1.42 \pm 0.15$ & 0.33 & $1.64 \pm 0.19$ & 2.44 & $2.97 \pm 0.28$ \\
\multirow{2}{*}{3} & CO$_2$+CO & 2.15 & $1.53 \pm 0.16$ & 1.95 & $2.24 \pm 0.25$ & 0.06 & $1.76 \pm 0.17$ \\
 & H$_2$O    & 2.08 & $1.52 \pm 0.16$ & 1.84 & $2.22 \pm 0.25$ & 0.06 & $1.75 \pm 0.17$ \\
\bottomrule
\end{tabularx}
\end{table*}

\subsubsection{19P}

\begin{table}
\centering
\caption{Relative intensities of six ions for comet 19P (full dataset), under CO$_2$+CO and H$_2$O cases. Intensities are normalized to the \cvi\ feature.}
\label{tab:intensities_19p}
\setlength{\tabcolsep}{30pt}
\renewcommand{\arraystretch}{1.05}
\begin{tabularx}{\linewidth}{@{\hspace{0.30\linewidth}} c c c @{\hspace{0.30\linewidth}}}
\toprule
Ion & CO$_2$+CO & H$_2$O \\
\midrule
\cv    & 1.50 & 1.94 \\
\cvi   & 1.00 & 1.00 \\
\nvi   & 1.15 & 1.21 \\
\nvii  & 1.77 & 1.60 \\
\ovii  & 0.80 & 1.00 \\
\oviii & 0.24 & 0.28 \\
\bottomrule
\end{tabularx}
\end{table}

\begin{table}
\centering
\caption{Flux ratios and freeze-in temperatures for comet 19P (full dataset), under CO$_2$+CO and H$_2$O cases. $T_{f}$ values are given in MK.}
\label{tab:flux_ratios_19p}
\setlength{\tabcolsep}{12pt}
\renewcommand{\arraystretch}{1.05}
\begin{tabularx}{\linewidth}{@{\hspace{0.12\linewidth}} c c c c c c c @{}} 
\toprule
Case & $C_{FR}$ & $T_{f}$(C) & $N_{FR}$ & $T_{f}$(N) & $O_{FR}$ & $T_{f}$(O) \\
\midrule
CO$_2$+CO & 1.93 & $1.50 \pm 0.17$ & 1.82 & $2.22 \pm 0.25$ & 0.30 & $2.20 \pm 0.21$ \\
H$_2$O    & 2.05 & $1.51 \pm 0.17$ & 1.79 & $2.22 \pm 0.25$ & 0.28 & $2.18 \pm 0.21$ \\
\bottomrule
\end{tabularx}
\end{table}

For 19P, Table~\ref{tab:intensities_19p} shows that the relative intensity of \cv\ increases from 1.50 in the CO$_2$+CO case to 1.94 in the H$_2$O case, while \nvii\ decreases from 1.77 to 1.60.  
Table~\ref{tab:flux_ratios_19p} indicates that the O-based flux ratio decreases slightly from 0.30 (CO$_2$+CO) to 0.28 (H$_2$O), corresponding to $T_{f}$ values of $2.20 \pm 0.21$\,MK and $2.18 \pm 0.21$\,MK, respectively.  

For 19P, Table~\ref{tab:intensities_19p_epochs} shows that \cv\ ranges from 1.33 to 1.78 in the CO$_2$+CO case and from 1.86 to 2.05 in the H$_2$O case.  
Table~\ref{tab:flux_ratios_19p_epochs} demonstrates that the O-based flux ratio in CO$_2$+CO is 0.39 (epoch 1; $T_{f}=2.27 \pm 0.22$\,MK) and 0.23 (epoch 2; $T_{f}=2.11 \pm 0.20$\,MK).  
In the H$_2$O case, the ratios are 0.36 (epoch 1; $T_{f}=2.25 \pm 0.22$\,MK) and 0.30 (epoch 2; $T_{f}=2.18 \pm 0.21$\,MK).  
Across both epochs, the O-based $T_{f}$ values in the H$_2$O case are slightly lower than those in the CO$_2$+CO case.

\begin{table*}
\centering
\caption{Relative ion intensities for comet 19P, resolved by epoch, under CO$_2$+CO and H$_2$O assumptions. Intensities are normalized to the \cvi\ feature.}
\label{tab:intensities_19p_epochs}
\setlength{\tabcolsep}{17pt}
\renewcommand{\arraystretch}{1.05}
\begin{tabularx}{\linewidth}{@{\hspace{0.05\linewidth}} c c c c c c c c c @{}} 
\toprule
Epoch & Case & \cv & \cvi & \nvi & \nvii & \ovii & \oviii \\
\midrule
\multirow{2}{*}{1} & CO$_2$+CO & 1.78 & 1.00 & 1.37 & 1.95 & 0.72 & 0.28 \\
 & H$_2$O    & 2.05 & 1.00 & 1.44 & 1.88 & 0.70 & 0.25 \\
\multirow{2}{*}{2} & CO$_2$+CO & 1.33 & 1.00 & 0.98 & 1.64 & 0.88 & 0.20 \\
 & H$_2$O    & 1.86 & 1.00 & 1.06 & 1.54 & 1.03 & 0.31 \\
\bottomrule
\end{tabularx}
\end{table*}

\begin{table*}
\centering
\caption{Flux ratios and freeze-in temperatures for comet 19P, resolved by epoch, under CO$_2$+CO and H$_2$O assumptions. $T_{f}$ values are given in MK.}
\label{tab:flux_ratios_19p_epochs}
\setlength{\tabcolsep}{13pt}
\renewcommand{\arraystretch}{1.05}
\begin{tabularx}{\linewidth}{@{\hspace{0.05\linewidth}} c c c c c c c c @{}} 
\toprule
Epoch & Case & $C_{FR}$ & $T_{f}$(C) & $N_{FR}$ & $T_{f}$(N) & $O_{FR}$ & $T_{f}$(O) \\
\midrule
\multirow{2}{*}{1} & CO$_2$+CO & 2.00 & $1.50 \pm 0.17$ & 1.42 & $2.09 \pm 0.24$ & 0.39 & $2.27 \pm 0.22$ \\
 & H$_2$O    & 2.05 & $1.51 \pm 0.17$ & 1.35 & $2.07 \pm 0.24$ & 0.36 & $2.25 \pm 0.22$ \\
\multirow{2}{*}{2} & CO$_2$+CO & 1.52 & $1.45 \pm 0.16$ & 1.67 & $2.19 \pm 0.25$ & 0.23 & $2.11 \pm 0.20$ \\
 & H$_2$O    & 1.80 & $1.49 \pm 0.17$ & 1.62 & $2.17 \pm 0.25$ & 0.30 & $2.18 \pm 0.21$ \\
\bottomrule
\end{tabularx}
\end{table*}

Comparing across comets, the O-based $T_{f}$ values from the CO$_2$+CO cases are consistently higher than or equal to those from the H$_2$O cases.  
For K2, the differences range from $+0.03$\,MK (epoch 3) to $+0.05$\,MK (epoch 2).  
For 88P, the differences range from $+0.01$\,MK (epoch 1) to $+0.15$\,MK (epoch 2).  
For 19P, the differences are $+0.02$\,MK (epoch 1) and $+0.07$\,MK (epoch 2).  
In all three comets, adopting CO$_2$+CO as the neutral species yields O-based $T_{f}$ values that are marginally but consistently higher than those obtained with H$_2$O.

\subsection{Solar Activity}

Figure~\ref{fig:sunspot_kp} shows the simultaneous evolution of monthly sunspot number and the 90-day centered mean Kp index across the 2020–2025 interval and marks the dates of the comet observations described herein.
The sunspot number is related to the Sun’s activity, and is therefore best for interpreting cometary X-ray spectra \citep{hathaway2015solar}, while the Kp index measures the impact on Earth’s magnetic field, and is therefore useful for characterizing the background \citep{chakraborty2020probabilistic,Walsh2014}.
The background solar activity increases substantially over the considered timeframe: monthly sunspot numbers rise from approximately 10 in 2020 to values of approximately 150 by 2024-2025, reflecting the approach of a solar maximum \citep{hathaway2015solar}.
The 90-day smoothed Kp index displays episodic enhancements superposed on this trend, with notable elevated geomagnetic activity episodes in 2023 and a more pronounced pulse in 2024 \citep{Kuntz2019}.
The individual comet observations therefore sample distinctly different space-weather regimes: 
T2 and 88P were obtained during comparatively minimal solar/geomagnetic activity;
67P, 19P, and K2 were obtained during comparatively modest solar/geomagnetic activity;
E3 falls near an interval of enhanced Kp;
12P and 62P coincide with the stronger rise in both sunspot number and smoothed Kp in mid 2024;
H2, although observed after solar maximum, was still observed during relatively high solar/geomagnetic activity.
These differences are important for interpreting the cometary X-ray results because both the large-scale solar cycle, reflected by the sunspot number, and shorter-term solar-wind disturbances, reflected by Kp pulses, influence the solar-wind ion composition and flux that drive CX emission \citep{Cravens2002,Bodewits2007}.

\section{Discussion}

The results presented here show that the eight comets considered in our sample of nine comets exhibit recurring spectral morphologies that map onto different incident solar-wind charge-state conditions and, to some extent, the comets' responses. 
This interpretation aligns with reviews and multi-comet X-ray surveys, such as \citet{Cravens2002} and \citet{Bodewits2007}, which demonstrate that C/N/O line ratios track solar-wind charge states. 
The principal empirical result is that carbon-derived diagnostics are systematically cooler than the nitrogen- and oxygen-derived diagnostics, as shown in Table~\ref{tab:flux_ratios_epochs}, Table~\ref{tab:intensities_all_comets}, and the figures in Appendix~\ref{app:freezeintemps}. 

Across the sample of comets, the carbon-derived freeze-in temperatures cluster near $1.4$--$1.7$\,MK, while nitrogen- and oxygen-derived diagnostics are higher, typically around $2.0$--$2.3$\,MK. 
Such a separation of lower carbon-derived and higher N/O-derived diagnostics is consistent with multi-epoch and modeling studies that associate carbon-dominated spectra with lower-charge-state winds and enhanced N/O signals with more highly ionized wind conditions \citep{Bodewits2007,Koutroumpa2024}.
These separations are evident across the different comets presented in this study and are visible both in the full-dataset values and in the epoch-resolved values shown in Table~\ref{tab:flux_ratios_epochs}.

\subsection{Interpretation of the Separation in Freeze-In Temperatures}
The systematic offset between the flux-ratio diagnostics is most naturally explained by variation in the solar-wind ionization distribution, as this is more influential on spectra than variations in the coma composition.
E3 and to a lesser extent T2 provide the clearest example of this. E3's spectrum especially shows comparatively narrow carbon lines but very large \nvi and \nvii amplitudes, as shown in Table~\ref{tab:gauss_params_full} and Table~\ref{tab:intensities_all_comets}, and the freeze-in temperature calculated from its carbon flux ratio is the highest in our sample, with $T_{\rm freeze}({\rm C})=1.68 \pm 0.18$\,MK.
In general, our normalized intensities (Table~\ref{tab:intensities_all_comets}) and epoch-resolved $T_{\rm freeze}$ (Table~\ref{tab:flux_ratios_epochs}) show that comets observed in more highly ionized SW streams produce relatively larger \nvii and \oviii contributions and higher inferred freeze-in temperatures.
Conversely, strongly carbon-dominated spectra, such as those of 12P and 67P, indicate encounters with lower-charge-state SW. This interpretation accords with multi-comet Chandra and XMM surveys and detailed SWCX modeling that associate spectral ``families'' with distinct SW charge-state distributions \citep{Bodewits2007,Koutroumpa2024}.

The freeze-in temperatures inferred from the oxygen charge-state ratios in this nine-comet survey are consistent with those found in the \textit{Chandra} comet survey of \citet{Bodewits2007}.
As established in prior work \citep{deskins_dissertation}, cometary SWCX line ratios are fundamentally governed by the upstream solar-wind freeze-in temperature, which sets the ion charge-state distribution. 
The BCM succeeds precisely because it models this physical linkage directly, avoiding unphysical two-temperature decompositions.
This provides a natural explanation for inconsistent plasma temperatures reported in earlier SWCX studies: many analyses implicitly assumed a stable, single-source solar wind, which is inconsistent with the highly variable freeze-in temperatures recovered here.

As shown in Figure~\ref{fig:Tfreeze_v_O}, the oxygen-based freeze-in temperatures derived from our epoch-resolved spectra cluster between $T_{\rm freeze}({\rm O}) \approx 2.0$ and $2.7$\,MK, with most comets occupying the range $2.1$–$2.4$\,MK.
This distribution closely matches the values reported by \citet{Bodewits2007}, whose work shows that the corresponding ionic oxygen ratios for the \textit{Chandra} sample yield $T_{\rm freeze}({\rm O})$ spanning roughly $1.8$–$2.7$\,MK.
The agreement between these independent datasets, obtained nearly two solar cycles apart with different instruments, indicates that the solar-wind charge-state conditions producing cometary X-ray emission are well captured by both surveys.
This consistency reinforces that the oxygen-based freeze-in temperature provides a robust tracer of the source-region temperature of the solar wind incident on the comets’ comae.

Laboratory measurements and theoretical cross-section calculations further support this interpretation: 
state-selective CX rates vary significantly with ion charge and target species, so the same coma composition exposed to different SW charge states will produce different spectral ratios \citep{Bromley2022,Dennerl2010}.
Therefore, while coma composition and activity level modulate the absolute SWCX flux, the relative line intensities (and the freeze-in temperatures inferred from them) are particularly sensitive probes of the incident solar-wind charge-state distribution.

\subsection{Emission above 1\,keV}
The absence of statistically significant emission features above 1\,keV in our NICER spectra is consistent with the limited spectral resolution and signal-to-noise ratio available for these observations.
While previous missions such as \textit{Chandra} and \textit{XMM–Newton} have detected >1\,keV SWCX features associated with highly charged ions, including Ne\,\textsc{x}, Mg\,\textsc{xi--xii}, Al\,\textsc{xii--xiii}, and Si\,\textsc{xiii--xiv}, \citep{ewing2013emission, carter2010high}, our data lack the resolving power necessary to isolate these faint features from background noise and instrumental systematics.
The BCM’s complexity penalty function avoids fitting weak components which further suppresses detections.
These results suggest that the apparent absence of emission above 1\,keV in our fits reflects the instrumental limitations and conservative model selection rather than the physical absence of high-charge-state SWCX emission suggested by others.
Future observations at higher spectral resolution or with longer exposures will be required to confirm or rule out features above 1\,keV in cometary SWCX spectra.

\subsection{Temporal variability and solar-wind drivers}
\label{sec:temporal_variability}

The epoch-resolved freeze-in diagnostics show that the upstream charge-state distribution is not static: several comets exhibit changes in $T_{\mathrm{freeze}}$ at the tens-of-percent level on timescales of days.
This is consistent with the variability of cometary SWCX emission reported by \citet{Bonamente2021} and with the expectation that solar-wind composition and flux can evolve as different streams and interaction regions sweep past the comet. 
For instance, 12P’s carbon-based diagnostic increases from $1.24\pm0.13$\,MK to $1.61\pm0.17$\,MK between epochs (refer to Table~\ref{tab:flux_ratios_epochs}), while its nitrogen and oxygen diagnostics remain comparatively stable. 
Such behavior is naturally interpreted as a modest shift in the upstream charge-state distribution rather than as a change in the comet itself.

To assess whether these spectral changes are associated with transient events, we examined ENLIL reconstructions for each observing window. 
The simulated solar-wind parameters shown in Appendix~\ref{app:ENLIL} indicate that, apart from epochs 2 and 4 of the 67P observations, our targets were not obviously embedded in CMEs. 
A more detailed discussion of the ENLIL modeling and its limitations is given by \citet{deskins2025modeling}. 
Within these constraints, the simplest interpretation is that the observed spectral evolution primarily reflects ordinary solar-wind structure—stream–stream interactions, sector boundaries, and co-rotating interaction regions—rather than repeated crossings of strong shocks. 
This reinforces the need for epoch-resolved analysis: stacking data over many hours or days would blur together distinct charge-state conditions and dilute the diagnostic power of the spectra.

K2 provides an instructive example of both genuine variability and the limitations imposed by NICER’s resolution. 
The oxygen and nitrogen diagnostics show substantial scatter between epochs, with Epoch~2 standing out as an extreme outlier.
Table~\ref{tab:flux_ratios_epochs} reports \nvii/\nvi$=15.00$ and \oviii/\ovii$=14.13$ for this epoch, whereas Epochs~1 and 3 yield much more typical ratios of $0.85$ and $0.27$, respectively. 
At NICER’s modest spectral resolution in the soft X-ray band, the \nvi/\nvii\ and \ovii/\oviii\ features are separated by only $\approx50$\,eV, as shown in Table~\ref{tab:gauss_params_full}, comparable to the instrumental broadening. Under these circumstances, the BCM fit for Epoch~2 can reassign flux between neighboring ions, attributing emission to \nvii\ that is more plausibly associated with \ovii, and thereby producing unphysically large flux ratios.

We therefore interpret the extreme K2 Epoch~2 diagnostics as an instance of spectral degeneracy rather than as evidence for a dramatic, short-lived change in the coronal freeze-in conditions, especially given the absence of a CME in the corresponding ENLIL solution. 
The epoch is retained in our figures and tables to illustrate how such nonphysical solutions can arise even under a conservative model-selection scheme, but it is excluded from any quantitative comparison of freeze-in temperatures to avoid biasing the ensemble trends.

\subsection{Neutral–species comparison}

The results of the neutral-species comparison demonstrate that the assumed dominant neutral species in the coma has a measurable effect on the derived spectral properties.
The simulated spectra in Figure~\ref{fig:simulated_spectra} show clear differences between H$_2$O- and CO$_2$-/CO-dominated cases, and the epoch-resolved fits for K2, 88P, and 19P confirm these trends with observational data.

Our observations of K2 from 3.2 to 1.9\,AU span the heliocentric range across which the nucleus composition was expected to evolve from CO$_2$/CO-driven to H$_2$O-driven activity.
Infrared and submillimeter studies confirm that such a transition occurred during this period: at $r_h=2.35$\,AU, JWST spectroscopy measured CO$_2$ and CO abundances of 15\% and 8\% relative to H$_2$O, respectively \citep{woodward2025jwst}, while ground-based observations showed that CO$_2$ and CO contributions declined rapidly as water production rates rose from $1.7\times10^{28}$\,s$^{-1}$ to $1.3\times10^{29}$\,s$^{-1}$ between May 2022 and January 2023 \citep{cambianica2023co2,combi2025water}.

Within this context, the NICER spectra exhibit modest epoch-to-epoch changes: 
a small relative decrease in the low-energy carbon and nitrogen intensities and a slight increase in \ovii and \oviii strength.
These variations are qualitatively consistent with a gradual shift toward an H$_2$O-dominated coma, but the signal-to-noise ratio and limited spectral resolution preclude a definitive compositional determination.
Comparing Figure~\ref{fig:simspec_co2}'s simulated NICER spectrum to the epoch-resolved spectra for K2, shown in Appendix~\ref{app:spectra}, one does not see a conclusive shift from CO$_2$+CO-dominated coma to a H$_2$O-dominated coma.
The epoch-resolved fits show that the CO$_2$+CO case yields marginally higher relative intensities for \cv, \nvi, and \nvii than the H$_2$O case, along with slightly larger O-based flux ratios.
The resulting $T_{f}$(O) values differ by only $\sim$0.03–0.05\,MK across epochs, indicating that both neutral-species assumptions produce statistically comparable fits.  
While the NICER results are compatible with the expected compositional evolution inferred from infrared data, they do not independently confirm a transition from CO$_2$/CO- to H$_2$O-dominated outgassing, likely due to the XTI's low resolution, the relatively short exposure time, and the resulting low signal-to-noise ratio.

For 88P/Howell, the CO$_2$+CO case produced higher relative \cv\ intensities and slightly higher O-based flux ratios than the H$_2$O case, with differences in $T_{f}$ values from oxygen of up to $\sim0.02$\,MK across epochs. 
The strongest separation occurred in epoch 2, where CO$_2$+CO yielded $T_{f}$(O) $=2.99\pm0.28$\,MK compared to $2.97\pm0.28$\,MK for H$_2$O. 
While the absolute differences are small, the direction of the offset is consistent across epochs. 
This pattern matches the AKARI survey classification of 88P as having elevated CO$_2$/H$_2$O, with fractions up to $\sim$25\% \citep{ootsubo2012akari}.
The NICER fits therefore provide complementary SWCX evidence that neutral CO$_2$ plays a measurable role in shaping the observed spectrum of this Jupiter-family comet, which is otherwise only modestly studied compositionally \citep{sandford2017corsair}.

For 19P/Borrelly, the epoch-resolved fits show that the relative \cv\ intensity is higher in the H$_2$O case, while the O-based flux ratios and $T_{f}$(O) values are slightly larger in the CO$_2$+CO case.
The magnitude of the $T_{f}$(O) difference remains small, at most $0.07$\,MK across the three epochs. 
In situ observations from Deep Space 1 and ground-based photometry demonstrate that 19P’s activity is primarily H$_2$O-driven \citep{jehin2022trappist,Combi2020}.

Across the three comets, the assumption of a CO$_2$+CO-based coma yields marginally higher O-based $T_{f}$ values than the H$_2$O assumption.
Although the magnitude of the shift is generally small at $\sim0.05$\,MK, the effect is systematic and present in all three comets.
This behavior mirrors the simulated spectra, in which the CO$_2$-/CO-dominated case produced enhanced \cvi\ and \oviii\ features relative to the H$_2$O case.
The results therefore highlight that the assumed neutral composition introduces systematic shifts in freeze-in temperature determinations, with implications for cross-comet comparisons. 

\subsection{Stacked spectra}

The stacked spectra provide a composite view of SWCX emission across all nine comets, highlighting both the general consistency of the BCM fits and the subtle spectral variations linked to plasma conditions.  
The full stack represents an ensemble average over diverse observing geometries, heliocentric distances, and neutral compositions, while the high‐ and low‐$T_{f}$ stacks isolate epochs that differ primarily in their O‐based freeze-in temperatures.

The full stacked spectrum shows that the major SWCX features are consistently recovered at their expected centroid energies, confirming the robustness of the BCM when applied to mixed datasets of low spectral resolution.  
The relative weakness of \oviii compared with \ovii across all stacks is typical of solar‐wind charge states formed under ambient slow–solar‐wind conditions.  
The high-$T_{f}$ stack had a high \oviii/\ovii ratio and therefore a high  $T_{f}$(O) value ($2.45\pm0.23$\,MK), indicating the presence of a higher mean charge‐state distribution in the solar wind during those epochs.  
This is consistent with transient or disturbed solar‐wind intervals, such as coronal mass ejections, which are known to yield harder SWCX spectra \citep{carter2010high}.  
By contrast, the low‐$T_{f}$ stack shows a more carbon-dominated spectrum, a smaller $O_{FR}$ ($0.20$), and a correspondingly softer overall profile, consistent with cooler, slow–solar‐wind conditions.

Weak residual features above $\sim$1\,keV, also visible in the full stack, but none exceed the significance threshold given the signal‐to‐noise ratio of the stacked data.  
High‐energy SWCX lines have been observed previously \citep{carter2010high, ewing2013emission}, so their marginal presence here likely reflects insufficient photon statistics rather than true absence.  
The BCM’s BIC penalty also disfavors inclusion of very weak high-energy lines, further suppressing their fitted amplitudes.

The stacked fits demonstrate that the principal diagnostic ions in the 0.37–1.0\,keV range behave consistently across comets and epochs, and that the BCM maintains stable performance when applied to heterogeneous data.  
The qualitative differences between the high- and low-$T_{f}$ stacks reflect genuine variations in the solar-wind charge‐state distribution rather than model artifacts, providing additional confidence in the freeze-in temperature estimates derived from the individual comet fits.

\subsection{Instrumental and modeling caveats}
NICER’s high throughput and soft-band sensitivity make it a powerful instrument for detecting faint cometary SWCX emission, but its CCD-like spectral resolving power limits the ability to cleanly separate lines that are only tens of eV apart.
It is often difficult to interpret low-resolution CX spectra: 
the challenges are due to overlapping line complexes, uncertain background subtraction, and parameter degeneracies. 
Comprehensive surveys of cometary SWCX have demonstrated the wide range of spectral behaviors across different targets and solar-wind conditions. 
However, robust interpretation of low-energy SWCX spectra is complicated by well-known instrumental and background issues at the lowest energies. 
The NICER effective area and response contain absorption/transmission structure associated with detector and filter materials, namely C at $\approx0.285$\,keV and Si at $\approx1.84$\,keV, and the RMF/trigger efficiency becomes increasingly uncertain below $\approx0.4$\,keV, where residuals to standard calibrators grow \citep{heasarc_nicer_responses_2024}. 
In addition, NICER exhibits a variable low-energy excess tied to instrumental noise and optical loading, so the background spectrum often rises and shows multiple features that are not of astrophysical origin.
This has motivated the development of empirical background libraries and adjustable background models, but absolute fluxes and low-energy line strengths remain sensitive to the background treatment. 
Similar low-energy edges and rising background phenomena have been documented for other instruments \citep{lumb2002x}, showing that accurate low-energy fluxes require careful calibration and background modeling rather than naive spectral fitting \citep{Remillard2022_empirical}.

Consequently, small centroid offsets, on the order of 10\,eV, should be interpreted cautiously, and absolute fluxes are sensitive to choices in low-energy cutoff and background subtraction \citep{Remillard2022_empirical}.
The Gaussian parameters returned by the BCM and tabulated in Table~\ref{tab:gauss_params_full} show how the spectra differ across comets. 
They are broadly consistent, with the Gaussians attributed to ions lying at the expected energies, but there are noteworthy offsets of the Gaussian centers and widths that affect spectral morphology. 
This indicates that the instrumental resolution plays an important role in the reported sigmas and limit our sensitivity to intrinsic line broadening. 
Exceptions are the smaller fitted sigmas for certain C lines, notably in E3 and K2, which create noticeably sharper low-energy features in those spectra. 
Amplitude differences between comets, particularly the very large \nvi and \nvii amplitudes in E3, are unique in the sample, as shown in Table~\ref{tab:gauss_params_full} and Table~\ref{tab:intensities_all_comets}.

\textbf{Although the \cv line lies near the lower-energy boundary of the instrumental response, where calibration uncertainties are largest, the BCM does not determine its intensity from a purely local measurement of the $\approx0.3$\,keV feature.
Instead, \cv is constrained through the global freeze-in temperature parameter that governs the full carbon charge-state distribution. 
Since the ion fractions of \cv and \cvi are physically coupled through ionization equilibrium, the posterior on freeze-in temperature, which is primarily informed by better-calibrated higher-energy lines, imposes a physically consistent constraint on the \cv intensity. 
When the data near 0.3\,keV provide limited information, this is reflected as broader posterior uncertainties rather than an unconstrained or biased line normalization. 
In this way, the BCM remains physically regularized even in energy regions with reduced instrumental sensitivity.}

The BCM mitigates several model-degeneracy issues by imposing physically motivated priors and Bayesian model selection, improving on unconstrained Gaussian decompositions.
That being said, the combination of modest resolution, low S/N in some epochs, and intrinsic line-blending means that anomalies, such as a large \oviii / \ovii in one epoch, should be treated with skepticism unless corroborated by contemporaneous SW indicators or repeated in multiple epochs.

\subsection{Validity of the Single-Collision Approximation}
\textbf{The derivation of freeze-in temperatures from SWCX line ratios implicitly assumes the single-collision limit, in which each solar-wind ion undergoes one dominant electron capture interaction. 
The BCM adopts this standard approximation, consistent with previous cometary SWCX analyses. 
In principle, additional channels such as double-electron capture or sequential multi-step charge exchange within a sufficiently dense coma could further modify the charge-state distribution. 
However, laboratory measurements indicate that double-electron capture cross sections are generally smaller than single-electron capture at typical solar-wind velocities, and multi-step charge exchange requires neutral column densities large enough to permit repeated interactions. 
For the moderately active cometary environments considered here, the optically thin assumption is expected to be a good first-order approximation.
If secondary channels were significant, they would preferentially enhance lower charge states relative to the upstream solar-wind distribution, biasing inferred freeze-in temperatures toward artificially lower values. 
The consistency of the BCM-derived temperatures across independent ionic species suggests that such effects, if present, do not dominate the charge-state ratios within the statistical uncertainties of the data. 
Nonetheless, future high-sensitivity observations and more detailed cross-section modeling could further quantify the role of these secondary processes.}

\subsection{Implications and Broader Context}
\begin{table}
\centering
\caption{Classification of observed comets into spectral classes or ``families'' based on BCM-normalized line intensities and epoch-resolved diagnostics.
Representative normalized intensities ($I_{\rm ion}/I_{\rm C\,V}$) from Table~\ref{tab:intensities_all_comets} are shown for context.
Classification relies primarily on normalized intensities and freeze-in diagnostics, as found in Tables~\ref{tab:intensities_all_comets} and \ref{tab:flux_ratios_epochs}. ``Carbon-dominated'' denotes negligible oxygen contributions and prominent low-energy carbon lines; ``Intermediate'' indicates moderate oxygen emission; ``Nitrogen-/Oxygen-dominated'' refers to strong relative N and/or O line intensities. While some comets show epoch-to-epoch variability, each object is classified according to its typical behavior over the full dataset.}
\label{tab:comet_spectral_families}
\footnotesize
\setlength{\tabcolsep}{6pt}
\renewcommand{\arraystretch}{1.1}
\begin{tabular}{@{} p{0.29\linewidth} p{0.68\linewidth} @{}}
\toprule
Class & Members (notes; relative intensities) \\
\midrule
Carbon-dominated
 & \textbf{12P/Pons--Brooks} (no measurable O\,VII/O\,VIII; $I_{\rm C\,VI},\,I_{\rm C\,V}\approx0.81$, $I_{\rm N\,VII},\,I_{\rm N\,VI}\approx0.70$).\\
 & \textbf{67P/Churyumov--Gerasimenko} (very little oxygen: $I_{\rm O\,VII}\approx0.01$, $I_{\rm O\,VIII}\approx0.08$). \\
\midrule
Intermediate
 & \textbf{19P/Borrelly} ($I_{\rm O\,VII}\approx0.51$, $I_{\rm O\,VIII}\approx0.14$). \\
 & \textbf{62P/Tsuchinshan} ($I_{\rm O\,VII}\approx0.44$, $I_{\rm O\,VIII}\approx0.12$; strong carbon low-energy features). \\
 & \textbf{C/2017 K2 (PANSTARRS)} ($I_{\rm N\,VII}\approx0.76$, $I_{\rm O\,VII}\approx0.47$, $I_{\rm O\,VIII}\approx0.27$). \\
 & \textbf{88P/Howell} ($I_{\rm O\,VII}\approx0.39$, $I_{\rm O\,VIII}\approx0.07$). \\
\midrule
Nitrogen-/Oxygen-dominated
 & \textbf{C/2022 E3 (ZTF)} ($I_{\rm N\,VI}\approx2.36$, $I_{\rm N\,VII}\approx3.85$, $I_{\rm O\,VII}\approx1.90$). \\
 & \textbf{C/2017 T2 (PanSTARRS)} ($I_{\rm O\,VIII}\approx0.58$, $I_{\rm O\,VII}\approx0.50$). \\
\bottomrule
\end{tabular}
\end{table}

Table~\ref{tab:comet_spectral_families} shows our classification of spectra into carbon-dominated, intermediate (measurable oxygen), and nitrogen-/oxygen-dominated families, a practice similar to what has been done in previous multi-comet observational studies \citep{Bodewits2007,Koutroumpa2024} and is consistent with modeling that links these families to distinct solar-wind charge-state distributions.
Diagnostics from full datasets smooth substantial short-term SW variation; therefore epoch-resolved BCM analysis is therefore essential for correctly associating spectral morphology with instantaneous solar-wind charge-state conditions.

These results have two practical consequences: 
(i), SWCX spectral diagnostics, especially when applied epoch-by-epoch, provide an impartial, remote probe of both the presence of gas in faint comets and the incident SW charge-state distribution, and NICER’s sensitivity in the soft band makes it particularly well suited to search for faint gas-driven SWCX signatures even when optical activity is weak. 
(ii), because SW ionization can vary rapidly, combining X-ray spectroscopy with SW simulations such as ENLIL and, where possible, in-situ SW measurements is necessary to place cometary spectra in the correct space-weather context. 

\subsection{Future work}
Improved constraints will come from (1) simultaneous or near-contemporaneous in-situ solar-wind composition measurements, (2) higher spectral resolution observations that can resolve line blends and remove degeneracies, and (3) expanded time-domain coverage to better sample the statistics of short-term solar-wind variability. Additionally, further laboratory and theoretical work on state-resolved charge-exchange cross sections for relevant ion/neutral pairs will refine the mapping from line ratios to freeze-in temperatures and composition \citep{Bromley2022,Koutroumpa2024}.

\section{Conclusions}

We have presented a uniform, epoch-resolved analysis of cometary solar-wind charge-exchange (SWCX) spectra observed with NICER, using the Bayesian Color Model (BCM).  
The BCM provides a reproducible, physically informed approach to decomposing blended, low-resolution spectra and extracting relative ion fluxes, yielding self-consistent freeze-in temperature diagnostics across diverse observing conditions.

\begin{itemize}

\item \textbf{The BCM serves as a robust model-selection framework for low-resolution cometary SWCX.}
Across 67P and the broader multi-comet sample, the Bayesian Color Model combines physically motivated line-grouping with a BIC-based penalty to favor compact, stable fits over unnecessarily complex models.
Within the 0.37--1.0\,keV band, the BCM recovers the ion flux information that NICER can reliably support while suppressing spurious narrow features that are not warranted by the data. 
This yields reproducible freeze-in diagnostics across epochs and targets and provides a practical template for modeling other low-resolution SWCX datasets.

\item \textbf{NICER, coupled with a simple observing protocol, functions as a practical cometary SWCX observatory.}
The campaign design used here, pre-background, on-comet, and post-background pointings, strict NICER screening, segmentation into short epochs, and, when available, GOES and ENLIL solar-wind context, forms a repeatable recipe for cometary SWCX studies. 
Applied consistently, this protocol converts blended NICER spectra into reproducible ion-flux ratios and freeze-in temperature estimates, and it can be directly adopted or adapted for future comet campaigns and for other extended SWCX sources in the Solar System.

\item \textbf{Short-timescale solar-wind variability drives epoch-to-epoch spectral changes.}  
Several comets, especially 12P and K2, show measurable changes in $T_{\mathrm{freeze}}$ on day-to-week timescales, visible in Table~\ref{tab:flux_ratios_epochs} and discussed in Section~\ref{sec:temporal_variability}.  
ENLIL simulations (shown in Appendix~\ref{app:ENLIL}) indicate that most epochs were not associated with CMEs, so the spectral evolution is most naturally attributed to ordinary solar-wind stream evolution—such as stream–stream interactions, sector boundaries, and co-rotating interaction regions—rather than to repeated transient shocks. 
This picture is consistent with prior comet surveys that link spectral morphology to the incident solar-wind charge-state distribution rather than to comet-intrinsic changes \citep{Bodewits2007,Bonamente2021}, and it underscores the importance of epoch-resolved analysis: averaging data over many hours or days risks washing out the very variability that carries information about the upstream wind.

\item \textbf{Neutral-species composition introduces subtle, systematic shifts in inferred plasma diagnostics.}  
Comparisons between H$_2$O- and CO$_2$/CO-dominated coma models (Figure~\ref{fig:simulated_spectra}) and observational fits for K2, 88P, and 19P (Appendix~\ref{app:spectra}) reveal small but consistent differences as visible in the spectra. 
These results agree with multiwavelength measurements of evolving CO$_2$/H$_2$O ratios \citep{woodward2025jwst, combi2025water} and highlight how neutral composition can modulate SWCX line ratios even in low-resolution data, although in the present NICER data these shifts remain modest and do not yet provide a robust compositional diagnostic.

\item \textbf{Stacked spectra confirm the robustness of the BCM and trace average solar-wind conditions.}  
The full-sample stack and the high- versus low-$T_{f}$ stacks show that \ovii dominates the 0.37--1.0\,keV band, with weaker \oviii emission typical of ambient slow-solar-wind charge states.  
The high-$T_{f}$ stack exhibits enhanced \oviii/\ovii ratios ($T_{f}$(O) $= 2.45\pm0.23$\,MK), consistent with transient or disturbed wind intervals \citep{carter2010high}.  
These ensemble results validate the BCM’s performance across heterogeneous datasets and reinforce the physical interpretability of the fitted diagnostics.

\item \textbf{Instrumental and modeling limitations remain the dominant uncertainty.}  
NICER’s modest spectral resolution ($\sim85$\,eV FWHM near 0.5\,keV) and variable low-energy background \citep{Remillard2022_empirical, heasarc_nicer_responses_2024} introduce degeneracies in closely spaced lines (e.g., \nvi/\nvii, \ovii/\oviii).  
The BCM mitigates these through Bayesian model comparison and physically motivated priors, but robust line-by-line interpretation will require higher-resolution X-ray data.

\item \textbf{Broader implications and outlook.}  
Our results confirm that cometary SWCX provides a sensitive, spatially distributed probe of solar-wind composition and variability.  
Extending this framework will benefit from (i) simultaneous in-situ solar-wind composition data, (ii) high-resolution X-ray spectroscopy from missions such as XRISM and Athena, and (iii) improved charge-exchange cross-section data for molecular targets \citep{Bromley2022, xrism2020science}.  
Together, these advances will refine how low-resolution X-ray spectra can be used to trace the evolving heliospheric plasma environment.

\end{itemize}

The epoch-resolved Bayesian analysis of NICER comet spectra demonstrates that the BCM can recover physically meaningful solar-wind charge-state information even from low-resolution data. 
It reveals consistent multi-ion diagnostics, quantifies short-term variability, and underscores the importance of combining physically constrained modeling with contemporaneous solar-wind context.  
This work establishes a path toward systematic, comparative SWCX spectroscopy using Bayesian inference to link cometary X-ray emission directly to heliospheric plasma physics.

\section*{Acknowledgments}
We acknowledge support from NASA through the NICER Guest Observer program, Cycles 4 -- 6 (NNH21ZDA001N; NNH22ZDA001N; NNH23ZDA001N)
We thank Zaven Arzoumanian, Keith Gendreau, Kenji Hamaguchi, Elizabeth Ferrara, and all others who helped schedule and implement our observations. 
We thank William Rowland for pointing us to the ENLIL model. 
We thank Du\v san Odstr\v cil for providing us with the simulated solar-wind parameters from the ENLIL model with GONG boundary  conditions, evaluated at each comet. 
This research has made use of data and/or software provided by the High Energy Astrophysics Science Archive Research Center (HEASARC), which is a service of the Astrophysics Science Division at NASA/GSFC.
\section*{Data Availability}
The data supporting the findings of this study are openly available in the nicermastr:NICER Master Catalog within \href{https://heasarc.gsfc.nasa.gov/cgi-bin/W3Browse/w3catindex.pl}{HEASARC Browse}.
The raw and NASA-processed data can be accessed and downloaded from this repository by searching the nicermastr: NICER Master Catalog with the comet name in the name field.

Additionally, any supplementary information, figures, or materials integral to the understanding and replication of the reported results can be found in the Appendix section of this article. For further inquiries, please contact T. K. Deskins at \href{mailto:tkd0009@auburn.edu}{tkd0009@auburn.edu}.

\bibliography{references}

\appendix
\section{Observations Logs}
\label{app:obs-logs}

This appendix provides complete observing logs for all NICER XTI exposures used in this study. The tables list every individual exposure grouped into discrete epochs: an epoch collects observations obtained close in time and under similar solar-wind conditions so that epoch-resolved BCM analysis is meaningful. Each exposure is assigned to a set; by design each set contains two dedicated blank-sky (background) pointings and one on-target comet pointing, which we used to monitor and subtract NICER's variable low-energy background. The tables therefore enable an assessment of the background strategy, the on-target exposure cadence, and the temporal sampling used for epoch construction.

\begin{table}
\centering
\caption{Explanation of fields use in the observation logs, including units and formats.}
\label{tab:obslog_fields_updated}
\scriptsize
\begin{tabular}{@{} ll @{}}
\toprule
Field & Description \\
\midrule
\textbf{Obs Id} & Observation identifier (NICER ObsID). Integer/string as in mission logs. \\
\textbf{Nom RA} (deg) & Nominal spacecraft pointing right ascension (J2000), decimal degrees. Pointing used for the exposure, not target coordinates. \\
\textbf{Nom Dec} (deg) & Nominal spacecraft pointing declination (J2000), decimal degrees. \\
\textbf{Date start} & Observation start timestamp in UTC, ISO 8601 format (YYYY-MM-DDThh:mm:ss). \\
\textbf{Date stop} & Observation stop timestamp in UTC, ISO 8601 format (YYYY-MM-DDThh:mm:ss). \\
\textbf{time start} & Mission elapsed time at observation start, integer seconds (MET). Matches mission/event table units. \\
\textbf{time stop} & Mission elapsed time at observation stop, integer seconds (MET). \\
\textbf{Obj RA} (deg) & Target (comet) right ascension at the observation epoch (J2000), decimal degrees. \\
\textbf{Obj Dec} (deg) & Target (comet) declination at the observation epoch (J2000), decimal degrees. \\
\textbf{Obj r} (AU) & Heliocentric distance of the target at the observation time, in astronomical units. \\
\textbf{Obj delta} (AU) & Geocentric (observer) distance of the target at the observation time, in astronomical units. \\
\textbf{Phase angle} (deg) & Sun--target--observer phase angle at the observation time, degrees. \\
\textbf{BG} & Background designation for the exposure, boolean. \\
\textbf{n\_evt} & Number of cleaned XTI events used for analysis (counts after screening). \\
\textbf{t\_exp} (s) & Net screened exposure time in seconds. \\ 
\textbf{epoch} & Epoch label used in the epoch-resolved analysis. \\
\textbf{ISS Day} & Indicator of whether the observation took place during ISS day or ISS night. \\
\bottomrule
\end{tabular}
\end{table}

\begin{landscape}
\begin{table*}[htbp]
\centering
\tiny
\caption{Observation log: C/2017 T2 (PanSTARRS)}
\label{tab:obs_T2}
\begin{tabular}{@{} l r r l l r r r r r r r l r r l l @{}}
\toprule
Obs Id & Nom RA & Nom Dec & Date start & Date stop & time start & time stop & Obj RA & Obj Dec & Obj r & Obj delta & Phase angle & BG & n\_evt & t\_exp & epoch & ISS Day \\
 & (deg) & (deg) & (UTC) & (UTC) & (s) & (s) & (deg) & (deg) & (AU) & (AU) & (deg) & & & (s) & & \\
\midrule
3543040101 & 77.20  & 75.32 & 2020-04-23T15:40:24 & 2020-04-23T16:17:40 & 199,122,026 & 199,124,262 & 76.20  & 75.22 & 1.62 & 1.72 & 34.80 & true  & 9,765  & 2,236 &  & DAY \\
3543040201 & 77.20  & 75.32 & 2020-04-24T02:30:44 & 2020-04-24T03:07:40 & 199,161,046 & 199,163,262 & 77.18  & 75.32 & 1.62 & 1.72 & 34.83 & false & 7,730  & 2,216 & 1 & DAY \\
3543040301 & 77.21  & 75.32 & 2020-04-26T02:46:39 & 2020-04-26T03:07:00 & 199,334,801 & 199,336,022 & 81.77  & 75.69 & 1.62 & 1.72 & 34.94 & true  & 4,072  & 1,221 &  & DAY \\
3543140101 & 128.75 & 74.73 & 2020-05-12T20:53:17 & 2020-05-12T21:25:55 & 200,782,399 & 200,784,357 & 127.39 & 74.89 & 1.62 & 1.68 & 35.66 & true  & 3,048  & 1,958 &  & DAY \\
3543140201 & 128.75 & 74.73 & 2020-05-13T09:13:04 & 2020-05-13T09:48:46 & 200,826,786 & 200,828,928 & 128.72 & 74.73 & 1.62 & 1.68 & 35.67 & false & 5,275  & 2,140 & 1 & DAY \\
3543140301 & 128.76 & 74.73 & 2020-05-15T10:44:22 & 2020-05-15T10:59:12 & 201,005,064.76 & 201,005,954 & 133.88 & 74.02 & 1.62 & 1.67 & 35.72 & true  & 807    & 887.24&  & DAY \\
\bottomrule
\end{tabular}
\end{table*}
\begin{table*}[htbp]
\centering
\tiny
\caption{Observation log: 88P/Howell}
\label{tab:obs_88P}
\begin{tabular}{@{} l r r l l r r r r r r r l r r l l @{}}
\toprule
Obs Id & Nom RA & Nom Dec & Date start & Date stop & time start & time stop & Obj RA & Obj Dec & Obj r & Obj delta & Phase angle & BG & n\_evt & t\_exp & epoch & ISS Day \\
 & (deg) & (deg) & (UTC) & (UTC) & (s) & (s) & (deg) & (deg) & (AU) & (AU) & (deg) & & & (s) & & \\
\midrule
3543420101 & 233.60 & -22.35 & 2020-09-08T15:56:22 & 2020-09-08T16:25:39 & 211,046,184 & 211,047,941 & 231.96 & -21.90 & 1.37 & 1.36 & 43.41 & true  & 2,773  & 1,757 &  & DAY \\
3543420201 & 233.60 & -22.35 & 2020-09-10T15:45:07 & 2020-09-10T16:27:36 & 211,218,309 & 211,220,858 & 233.59 & -22.35 & 1.36 & 1.36 & 43.37 & false & 3,715  & 1,346 & 1 & DAY \\
3543420301 & 233.60 & -22.35 & 2020-09-12T15:47:07 & 2020-09-12T16:29:20 & 211,391,229 & 211,393,762 & 235.24 & -22.78 & 1.36 & 1.36 & 43.32 & true  & 4,360  & 1,524 &  & DAY \\
3543430101 & 234.72 & -22.65 & 2020-09-10T00:20:02 & 2020-09-10T00:52:41 & 211,162,804 & 211,164,763 & 233.06 & -22.20 & 1.37 & 1.36 & 43.39 & true  & 4,036  & 1,767 &  & DAY \\
3543430201 & 234.72 & -22.65 & 2020-09-12T00:17:26 & 2020-09-12T00:59:40 & 211,335,448 & 211,337,982 & 234.71 & -22.64 & 1.36 & 1.36 & 43.34 & false & 7,464  & 2,530 & 1 & DAY \\
3543430301 & 234.72 & -22.65 & 2020-09-14T00:41:39 & 2020-09-14T01:01:40 & 211,509,701 & 211,510,902 & 236.40 & -23.07 & 1.36 & 1.37 & 43.28 & true  & 3,297  & 1,201 &  & DAY \\
\bottomrule
\end{tabular}
\end{table*}
\begin{table*}[htbp]
\centering
\tiny
\caption{Observation log: 67P/Churyumov--Gerasimenko}
\label{tab:obs_67P}
\begin{tabular}{@{} l r r l l r r r r r r r l r r l l @{}}
\toprule
Obs Id & Nom RA & Nom Dec & Date start & Date stop & time start & time stop & Obj RA & Obj Dec & Obj r & Obj delta & Phase angle & BG & n\_evt & t\_exp & epoch & ISS Day \\
 & (deg) & (deg) & (UTC) & (UTC) & (s) & (s) & (deg) & (deg) & (AU) & (AU) & (deg) & & & (s) & & \\
\midrule
4623130101 & 118.33 & 26.68 & 2021-11-08T16:34:00 & 2021-11-08T16:49:40 & 247,854,990 & 247,855,782 & 116.60 & 26.63 & 1.21 & 0.42 & 49.02 & true  & 564    & 788   &  & NIGHT \\
4623130201 & 118.33 & 26.68 & 2021-11-10T16:37:09 & 2021-11-10T16:56:20 & 248,027,910 & 248,028,982 & 118.32 & 26.68 & 1.22 & 0.42 & 48.65 & false & 1,844  & 1,068 & 1 & NIGHT \\
4623130301 & 118.33 & 26.68 & 2021-11-12T16:35:40 & 2021-11-12T16:58:00 & 248,200,817 & 248,201,882 & 119.97 & 26.72 & 1.22 & 0.42 & 48.23 & true  & 1,046  & 1,052 &  & NIGHT \\
4623140101 & 119.30 & 26.71 & 2021-11-09T20:27:38 & 2021-11-09T20:46:20 & 247,955,409 & 247,956,382 & 117.61 & 26.66 & 1.21 & 0.42 & 48.82 & true  & 1,033  & 967   &  & NIGHT \\
4623140201 & 119.30 & 26.71 & 2021-11-11T22:02:24 & 2021-11-11T22:22:40 & 248,133,908 & 248,134,962 & 119.35 & 26.71 & 1.22 & 0.42 & 48.40 & false & 1,329  & 1,052 & 1 & NIGHT \\
4623140301 & 119.30 & 26.71 & 2021-11-13T22:06:21 & 2021-11-13T22:24:40 & 248,306,869 & 248,307,882 & 120.95 & 26.75 & 1.22 & 0.42 & 47.93 & true  & 880    & 1,011 &  & NIGHT \\
\bottomrule
\end{tabular}
\end{table*}
\begin{table*}[htbp]
\centering
\tiny
\caption{Observation log: 19P/Borrelly}
\label{tab:obs_19P}
\begin{tabular}{@{} l r r l l r r r r r r r l r r l l @{}}
\toprule
Obs Id & Nom RA & Nom Dec & Date start & Date stop & time start & time stop & Obj RA & Obj Dec & Obj r & Obj delta & Phase angle & BG & n\_evt & t\_exp & epoch & ISS Day \\
 & (deg) & (deg) & (UTC) & (UTC) & (s) & (s) & (deg) & (deg) & (AU) & (AU) & (deg) & & & (s) & & \\
\midrule
4623290101 & 13.23 & -6.06  & 2022-01-15T19:07:42 & 2022-01-15T19:09:04 & 253,739,264 & 253,739,346 & 12.14 & -7.53 & 1.32 & 1.21 & 45.41 & true  & 230    & 82    &  & DAY \\
4623290201 & 13.23 & -6.06  & 2022-01-17T19:03:28 & 2022-01-17T19:09:59 & 253,911,810 & 253,912,201 & 13.23 & -6.06 & 1.32 & 1.22 & 45.43 & false & 1,175  & 391   & 1 & DAY \\
4623290301 & 13.23 & -6.06  & 2022-01-19T19:04:28 & 2022-01-19T19:11:01 & 254,084,670 & 254,085,063 & 14.34 & -4.60 & 1.32 & 1.22 & 45.43 & true  & 1,698  & 393   &  & DAY \\
4623300101 & 13.80 & -5.31  & 2022-01-16T19:57:19 & 2022-01-16T19:57:19 & 253,828,641.94 & 253,828,641.94 & 12.70 & -6.77 & 1.32 & 1.21 & 45.42 & true  & 0      & 0     &  & DAY \\
4623300201 & 13.80 & -5.31  & 2022-01-18T19:50:30 & 2022-01-18T19:59:54 & 254,001,032 & 254,001,596 & 13.80 & -5.31 & 1.32 & 1.22 & 45.43 & false & 2,296  & 564   & 1 & DAY \\
4623300301 & 13.80 & -5.31  & 2022-01-20T19:51:29 & 2022-01-20T20:15:19 & 254,173,891 & 254,175,321 & 14.92 & -3.85 & 1.31 & 1.23 & 45.42 & true  & 6,147  & 1,321 &  & DAY \\
\bottomrule
\end{tabular}
\end{table*}
\begin{table*}[htbp]
\centering
\tiny
\caption{Observation log: C/2017 K2 (PANSTARRS)}
\label{tab:obs_K2}
\begin{tabular}{@{} l r r l l r r r r r r r l r r l l @{}}
\toprule
Obs Id & Nom RA & Nom Dec & Date start & Date stop & time start & time stop & Obj RA & Obj Dec & Obj r & Obj delta & Phase angle & BG & n\_evt & t\_exp & epoch & ISS Day \\
 & (deg) & (deg) & (UTC) & (UTC) & (s) & (s) & (deg) & (deg) & (AU) & (AU) & (deg) & & & (s) & & \\
\midrule
5555030101 & 275.64 & 9.56  & 2022-05-29T13:40:11 & 2022-05-29T14:12:21 & 265,297,213 & 265,299,143 & 276.35 & 9.84  & 3.05 & 2.24 & 13.14 & true  & 3,383  & 1,930 &  & DAY \\
5555030201 & 275.64 & 9.56  & 2022-05-31T13:47:33 & 2022-05-31T14:17:20 & 265,470,455 & 265,472,242 & 275.64 & 9.56  & 3.04 & 2.20 & 12.77 & false & 2,302  & 1,787 & 1 & NIGHT \\
5555030301 & 275.64 & 9.56  & 2022-06-02T13:44:19 & 2022-06-02T13:53:40 & 265,643,061 & 265,643,622 & 274.90 & 9.26  & 3.02 & 2.17 & 12.39 & true  & 494    & 554   &  & DAY \\
5555040101 & 275.57 & 9.53  & 2022-05-29T18:18:45 & 2022-05-29T18:35:20 & 265,313,927 & 265,314,922 & 276.28 & 9.81  & 3.05 & 2.23 & 13.11 & true  & 2,349  & 995   &  & DAY \\
5555040201 & 275.57 & 9.53  & 2022-05-31T18:21:05 & 2022-05-31T18:52:19 & 265,486,867 & 265,488,741 & 275.57 & 9.53  & 3.03 & 2.20 & 12.73 & false & 39,088 & 1,872 & 1 & DAY \\
5555040301 & 275.57 & 9.53  & 2022-06-02T18:20:33 & 2022-06-02T18:32:08 & 265,659,635 & 265,660,330 & 274.83 & 9.23  & 3.02 & 2.16 & 12.35 & true  & 562    & 695   &  & DAY \\
\bottomrule
\end{tabular}
\end{table*}
\begin{table*}[htbp]
\centering
\tiny
\caption{Observation log: C/2022 E3 (ZTF)}
\label{tab:obs_E3}
\begin{tabular}{@{} l r r l l r r r r r r r l r r l l @{}}
\toprule
Obs Id & Nom RA & Nom Dec & Date start & Date stop & time start & time stop & Obj RA & Obj Dec & Obj r & Obj delta & Phase angle & BG & n\_evt & t\_exp & epoch & ISS Day \\
 & (deg) & (deg) & (UTC) & (UTC) & (s) & (s) & (deg) & (deg) & (AU) & (AU) & (deg) & & & (s) & & \\
\midrule
6722070101 & 71.48 & -5.68 & 2023-03-19T02:20:40 & 2023-03-19T02:57:00 & 290,658,327.46 & 290,660,114 & 71.21 & -5.23 & 1.52 & 1.47 & 38.87 & true  & 9,561  & 926.54 &  & DAY \\
6722070201 & 71.48 & -5.68 & 2023-03-21T02:23:55 & 2023-03-21T02:58:20 & 290,831,224    & 290,833,065 & 71.48 & -5.68 & 1.54 & 1.53 & 37.92 & false & 17,871 & 1,838  & 1 & DAY \\
6722070301 & 71.48 & -5.68 & 2023-03-23T02:24:42 & 2023-03-23T02:59:40 & 291,004,107    & 291,005,969 & 71.75 & -6.09 & 1.56 & 1.58 & 36.98 & true  & 23,654 & 1,851  &  & DAY \\
6722080101 & 71.49 & -5.69 & 2023-03-19T03:55:10 & 2023-03-19T04:30:00 & 290,663,927    & 290,665,694 & 71.22 & -5.25 & 1.52 & 1.47 & 38.84 & true  & 11,426 & 907    &  & DAY \\
6722080201 & 71.49 & -5.69 & 2023-03-21T03:56:14 & 2023-03-21T04:31:20 & 290,836,788    & 290,838,642 & 71.49 & -5.69 & 1.54 & 1.53 & 37.89 & false & 18,463 & 1,854  & 1 & DAY \\
6722080301 & 71.49 & -5.69 & 2023-03-23T03:57:56 & 2023-03-23T04:32:20 & 291,009,687    & 291,011,533 & 71.76 & -6.10 & 1.56 & 1.58 & 36.95 & true  & 24,154 & 1,846  &  & DAY \\
\bottomrule
\end{tabular}
\end{table*}
\begin{table*}[htbp]
\centering
\tiny
\caption{Observation log: 62P/Tsuchinshan}
\label{tab:obs_62P}
\begin{tabular}{@{} l r r l l r r r r r r r l r r l l @{}}
\toprule
Obs Id & Nom RA & Nom Dec & Date start & Date stop & time start & time stop & Obj RA & Obj Dec & Obj r & Obj delta & Phase angle & BG & n\_evt & t\_exp & epoch & ISS Day \\
 & (deg) & (deg) & (UTC) & (UTC) & (s) & (s) & (deg) & (deg) & (AU) & (AU) & (deg) & & & (s) & & \\
\midrule
7663510101 & 190.29 & 10.02 & 2024-02-20T02:18:51 & 2024-02-20T02:30:00 & 319,861,533 & 319,861,772 & 190.21 & 10.06 & 1.43 & 0.52 & 24.45 & true  & 265    & 237   &  & NIGHT \\
7663510201 & 190.29 & 10.02 & 2024-02-18T02:26:36 & 2024-02-18T02:29:53 & 319,688,798 & 319,688,995 & 190.29 & 10.02 & 1.42 & 0.51 & 25.81 & false & 329    & 197   & 1 & DAY \\
7663510301 & 190.29 & 10.02 & 2024-02-20T02:25:31 & 2024-02-20T02:29:30 & 319,861,533 & 319,861,772 & 190.21 & 10.06 & 1.43 & 0.52 & 24.44 & true  & 265    & 237   &  & NIGHT \\
7663540101 & 190.28 & 10.02 & 2024-02-20T04:09:04 & 2024-02-20T04:14:20 & 319,867,783 & 319,868,019 & 190.20 & 10.06 & 1.43 & 0.52 & 24.39 & true  & 238    & 234   &  & NIGHT \\
7663540201 & 190.28 & 10.02 & 2024-02-18T04:10:19 & 2024-02-18T04:13:41 & 319,695,021 & 319,695,223 & 190.28 & 10.02 & 1.42 & 0.51 & 25.76 & false & 243    & 202   & 1 & NIGHT \\
7663540301 & 190.28 & 10.02 & 2024-02-20T04:09:44 & 2024-02-20T04:13:37 & 319,867,786 & 319,868,019 & 190.20 & 10.06 & 1.43 & 0.52 & 24.39 & true  & 238    & 233   &  & NIGHT \\
\bottomrule
\end{tabular}
\end{table*}
\begin{table*}[htbp]
\centering
\tiny
\caption{Observation log: 12P/Pons--Brooks}
\label{tab:obs_12P}
\begin{tabular}{@{} l r r l l r r r r r r r l r r l l @{}}
\toprule
Obs Id & Nom RA & Nom Dec & Date start & Date stop & time start & time stop & Obj RA & Obj Dec & Obj r & Obj delta & Phase angle & BG & n\_evt & t\_exp & epoch & ISS Day \\
 & (deg) & (deg) & (UTC) & (UTC) & (s) & (s) & (deg) & (deg) & (AU) & (AU) & (deg) & & & (s) & & \\
\midrule
7663160101 & 346.54 & 36.90 & 2024-02-25T14:17:27 & 2024-02-25T14:23:40 & 320,336,348 & 320,336,615 & 349.04 & 36.58 & 1.27 & 1.72 & 34.83 & true  & 1,692  & 267   &  & NIGHT \\
7663160201 & 346.54 & 36.90 & 2024-02-23T14:19:07 & 2024-02-23T14:22:45 & 320,163,549 & 320,163,767 & 346.54 & 36.90 & 1.29 & 1.73 & 34.58 & false & 408    & 218   & 1 & DAY \\
7663160301 & 346.54 & 36.90 & 2024-02-25T14:19:06 & 2024-02-25T14:23:32 & 320,336,348 & 320,336,614 & 349.04 & 36.58 & 1.27 & 1.72 & 34.83 & true  & 1,660  & 266   &  & NIGHT \\
7663170101 & 346.62 & 36.89 & 2024-02-21T15:53:07 & 2024-02-21T15:56:00 & 319,996,389 & 319,996,562 & 344.14 & 37.17 & 1.32 & 1.74 & 34.32 & true  & 108    & 173   &  & DAY \\
7663170201 & 346.62 & 36.89 & 2024-02-23T15:51:45 & 2024-02-23T15:55:38 & 320,169,107 & 320,169,340 & 346.62 & 36.89 & 1.29 & 1.73 & 34.59 & false & 241    & 233   & 1 & DAY \\
7663170301 & 346.62 & 36.89 & 2024-02-25T15:51:46 & 2024-02-25T15:56:29 & 320,341,908 & 320,342,191 & 349.12 & 36.57 & 1.27 & 1.71 & 34.84 & true  & 641    & 283   &  & DAY \\
\bottomrule
\end{tabular}
\end{table*}
\begin{table*}[htbp]
\centering
\tiny
\caption{Observation log: P/2010 H2 (Vales)}
\label{tab:obs_H2}
\begin{tabular}{@{} l r r l l r r r r r r r l r r l l @{}}
\toprule
Obs Id & Nom RA & Nom Dec & Date start & Date stop & time start & time stop & Obj RA & Obj Dec & Obj r & Obj delta & Phase angle & BG & n\_evt & t\_exp & epoch & ISS Day \\
 & (deg) & (deg) & (UTC) & (UTC) & (s) & (s) & (deg) & (deg) & (AU) & (AU) & (deg) & & & (s) & & \\
\midrule
8555280101 & 204.52 & 6.25 & 2025-03-31T21:47:09 & 2025-03-31T21:56:34 & 354,923,231 & 354,923,796 & 204.88 & 6.19 & 3.08 & 2.11 & 5.79 & true  & 669   & 565   &  & NIGHT \\
8555280201 & 204.52 & 6.26 & 2025-04-02T21:50:53 & 2025-04-02T22:05:40 & 355,096,255 & 355,097,142 & 204.52 & 6.26 & 3.08 & 2.11 & 5.43 & false & 2,047 & 451   & 1 & NIGHT \\
8555280301 & 204.52 & 6.26 & 2025-04-04T21:51:56 & 2025-04-04T21:53:54 & 355,269,118 & 355,269,236 & 204.16 & 6.32 & 3.08 & 2.10 & 5.14 & true  & 709   & 118   &  & NIGHT \\
8555290101 & 204.51 & 6.26 & 2025-03-31T23:20:09 & 2025-03-31T23:29:08 & 354,928,811 & 354,929,350 & 204.87 & 6.19 & 3.08 & 2.11 & 5.77 & true  & 649   & 537   &  & NIGHT \\
8555290201 & 204.51 & 6.26 & 2025-04-02T23:37:54 & 2025-04-02T23:44:00 & 355,102,676 & 355,103,042 & 204.51 & 6.26 & 3.08 & 2.11 & 5.41 & false & 2,240 & 364   & 1 & NIGHT \\
8555290301 & 204.51 & 6.26 & 2025-04-04T23:24:19 & 2025-04-04T23:24:19 & 355,274,661.05 & 355,274,661.05 & 204.15 & 6.32 & 3.08 & 2.10 & 5.13 & true  & 0     & 0     &  & NIGHT \\
8555350101 & 204.44 & 6.27 & 2025-04-01T08:36:24 & 2025-04-01T08:46:54 & 354,962,186 & 354,962,816 & 204.80 & 6.20 & 3.08 & 2.11 & 5.70 & true  & 752   & 626   &  & NIGHT \\
8555350201 & 204.44 & 6.27 & 2025-04-03T08:43:09 & 2025-04-03T09:06:40 & 355,135,391 & 355,136,802 & 204.44 & 6.27 & 3.08 & 2.11 & 5.36 & false & 5,214 & 1,377 & 1 & NIGHT \\
8555350301 & 204.45 & 6.27 & 2025-04-05T09:01:14 & 2025-04-05T09:08:01 & 355,309,276 & 355,309,683 & 204.08 & 6.33 & 3.08 & 2.10 & 5.09 & true  & 1,534 & 407   &  & NIGHT \\
\bottomrule
\end{tabular}
\end{table*}
\end{landscape}

\section{Spectra}
\label{app:spectra}
This appendix section presents the Bayesian Color Model (BCM) spectral fits for each comet in the survey, showing both spectra from individual observing-epochs and the corresponding full-dataset spectrum and model. 
The BCM utilizes Bayesian principles to quantitatively determine optimal parameters for fitting the low-resolution data.
By iterating through peak combinations associated with C~V, C~VI, N~VI, N~VII, O~VII, O~VIII, Ne~IX, Ne~X, Mg~XI, Mg~XII, Al~XII, Al~XIII, Si~XIII, and Si~XIV, the model selects the combination with the best fit and minimal complexity, as indicated by the minimum BIC.
Readers may consult Table~\ref{tab:gauss_params_full} for the numerical Gaussian centroids, widths, and amplitudes, and Table~\ref{tab:flux_ratios_epochs} for the epoch-resolved flux ratios and inferred freeze-in temperatures that accompany each epochal fit.
Figures are ordered to match the comet list in the main text so that epochal behavior and the full-dataset morphology can be compared easily across the sample.
These plots illustrate both reproducible spectral families and the epoch-to-epoch variations, and they provide the visual basis for the quantitative diagnostics reported in the paper.


\begin{figure*}
    \centering
    \begin{subcaptionblock}[b]{0.47\textwidth}
        \includegraphics[width=\textwidth,trim={0.35cm 0.25cm 0.2cm 1.35cm},clip]{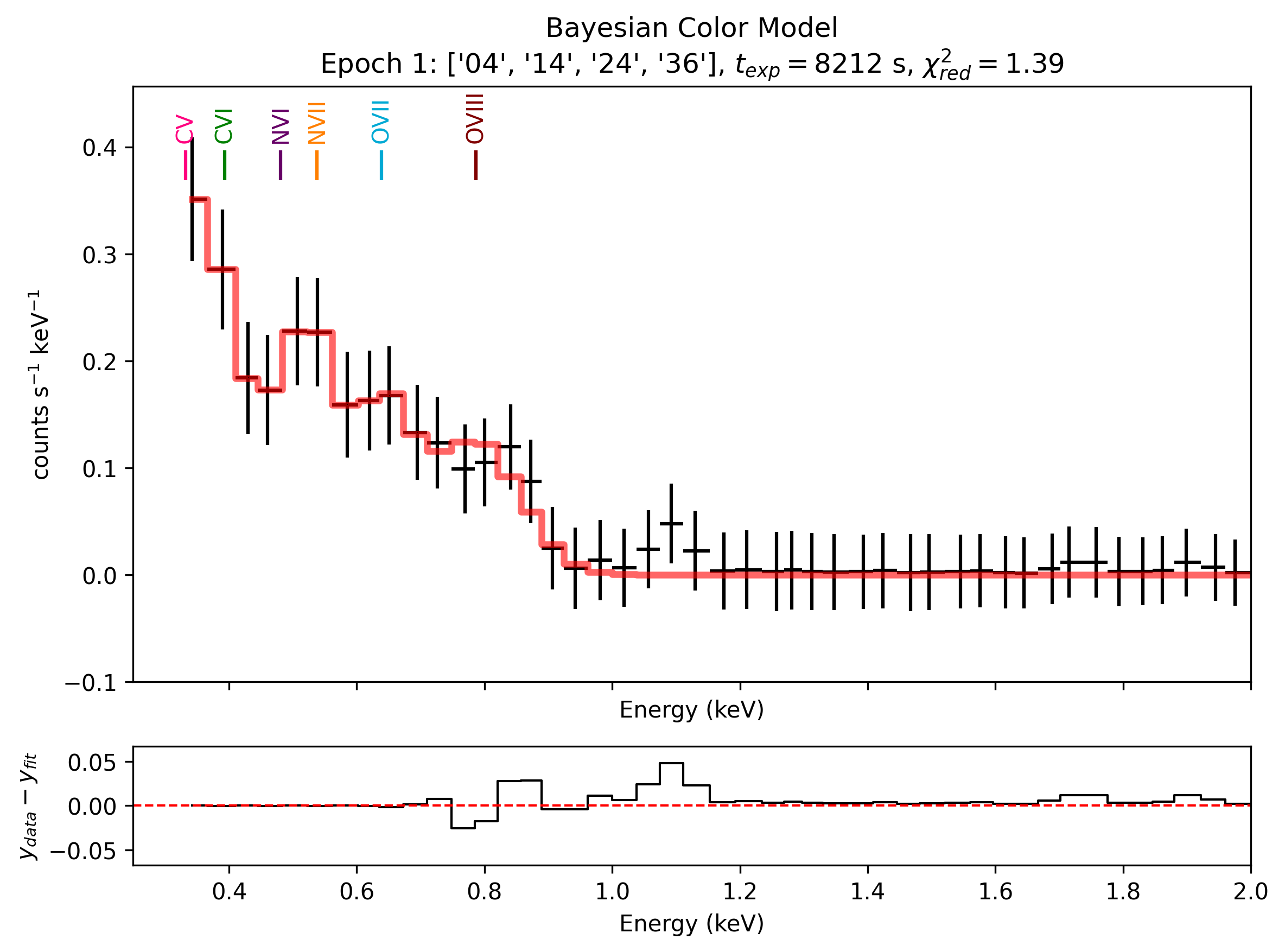}
        \caption{Epoch 1 ($t_{exp}=8212$\,ks).}
        \label{fig:spec_t2_e1}
    \end{subcaptionblock}
    \hfill
    \begin{subcaptionblock}[b]{0.47\textwidth}
        \includegraphics[width=\textwidth,trim={0.35cm 0.25cm 0.2cm 1.35cm},clip]{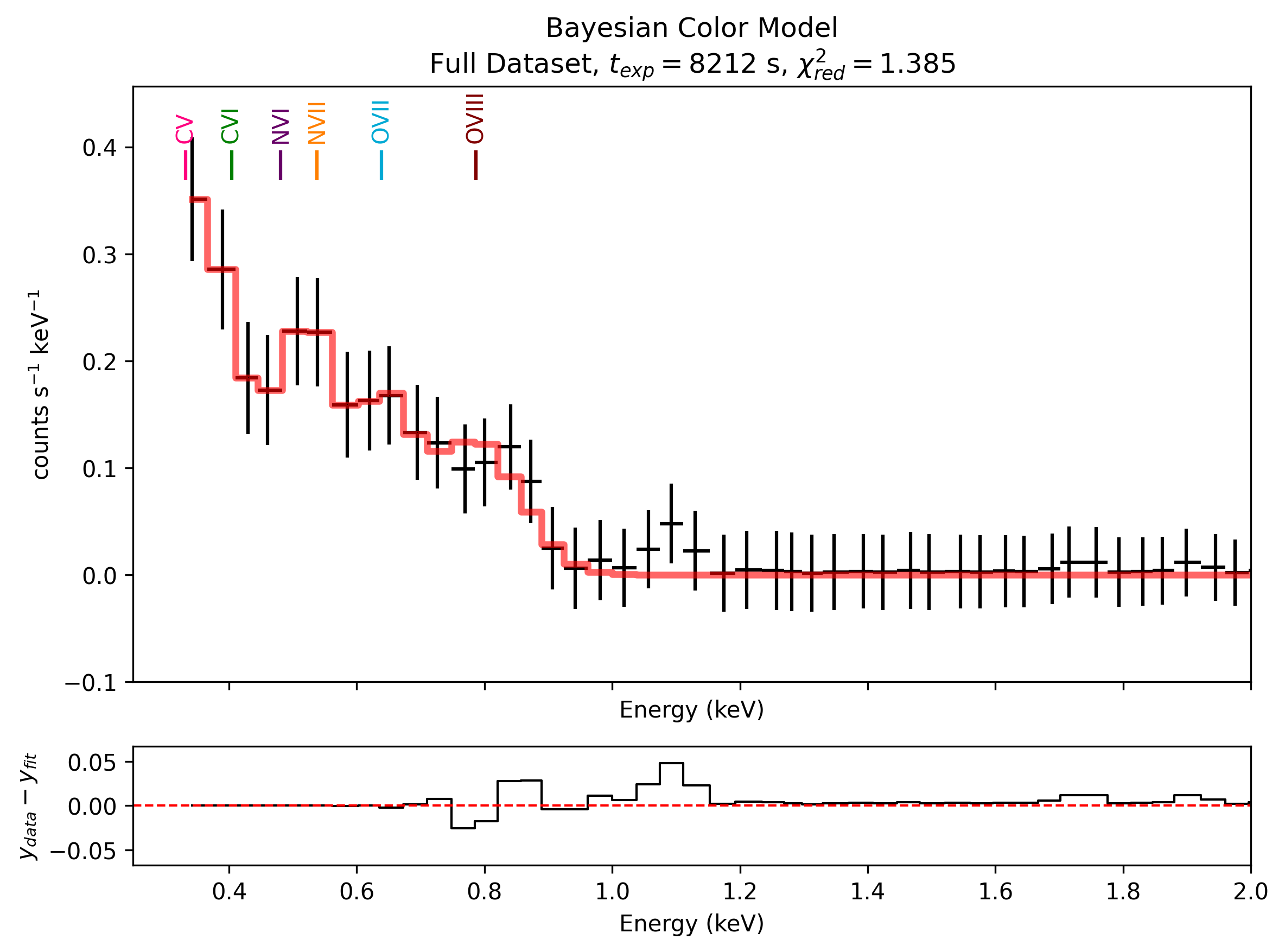}
        \caption{Full dataset ($t_{exp}=8212$\,ks).}
        \label{fig:spec_t2_full}
    \end{subcaptionblock}

    \caption{X-ray spectra from the interaction between the solar wind and the atmosphere of C/2017 T2 fitted with the Bayesian Color Model (BCM).}
    \label{fig:app-spec-t2}
\end{figure*}


\begin{figure*}
    \centering
    \begin{subcaptionblock}[b]{0.47\textwidth}
        \includegraphics[width=\textwidth,trim={0.35cm 0.25cm 0.2cm 1.35cm},clip]{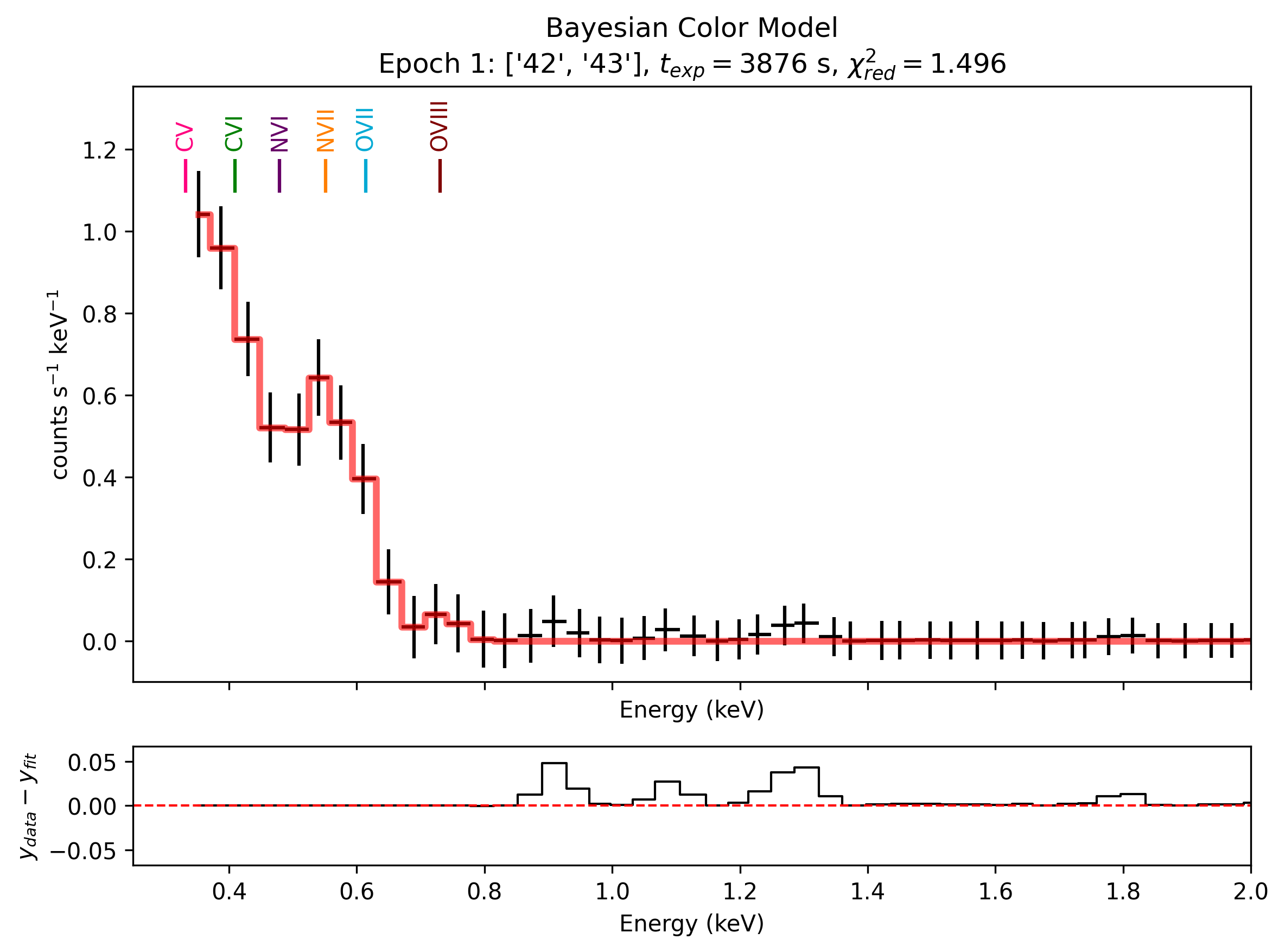}
        \caption{Epoch 1 ($t_{exp}=3876$\,ks).}
        \label{fig:spec_88p_e1}
    \end{subcaptionblock}
    \hfill
    \begin{subcaptionblock}[b]{0.47\textwidth}
        \includegraphics[width=\textwidth,trim={0.35cm 0.25cm 0.2cm 1.35cm},clip]{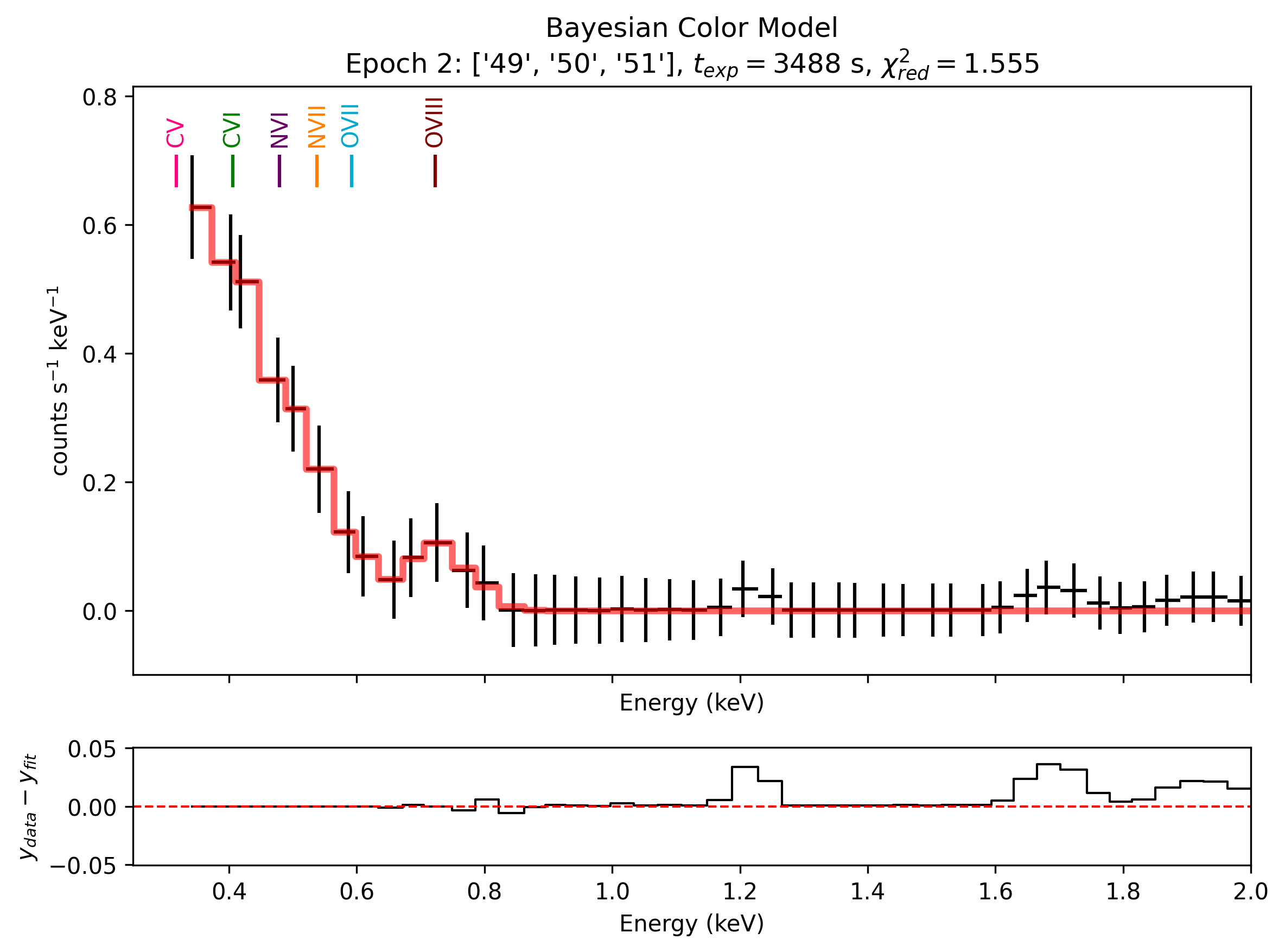}
        \caption{Epoch 2 ($t_{exp}=3488$\,ks).}
        \label{fig:spec_88p_e2}
    \end{subcaptionblock}
    \hfill
    \begin{subcaptionblock}[b]{0.47\textwidth}
        \includegraphics[width=\textwidth,trim={0.35cm 0.25cm 0.2cm 1.35cm},clip]{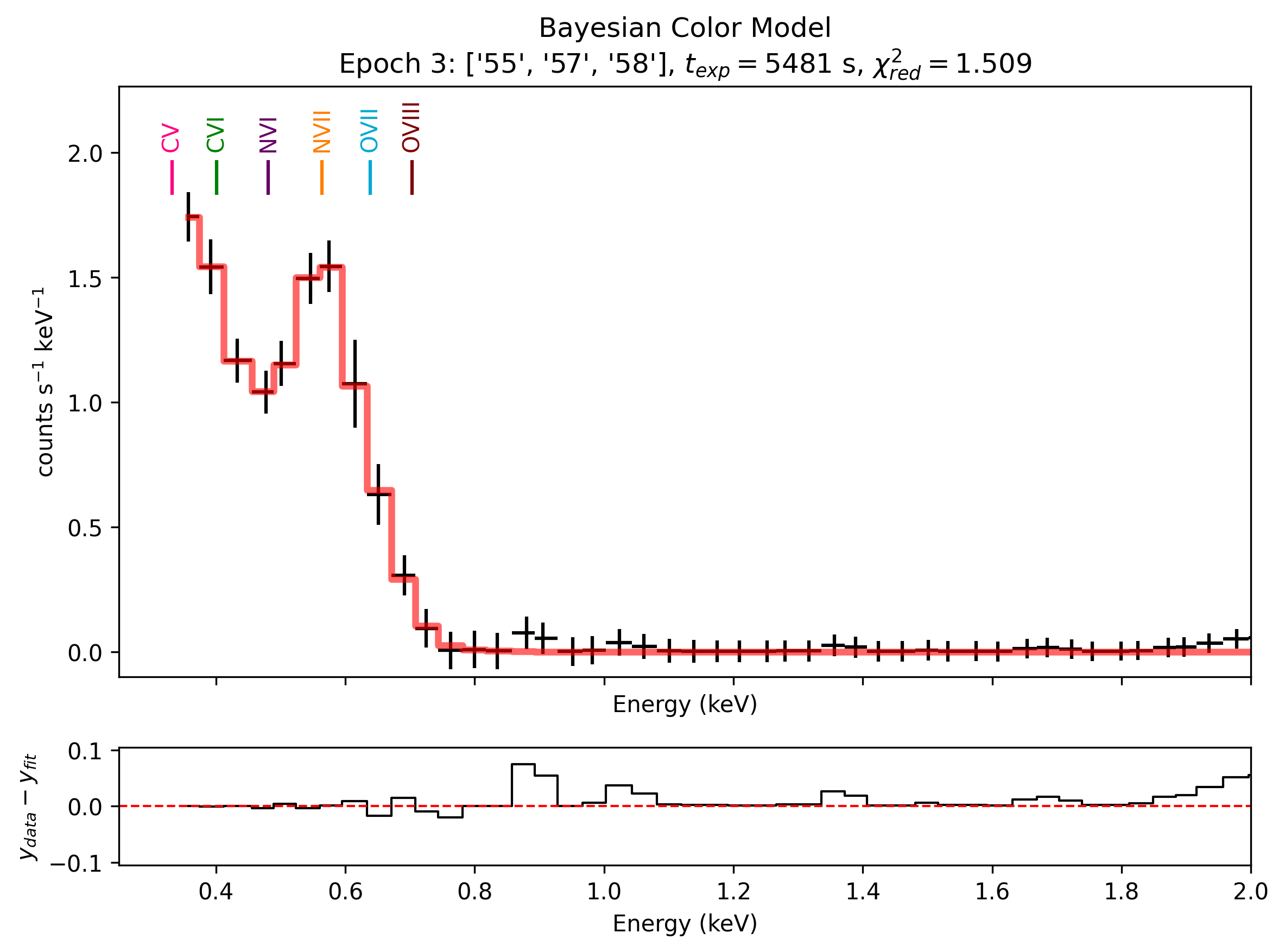}
        \caption{Epoch 3 ($t_{exp}=5481$\,ks).}
        \label{fig:spec_88p_e3}
    \end{subcaptionblock}
    \hfill
    \begin{subcaptionblock}[b]{0.47\textwidth}
        \includegraphics[width=\textwidth,trim={0.35cm 0.25cm 0.2cm 1.35cm},clip]{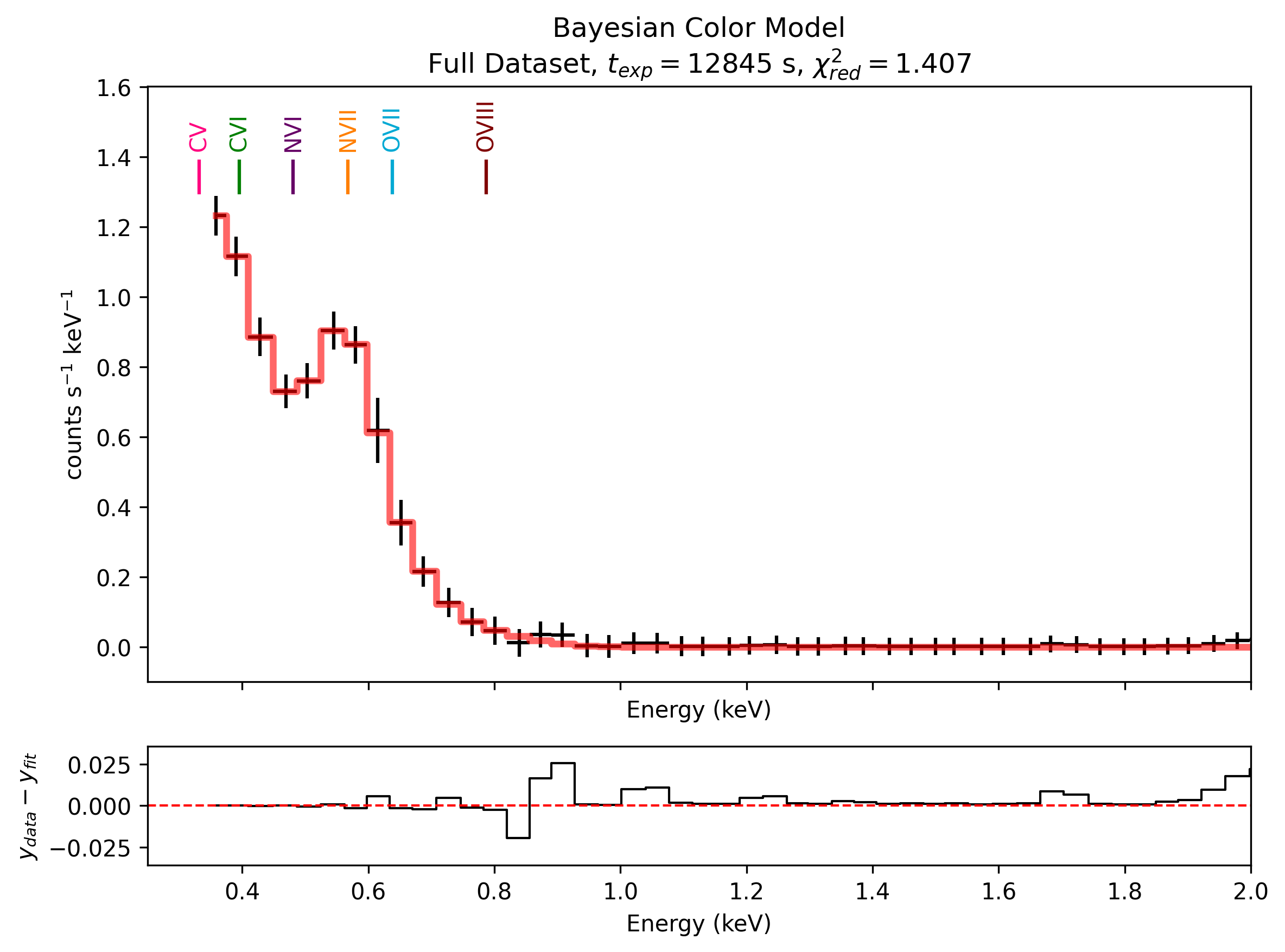}
        \caption{Full dataset ($t_{exp}=12845$\,ks).}
        \label{fig:spec_88p_full}
    \end{subcaptionblock}

    \caption{X-ray spectra from the interaction between the solar wind and the atmosphere of 88P/Howell fitted with the Bayesian Color Model (BCM).}
    \label{fig:app-spec-88p}
\end{figure*}


\begin{figure*}
    \centering
    \begin{subcaptionblock}[b]{0.47\textwidth}
        \includegraphics[width=\textwidth,trim={0.35cm 0.25cm 0.2cm 1.35cm},clip]{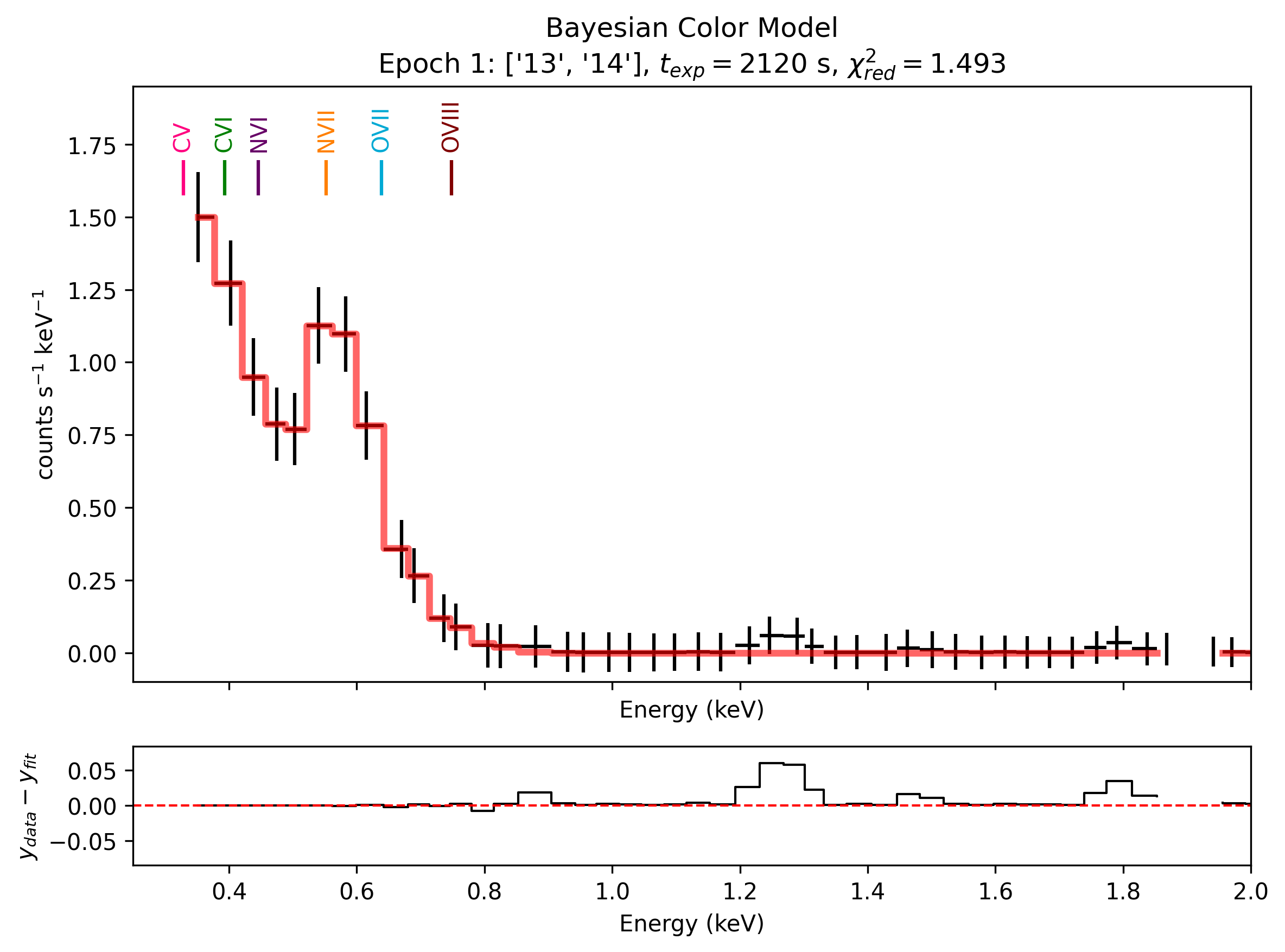}
        \caption{Epoch 1 ($t_{exp}=2120$\,ks).}
        \label{fig:spec_67p_e1}
    \end{subcaptionblock}
    \hfill
    \begin{subcaptionblock}[b]{0.47\textwidth}
        \includegraphics[width=\textwidth,trim={0.35cm 0.25cm 0.2cm 1.35cm},clip]{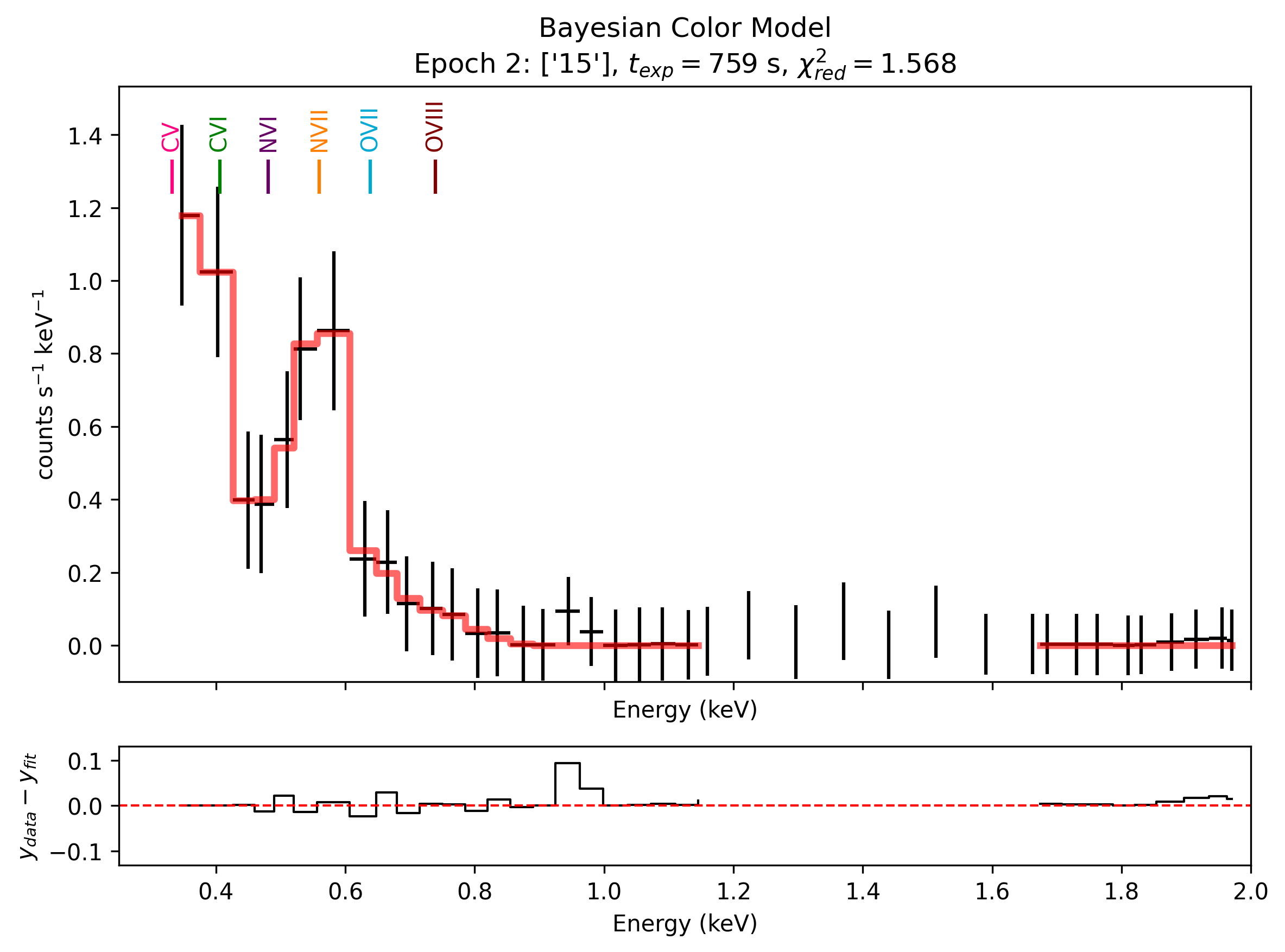}
        \caption{Epoch 2 ($t_{exp}=759$\,ks).}
        \label{fig:spec_67p_e2}
    \end{subcaptionblock}
    \hfill
    \begin{subcaptionblock}[b]{0.47\textwidth}
        \includegraphics[width=\textwidth,trim={0.35cm 0.25cm 0.2cm 1.35cm},clip]{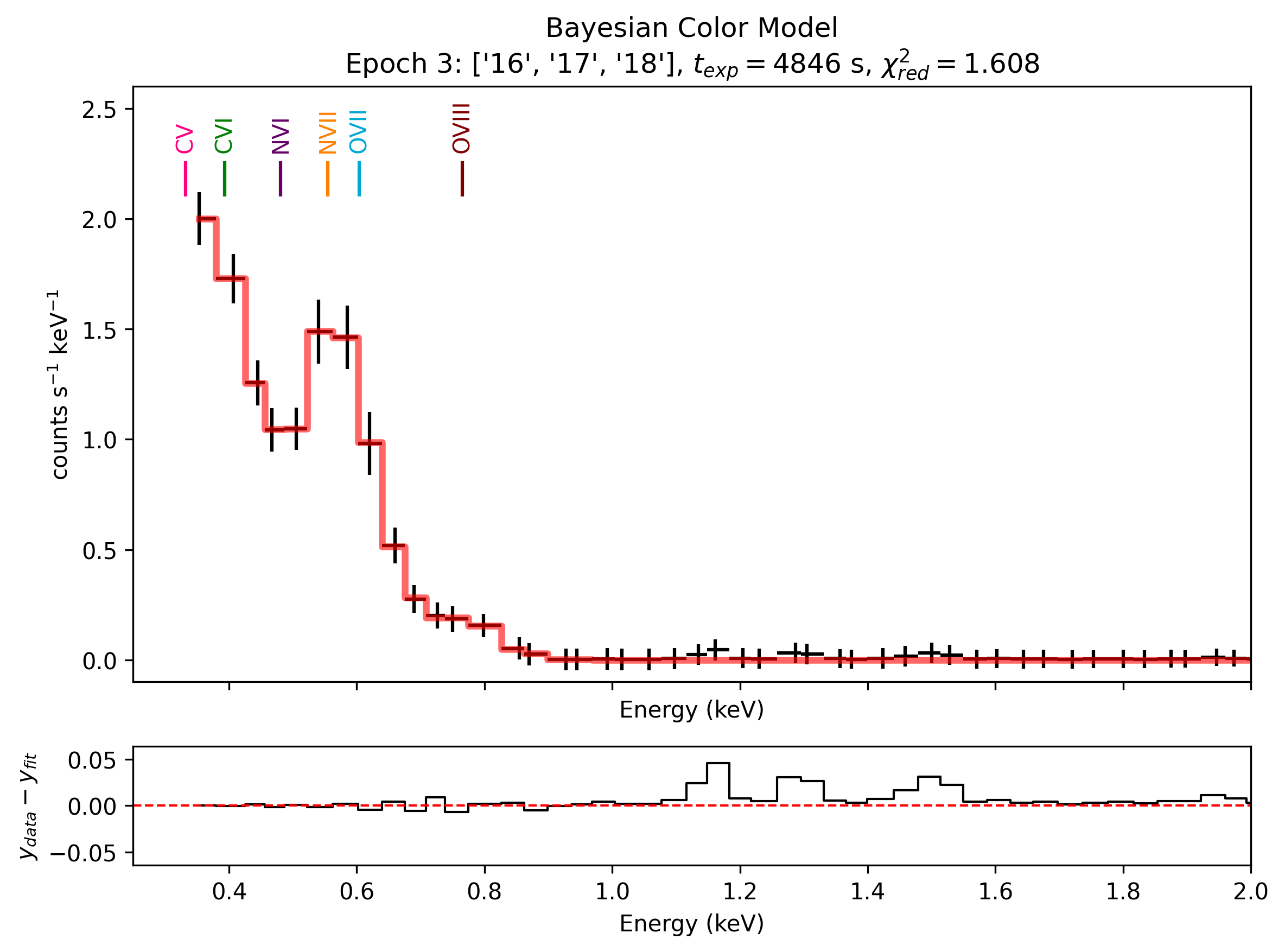}
        \caption{Epoch 3 ($t_{exp}=4846$\,ks).}
        \label{fig:spec_67p_e3}
    \end{subcaptionblock}
    \hfill
    \begin{subcaptionblock}[b]{0.47\textwidth}
        \includegraphics[width=\textwidth,trim={0.35cm 0.25cm 0.2cm 1.35cm},clip]{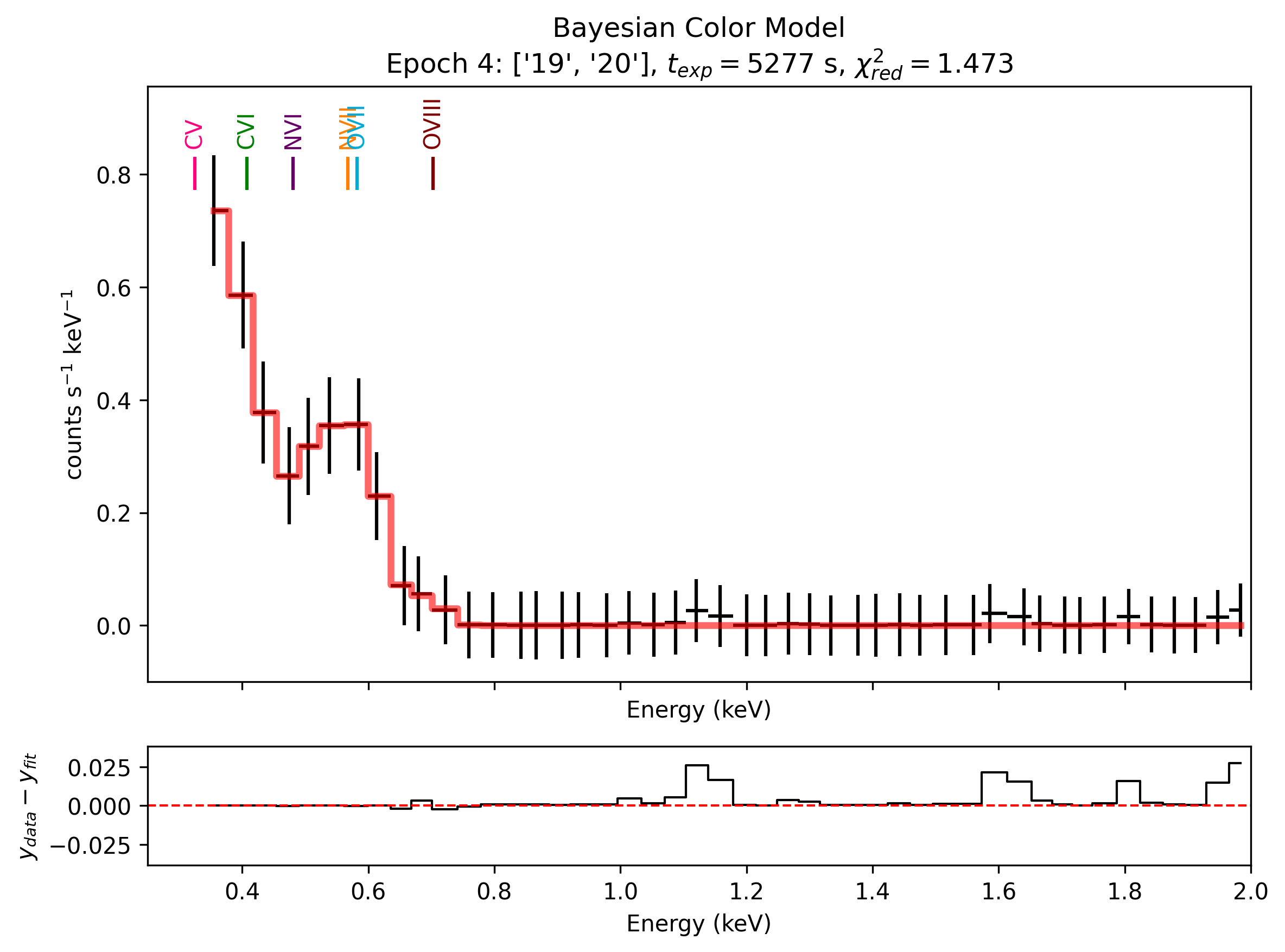}
        \caption{Epoch 4 ($t_{exp}=5277$\,ks).}
        \label{fig:spec_67p_e4}
    \end{subcaptionblock}
    \hfill
    \begin{subcaptionblock}[b]{0.47\textwidth}
        \includegraphics[width=\textwidth,trim={0.35cm 0.25cm 0.2cm 1.35cm},clip]{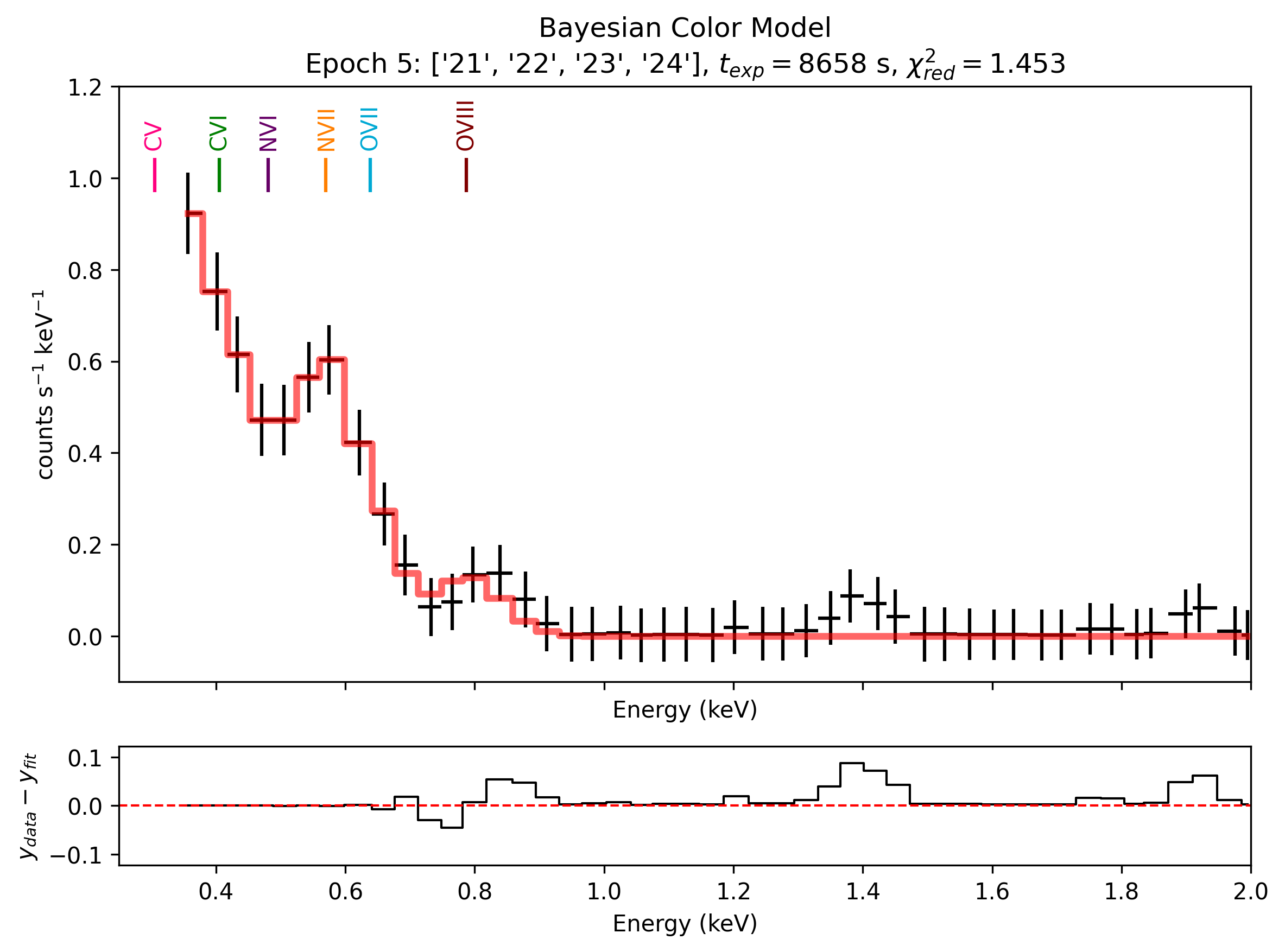}
        \caption{Epoch 5 ($t_{exp}=8658$\,ks).}
        \label{fig:spec_67p_e5}
    \end{subcaptionblock}
    \hfill
    \begin{subcaptionblock}[b]{0.47\textwidth}
        \includegraphics[width=\textwidth,trim={0.35cm 0.25cm 0.2cm 1.35cm},clip]{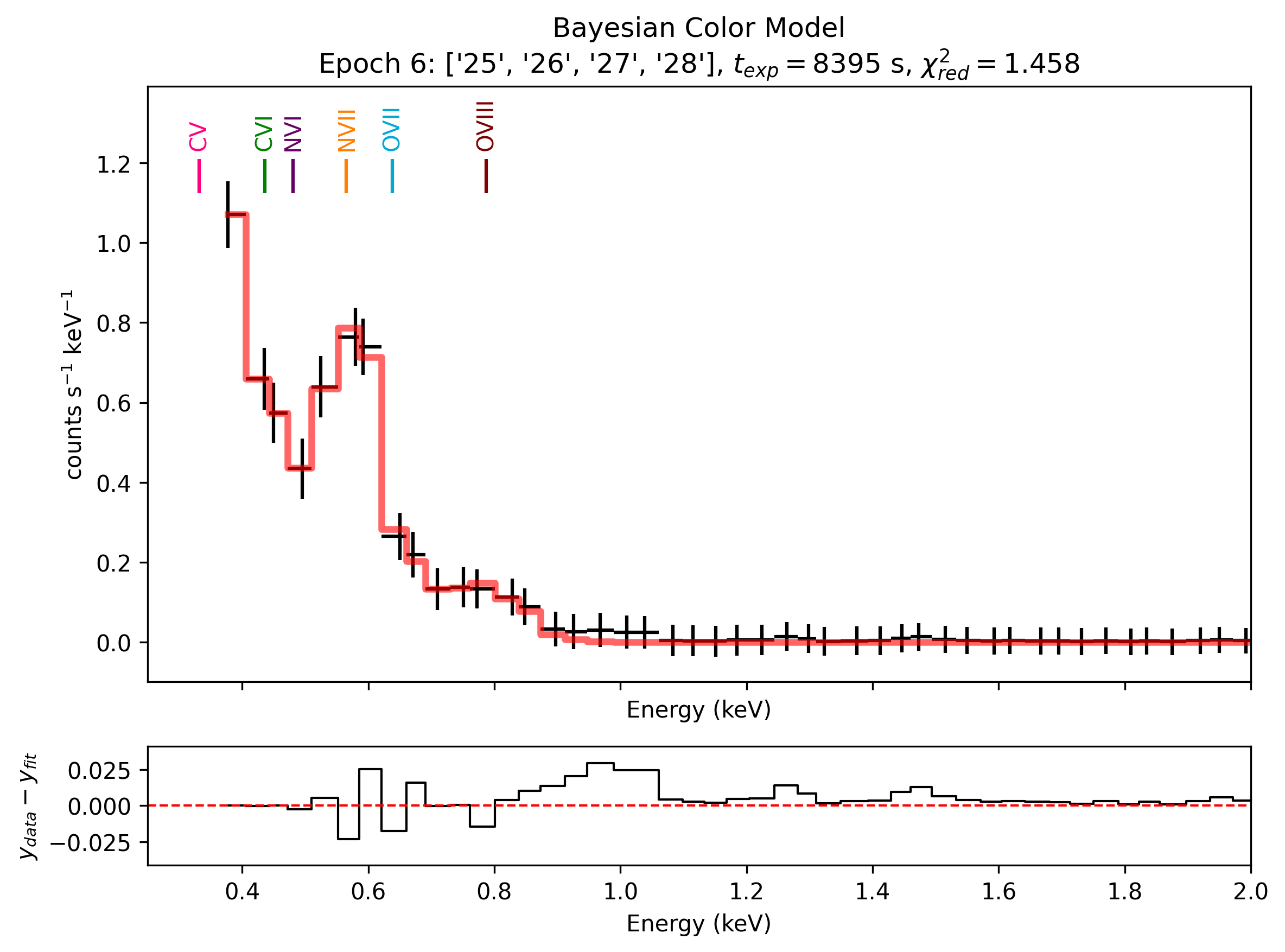}
        \caption{Epoch 6 ($t_{exp}=8395$\,ks).}
        \label{fig:spec_67p_e6}
    \end{subcaptionblock}

    \caption{X-ray spectra from the interaction between the solar wind and the atmosphere of 67P/Churyumov–Gerasimenko fitted with the Bayesian Color Model (BCM).}
    \label{fig:app-spec-67p-epochs}
\end{figure*}

\begin{figure*}
    \centering
    \begin{subcaptionblock}[b]{0.47\textwidth}
        \includegraphics[width=\textwidth,trim={0.35cm 0.25cm 0.2cm 1.35cm},clip]{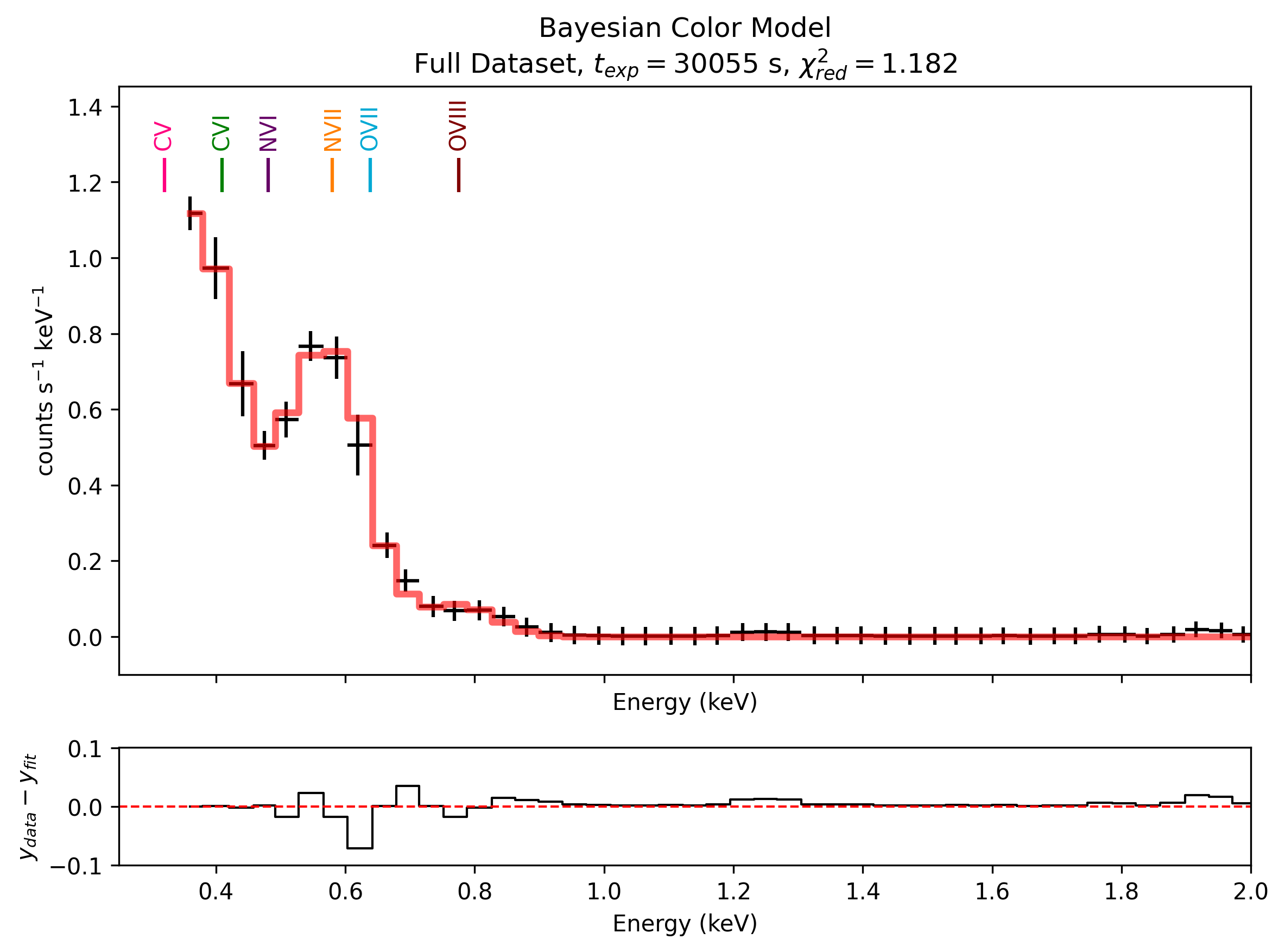}
        \caption{Full dataset ($t_{exp}=30055$\,ks).}
        \label{fig:spec_67p_full}
    \end{subcaptionblock}

    \caption{X-ray spectra from the interaction between the solar wind and the atmosphere of 67P/Churyumov–Gerasimenko fitted with the Bayesian Color Model (BCM).}
    \label{fig:app-spec-67p-full}
\end{figure*}


\begin{figure*}
    \centering
    \begin{subcaptionblock}[b]{0.47\textwidth}
        \includegraphics[width=\textwidth,trim={0.35cm 0.25cm 0.2cm 1.35cm},clip]{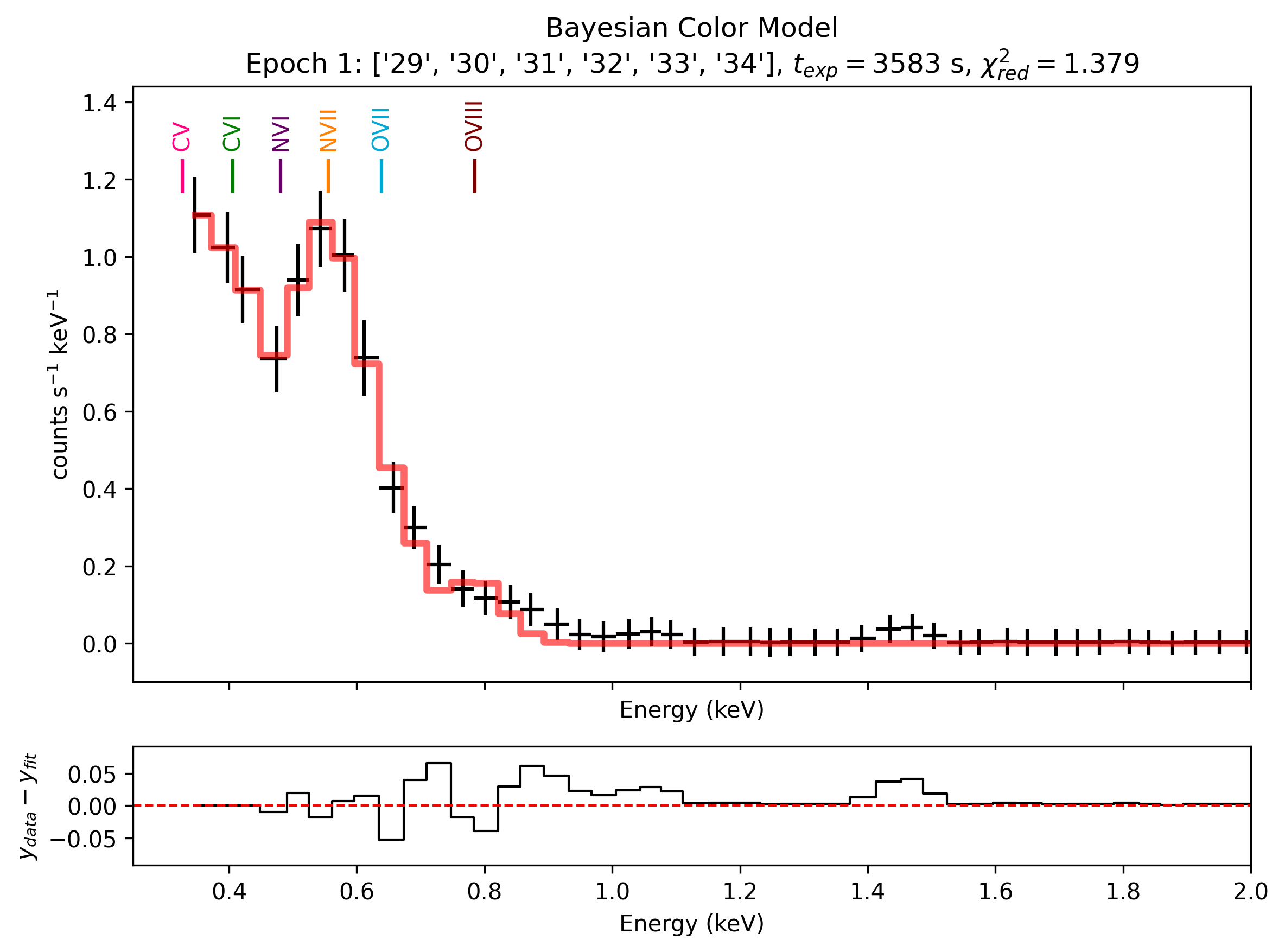}
        \caption{Epoch 1 ($t_{exp}=3583$\,ks).}
        \label{fig:spec_19p_e1}
    \end{subcaptionblock}
    \hfill
    \begin{subcaptionblock}[b]{0.47\textwidth}
        \includegraphics[width=\textwidth,trim={0.35cm 0.25cm 0.2cm 1.35cm},clip]{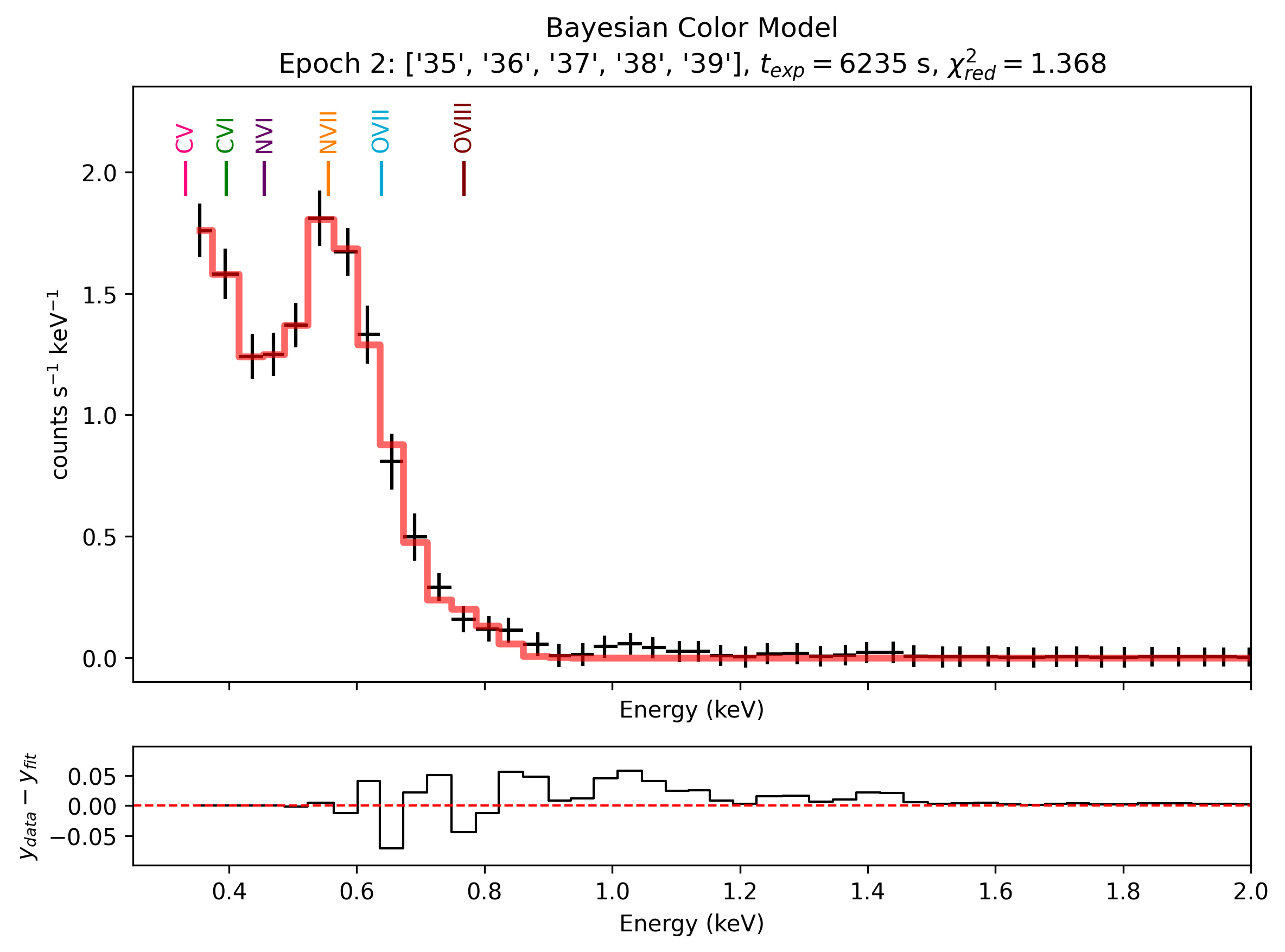}
        \caption{Epoch 2 ($t_{exp}=6235$\,ks).}
        \label{fig:spec_19p_e2}
    \end{subcaptionblock}
    \hfill
    \begin{subcaptionblock}[b]{0.47\textwidth}
        \includegraphics[width=\textwidth,trim={0.35cm 0.25cm 0.2cm 1.35cm},clip]{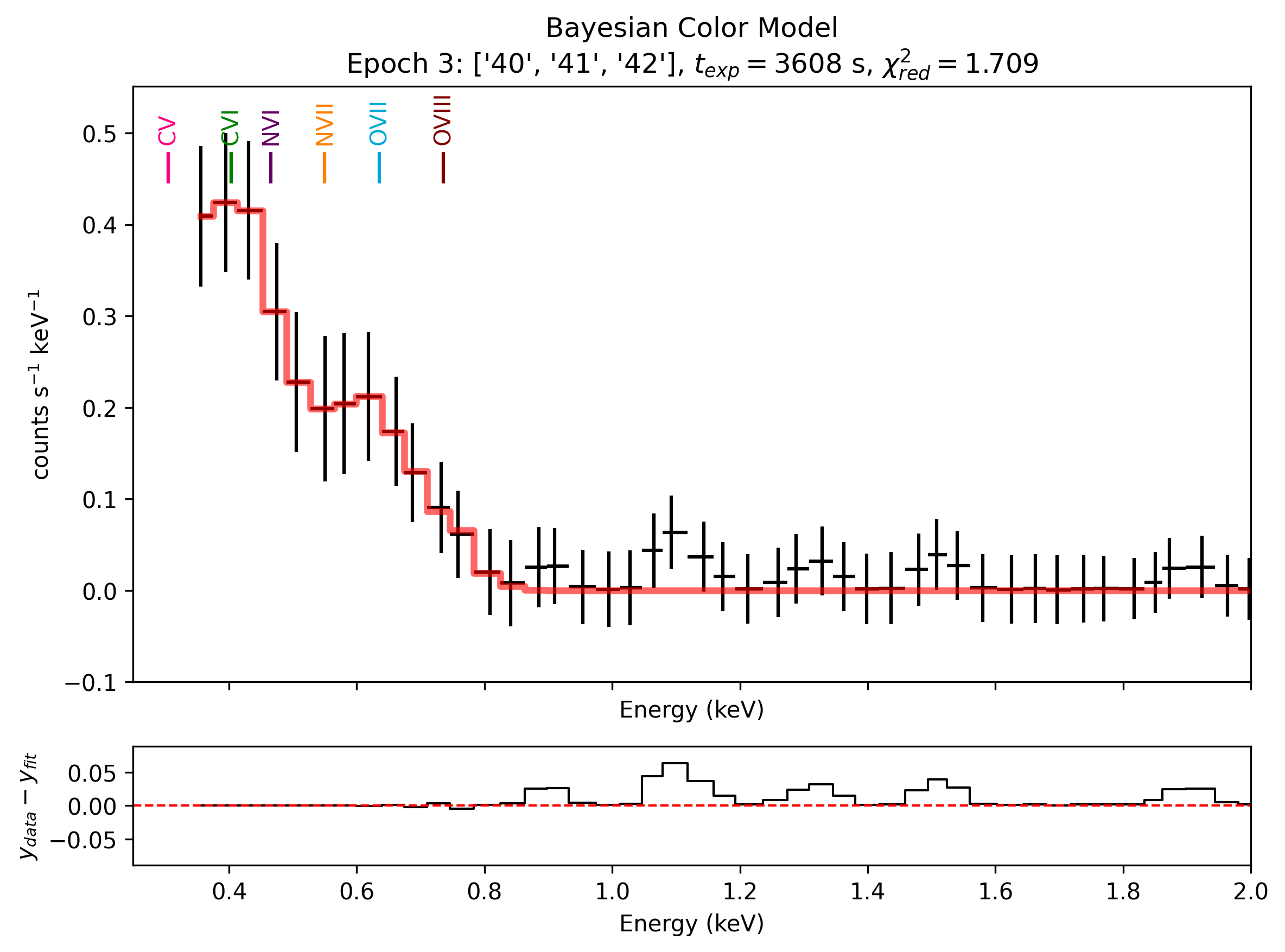}
        \caption{Epoch 3 ($t_{exp}=3608$\,ks).}
        \label{fig:spec_19p_e3}
    \end{subcaptionblock}
    \hfill
    \begin{subcaptionblock}[b]{0.47\textwidth}
        \includegraphics[width=\textwidth,trim={0.35cm 0.25cm 0.2cm 1.35cm},clip]{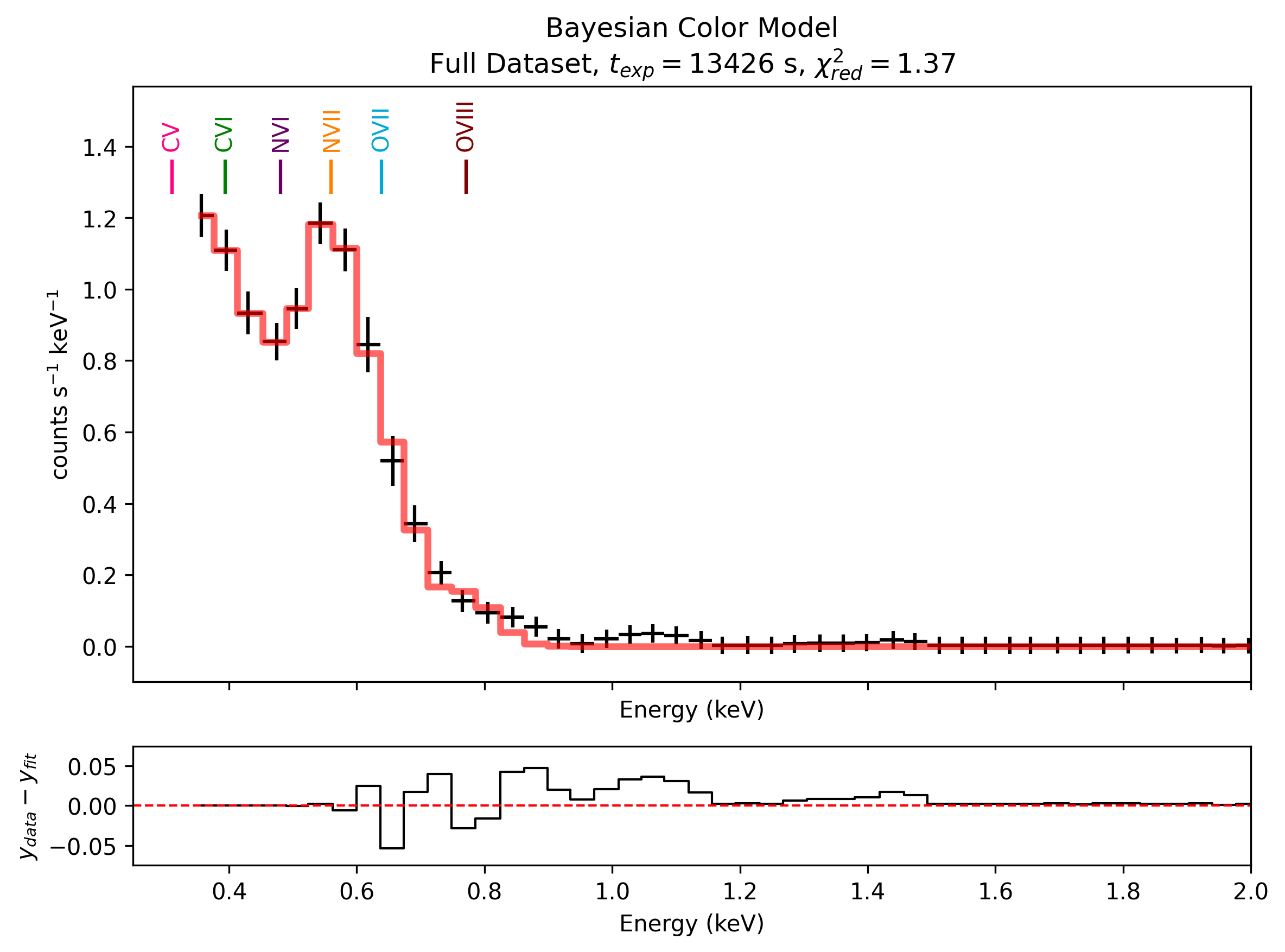}
        \caption{Full dataset ($t_{exp}=13426$\,ks).}
        \label{fig:spec_19p_full}
    \end{subcaptionblock}

    \caption{X-ray spectra from the interaction between the solar wind and the atmosphere of 19P/Borrelly fitted with the Bayesian Color Model (BCM).}
    \label{fig:app-spec-19p}
\end{figure*}


\begin{figure*}
    \centering
    \begin{subcaptionblock}[b]{0.47\textwidth}
        \includegraphics[width=\textwidth,trim={0.35cm 0.25cm 0.2cm 1.35cm},clip]{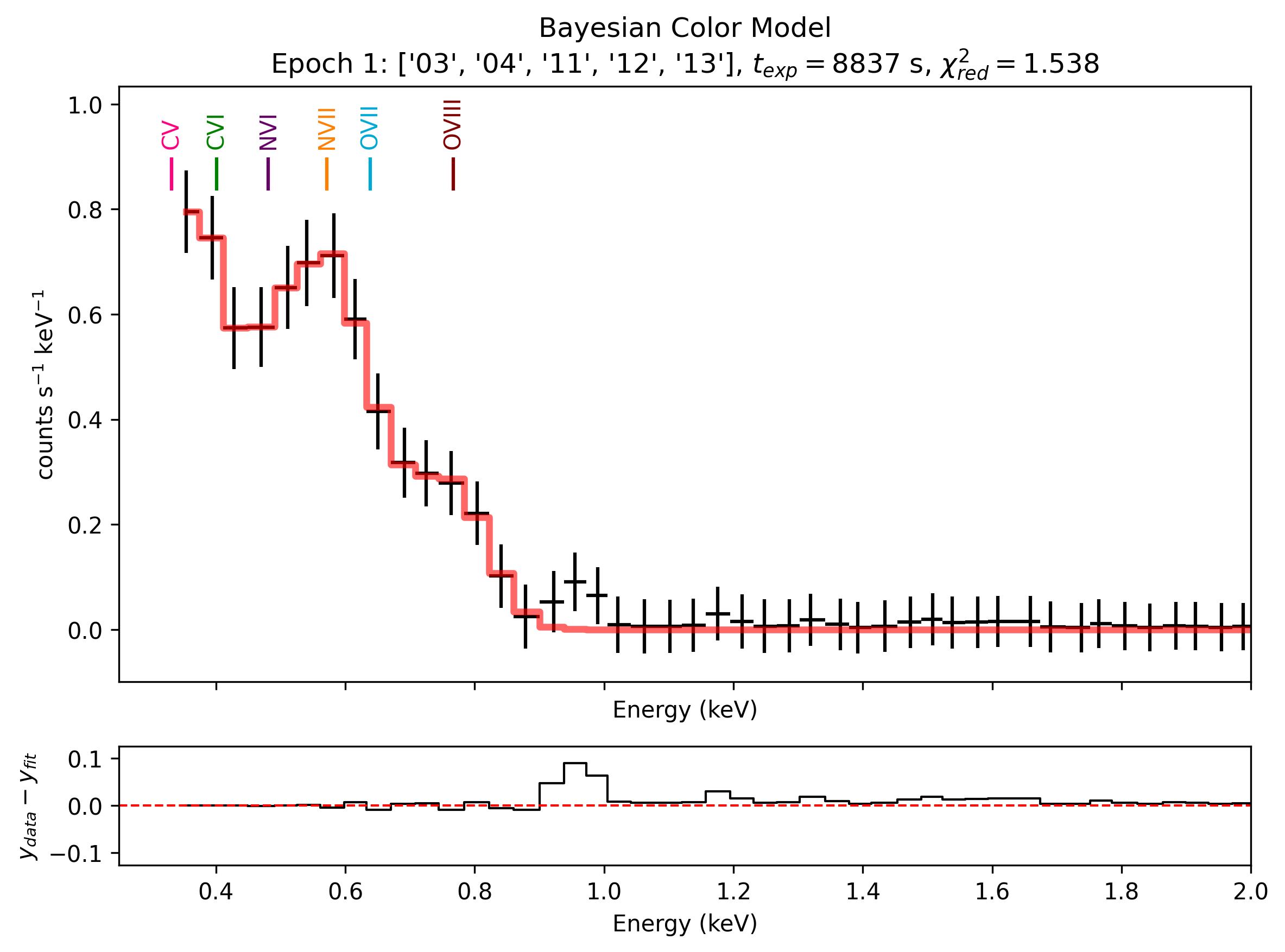}
        \caption{Epoch 1 ($t_{exp}=8837$\,ks).}
        \label{fig:spec_k2_e1}
    \end{subcaptionblock}
    \hfill
    \begin{subcaptionblock}[b]{0.47\textwidth}
        \includegraphics[width=\textwidth,trim={0.35cm 0.25cm 0.2cm 1.35cm},clip]{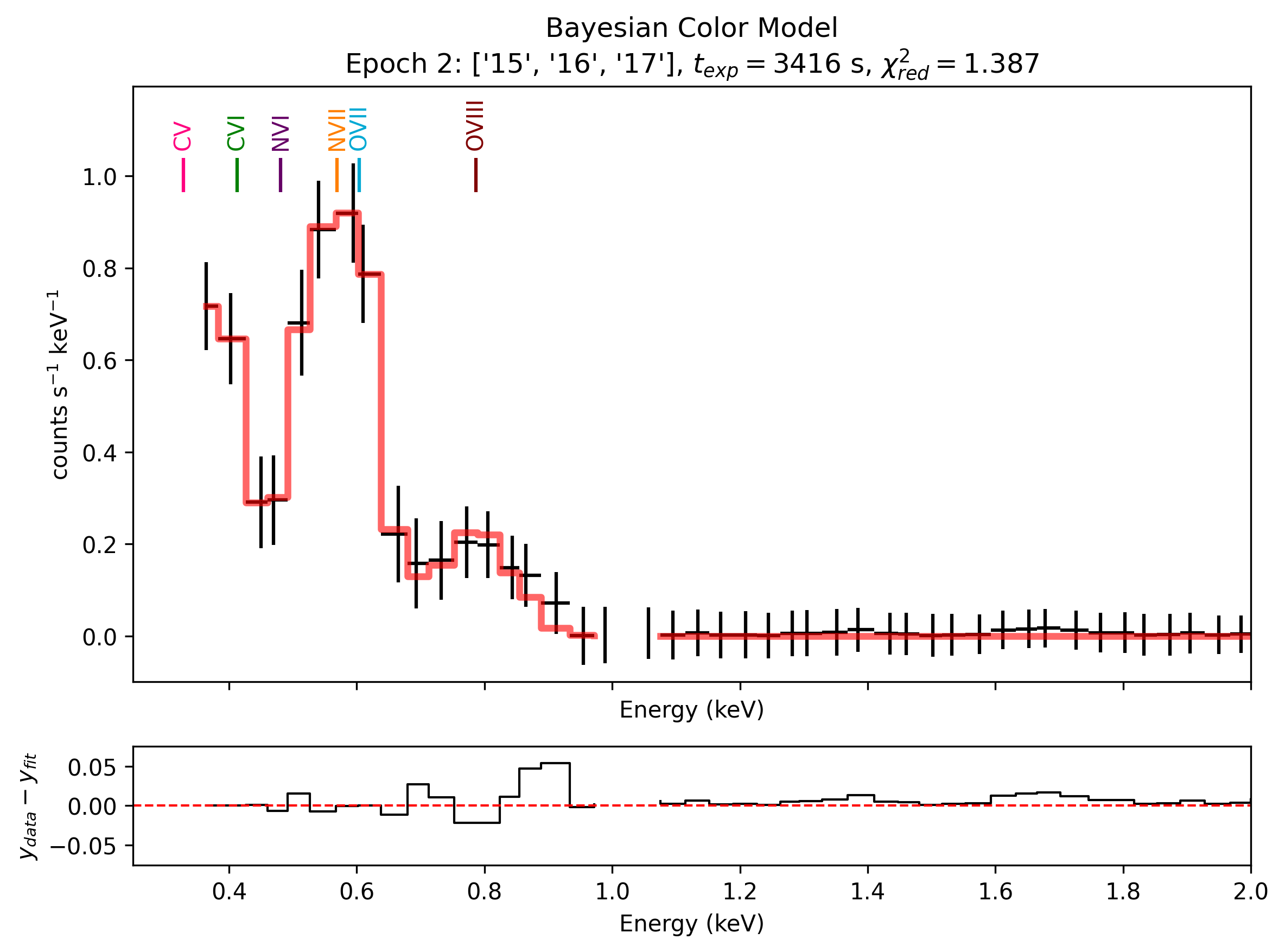}
        \caption{Epoch 2 ($t_{exp}=3416$\,ks).}
        \label{fig:spec_k2_e2}
    \end{subcaptionblock}

    \begin{subcaptionblock}[b]{0.47\textwidth}
        \includegraphics[width=\textwidth,trim={0.35cm 0.25cm 0.2cm 1.35cm},clip]{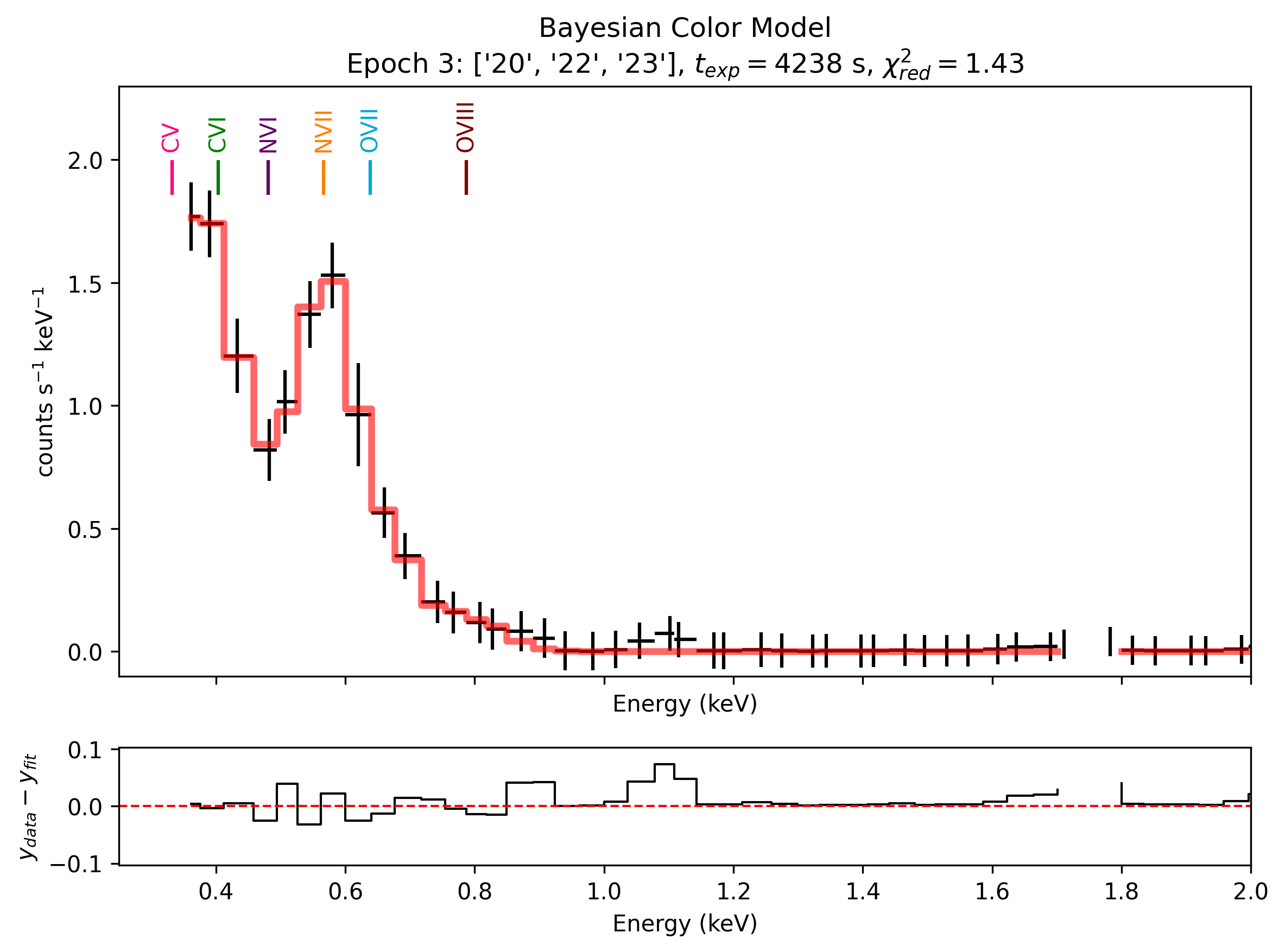}
        \caption{Epoch 3 ($t_{exp}=4238$\,ks).}
        \label{fig:spec_k2_e3}
    \end{subcaptionblock}
    \hfill
    \begin{subcaptionblock}[b]{0.47\textwidth}
        \includegraphics[width=\textwidth,trim={0.35cm 0.25cm 0.2cm 1.35cm},clip]{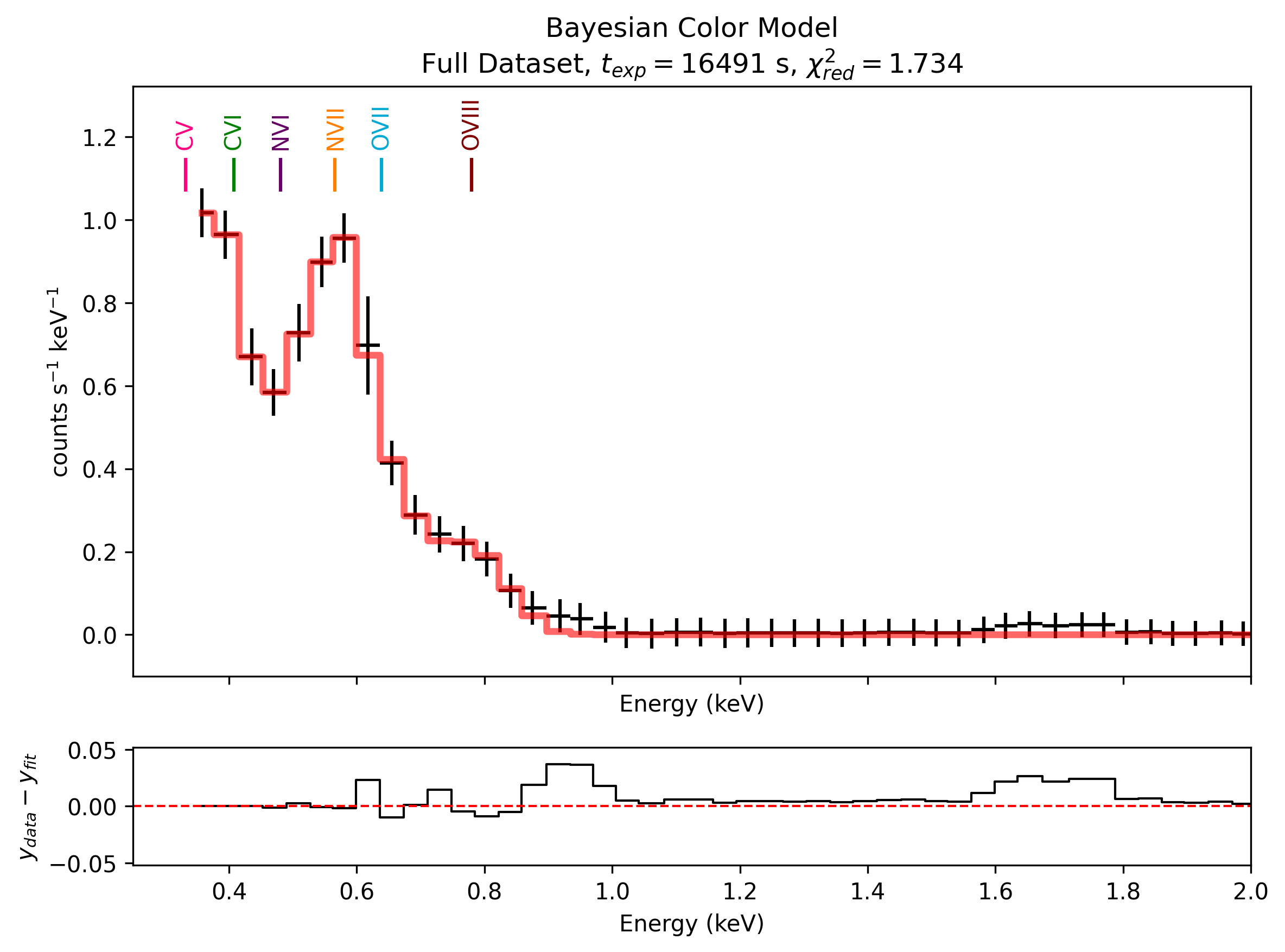}
        \caption{Full dataset ($t_{exp}=16491$\,ks).}
        \label{fig:spec_k2_full}
    \end{subcaptionblock}

    \caption{X-ray spectra from the interaction between the solar wind and the atmosphere of C/2017 K2 (PANSTARRS) fitted with the Bayesian Color Model (BCM).}
    \label{fig:app-spec-k2}
\end{figure*}


\begin{figure*}
    \centering
    \begin{subcaptionblock}[b]{0.47\textwidth}
        \includegraphics[width=\textwidth,trim={0.35cm 0.25cm 0.2cm 1.35cm},clip]{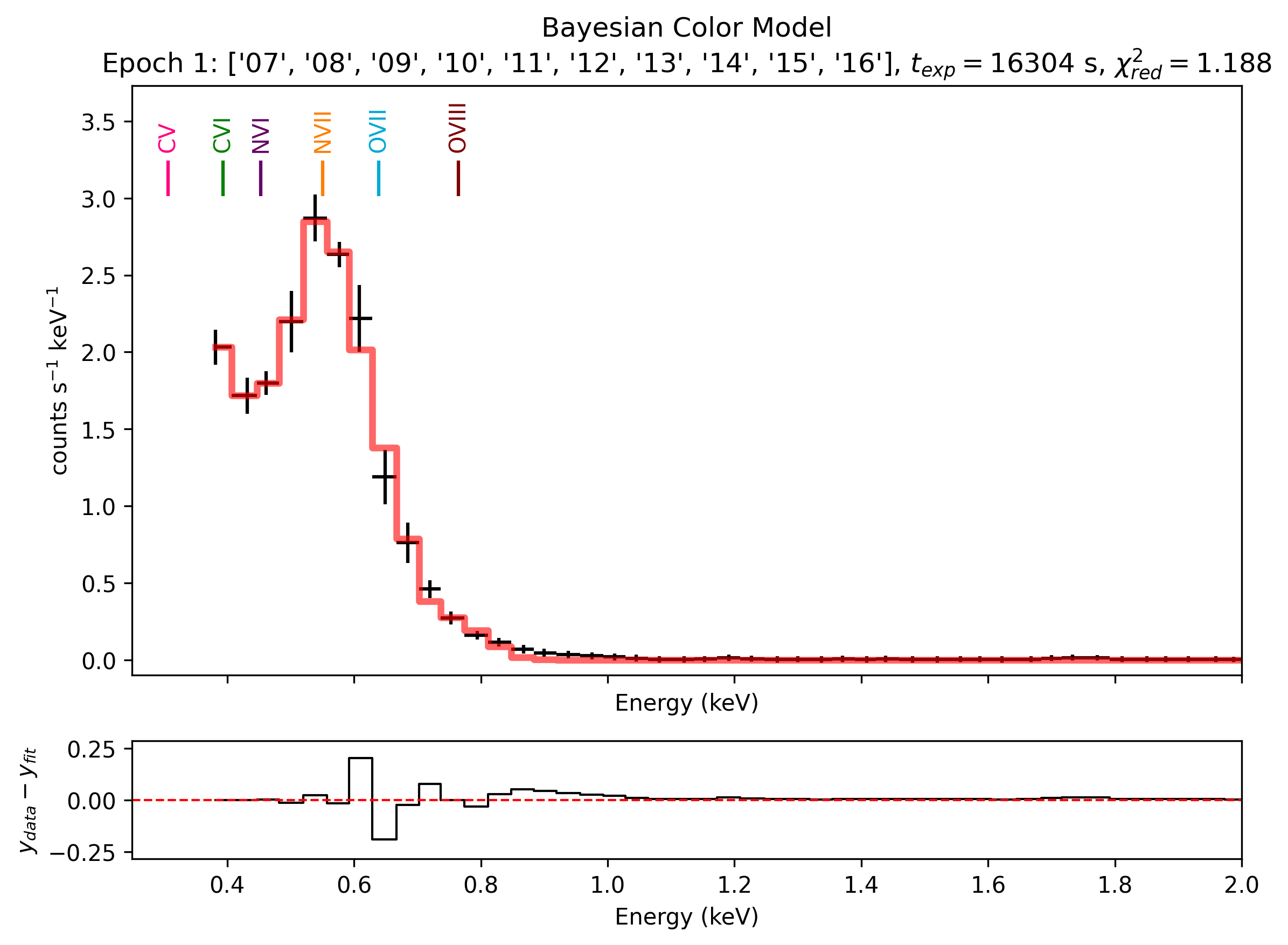}
        \caption{Epoch 1 ($t_{exp}=16304$\,ks).}
        \label{fig:spec_e3_e1}
    \end{subcaptionblock}
    \begin{subcaptionblock}[b]{0.47\textwidth}
        \includegraphics[width=\textwidth,trim={0.35cm 0.25cm 0.2cm 1.35cm},clip]{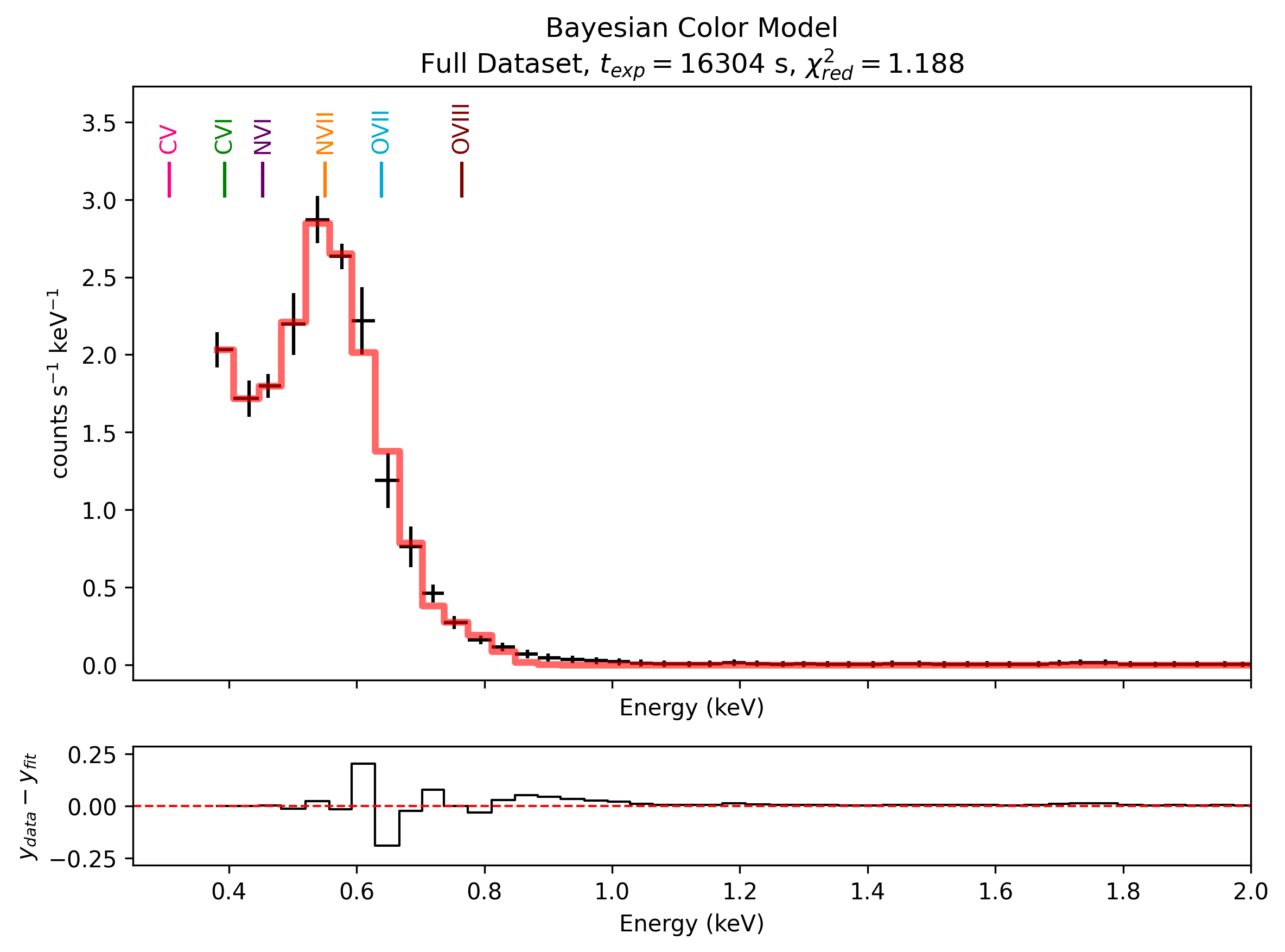}
        \caption{Full dataset ($t_{exp}=16304$\,ks).}
        \label{fig:spec_e3_full}
    \end{subcaptionblock}

    \caption{X-ray spectra from the interaction between the solar wind and the atmosphere of C/2022 E3 (ZTF) fitted with the Bayesian Color Model (BCM).}
    \label{fig:app-spec-e3}
\end{figure*}


\begin{figure*}
    \centering
    \begin{subcaptionblock}[b]{0.47\textwidth}
        \includegraphics[width=\textwidth,trim={0.35cm 0.25cm 0.2cm 1.35cm},clip]{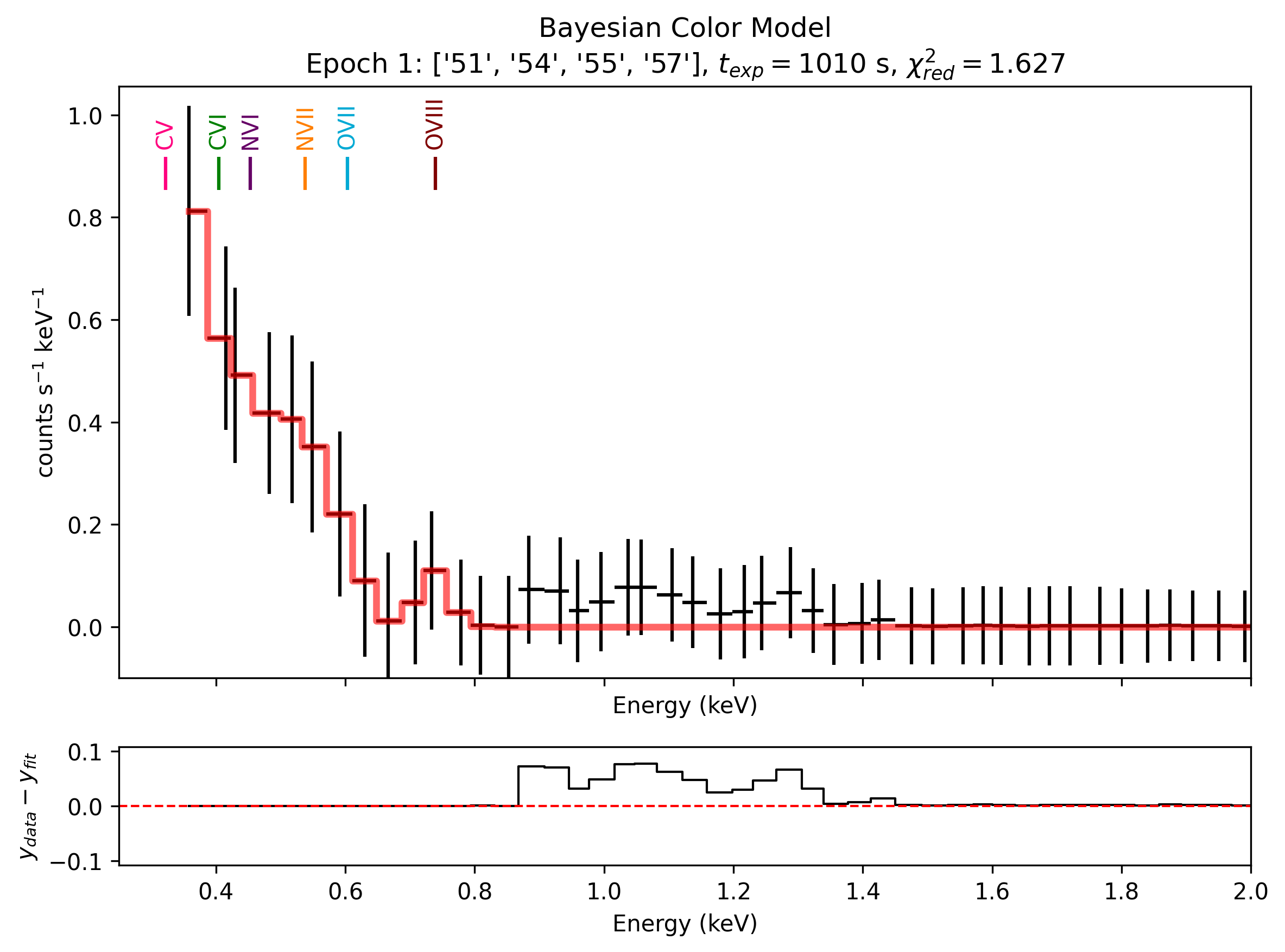}
        \caption{Epoch 1 ($t_{exp}=1010$\,ks).}
        \label{fig:spec_62p_e1}
    \end{subcaptionblock}
    \hfill
    \begin{subcaptionblock}[b]{0.47\textwidth}
        \includegraphics[width=\textwidth,trim={0.35cm 0.25cm 0.2cm 1.35cm},clip]{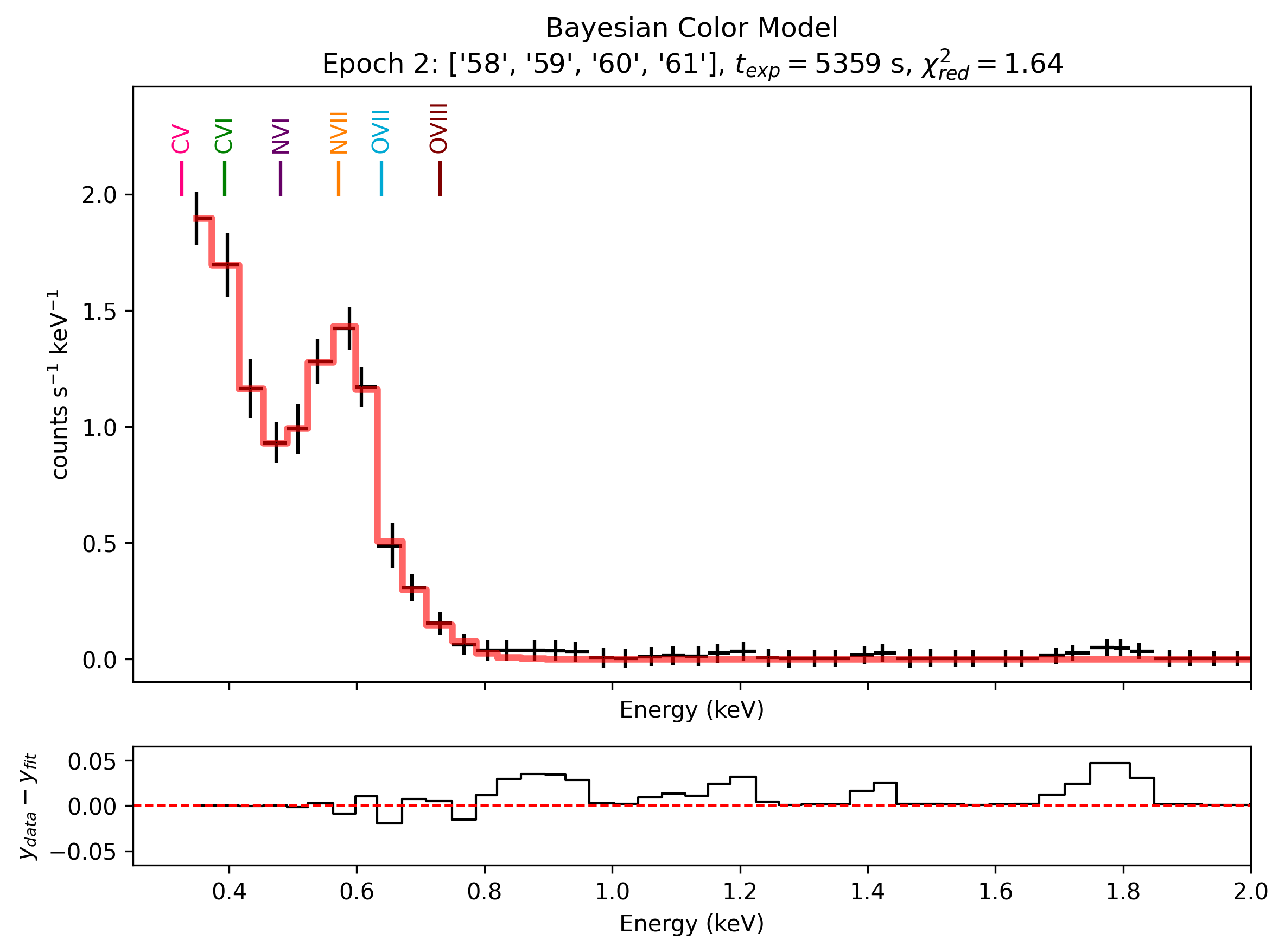}
        \caption{Epoch 2 ($t_{exp}=5359$\,ks).}
        \label{fig:spec_62p_e2}
    \end{subcaptionblock}
    \hfill
    \begin{subcaptionblock}[b]{0.47\textwidth}
        \includegraphics[width=\textwidth,trim={0.35cm 0.25cm 0.2cm 1.35cm},clip]{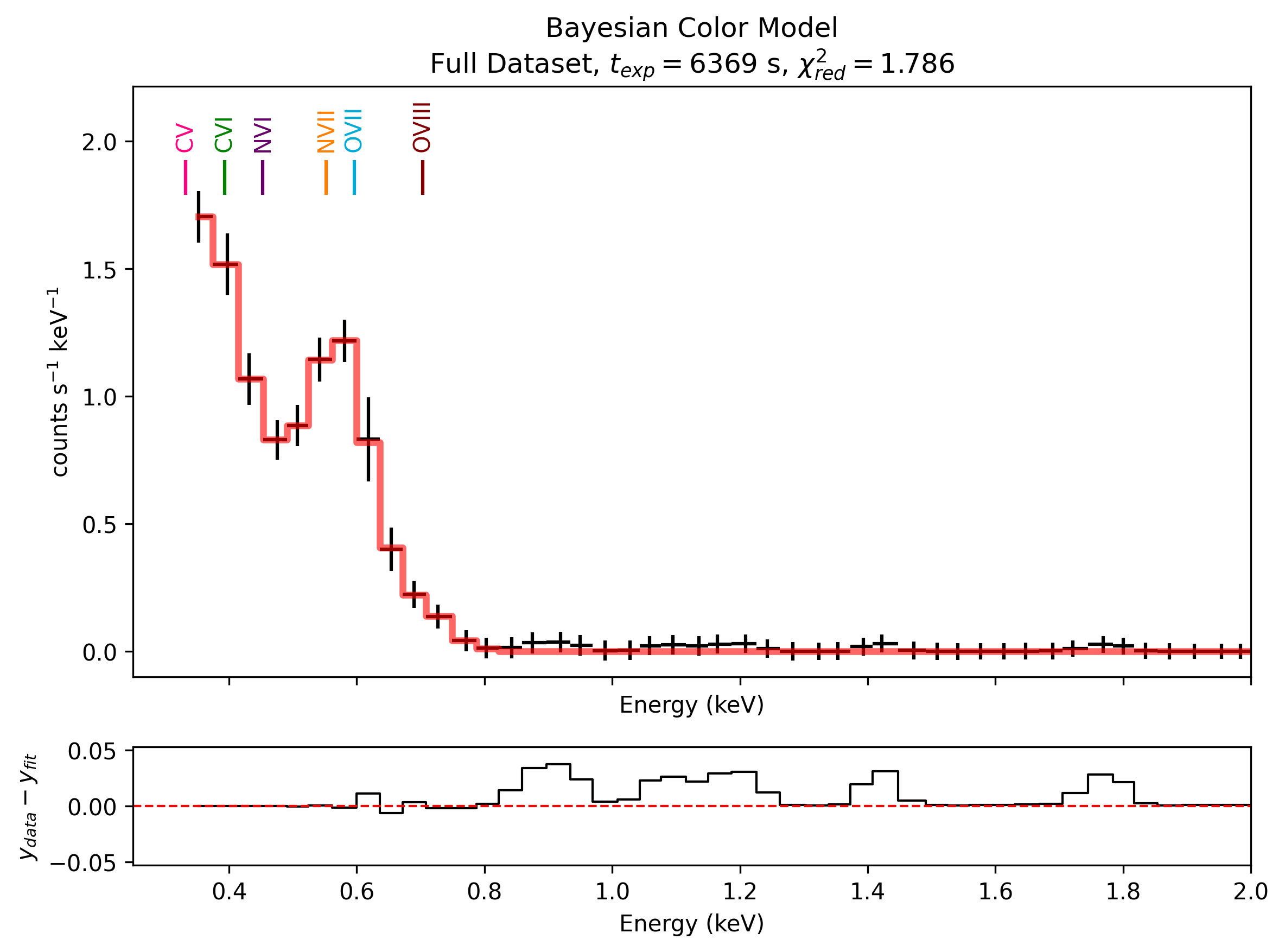}
        \caption{Full dataset ($t_{exp}=6369$\,ks).}
        \label{fig:spec_62p_full}
    \end{subcaptionblock}

    \caption{X-ray spectra from the interaction between the solar wind and the atmosphere of 62P/Tsuchinshan fitted with the Bayesian Color Model (BCM).}
    \label{fig:app-spec-62p}
\end{figure*}


\begin{figure*}
    \centering
    \begin{subcaptionblock}[b]{0.47\textwidth}
        \includegraphics[width=\textwidth,trim={0.35cm 0.25cm 0.2cm 1.35cm},clip]{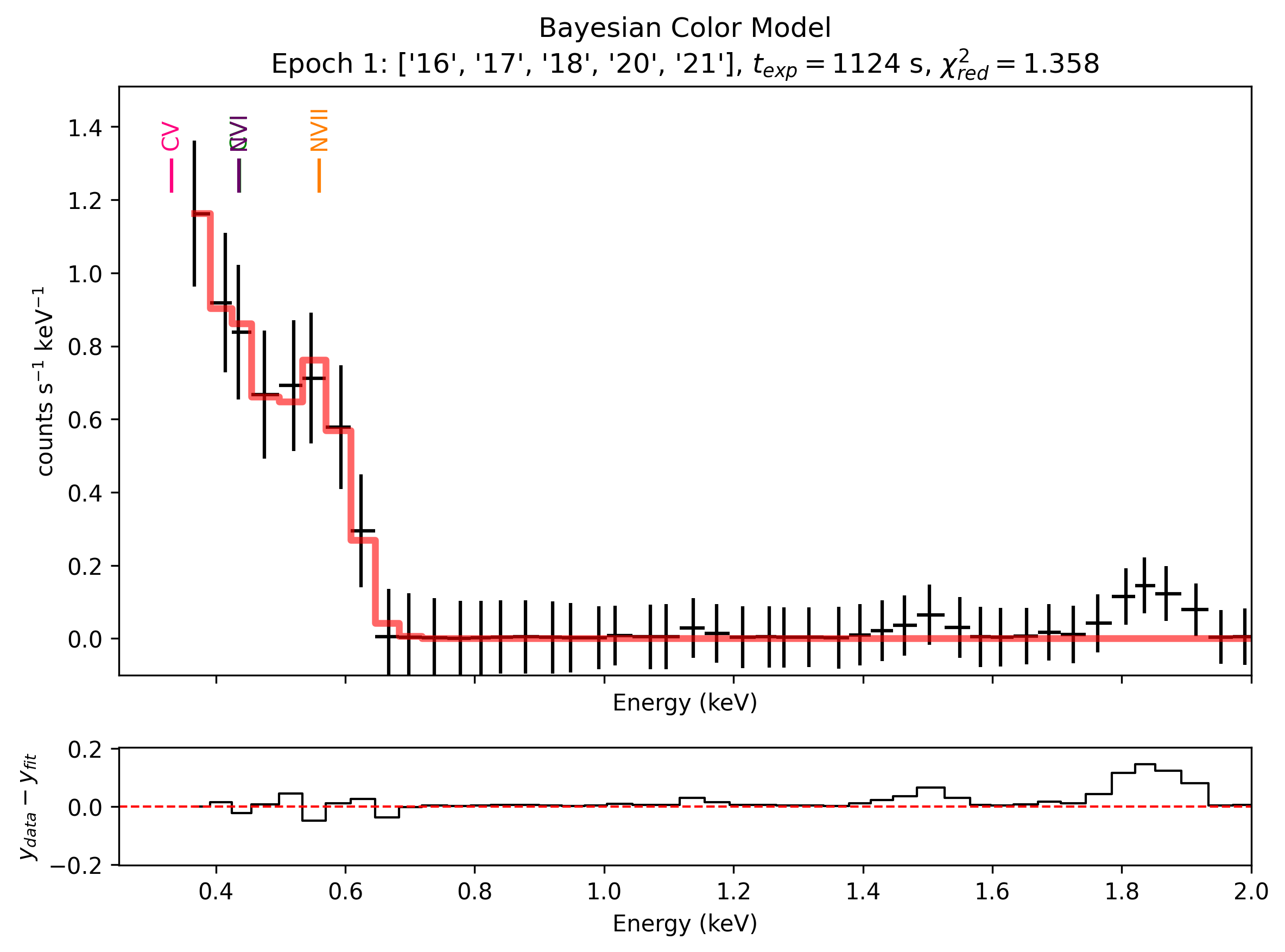}
        \caption{Epoch 1 ($t_{exp}=1124$\,ks).}
        \label{fig:spec_12p_e1}
    \end{subcaptionblock}
    \hfill
    \begin{subcaptionblock}[b]{0.47\textwidth}
        \includegraphics[width=\textwidth,trim={0.35cm 0.25cm 0.2cm 1.35cm},clip]{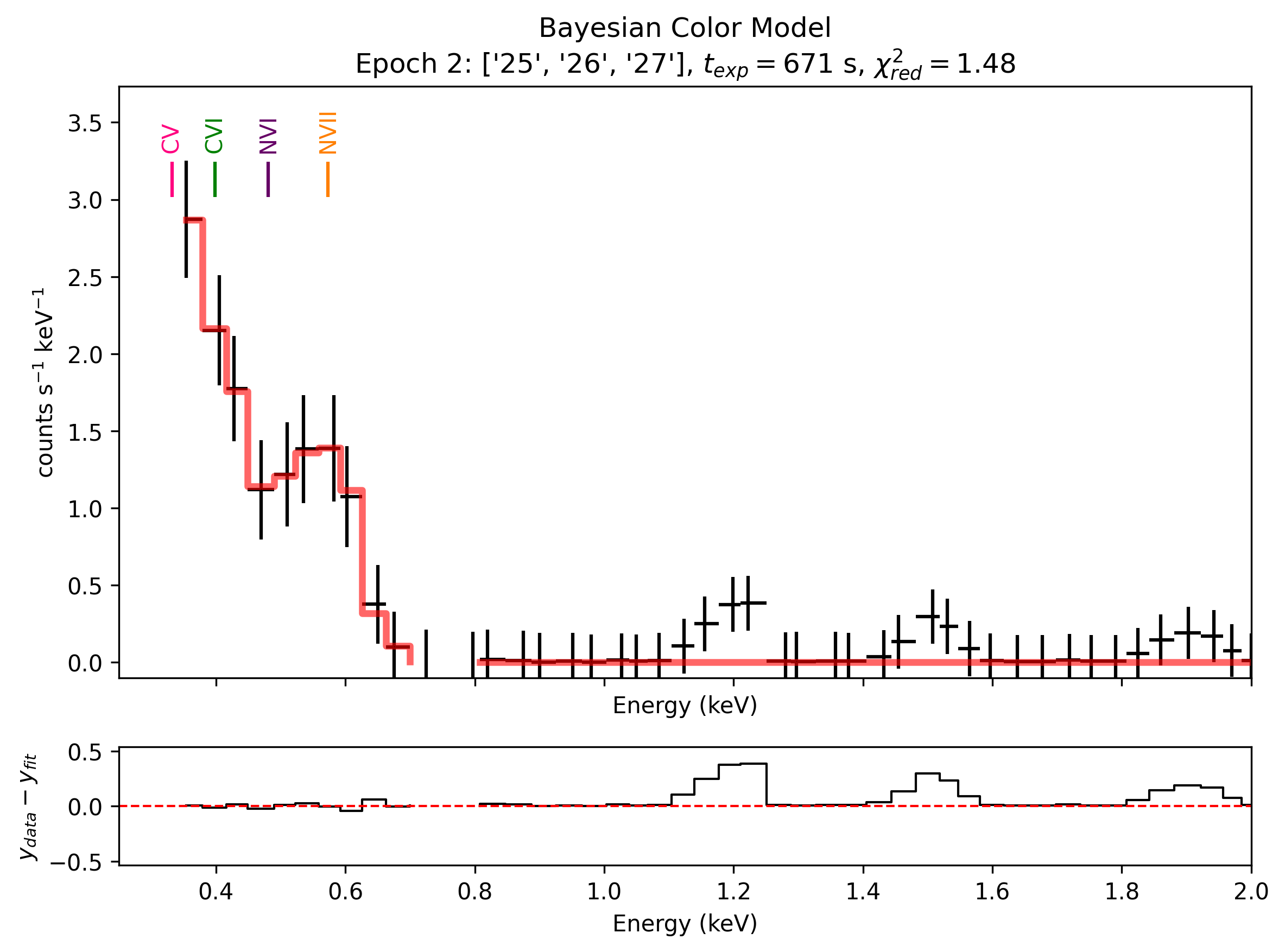}
        \caption{Epoch 2 ($t_{exp}=671$\,ks).}
        \label{fig:spec_12p_e2}
    \end{subcaptionblock}
    \hfill
    \begin{subcaptionblock}[b]{0.47\textwidth}
        \includegraphics[width=\textwidth,trim={0.35cm 0.25cm 0.2cm 1.35cm},clip]{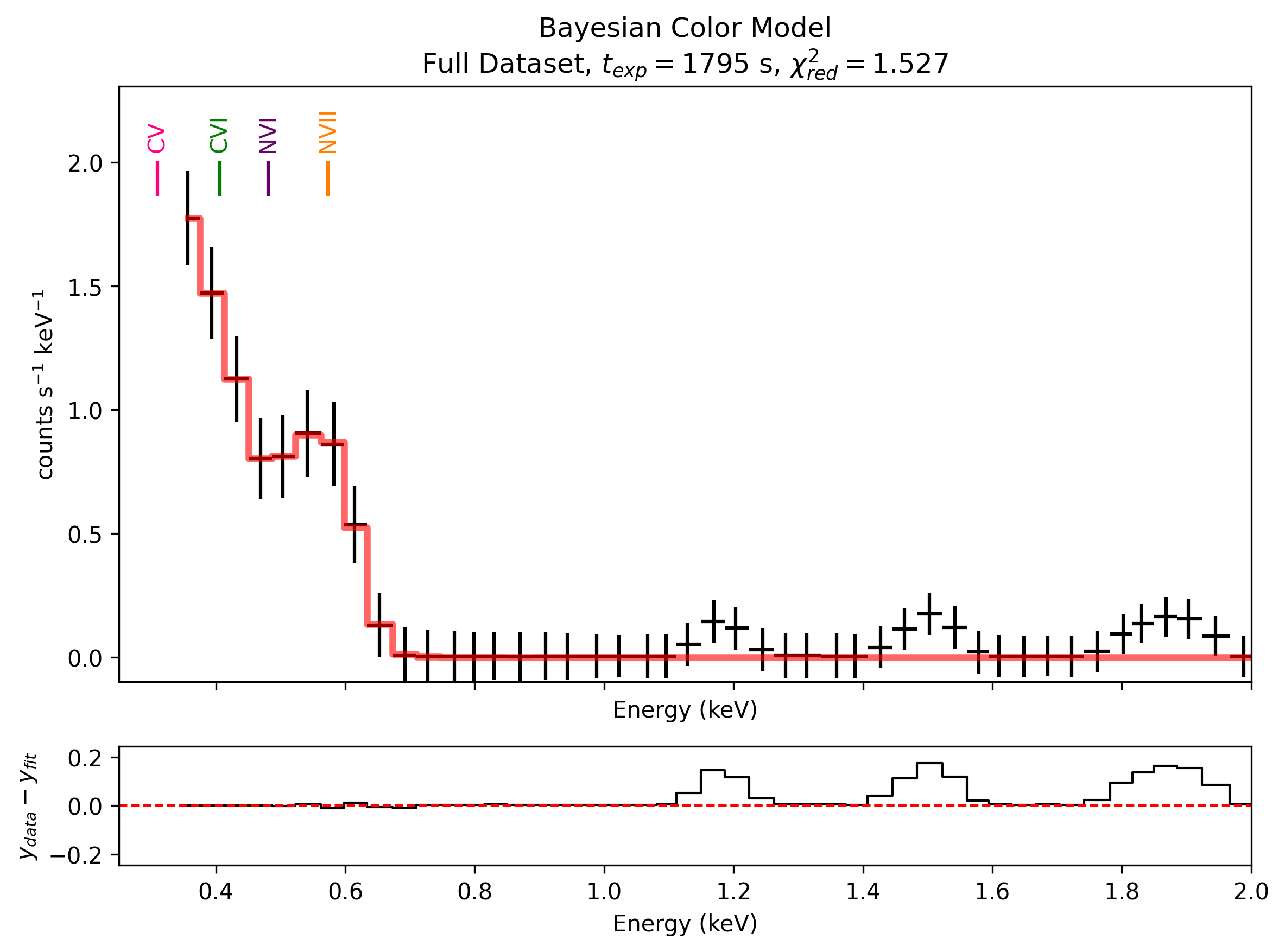}
        \caption{Full dataset ($t_{exp}=1795$\,ks).}
        \label{fig:spec_12p_full}
    \end{subcaptionblock}

    \caption{X-ray spectra from the interaction between the solar wind and the atmosphere of 12P/Pons–Brooks fitted with the Bayesian Color Model (BCM).}
    \label{fig:app-spec-12p}
\end{figure*}


\begin{figure*}
    \centering
    \begin{subcaptionblock}[b]{0.47\textwidth}
        \includegraphics[width=\textwidth,trim={0.35cm 0.25cm 0.2cm 1.35cm},clip]{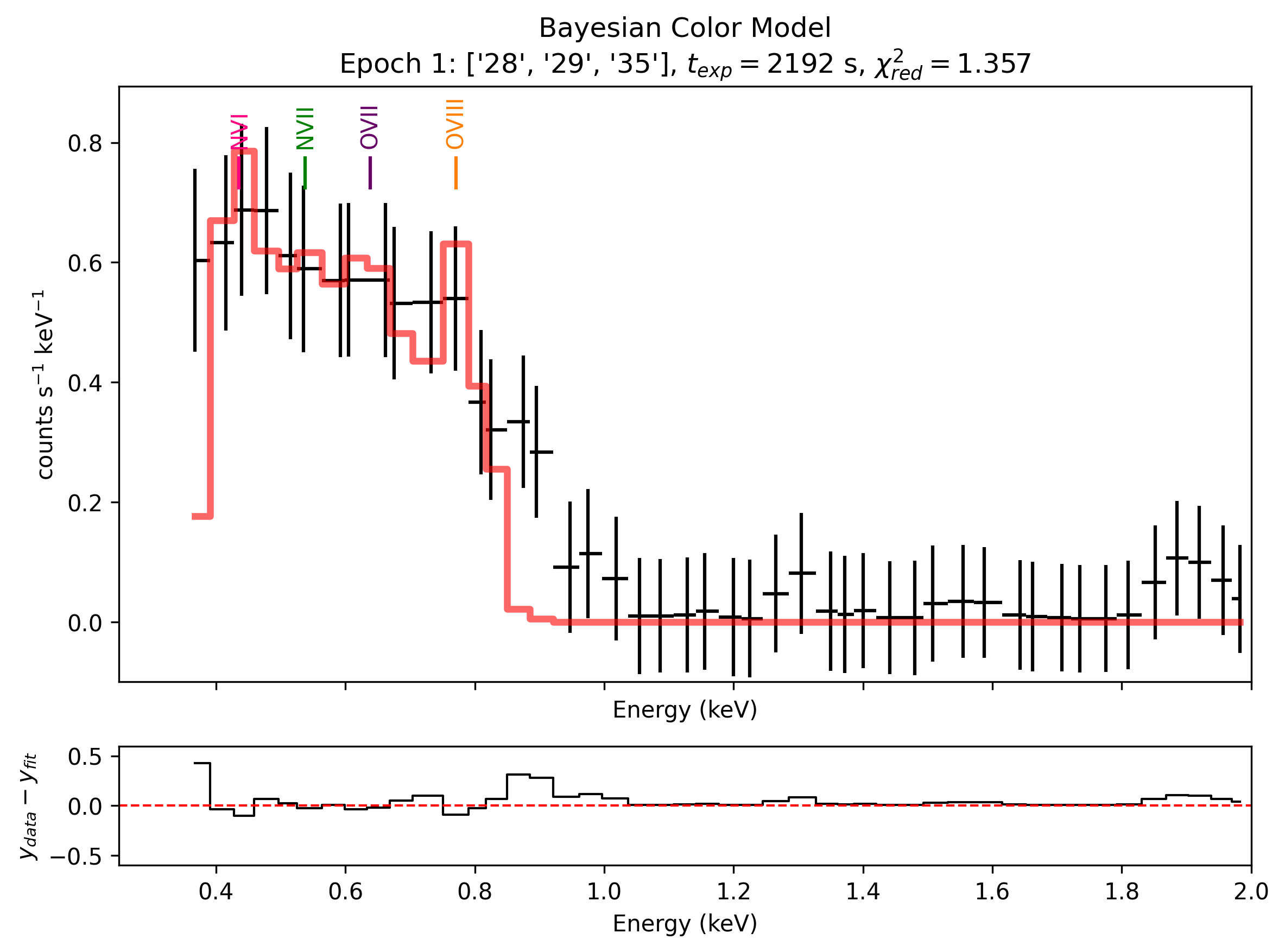}
        \caption{Epoch 1 ($t_{exp}=2192$\,ks).}
        \label{fig:spec_h2_e1}
    \end{subcaptionblock}
    \hfill
    \begin{subcaptionblock}[b]{0.47\textwidth}
        \includegraphics[width=\textwidth,trim={0.35cm 0.25cm 0.2cm 1.35cm},clip]{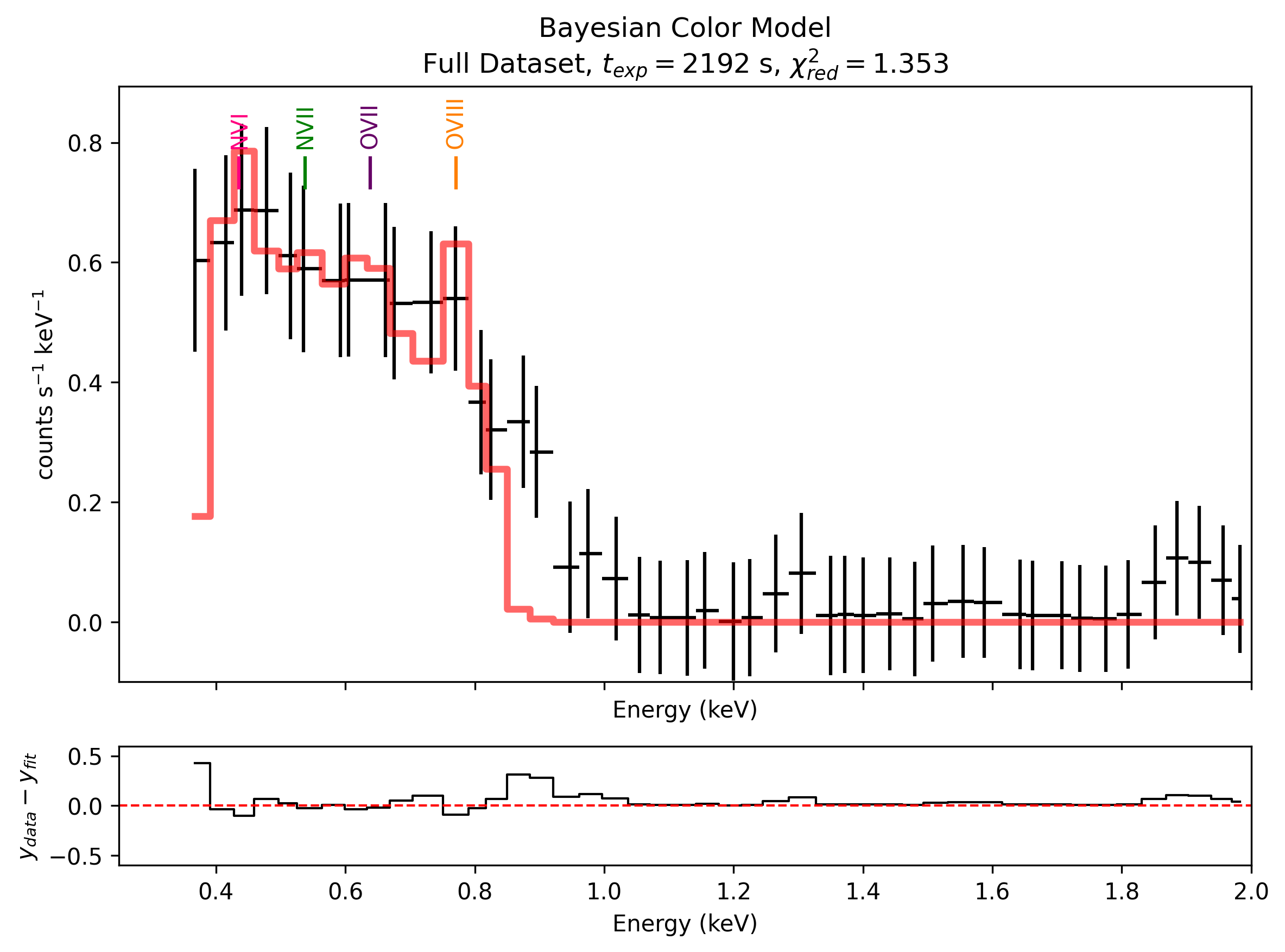}
        \caption{Full dataset ($t_{exp}=2192$\,ks).}
        \label{fig:spec_h2_full}
    \end{subcaptionblock}

    \caption{X-ray spectra from the interaction between the solar wind and the atmosphere of P/2010 H2 (Vales) fitted with the Bayesian Color Model (BCM). Given the low quality of the spectra, H2 is not considered in the results.}
    \label{fig:app-spec-h2}
\end{figure*}

\section{Freeze-In Temperatures}
\label{app:freezeintemps}
This section presents epoch-resolved freeze-in temperatures inferred from BCM flux ratios for each comet in our study. Each figure corresponds to one comet and shows $T_{\rm freeze}$ derived from \cvi\ / \cv (blue), \nvii\ / \nvi (green) and \oviii\ / \ovii (red) at the epoch mid-times listed in Table~\ref{tab:flux_ratios_epochs}; vertical error bars are the $1\sigma$ uncertainties propagated from the line-fit errors and solid lines join consecutive epochs to guide the eye. Readers may consult Table~\ref{tab:flux_ratios_epochs} for numeric values.

\begin{figure*}
    \centering
    \includegraphics[width=0.65\linewidth,trim={0cm 0cm 0cm 0cm},clip]{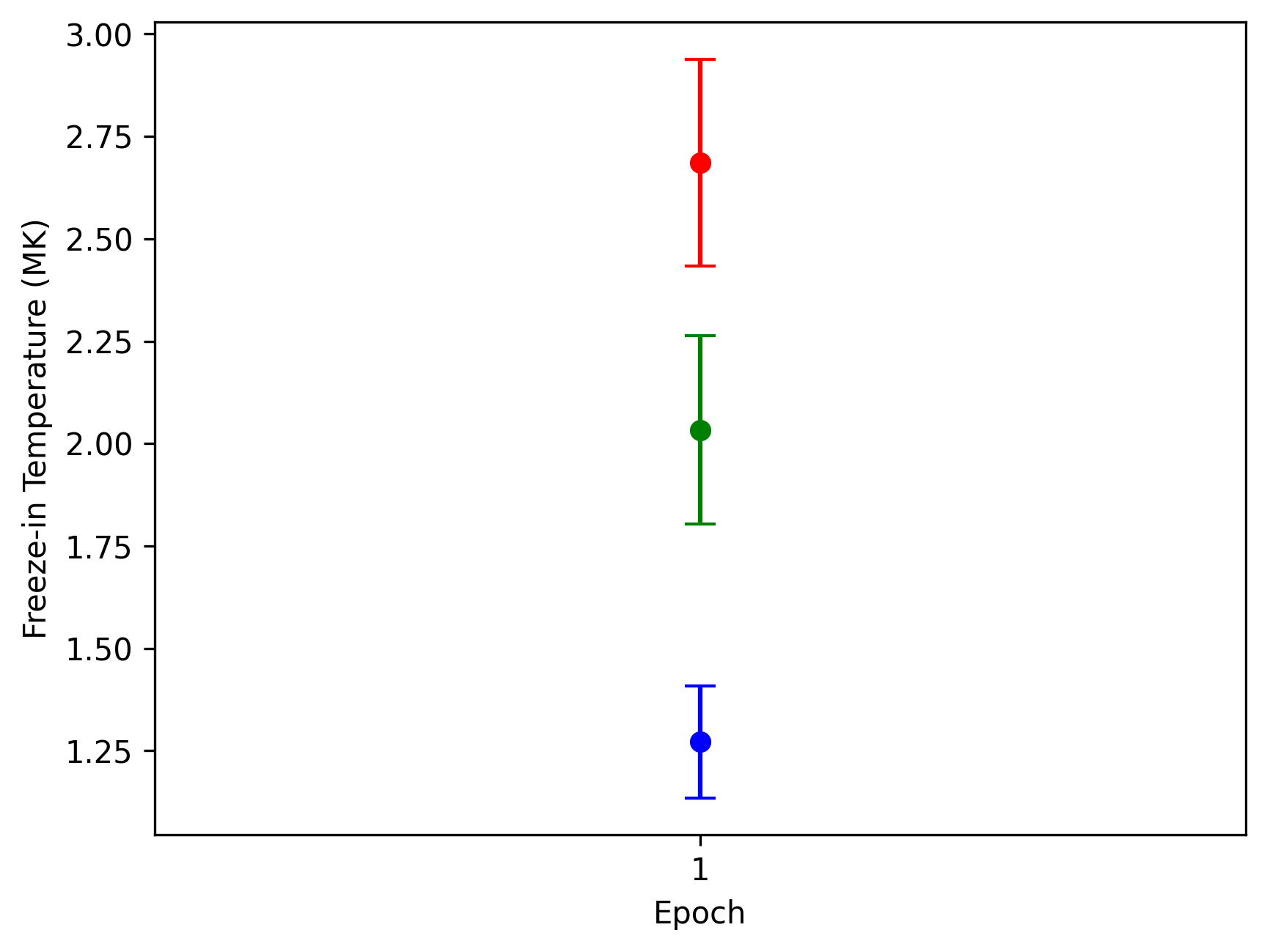}
    \caption{Freeze-in temperatures for comet C/2017 T2 inferred from flux ratios, plotted versus observing epoch. Blue shows $T_{\rm freeze}$ derived from \cvi\ / \cv, green is from \nvii\ / \nvi, and red is from \oviii\ / \ovii; vertical bars are $1\sigma$ uncertainties propagated from the line-fit errors.}
    \label{fig:t2_freezeintemps}
\end{figure*}

\begin{figure*}
    \centering
    \includegraphics[width=0.65\linewidth,trim={0cm 0cm 0cm 0cm},clip]{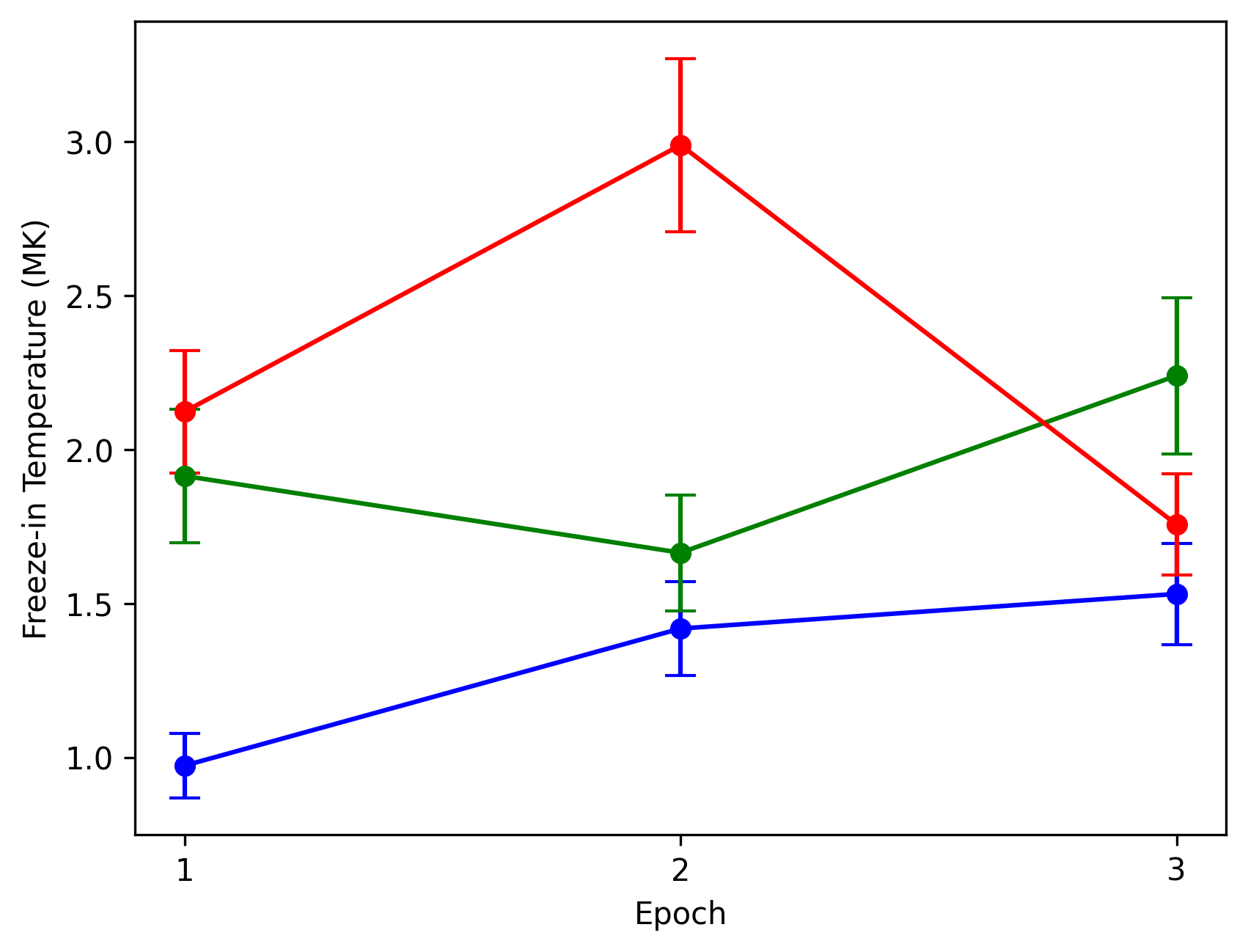}
    \caption{Freeze-in temperatures for comet 88P/Howell inferred from flux ratios, plotted versus observing epoch. Blue shows $T_{\rm freeze}$ derived from \cvi\ / \cv, green is from \nvii\ / \nvi, and red is from \oviii\ / \ovii; vertical bars are $1\sigma$ uncertainties propagated from the line-fit errors.}
    \label{fig:88p_freezeintemps}
\end{figure*}

\begin{figure*}
    \centering
    \includegraphics[width=0.65\linewidth,trim={0cm 0cm 0cm 0cm},clip]{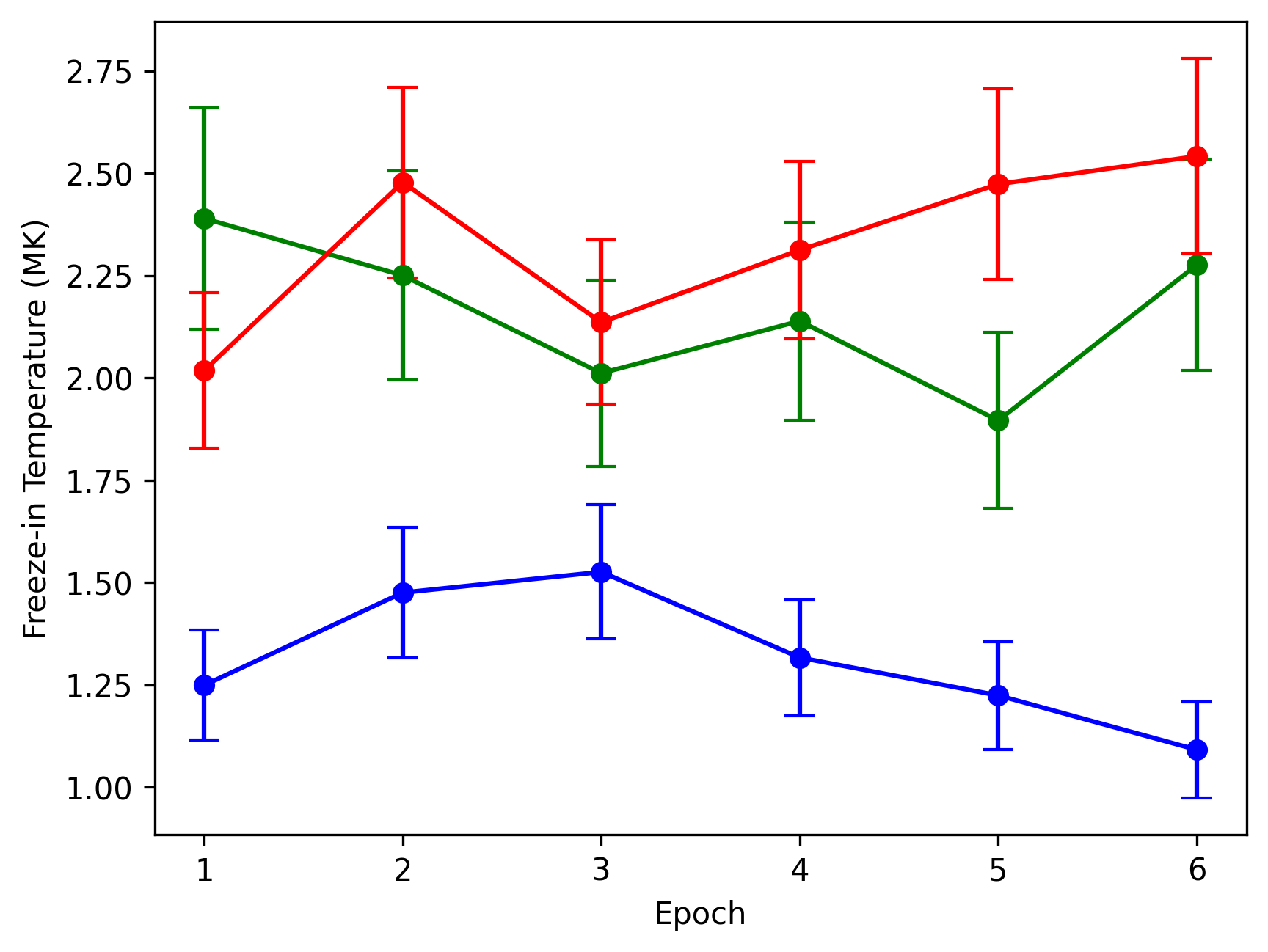}
    \caption{Freeze-in temperatures for comet 67P/Churyumov–Gerasimenko inferred from flux ratios, plotted versus observing epoch. Blue shows $T_{\rm freeze}$ derived from \cvi\ / \cv, green is from \nvii\ / \nvi, and red is from \oviii\ / \ovii; vertical bars are $1\sigma$ uncertainties propagated from the line-fit errors.}
    \label{fig:67p_freezeintemps}
\end{figure*}

\begin{figure*}
    \centering
    \includegraphics[width=0.65\linewidth,trim={0cm 0cm 0cm 0cm},clip]{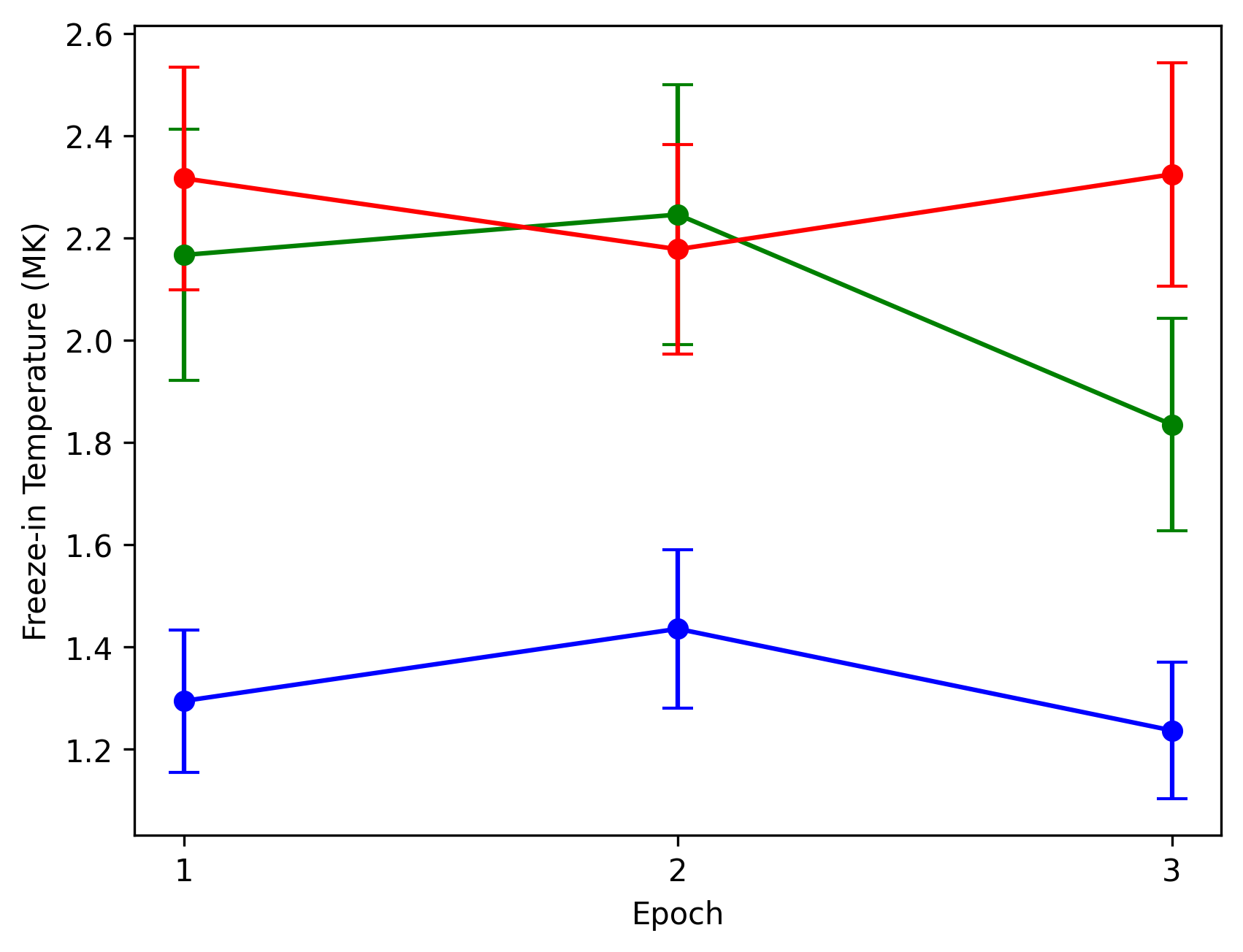}
    \caption{Freeze-in temperatures for comet 19P/Borrelly inferred from flux ratios, plotted versus observing epoch. Blue shows $T_{\rm freeze}$ derived from \cvi\ / \cv, green is from \nvii\ / \nvi, and red is from \oviii\ / \ovii; vertical bars are $1\sigma$ uncertainties propagated from the line-fit errors.}
    \label{fig:19P_freezeintemps}
\end{figure*}

\begin{figure*}
    \centering
    \includegraphics[width=0.65\linewidth,trim={0cm 0cm 0cm 0cm},clip]{Plots/K2_freezeInTemp_epochs.png}
    \caption{Freeze-in temperatures for comet C/2017 K2 (PANSTARRS) inferred from flux ratios, plotted versus observing epoch. Blue shows $T_{\rm freeze}$ derived from \cvi\ / \cv, green is from \nvii\ / \nvi, and red is from \oviii\ / \ovii; vertical bars are $1\sigma$ uncertainties propagated from the line-fit errors.}
    \label{fig:K2_freezeintemps}
\end{figure*}

\begin{figure*}
    \centering
    \includegraphics[width=0.65\linewidth,trim={0cm 0cm 0cm 0cm},clip]{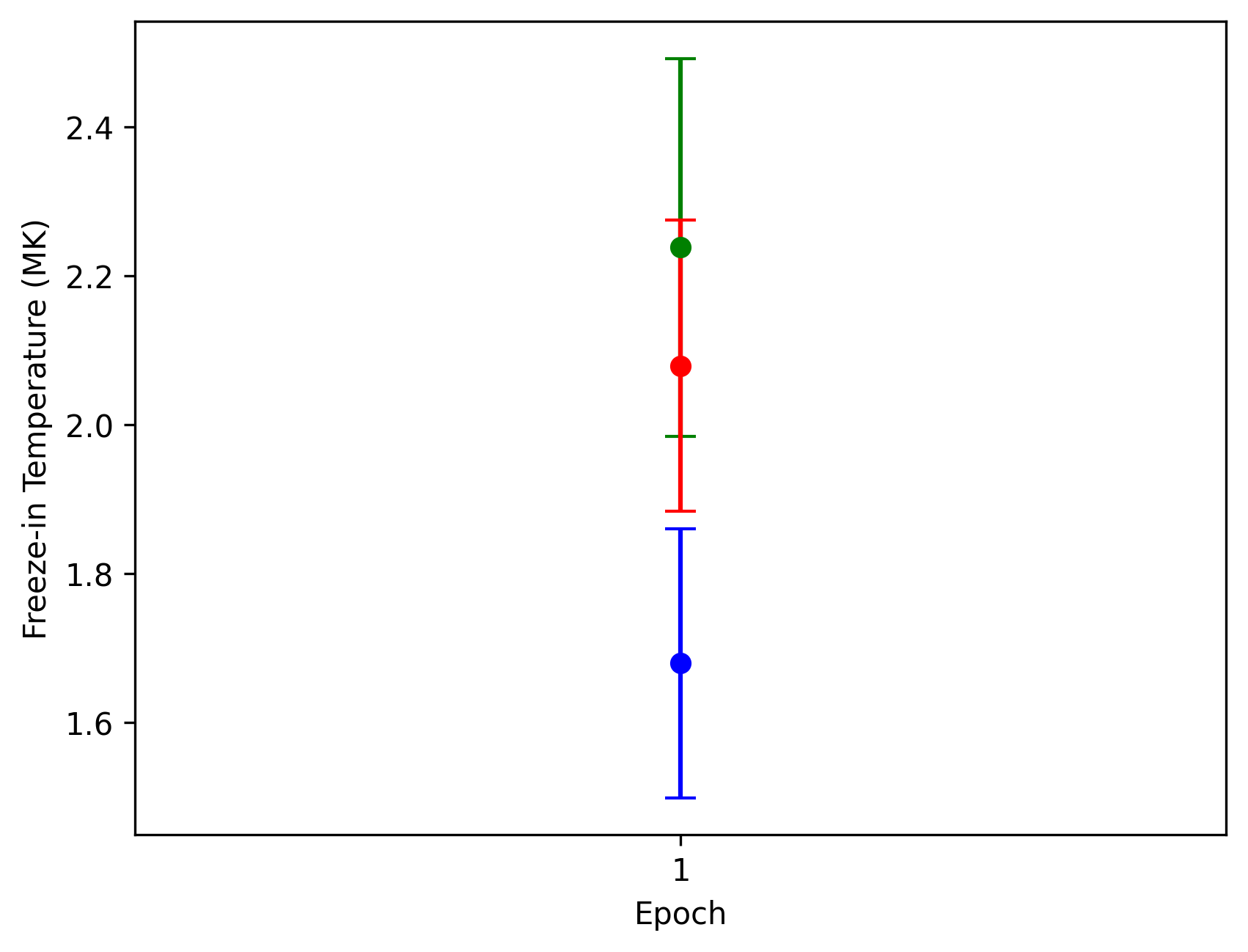}
    \caption{Freeze-in temperatures for comet C/2022 E3 (ZTF) inferred from flux ratios, plotted versus observing epoch. Blue shows $T_{\rm freeze}$ derived from \cvi\ / \cv, green is from \nvii\ / \nvi, and red is from \oviii\ / \ovii; vertical bars are $1\sigma$ uncertainties propagated from the line-fit errors.}
    \label{fig:E3_freezeintemps}
\end{figure*}

\begin{figure*}
    \centering
    \includegraphics[width=0.65\linewidth,trim={0cm 0cm 0cm 0cm},clip]{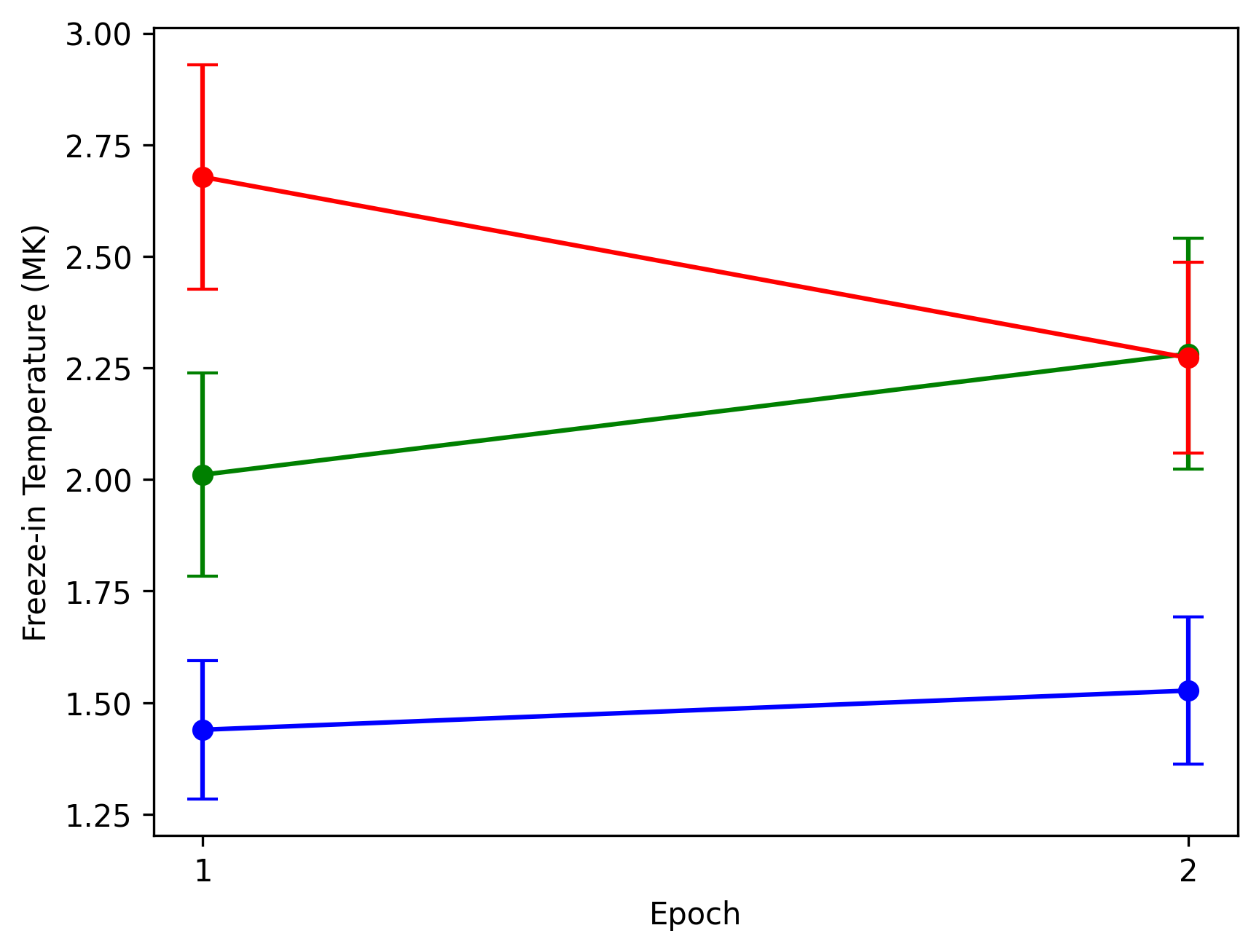}
    \caption{Freeze-in temperatures for comet 62P/Tsuchinshan inferred from flux ratios, plotted versus observing epoch. Blue shows $T_{\rm freeze}$ derived from \cvi\ / \cv, green is from \nvii\ / \nvi, and red is from \oviii\ / \ovii; vertical bars are $1\sigma$ uncertainties propagated from the line-fit errors.}
    \label{fig:62P_freezeintemps}
\end{figure*}

\begin{figure*}
    \centering
    \includegraphics[width=0.65\linewidth,trim={0cm 0cm 0cm 0cm},clip]{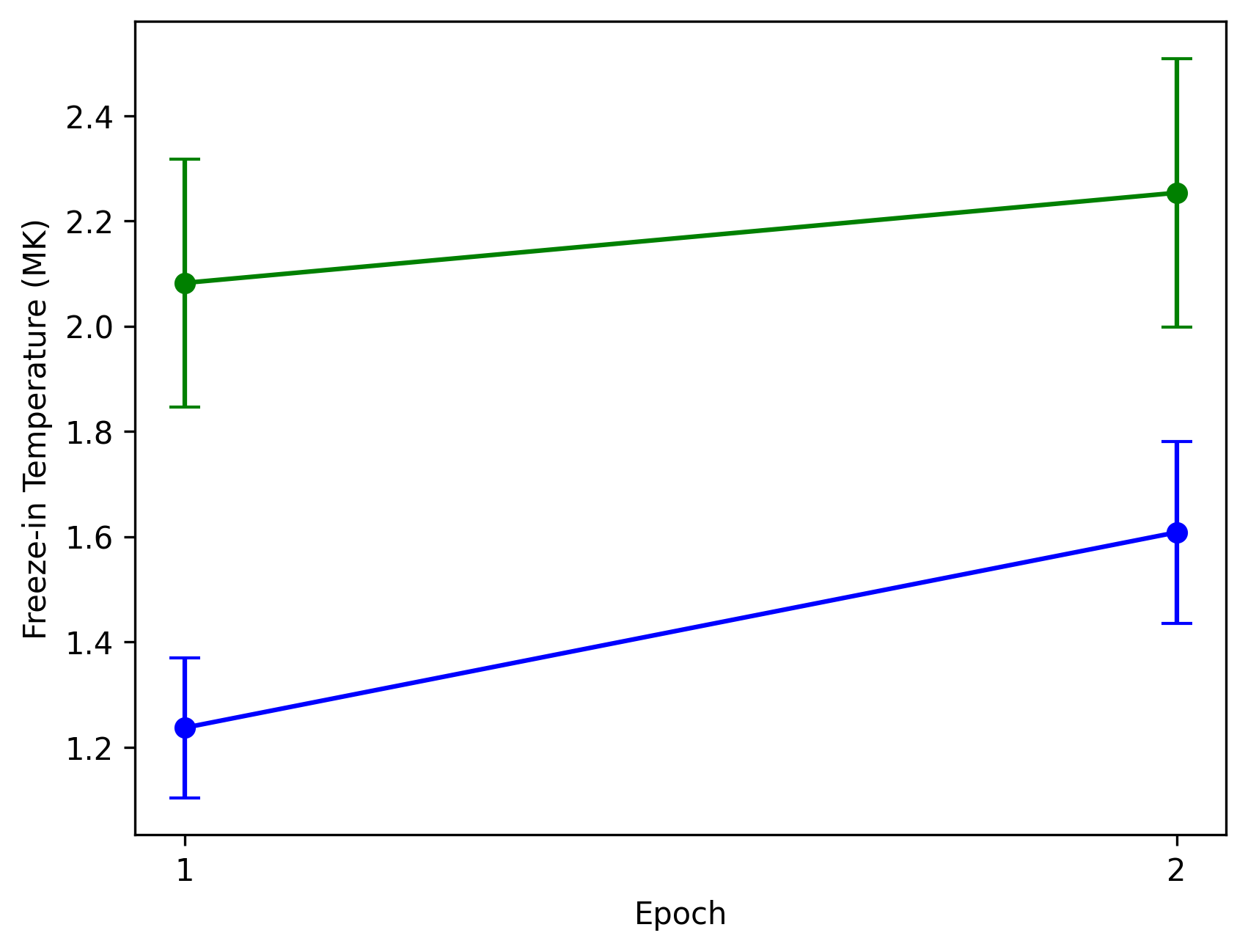}
    \caption{Freeze-in temperatures for comet 12P/Pons-Brooks inferred from flux ratios, plotted versus observing epoch. Blue shows $T_{\rm freeze}$ derived from \cvi\ / \cv and green is from \nvii\ / \nvi. There is no freeze-in temperature derived from oxygen, as the oxygen pair was not selected by the BCM model for 12P. Vertical bars are $1\sigma$ uncertainties propagated from the line-fit errors.}
    \label{fig:12P_freezeintemps}
\end{figure*}

\section{ENLIL}
\label{app:ENLIL}
This section presents simulated solar-wind parameters from the ENLIL model, using GONG boundary conditions, evaluated at each comet's position. The data includes time-series of the solar wind’s velocity ($V_r$), density ($N$), temperature ($T$), and magnetic field intensity ($|B|$). Key features such as K Shocks (marked by vertical blue lines) and coronal mass ejections (CMEs) (highlighted in yellow) are annotated to show their timing and impact on the solar wind conditions. The variability in these parameters is linked to shock and CME events, as well as changes in the interplanetary magnetic field (IMF) polarity, providing insight into the dynamic nature of the solar wind during this period.

\begin{figure}
    \centering
    \includegraphics[width=0.9\linewidth,trim={0.2cm .8cm 1cm 1.5cm},clip]{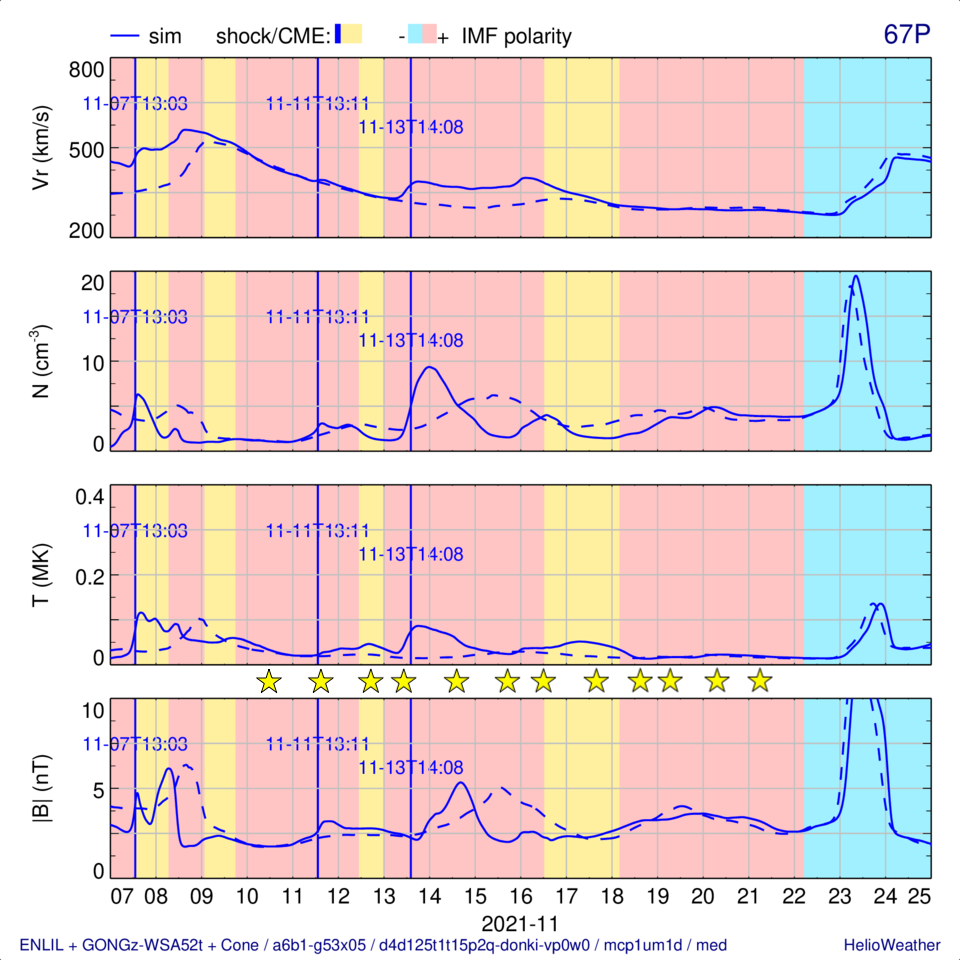}
    \caption{Simulated solar-wind parameters from the ENLIL model with GONG boundary conditions, evaluated at the position of Comet 67P Churyumov-Gerasimenko in November 2021. The horizontal panels display the velocity ($V_r$), density ($N$), temperature ($T$), and magnetic field intensity ($|B|$) of the solar wind. K Shocks are marked by vertical blue lines with annotated times, while coronal mass ejections are highlighted in yellow. Times accompanying vertical blue lines denote shocks. Variations in the solar-wind parameters coincide with shock/CME events and IMF polarity changes, suggesting dynamic solar wind conditions. Each observation is indicated by a star in the figure. Figure provided by Du\v san Odstr\v cil.}
    \label{fig:67p-11-21}
\end{figure}

\begin{figure}
    \centering
    \includegraphics[width=0.9\linewidth,trim={0.2cm .8cm 1cm 1.5cm},clip]{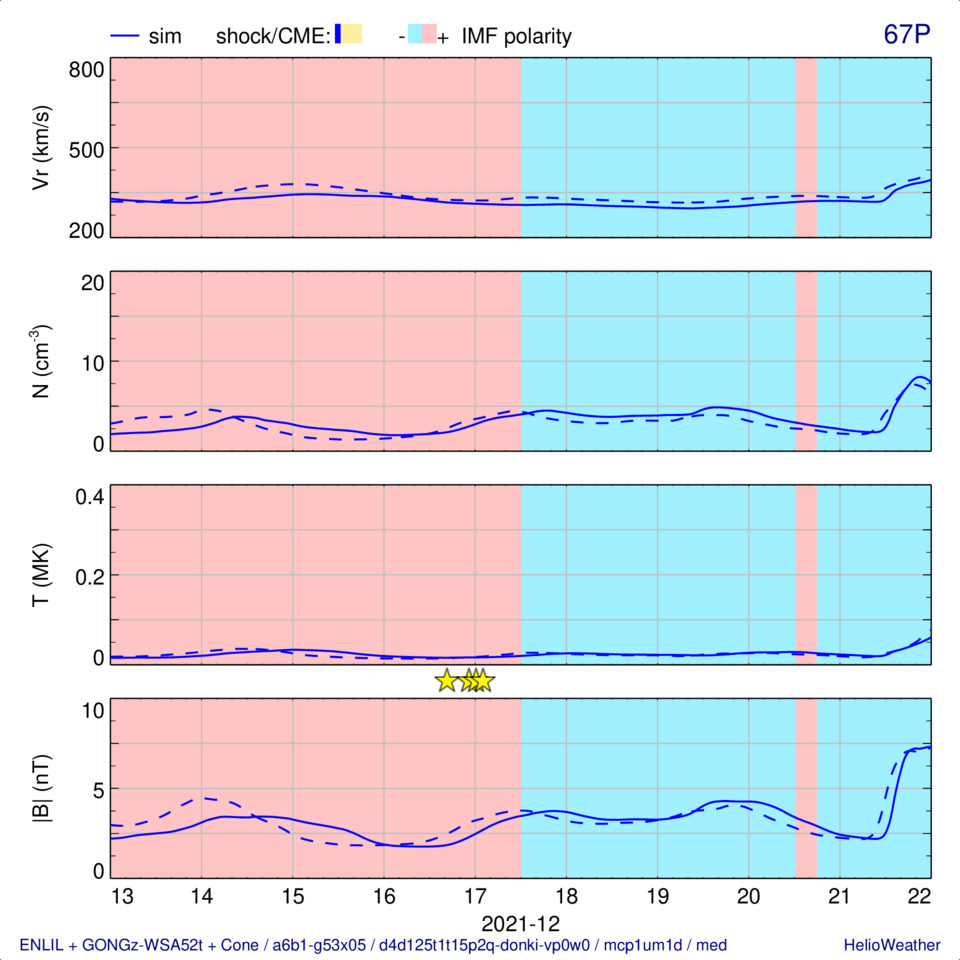}
    \caption{Simulated solar-wind parameters from the ENLIL model with GONG boundary conditions, evaluated at the position of Comet 67P Churyumov-Gerasimenko in December 2021. The horizontal panels display the velocity ($V_r$), density ($N$), temperature ($T$), and magnetic field intensity ($|B|$) of the solar wind. K Shocks are marked by vertical blue lines with annotated times, while coronal mass ejections are highlighted in yellow. Times accompanying vertical blue lines denote shocks. Variations in the solar-wind parameters coincide with shock/CME events and IMF polarity changes, suggesting dynamic solar wind conditions. Each observation is indicated by a star in the figure. Figure provided by Du\v san Odstr\v cil.}
    \label{fig:67p-12-21}
\end{figure}

\begin{figure*}
    \centering
    \includegraphics[width=0.9\linewidth,trim={0.2cm .8cm 1cm 1.5cm},clip]{Plots/ENLIL/t22z-a6b1-d4t1vrp2q-cmes-donki-2022_evo5o1-vr-den-tem-btot-19p_cxp20y_s2w1d_mcp1um1d_low_202201.png}
    \caption{Simulated solar-wind parameters from the ENLIL model with GONG boundary conditions, evaluated at the position of Comet 19P/Borrelly in January 2022. The horizontal panels display the velocity ($V_r$), density ($N$), temperature ($T$), and magnetic field intensity ($|B|$) of the solar wind. K Shocks are marked by vertical blue lines with annotated times, while coronal mass ejections are highlighted in yellow. Times accompanying vertical blue lines denote shocks. Variations in the solar-wind parameters coincide with shock/CME events and IMF polarity changes, suggesting dynamic solar wind conditions. Each observation is indicated by a star in the figure. Figure provided by Du\v san Odstr\v cil.}
    \label{fig:19P_01_2022}
\end{figure*}

\begin{figure}
    \centering
    \includegraphics[width=0.9\linewidth,trim={0.2cm .8cm 1cm 1.5cm},clip]{Plots/ENLIL/t22z-a6b1-d4t1vrp2q-cmes-donki-2022_evo5o1-vr-den-tem-btot-19p_cxp20y_s2w1d_mcp1um1d_low_202202.png}
    \caption{Simulated solar-wind parameters from the ENLIL model with GONG boundary conditions, evaluated at the position of Comet 19P/Borrelly in February 2022. The horizontal panels display the velocity ($V_r$), density ($N$), temperature ($T$), and magnetic field intensity ($|B|$) of the solar wind. K Shocks are marked by vertical blue lines with annotated times, while coronal mass ejections are highlighted in yellow. Times accompanying vertical blue lines denote shocks. Variations in the solar-wind parameters coincide with shock/CME events and IMF polarity changes, suggesting dynamic solar wind conditions. Each observation is indicated by a star in the figure. Figure provided by Du\v san Odstr\v cil.}
    \label{fig:19P_02_2022}
\end{figure}

\begin{figure}
    \centering
    \includegraphics[width=0.9\linewidth,trim={0.2cm .8cm 1cm 1.5cm},clip]{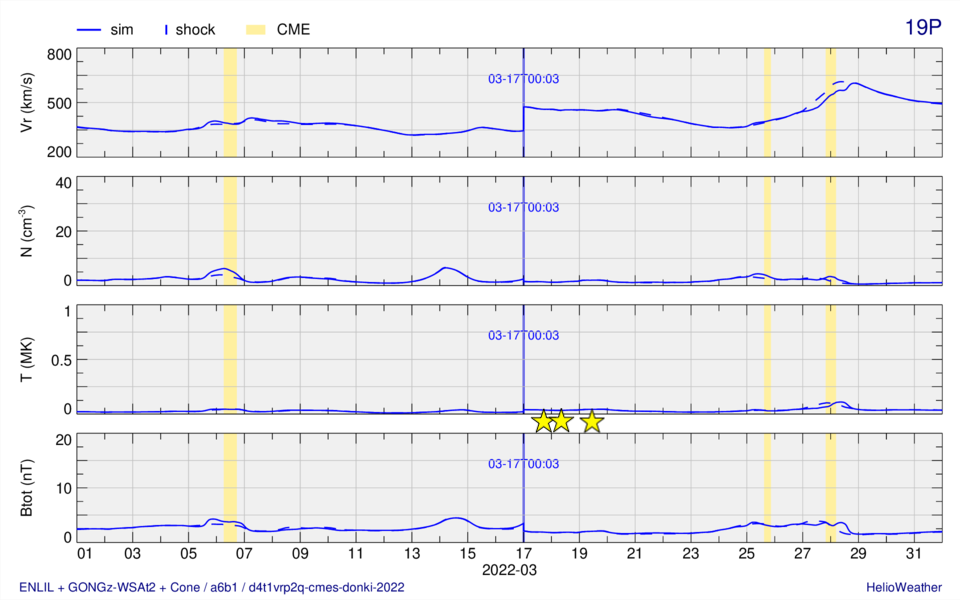}
    \caption{Simulated solar-wind parameters from the ENLIL model with GONG boundary conditions, evaluated at the position of Comet 19P/Borrelly in March 2022. The horizontal panels display the velocity ($V_r$), density ($N$), temperature ($T$), and magnetic field intensity ($|B|$) of the solar wind. K Shocks are marked by vertical blue lines with annotated times, while coronal mass ejections are highlighted in yellow. Times accompanying vertical blue lines denote shocks. Variations in the solar-wind parameters coincide with shock/CME events and IMF polarity changes, suggesting dynamic solar wind conditions. Each observation is indicated by a star in the figure. Figure provided by Du\v san Odstr\v cil.}
    \label{fig:19P_03_2022}
\end{figure}

\begin{figure}
    \centering
    \includegraphics[width=0.9\linewidth,trim={0.2cm .8cm 1cm 1.5cm},clip]{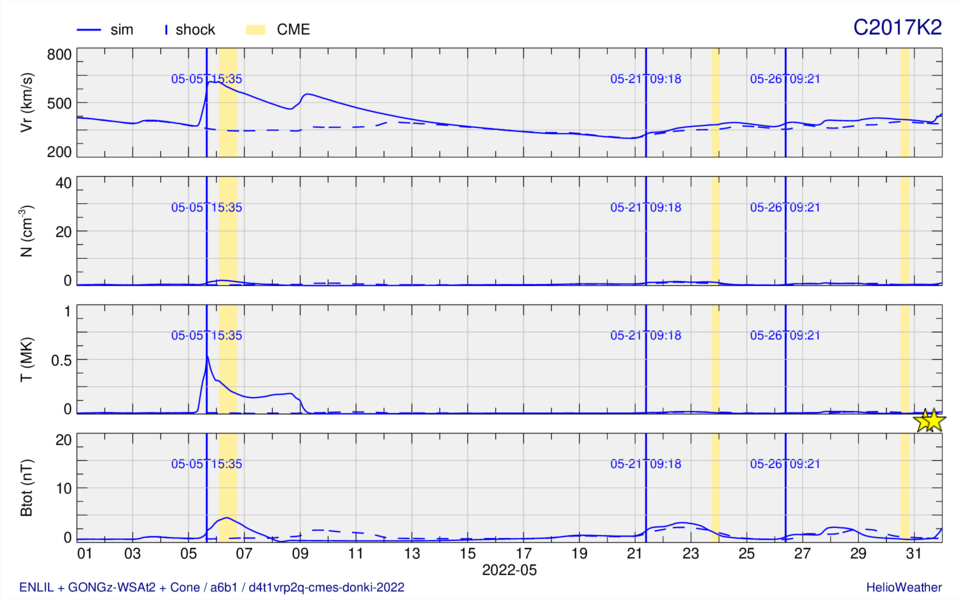}
    \caption{Simulated solar-wind parameters from the ENLIL model with GONG boundary conditions, evaluated at the position of Comet C/2017 K2 (PANSTARRS) in May 2022. The horizontal panels display the velocity ($V_r$), density ($N$), temperature ($T$), and magnetic field intensity ($|B|$) of the solar wind. K Shocks are marked by vertical blue lines with annotated times, while coronal mass ejections are highlighted in yellow. Times accompanying vertical blue lines denote shocks. Variations in the solar-wind parameters coincide with shock/CME events and IMF polarity changes, suggesting dynamic solar wind conditions. Each observation is indicated by a star in the figure. Figure provided by Du\v san Odstr\v cil.}
    \label{fig:K2_05_2022}
\end{figure}

\begin{figure}
    \centering
    \includegraphics[width=0.9\linewidth,trim={0.2cm .8cm 1cm 1.5cm},clip]{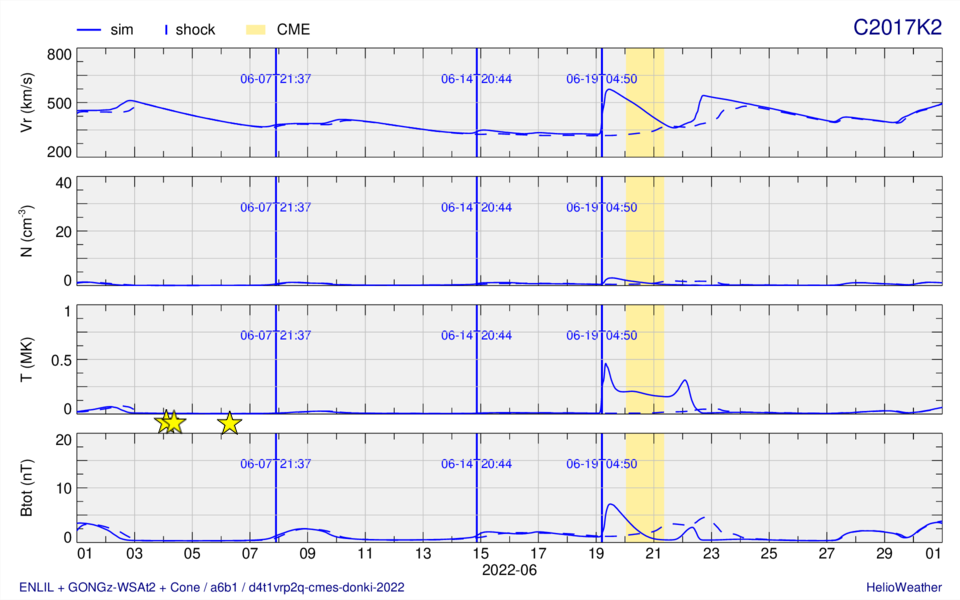}
    \caption{Simulated solar-wind parameters from the ENLIL model with GONG boundary conditions, evaluated at the position of Comet C/2017 K2 (PANSTARRS) in June 2022. The horizontal panels display the velocity ($V_r$), density ($N$), temperature ($T$), and magnetic field intensity ($|B|$) of the solar wind. K Shocks are marked by vertical blue lines with annotated times, while coronal mass ejections are highlighted in yellow. Times accompanying vertical blue lines denote shocks. Variations in the solar-wind parameters coincide with shock/CME events and IMF polarity changes, suggesting dynamic solar wind conditions. Each observation is indicated by a star in the figure. Figure provided by Du\v san Odstr\v cil.}
    \label{fig:K2_06_2022}
\end{figure}

\begin{figure}
    \centering
    \includegraphics[width=0.9\linewidth,trim={0.2cm .8cm 1cm 1.5cm},clip]{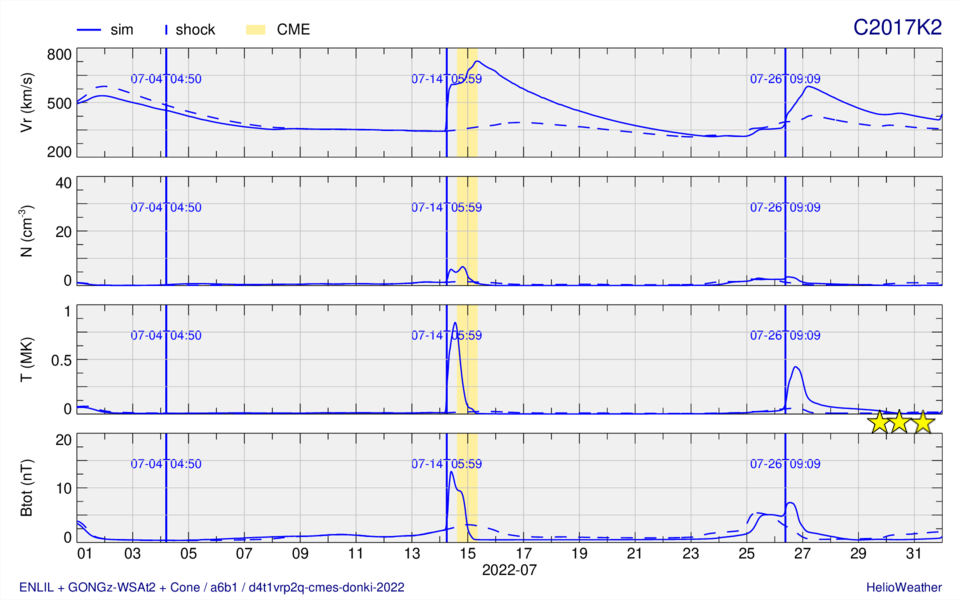}
    \caption{Simulated solar-wind parameters from the ENLIL model with GONG boundary conditions, evaluated at the position of Comet C/2017 K2 (PANSTARRS) in July 2022. The horizontal panels display the velocity ($V_r$), density ($N$), temperature ($T$), and magnetic field intensity ($|B|$) of the solar wind. K Shocks are marked by vertical blue lines with annotated times, while coronal mass ejections are highlighted in yellow. Times accompanying vertical blue lines denote shocks. Variations in the solar-wind parameters coincide with shock/CME events and IMF polarity changes, suggesting dynamic solar wind conditions. Each observation is indicated by a star in the figure. Figure provided by Du\v san Odstr\v cil.}
    \label{fig:K2_07_2022}
\end{figure}

\begin{figure}
    \centering
    \includegraphics[width=0.9\linewidth,trim={0.2cm .8cm 1cm 1.5cm},clip]{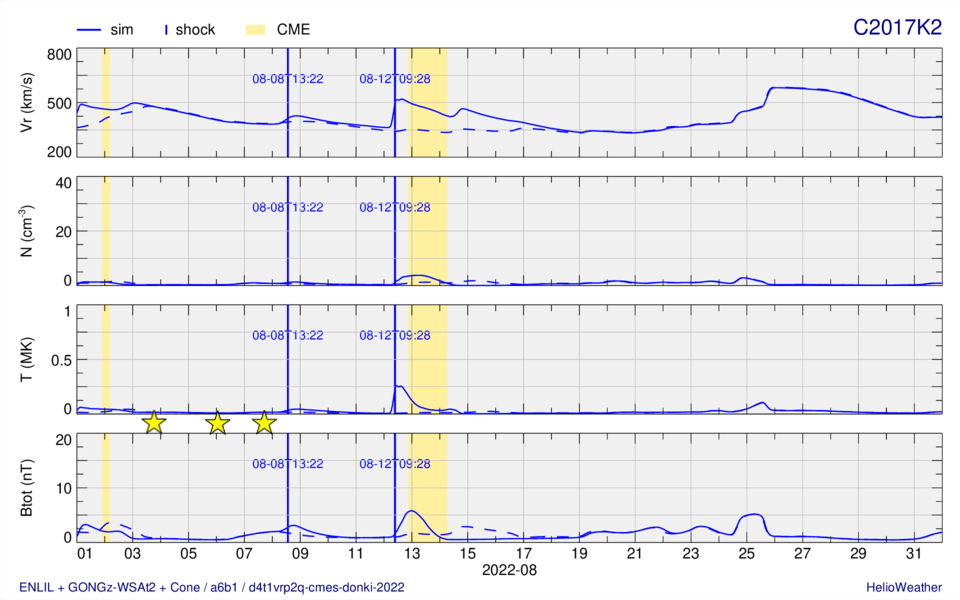}
    \caption{Simulated solar-wind parameters from the ENLIL model with GONG boundary conditions, evaluated at the position of Comet C/2017 K2 (PANSTARRS) in August 2022. The horizontal panels display the velocity ($V_r$), density ($N$), temperature ($T$), and magnetic field intensity ($|B|$) of the solar wind. K Shocks are marked by vertical blue lines with annotated times, while coronal mass ejections are highlighted in yellow. Times accompanying vertical blue lines denote shocks. Variations in the solar-wind parameters coincide with shock/CME events and IMF polarity changes, suggesting dynamic solar wind conditions. Each observation is indicated by a star in the figure. Figure provided by Du\v san Odstr\v cil.}
    \label{fig:K2_08_2022}
\end{figure}

\begin{figure}
    \centering
    \includegraphics[width=0.9\linewidth,trim={0.2cm .8cm 1cm 1.5cm},clip]{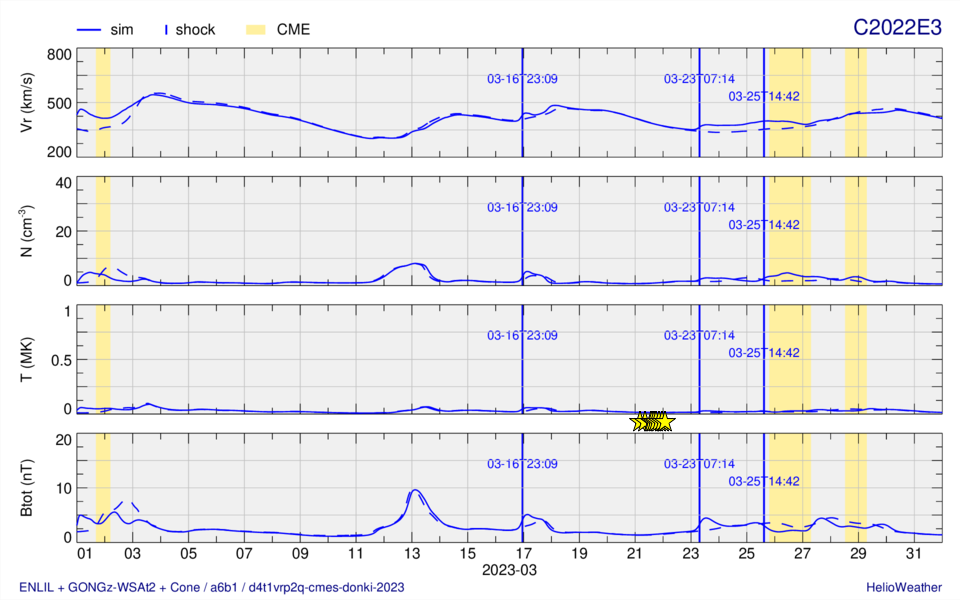}
    \caption{Simulated solar-wind parameters from the ENLIL model with GONG boundary conditions, evaluated at the position of Comet C/2022 E3 (ZTF) in March 2023. The horizontal panels display the velocity ($V_r$), density ($N$), temperature ($T$), and magnetic field intensity ($|B|$) of the solar wind. K Shocks are marked by vertical blue lines with annotated times, while coronal mass ejections are highlighted in yellow. Times accompanying vertical blue lines denote shocks. Variations in the solar-wind parameters coincide with shock/CME events and IMF polarity changes, suggesting dynamic solar wind conditions. Each observation is indicated by a star in the figure. Figure provided by Du\v san Odstr\v cil.}
    \label{fig:E3_03_2023}
\end{figure}

\begin{figure}
    \centering
    \includegraphics[width=0.9\linewidth,trim={0.2cm .8cm 1cm 1.5cm},clip]{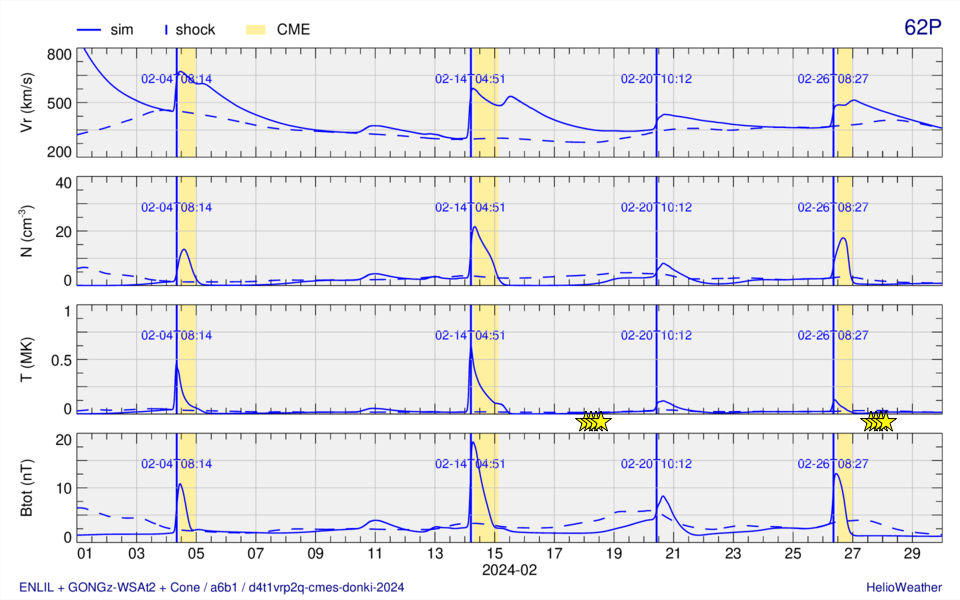}
    \caption{Simulated solar-wind parameters from the ENLIL model with GONG boundary conditions, evaluated at the position of Comet 62P/Tsuchinshan in February 2024. The horizontal panels display the velocity ($V_r$), density ($N$), temperature ($T$), and magnetic field intensity ($|B|$) of the solar wind. K Shocks are marked by vertical blue lines with annotated times, while coronal mass ejections are highlighted in yellow. Times accompanying vertical blue lines denote shocks. Variations in the solar-wind parameters coincide with shock/CME events and IMF polarity changes, suggesting dynamic solar wind conditions. Each observation is indicated by a star in the figure. Figure provided by Du\v san Odstr\v cil.}
    \label{fig:62P_02_2024}
\end{figure}


\begin{figure}
    \centering
    \includegraphics[width=0.9\linewidth,trim={0.2cm .8cm 1cm 1.5cm},clip]{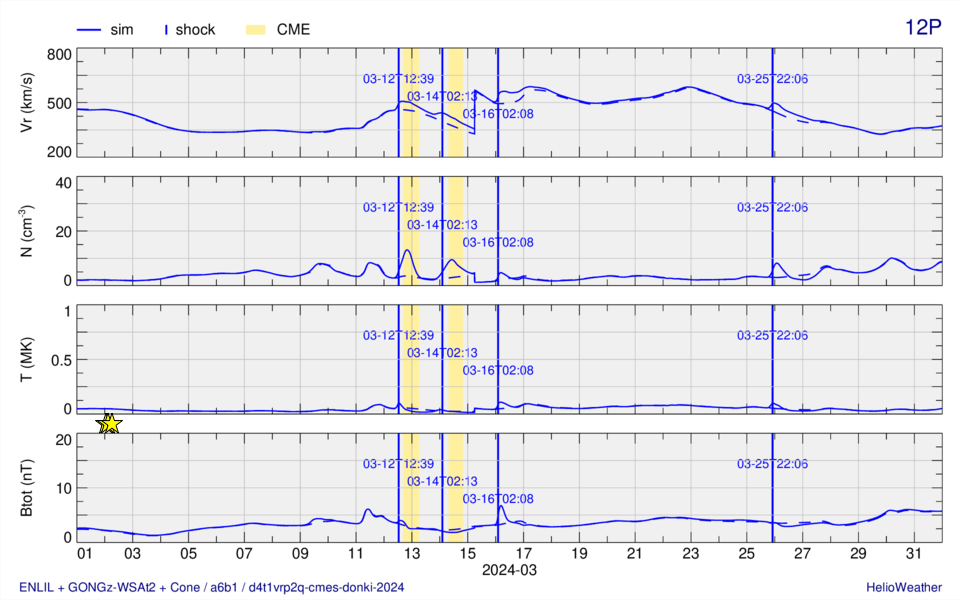}
    \caption{Simulated solar-wind parameters from the ENLIL model with GONG boundary conditions, evaluated at the position of Comet 12P/Pons-Brooks in March 2024. The horizontal panels display the velocity ($V_r$), density ($N$), temperature ($T$), and magnetic field intensity ($|B|$) of the solar wind. K Shocks are marked by vertical blue lines with annotated times, while coronal mass ejections are highlighted in yellow. Times accompanying vertical blue lines denote shocks. Variations in the solar-wind parameters coincide with shock/CME events and IMF polarity changes, suggesting dynamic solar wind conditions. Each observation is indicated by a star in the figure. Figure provided by Du\v san Odstr\v cil.}
    \label{fig:12P_03_2024}
\end{figure}

\section{GOES X-ray Fluxes}
\label{app:GOES}
The figures presented in this section show the evolution of solar X‑ray activity as measured by Geostationary Operational Environmental Satellites (GOES) in the 1.6–12.4\,keV band, with raw, high‑cadence data in blue and a 60-s rolling mean in orange. 
Each figure covers a single epoch for each comet.
These figures illustrate both the day‑to‑day variability and shorter‑term fluctuations in solar X‑ray emission over the duration of the observing campaign.


\begin{landscape}
\begin{figure}
  \centering
  \includegraphics[width=\linewidth,trim={0.35cm 0.25cm 0.25cm 1.4cm},clip]{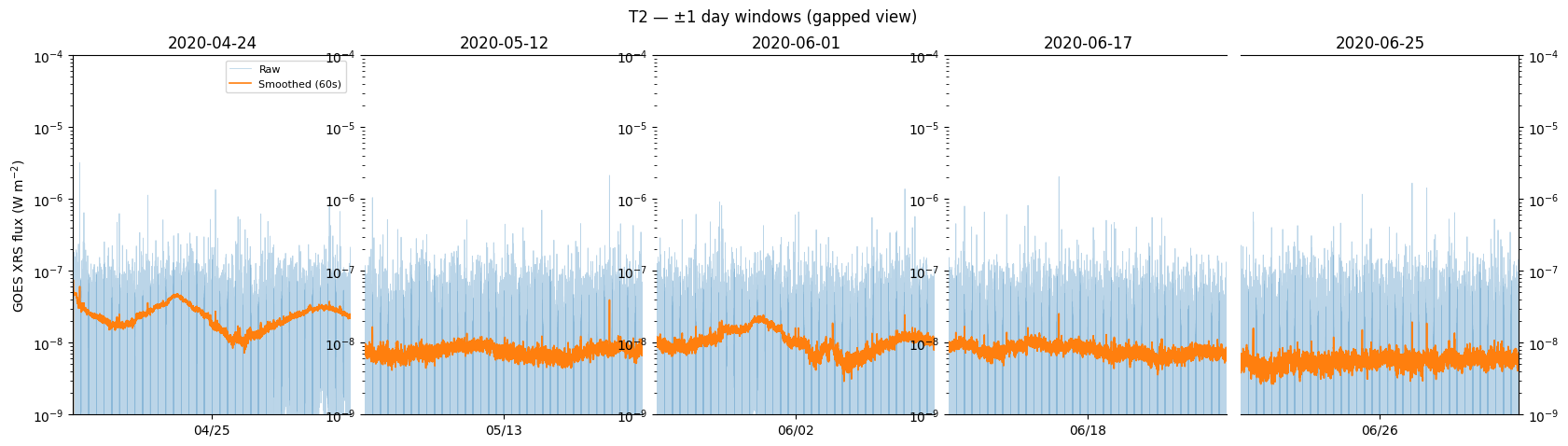}
  \caption{Geostationary Operational Environmental Satellites (GOES) X-ray fluxes over the period of our observations for Comet C/2017 T2. Raw data corresponding to the $1.6-12.4$\,keV energy range is shown in blue in the background and a rolling 60-second mean is shown in orange in the foreground. Our observations were made during each of the periods indicated in the figure.}
  \label{fig:T2-goes-e1}
\end{figure}
\end{landscape}

\begin{figure}
  \centering
  \includegraphics[width=\linewidth,trim={0.35cm 0.25cm 0.25cm 1.1cm},clip]{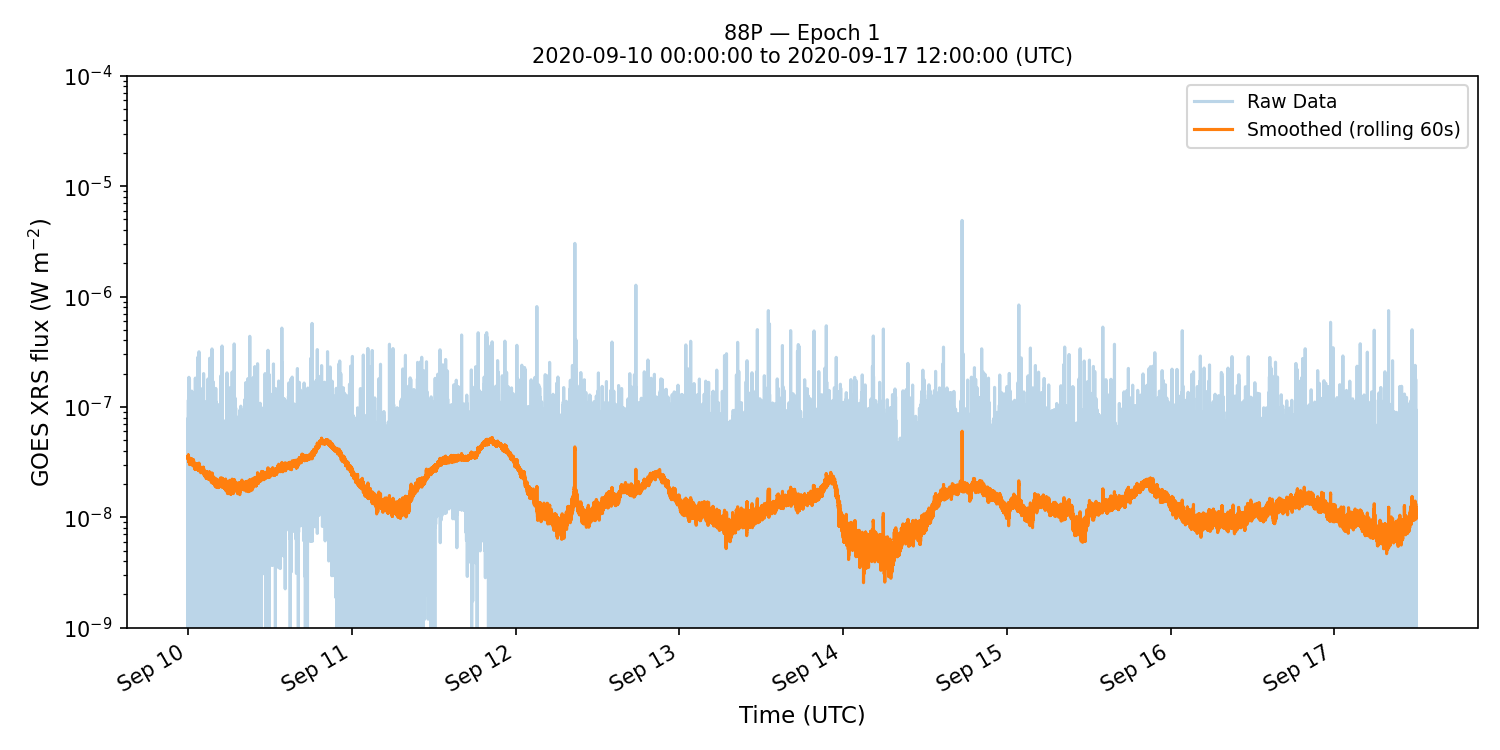}
  \caption{Geostationary Operational Environmental Satellites (GOES) X-ray fluxes over the period of our observations for Comet 88P during epoch 1. Raw data corresponding to the $1.6-12.4$\,keV energy range is shown in blue in the background and a rolling 60-second mean is shown in orange in the foreground.}
  \label{fig:88p-goes-e1}
\end{figure}

\begin{figure}
  \centering
  \includegraphics[width=\linewidth,trim={0.35cm 0.25cm 0.25cm 1.1cm},clip]{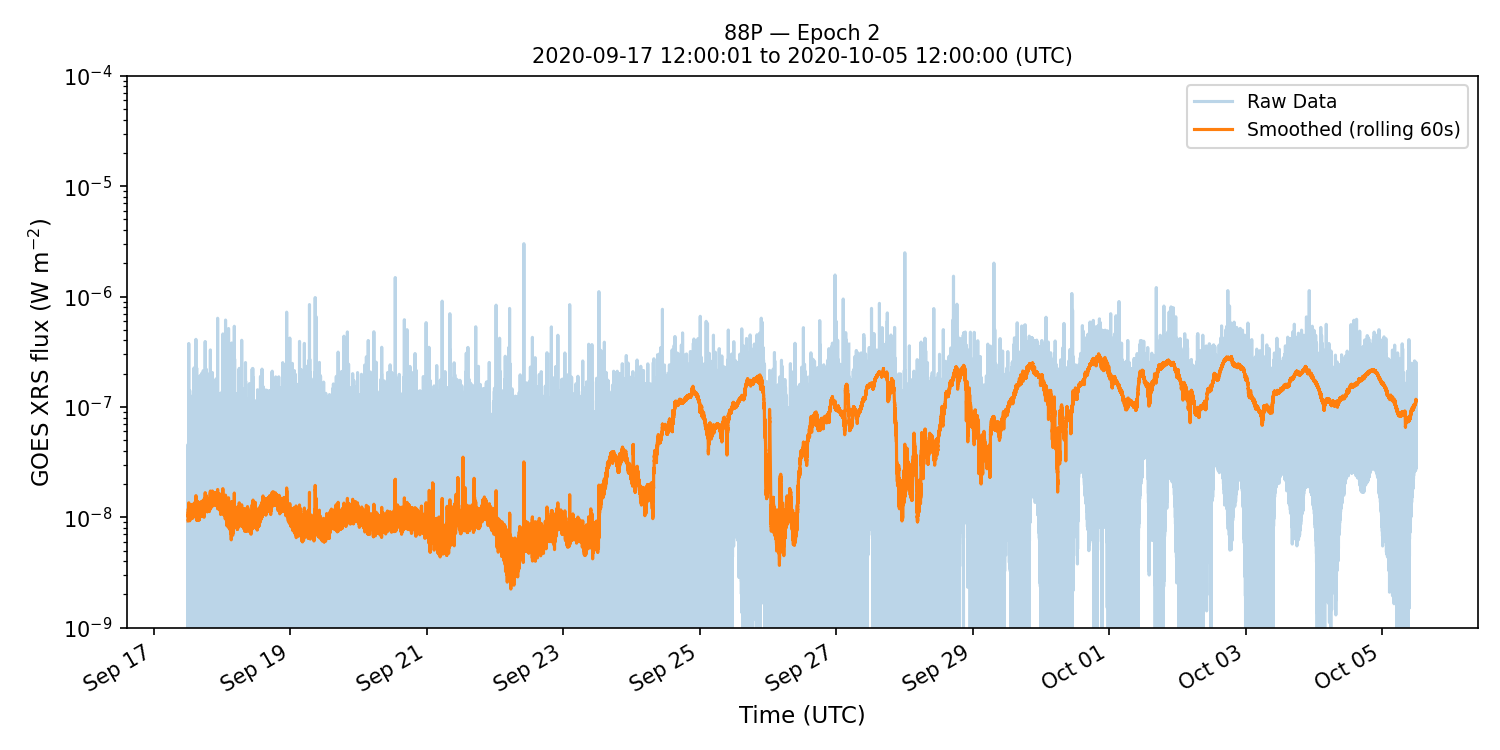}
  \caption{Geostationary Operational Environmental Satellites (GOES) X-ray fluxes over the period of our observations for Comet 88P during epoch 2. Raw data corresponding to the $1.6-12.4$\,keV energy range is shown in blue in the background and a rolling 60-second mean is shown in orange in the foreground.}
  \label{fig:88p-goes-e2}
\end{figure}

\begin{figure}
  \centering
  \includegraphics[width=\linewidth,trim={0.35cm 0.25cm 0.25cm 1.1cm},clip]{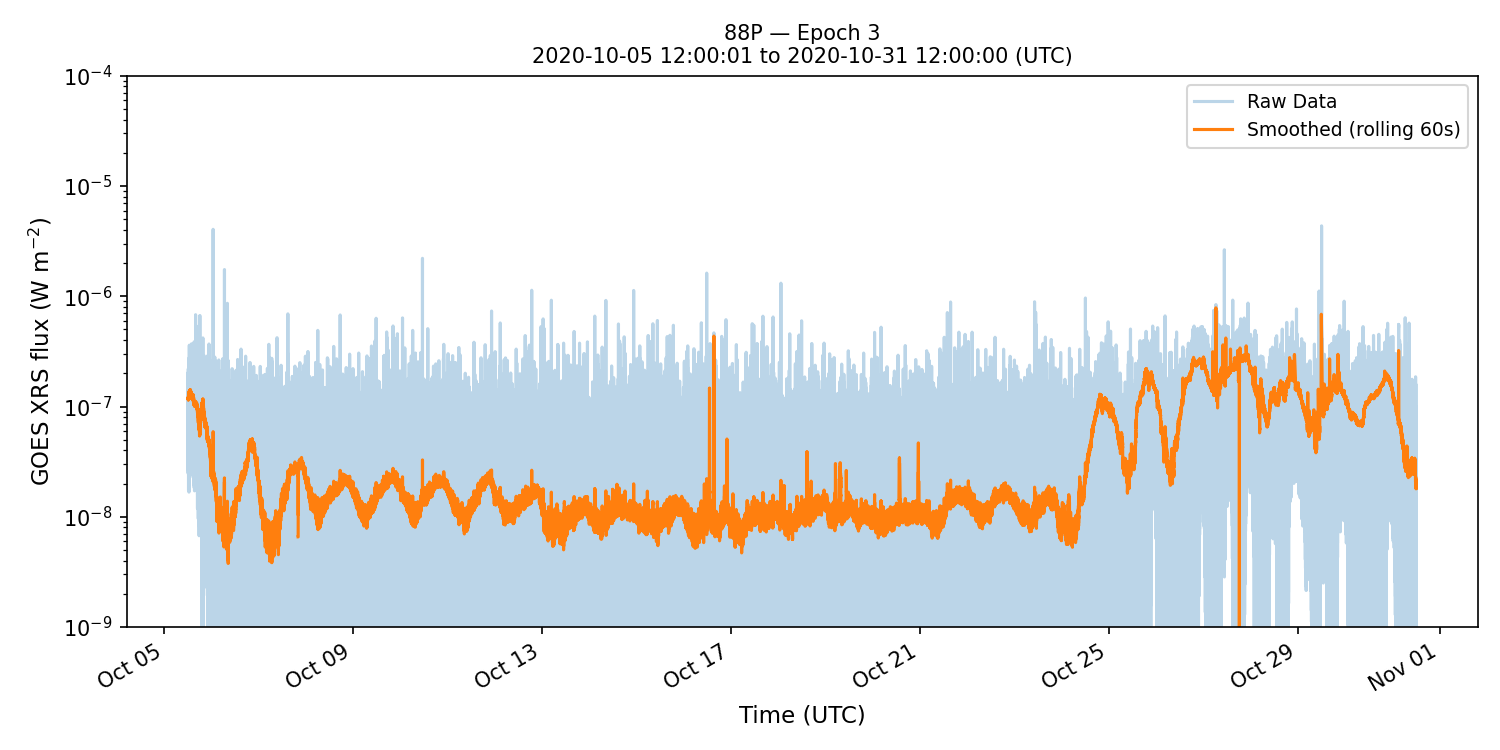}
  \caption{Geostationary Operational Environmental Satellites (GOES) X-ray fluxes over the period of our observations for Comet 88P during epoch 3. Raw data corresponding to the $1.6-12.4$\,keV energy range is shown in blue in the background and a rolling 60-second mean is shown in orange in the foreground.}
  \label{fig:88p-goes-e3}
\end{figure}

\begin{figure}
  \centering
  \includegraphics[width=\linewidth,trim={0.35cm 0.25cm 0.25cm 1.1cm},clip]{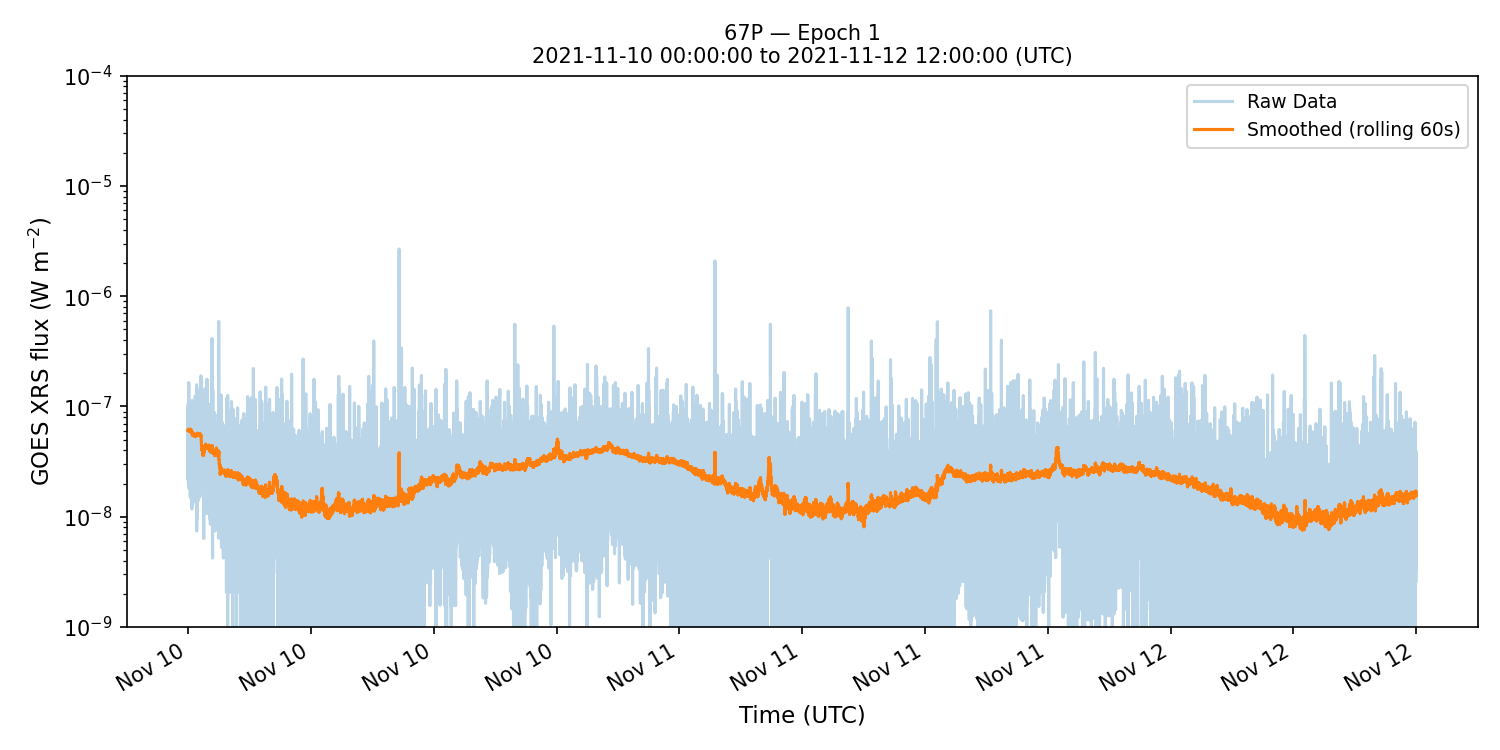}
  \caption{Geostationary Operational Environmental Satellites (GOES) X-ray fluxes over the period of our observations for Comet 67P during epoch 1. Raw data corresponding to the $1.6-12.4$\,keV energy range is shown in blue in the background and a rolling 60-second mean is shown in orange in the foreground.}
  \label{fig:67p-goes-e1}
\end{figure}

\begin{figure}
  \centering
  \includegraphics[width=\linewidth,trim={0.35cm 0.25cm 0.25cm 1.1cm},clip]{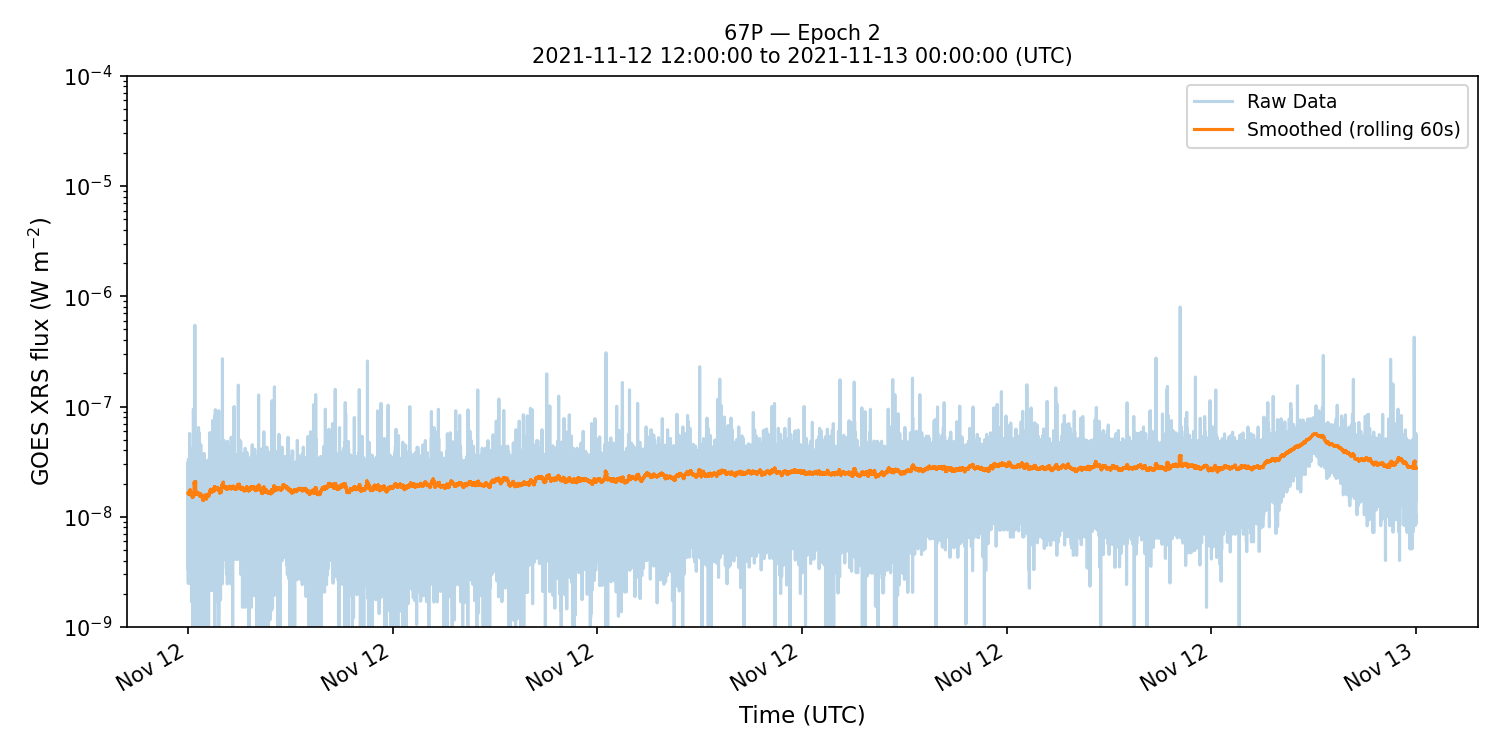}
  \caption{Geostationary Operational Environmental Satellites (GOES) X-ray fluxes over the period of our observations for Comet 67P during epoch 2. Raw data corresponding to the $1.6-12.4$\,keV energy range is shown in blue in the background and a rolling 60-second mean is shown in orange in the foreground.}
  \label{fig:67p-goes-e2}
\end{figure}

\begin{figure}
  \centering
  \includegraphics[width=\linewidth,trim={0.35cm 0.25cm 0.25cm 1.1cm},clip]{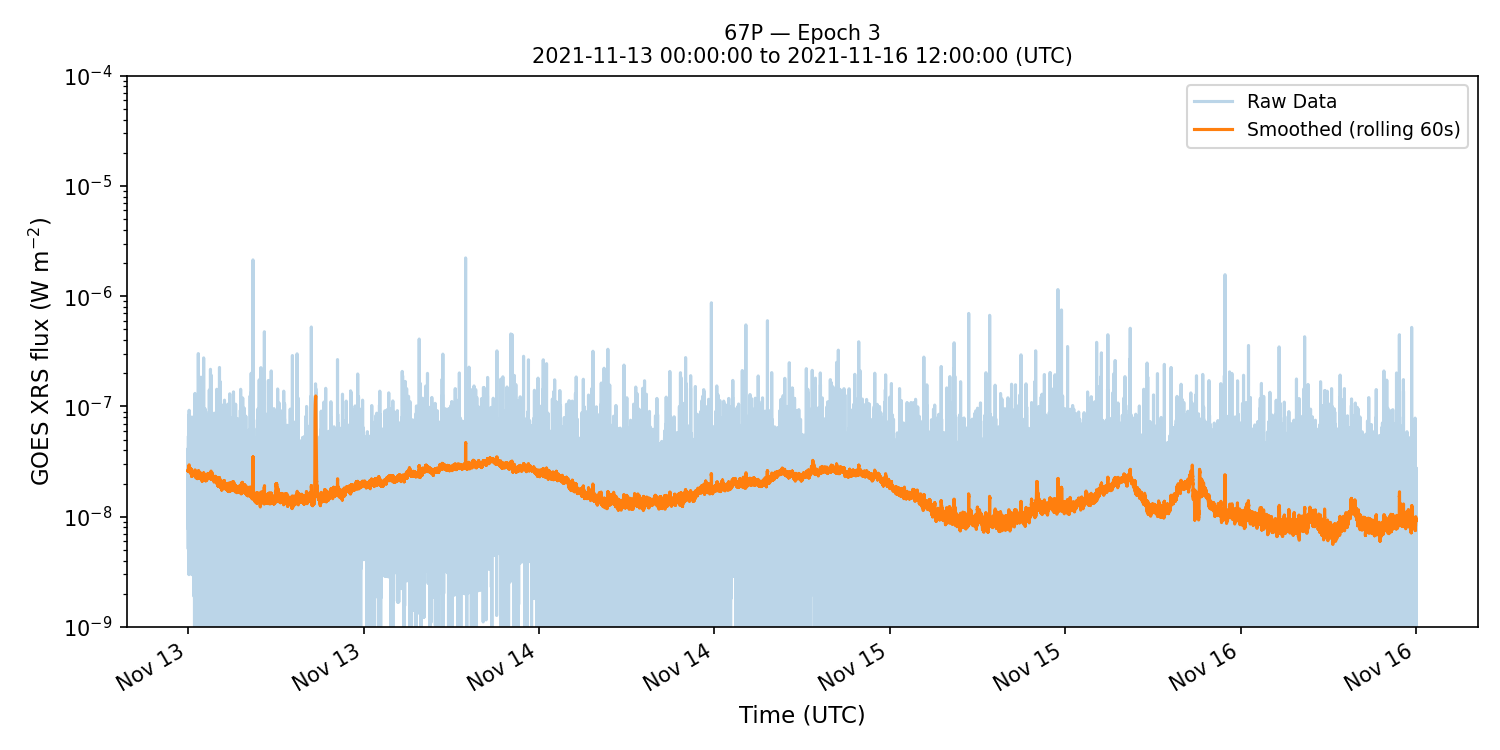}
  \caption{Geostationary Operational Environmental Satellites (GOES) X-ray fluxes over the period of our observations for Comet 67P during epoch 3. Raw data corresponding to the $1.6-12.4$\,keV energy range is shown in blue in the background and a rolling 60-second mean is shown in orange in the foreground.}
  \label{fig:67p-goes-e3}
\end{figure}

\begin{figure}
  \centering
  \includegraphics[width=\linewidth,trim={0.35cm 0.25cm 0.25cm 1.1cm},clip]{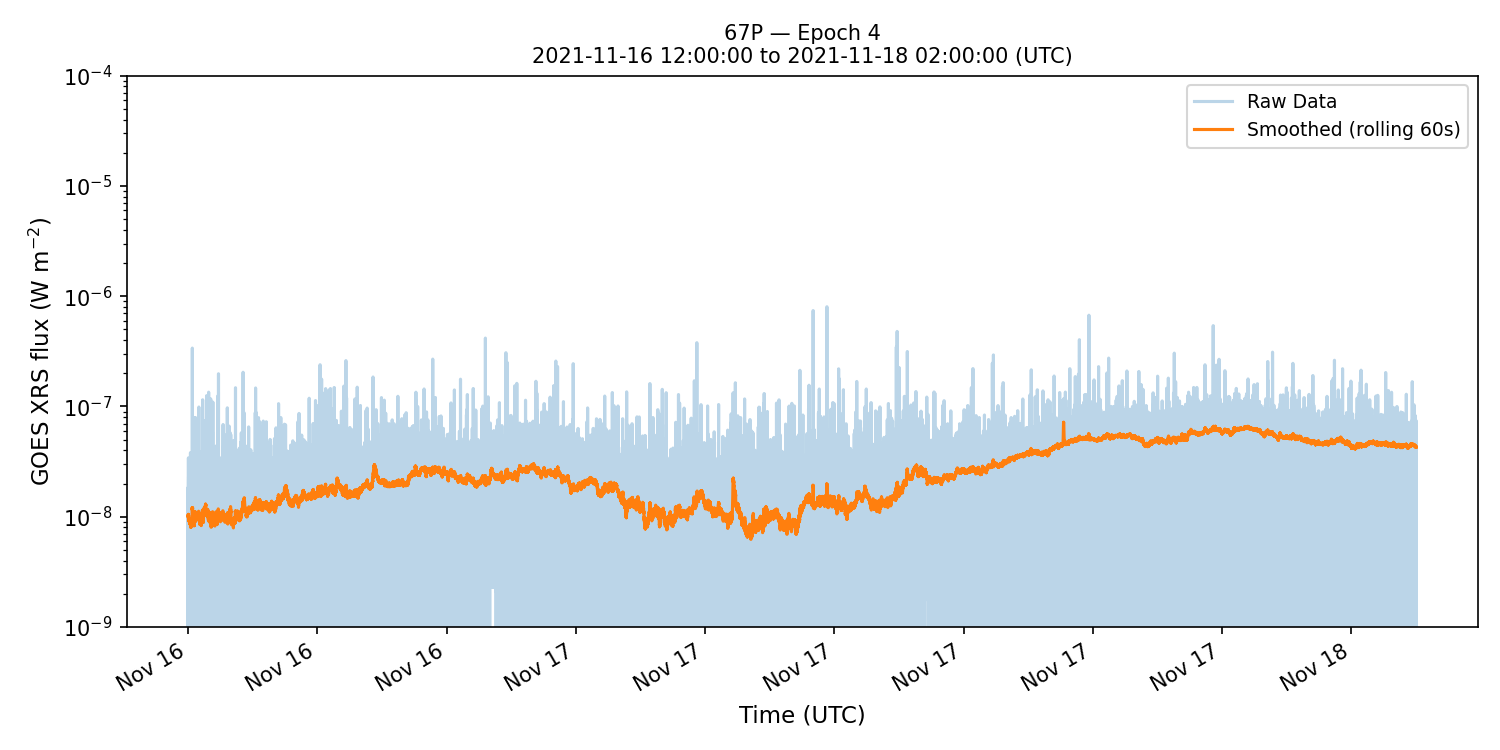}
  \caption{Geostationary Operational Environmental Satellites (GOES) X-ray fluxes over the period of our observations for Comet 67P during epoch 4. Raw data corresponding to the $1.6-12.4$\,keV energy range is shown in blue in the background and a rolling 60-second mean is shown in orange in the foreground.}
  \label{fig:67p-goes-e4}
\end{figure}

\begin{figure}
  \centering
  \includegraphics[width=\linewidth,trim={0.35cm 0.25cm 0.25cm 1.1cm},clip]{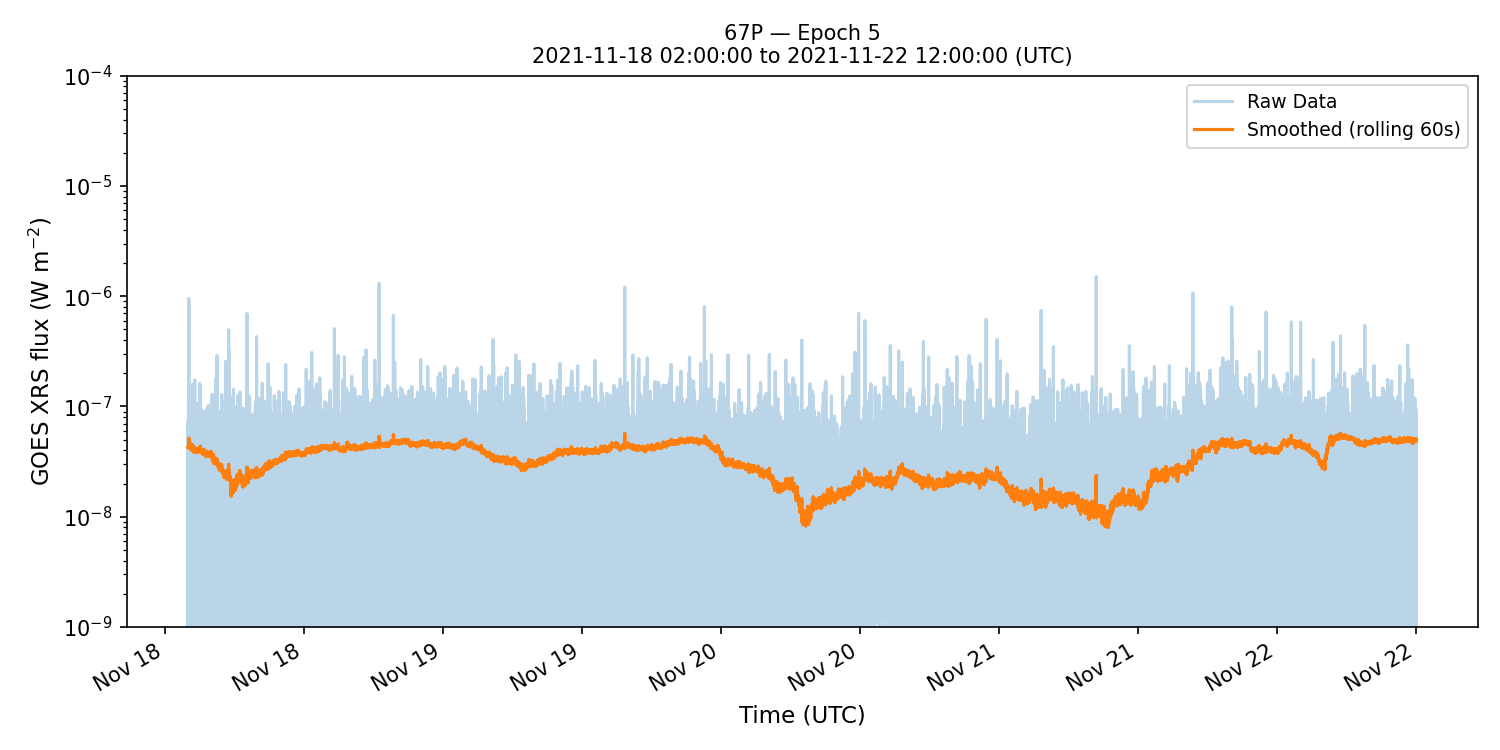}
  \caption{Geostationary Operational Environmental Satellites (GOES) X-ray fluxes over the period of our observations for Comet 67P during epoch 5. Raw data corresponding to the $1.6-12.4$\,keV energy range is shown in blue in the background and a rolling 60-second mean is shown in orange in the foreground.}
  \label{fig:67p-goes-e5}
\end{figure}

\begin{figure}
  \centering
  \includegraphics[width=\linewidth,trim={0.35cm 0.25cm 0.25cm 1.1cm},clip]{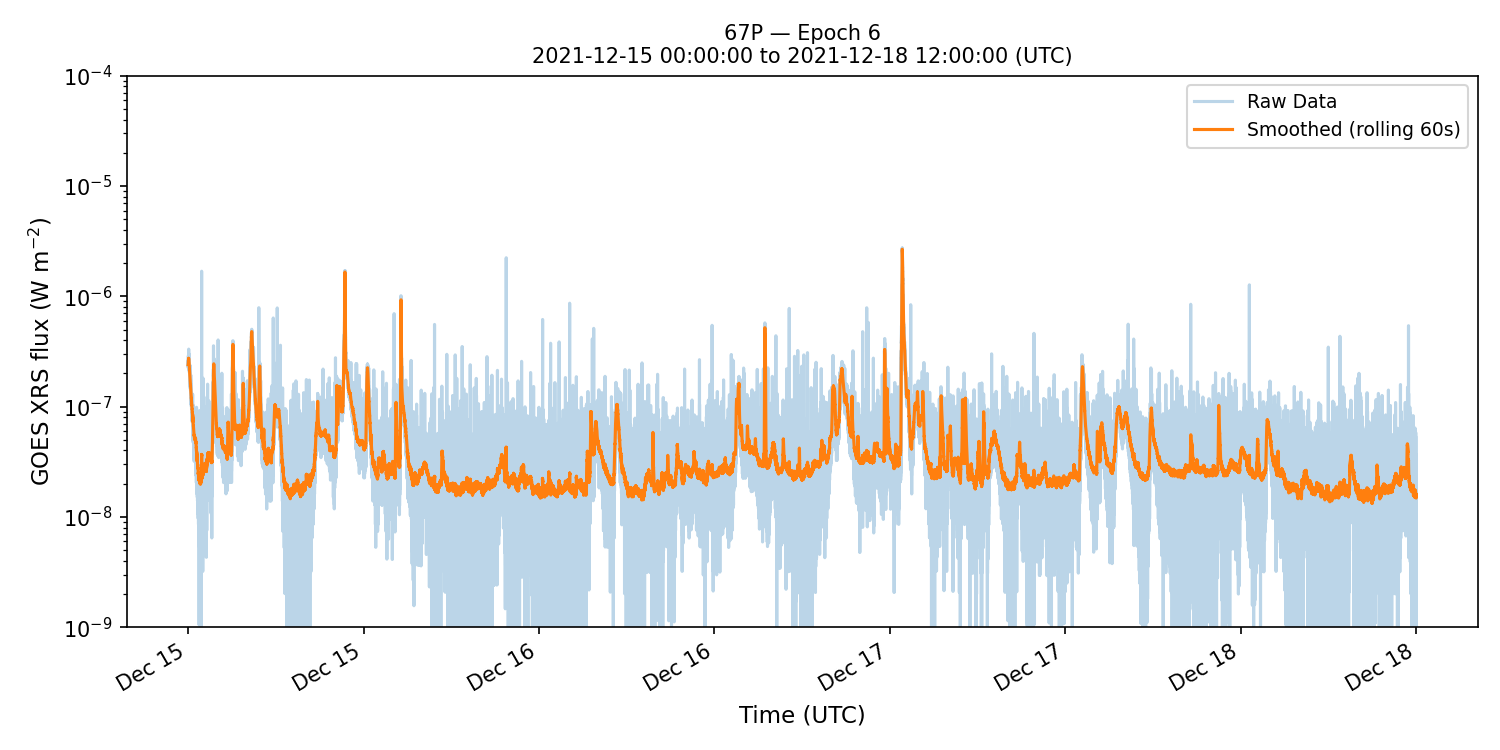}
  \caption{Geostationary Operational Environmental Satellites (GOES) X-ray fluxes over the period of our observations for Comet 67P during epoch 6. Raw data corresponding to the $1.6-12.4$\,keV energy range is shown in blue in the background and a rolling 60-second mean is shown in orange in the foreground.}
  \label{fig:67p-goes-e6}
\end{figure}

\begin{figure}
  \centering
  \includegraphics[width=\linewidth,trim={0.35cm 0.25cm 0.25cm 1.1cm},clip]{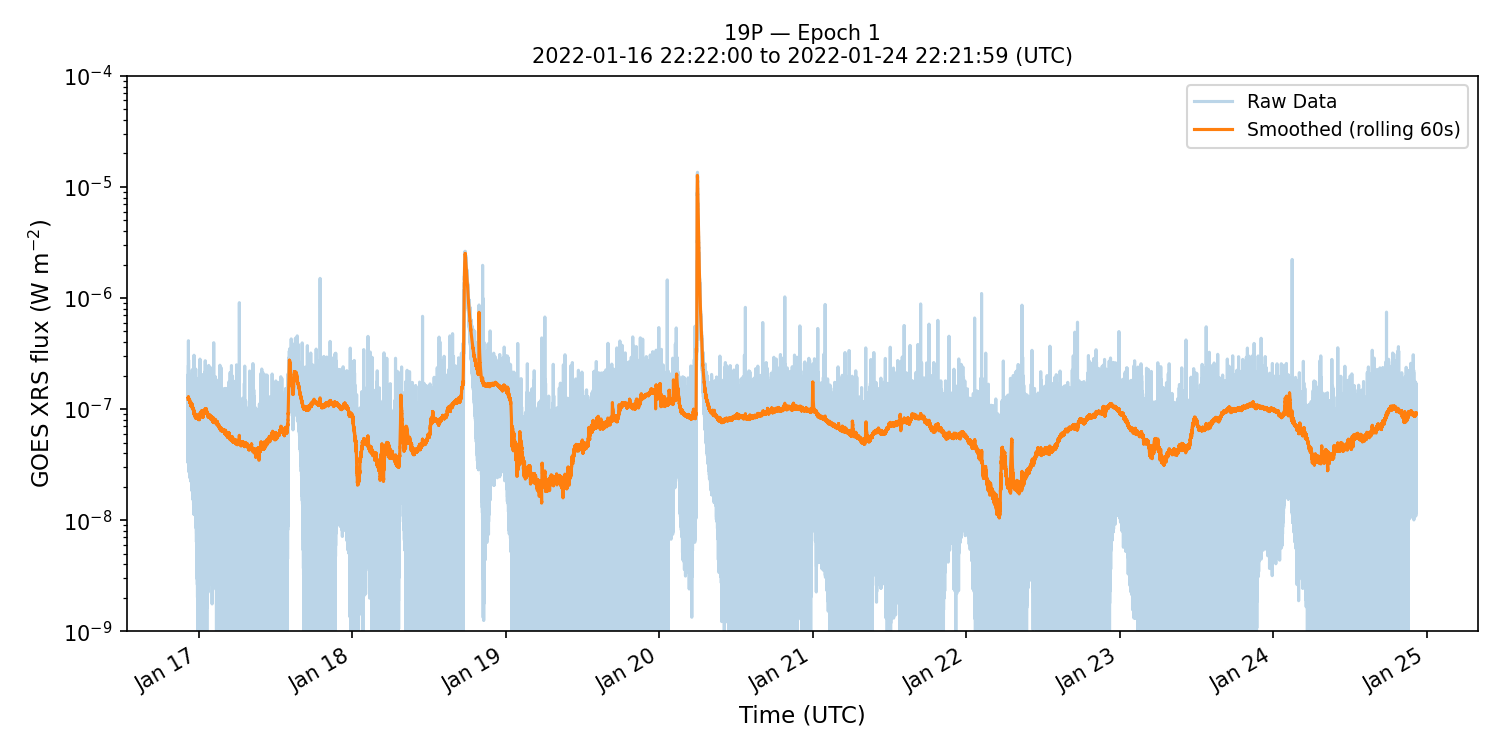}
  \caption{Geostationary Operational Environmental Satellites (GOES) X-ray fluxes over the period of our observations for Comet 19P during epoch 1. Raw data corresponding to the $1.6-12.4$\,keV energy range is shown in blue in the background and a rolling 60-second mean is shown in orange in the foreground.}
  \label{fig:19p-goes-e1}
\end{figure}

\begin{figure}
  \centering
  \includegraphics[width=\linewidth,trim={0.35cm 0.25cm 0.25cm 1.1cm},clip]{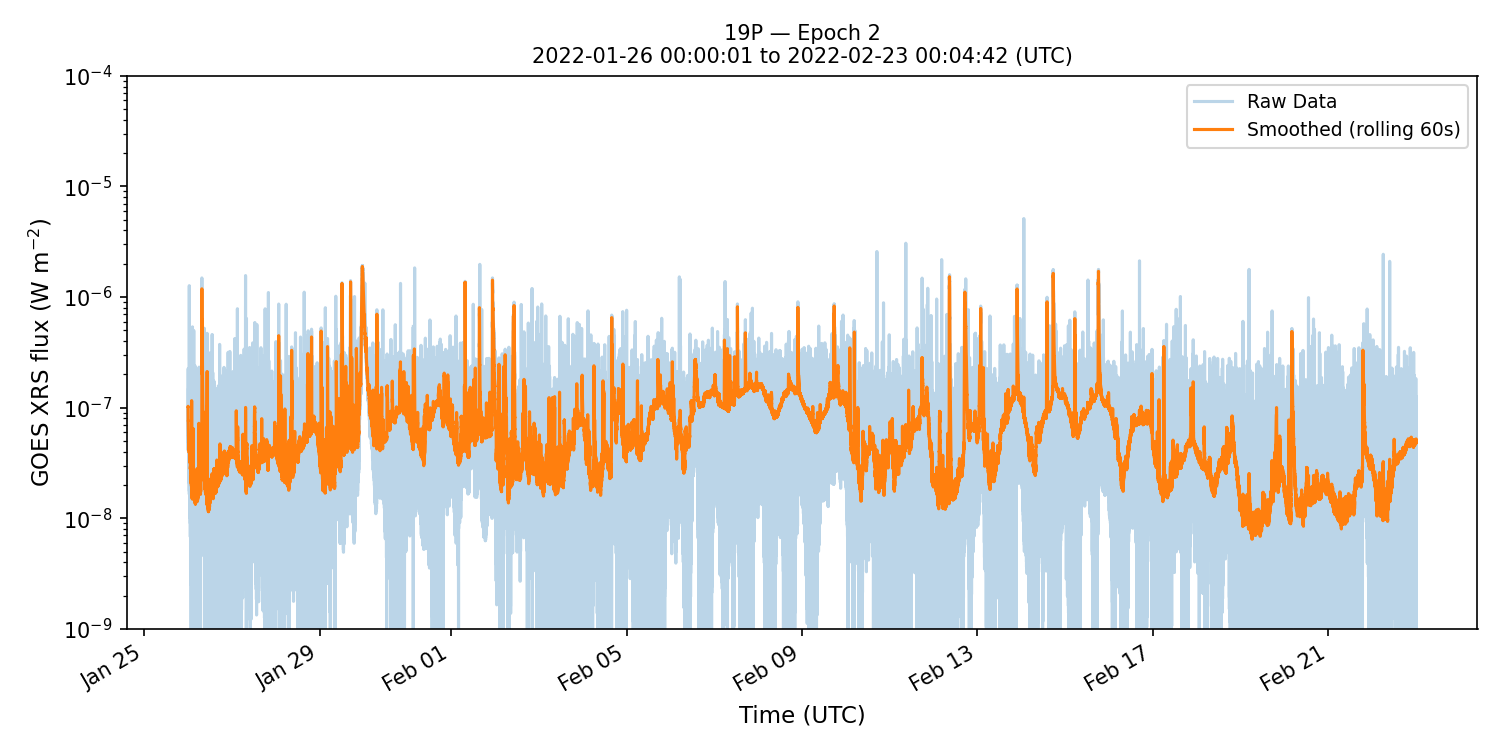}
  \caption{Geostationary Operational Environmental Satellites (GOES) X-ray fluxes over the period of our observations for Comet 19P during epoch 2. Raw data corresponding to the $1.6-12.4$\,keV energy range is shown in blue in the background and a rolling 60-second mean is shown in orange in the foreground.}
  \label{fig:19p-goes-e2}
\end{figure}

\begin{figure}
  \centering
  \includegraphics[width=\linewidth,trim={0.35cm 0.25cm 0.25cm 1.1cm},clip]{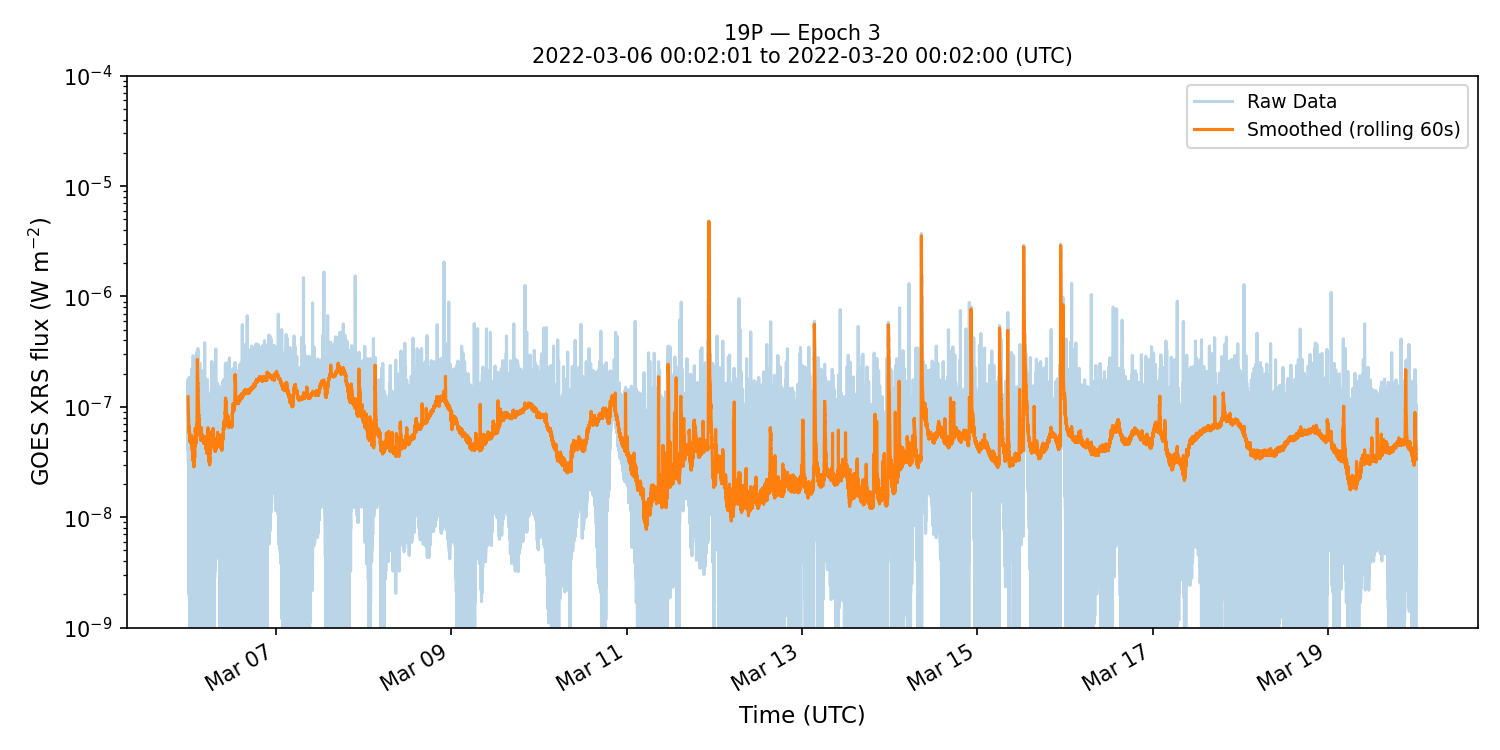}
  \caption{Geostationary Operational Environmental Satellites (GOES) X-ray fluxes over the period of our observations for Comet 19P during epoch 3. Raw data corresponding to the $1.6-12.4$\,keV energy range is shown in blue in the background and a rolling 60-second mean is shown in orange in the foreground.}
  \label{fig:19p-goes-e3}
\end{figure}

\begin{figure}
  \centering
  \includegraphics[width=\linewidth,trim={0.35cm 0.25cm 0.25cm 1.1cm},clip]{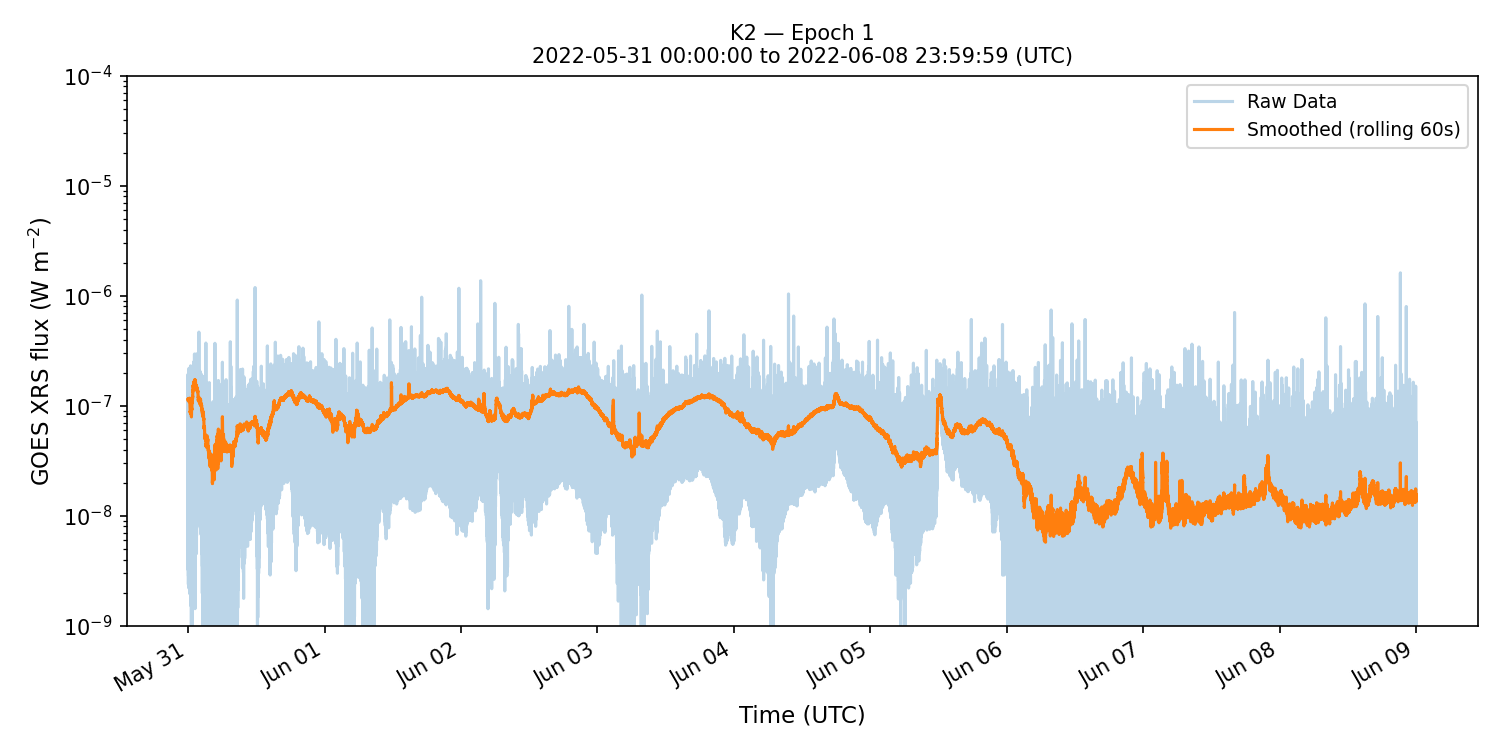}
  \caption{Geostationary Operational Environmental Satellites (GOES) X-ray fluxes over the period of our observations for Comet C/2017 K2 during epoch 1. Raw data corresponding to the $1.6-12.4$\,keV energy range is shown in blue in the background and a rolling 60-second mean is shown in orange in the foreground.}
  \label{fig:k2-goes-e1}
\end{figure}

\begin{figure}
  \centering
  \includegraphics[width=\linewidth,trim={0.35cm 0.25cm 0.25cm 1.1cm},clip]{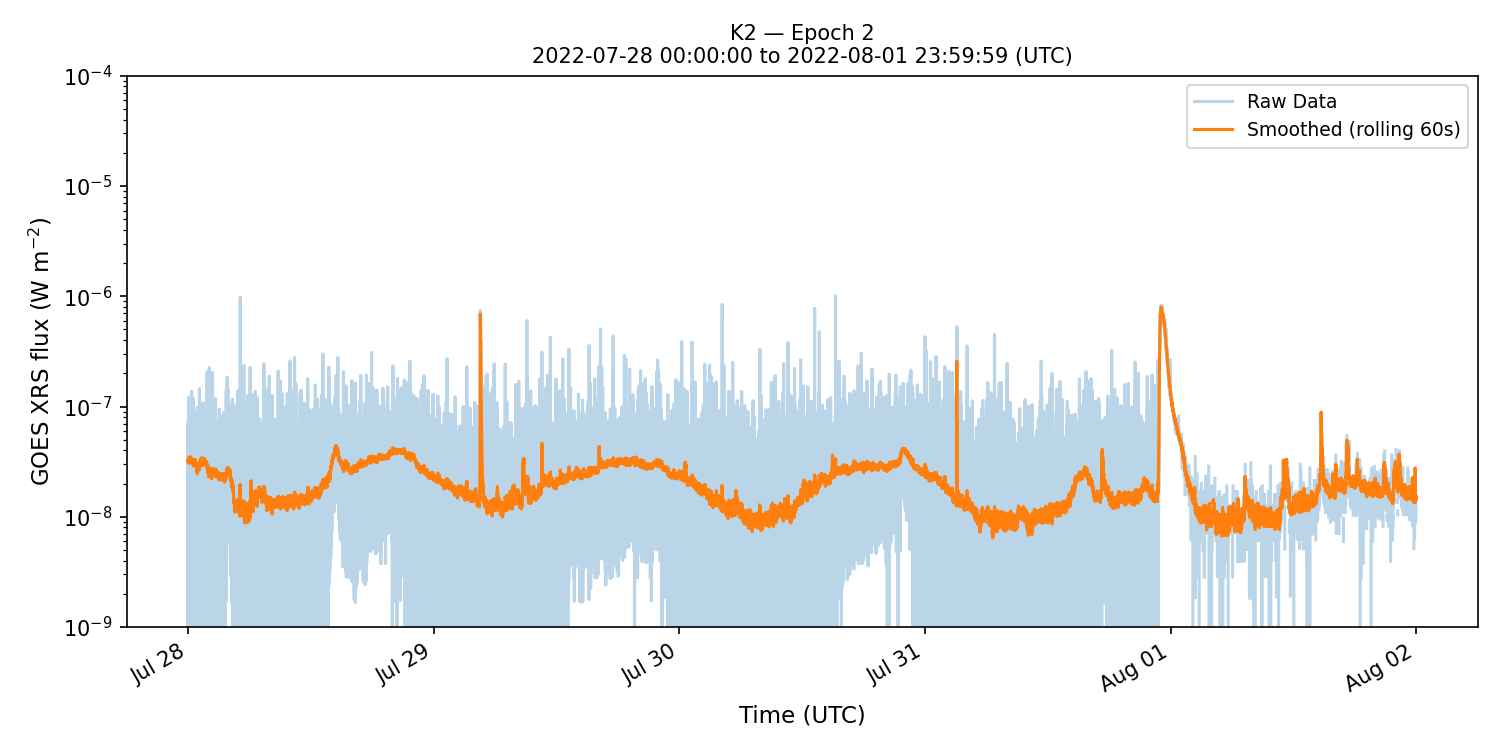}
  \caption{Geostationary Operational Environmental Satellites (GOES) X-ray fluxes over the period of our observations for Comet C/2017 K2 during epoch 2. Raw data corresponding to the $1.6-12.4$\,keV energy range is shown in blue in the background and a rolling 60-second mean is shown in orange in the foreground.}
  \label{fig:k2-goes-e2}
\end{figure}

\begin{figure}
  \centering
  \includegraphics[width=\linewidth,trim={0.35cm 0.25cm 0.25cm 1.1cm},clip]{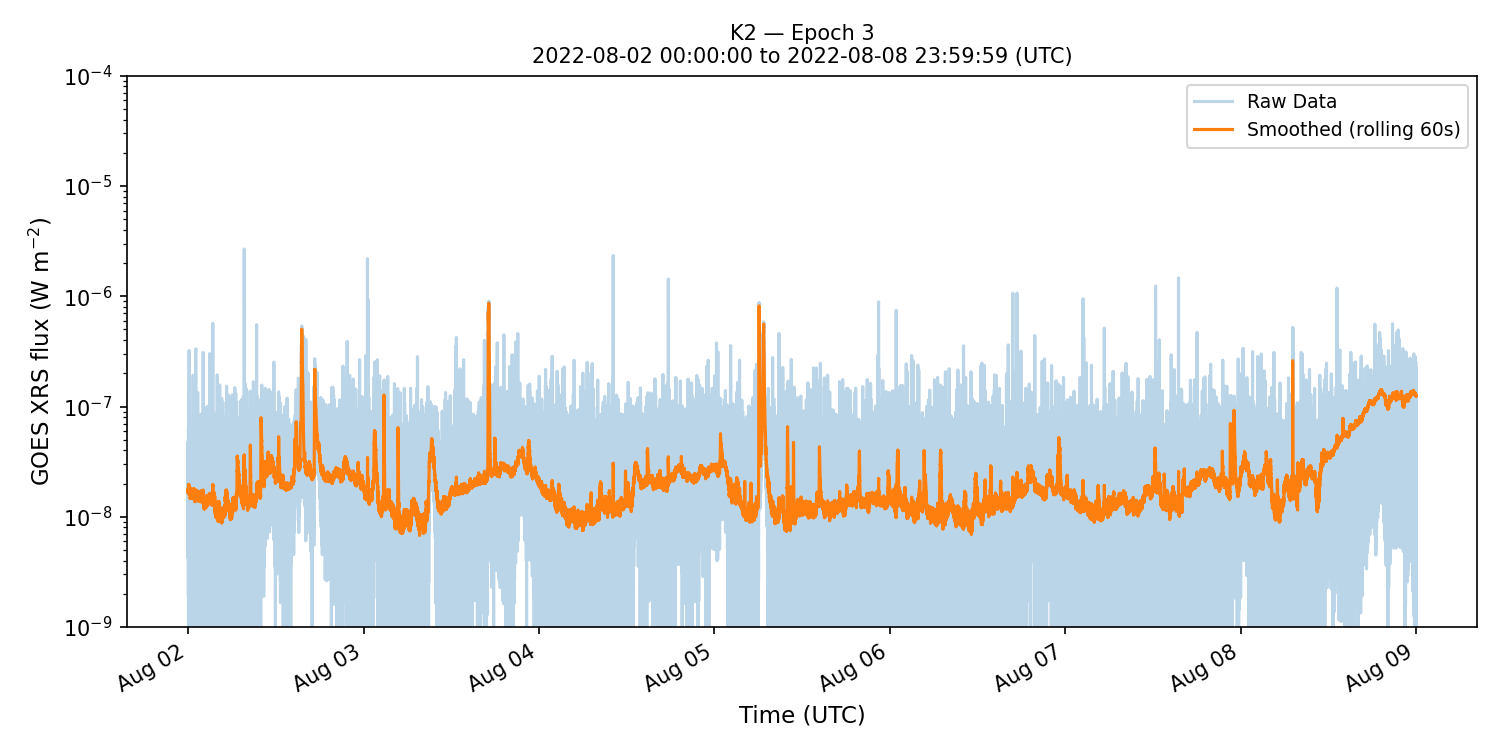}
  \caption{Geostationary Operational Environmental Satellites (GOES) X-ray fluxes over the period of our observations for Comet C/2017 K2 during epoch 3. Raw data corresponding to the $1.6-12.4$\,keV energy range is shown in blue in the background and a rolling 60-second mean is shown in orange in the foreground.}
  \label{fig:k2-goes-e3}
\end{figure}

\begin{figure}
  \centering
  \includegraphics[width=\linewidth,trim={0.35cm 0.25cm 0.25cm 1.1cm},clip]{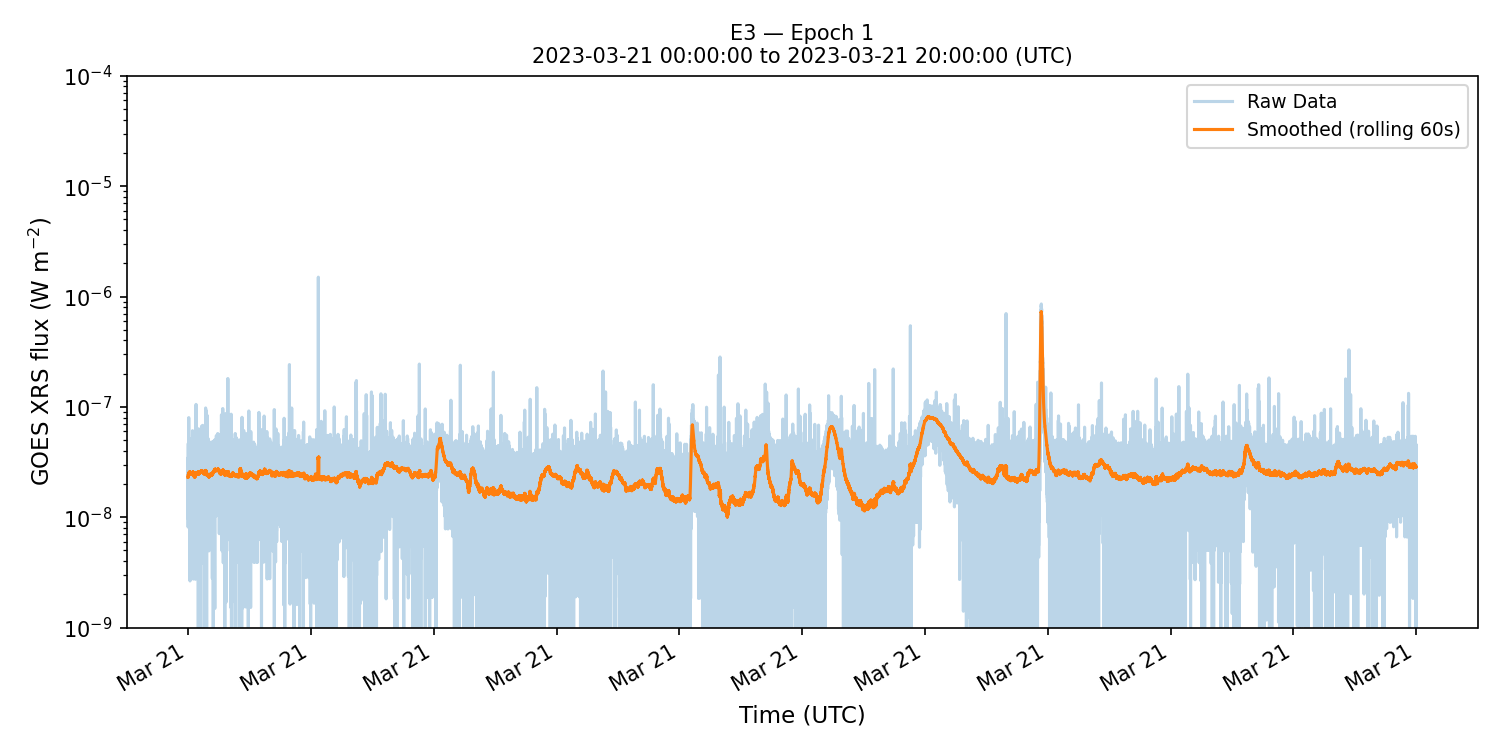}
  \caption{Geostationary Operational Environmental Satellites (GOES) X-ray fluxes over the period of our observations for Comet C/2022 E3 during epoch 1. Raw data corresponding to the $1.6-12.4$\,keV energy range is shown in blue in the background and a rolling 60-second mean is shown in orange in the foreground.}
  \label{fig:e3-goes-e1}
\end{figure}

\begin{figure}
  \centering
  \includegraphics[width=\linewidth,trim={0.35cm 0.25cm 0.25cm 1.1cm},clip]{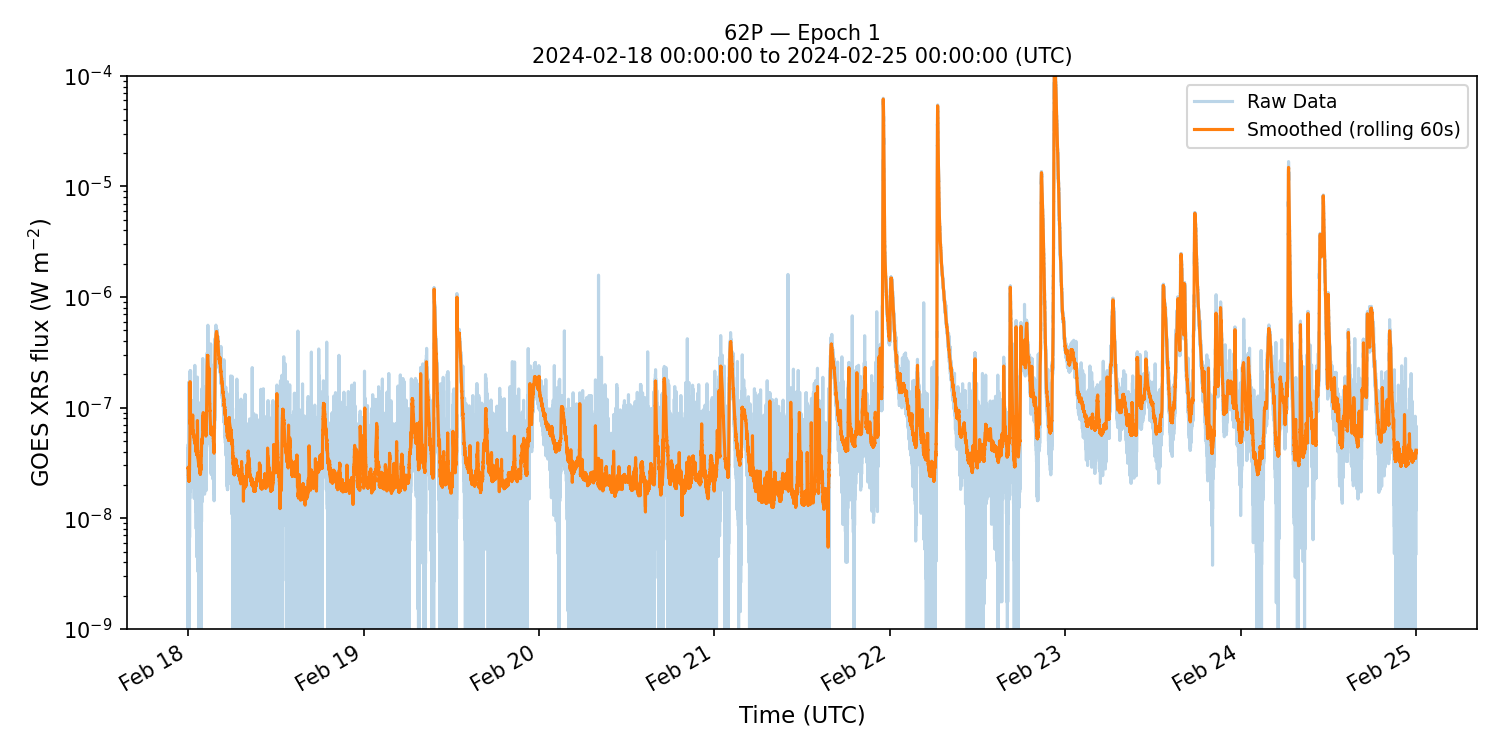}
  \caption{Geostationary Operational Environmental Satellites (GOES) X-ray fluxes over the period of our observations for Comet 62P during epoch 1. Raw data corresponding to the $1.6-12.4$\,keV energy range is shown in blue in the background and a rolling 60-second mean is shown in orange in the foreground.}
  \label{fig:62p-goes-e1}
\end{figure}

\begin{figure}
  \centering
  \includegraphics[width=\linewidth,trim={0.35cm 0.25cm 0.25cm 1.1cm},clip]{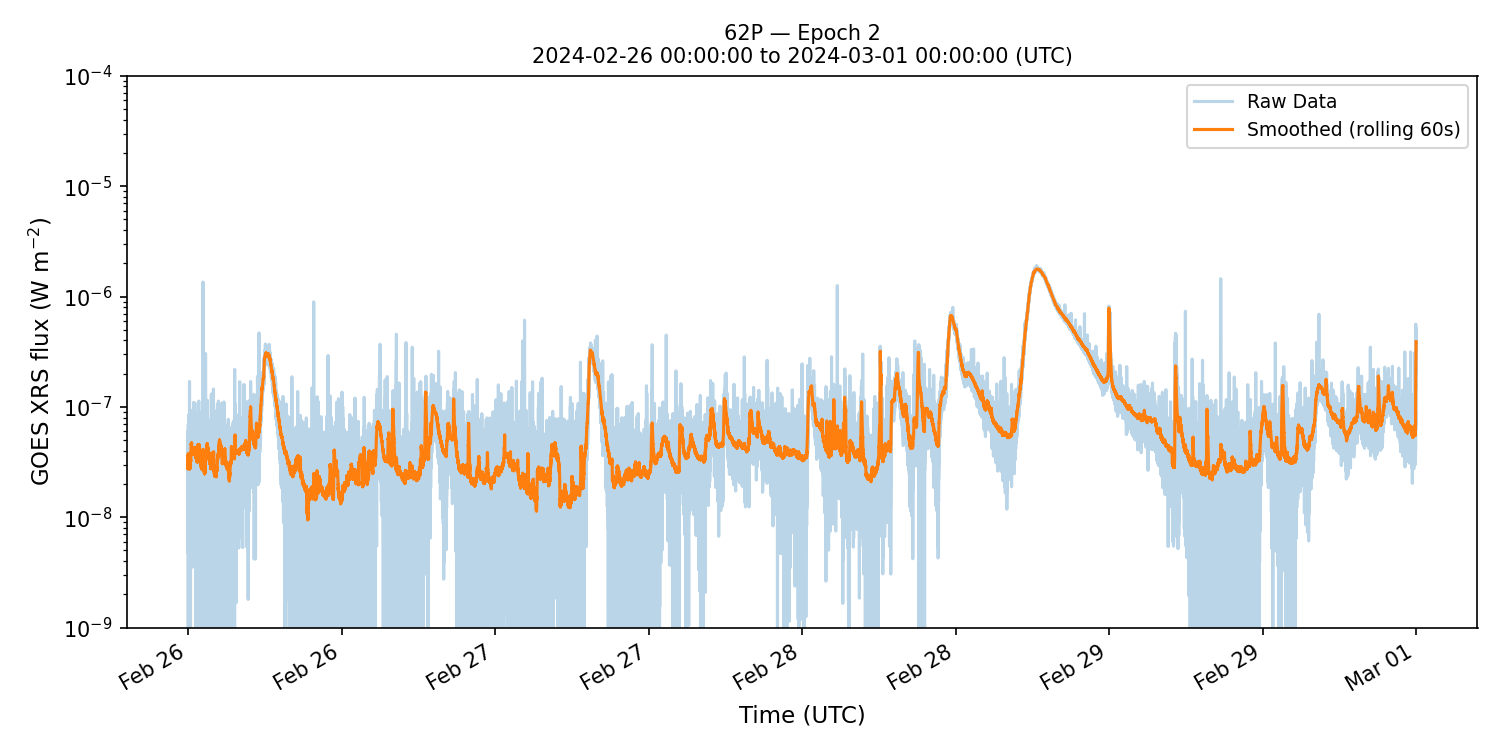}
  \caption{Geostationary Operational Environmental Satellites (GOES) X-ray fluxes over the period of our observations for Comet 62P during epoch 2. Raw data corresponding to the $1.6-12.4$\,keV energy range is shown in blue in the background and a rolling 60-second mean is shown in orange in the foreground.}
  \label{fig:62p-goes-e2}
\end{figure}

\begin{figure}
  \centering
  \includegraphics[width=\linewidth,trim={0.35cm 0.25cm 0.25cm 1.1cm},clip]{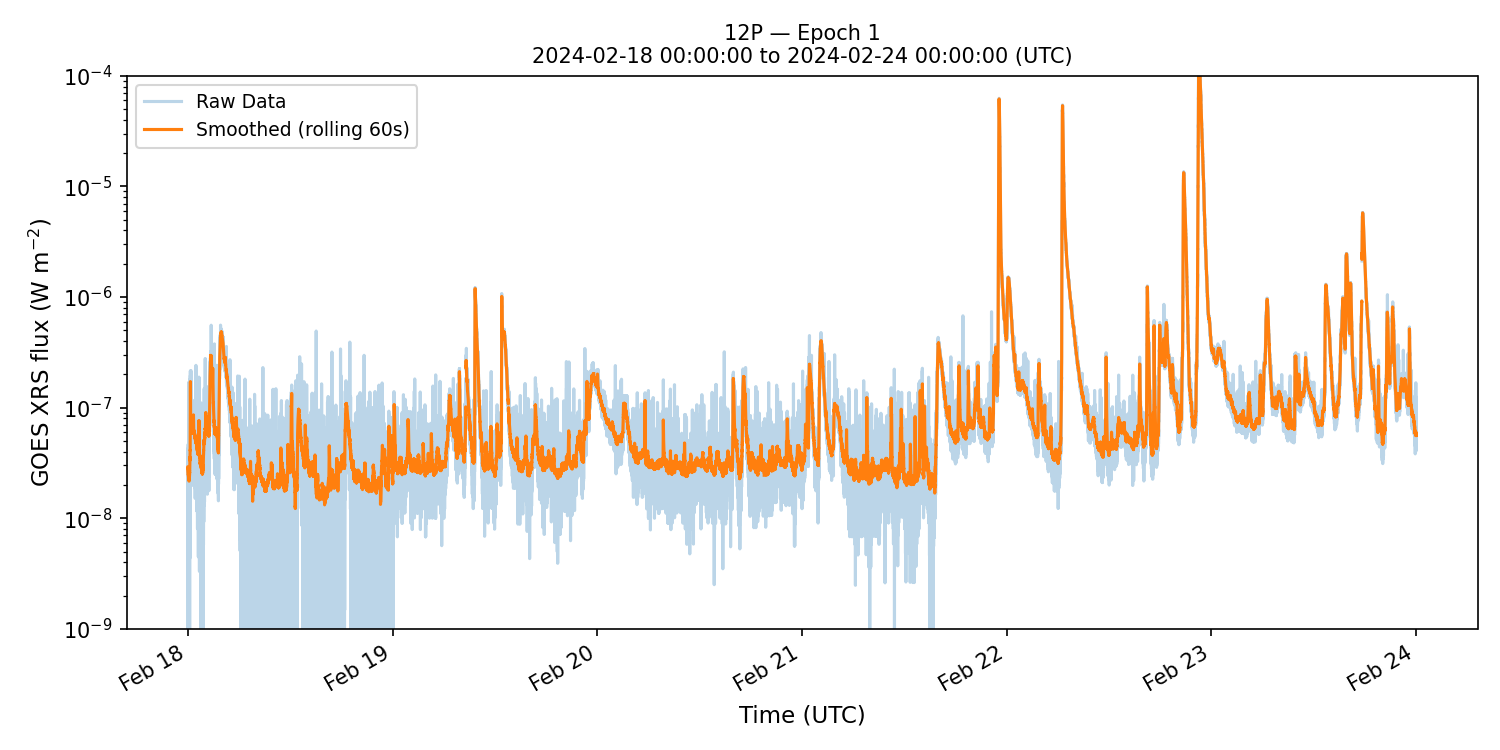}
  \caption{Geostationary Operational Environmental Satellites (GOES) X-ray fluxes over the period of our observations for Comet 12P during epoch 1. Raw data corresponding to the $1.6-12.4$\,keV energy range is shown in blue in the background and a rolling 60-second mean is shown in orange in the foreground.}
  \label{fig:12p-goes-e1}
\end{figure}

\begin{figure}
  \centering
  \includegraphics[width=\linewidth,trim={0.35cm 0.25cm 0.25cm 1.1cm},clip]{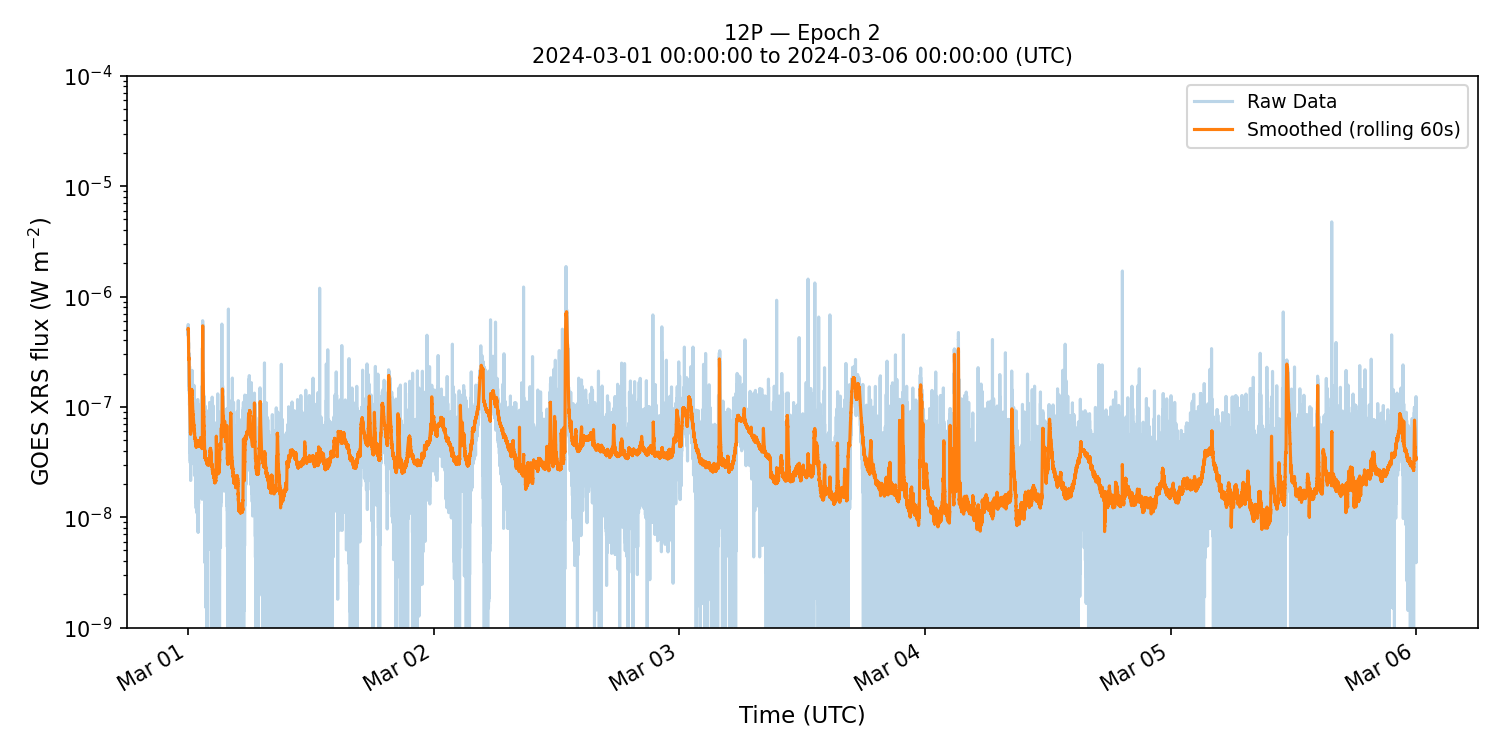}
  \caption{Geostationary Operational Environmental Satellites (GOES) X-ray fluxes over the period of our observations for Comet 12P during epoch 2. Raw data corresponding to the $1.6-12.4$\,keV energy range is shown in blue in the background and a rolling 60-second mean is shown in orange in the foreground.}
  \label{fig:12p-goes-e2}
\end{figure}

\begin{figure}
  \centering
  \includegraphics[width=\linewidth,trim={0.35cm 0.25cm 0.25cm 1.1cm},clip]{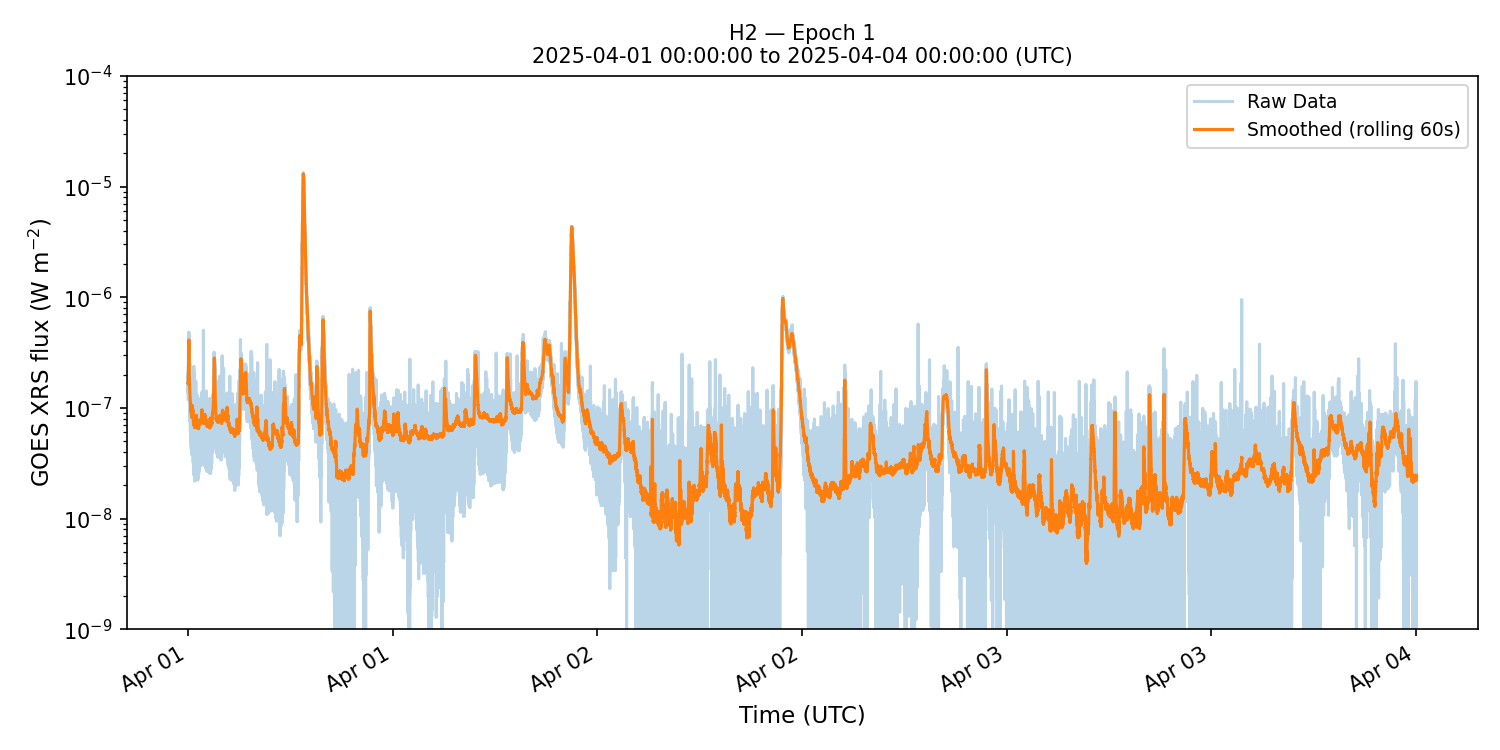}
  \caption{Geostationary Operational Environmental Satellites (GOES) X-ray fluxes over the period of our observations for Comet P/2010 H2 during epoch 1. Raw data corresponding to the $1.6-12.4$\,keV energy range is shown in blue in the background and a rolling 60-second mean is shown in orange in the foreground.}
  \label{fig:h2-goes-e1}
\end{figure}

\section{Neutral–Species Model Spectra}
\label{app:neutral_species_spectra}

This appendix presents model spectra illustrating the expected differences between SWCX emission with H$_2$O and CO$_2$+CO as the dominant neutral targets.  
For each case, synthetic spectra were generated using line energies and normalized line ratios from the \texttt{Kronos} charge–exchange database \citep{cumbee2017charge}, calculated at a collision energy of 987.3\,eV\,u$^{-1}$, corresponding to a collision velocity of 434.8\,km\,s$^{-1}$.  
This velocity is typical of the slow solar wind near 1\,AU, making the resulting line ratios representative of heliospheric and cometary SWCX conditions.  
Each ion group is represented by Gaussian profiles at the energies and relative strengths tabulated in the corresponding Kronos datasets.  
All intensities are normalized to the strongest line within each ion species to highlight differences in spectral morphology rather than absolute flux.  
Figure~\ref{fig:simspec_h2o} and Figure~\ref{fig:simspec_co2} show the NICER simulated spectra for the two neutral cases.



\begin{figure*}
    \centering
    \begin{subcaptionblock}[b]{0.47\textwidth}
        \includegraphics[width=\textwidth,trim={0.35cm 0.25cm 0.2cm 1.35cm},clip]{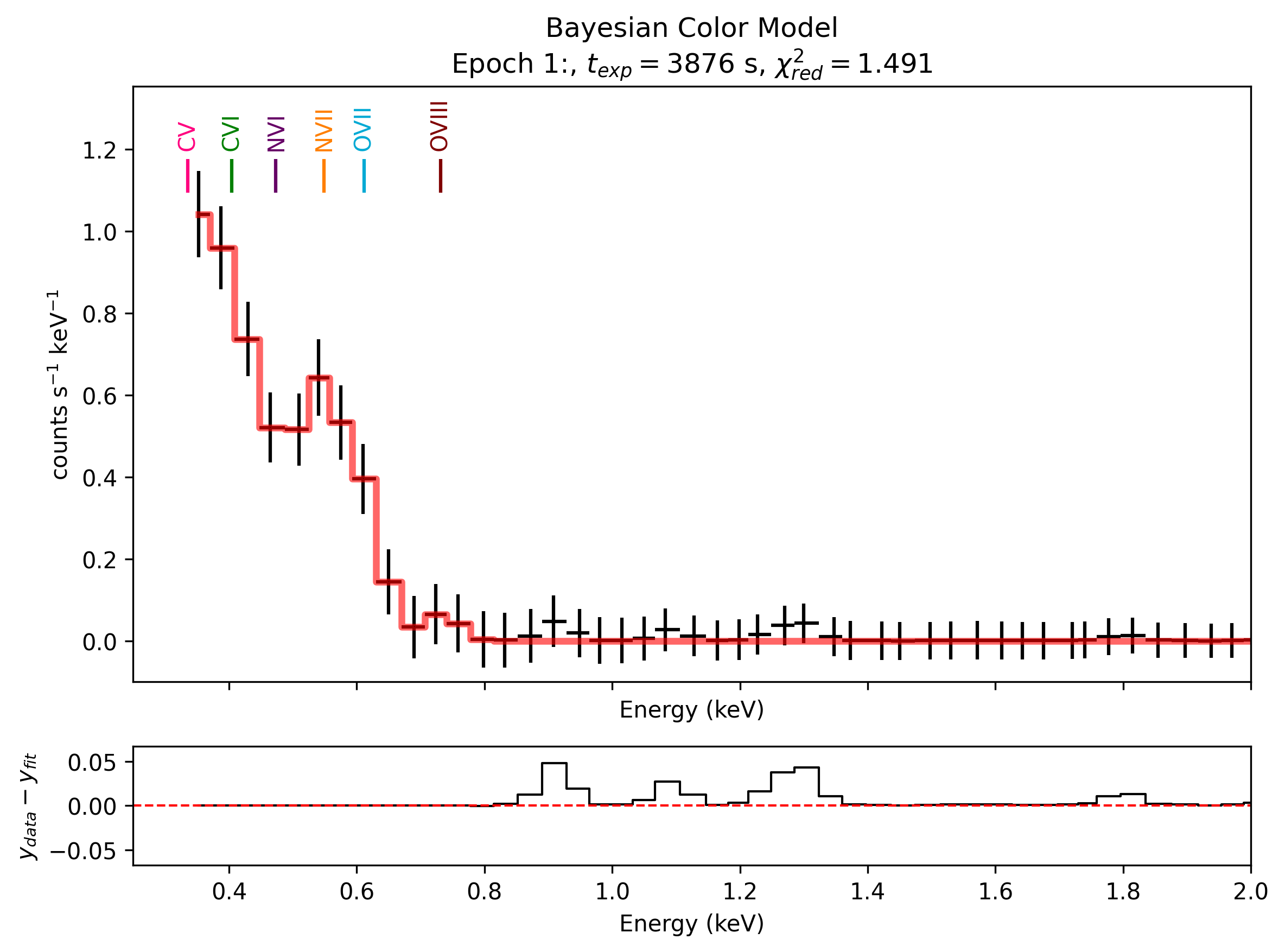}
        \caption{Epoch 1 ($t_{exp}=3876$\,ks).}
        \label{fig:spec_88p_e1-co2}
    \end{subcaptionblock}
    \hfill
    \begin{subcaptionblock}[b]{0.47\textwidth}
        \includegraphics[width=\textwidth,trim={0.35cm 0.25cm 0.2cm 1.35cm},clip]{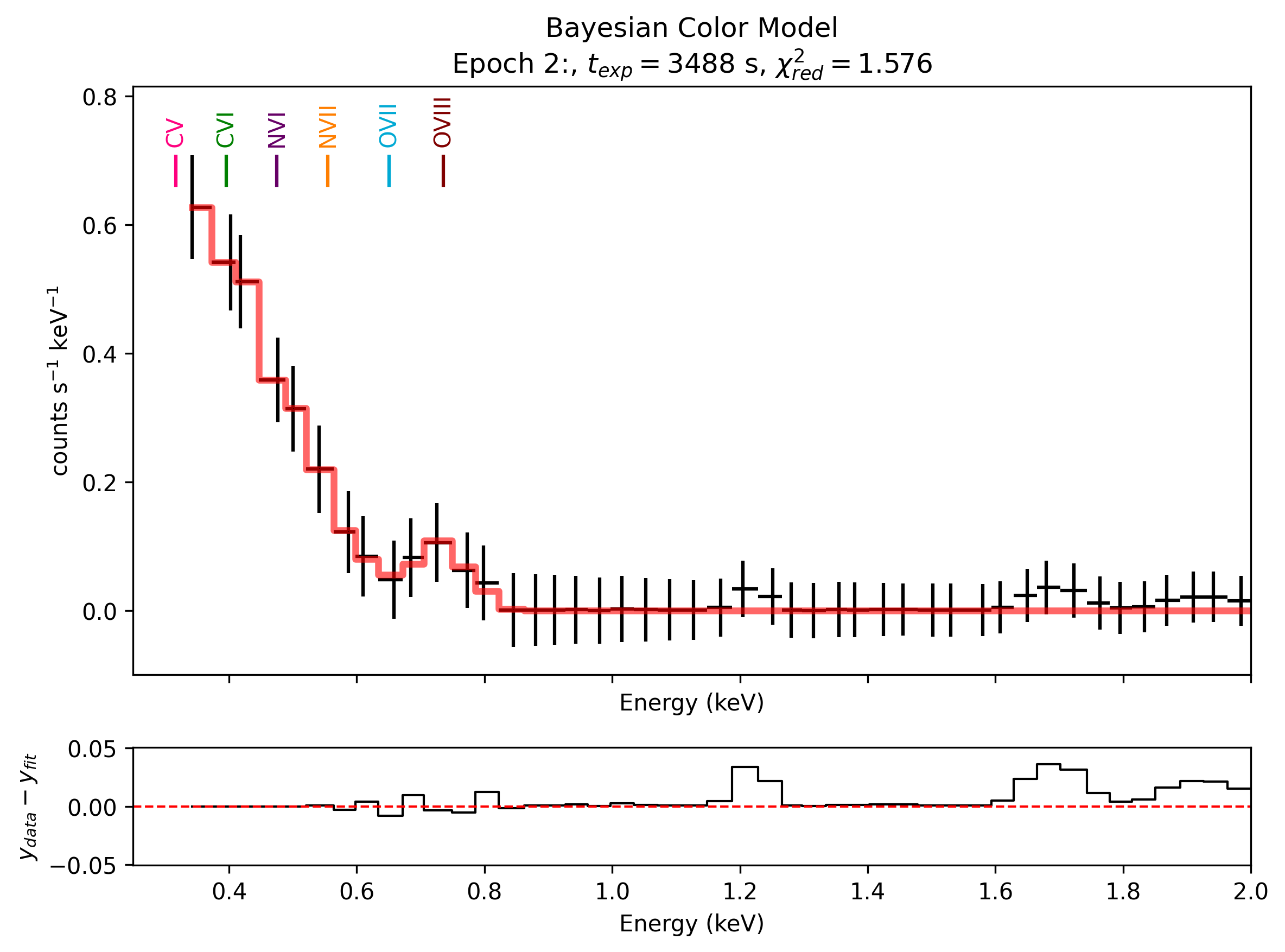}
        \caption{Epoch 2 ($t_{exp}=3488$\,ks).}
        \label{fig:spec_88p_e2-co2}
    \end{subcaptionblock}
    \hfill
    \begin{subcaptionblock}[b]{0.47\textwidth}
        \includegraphics[width=\textwidth,trim={0.35cm 0.25cm 0.2cm 1.35cm},clip]{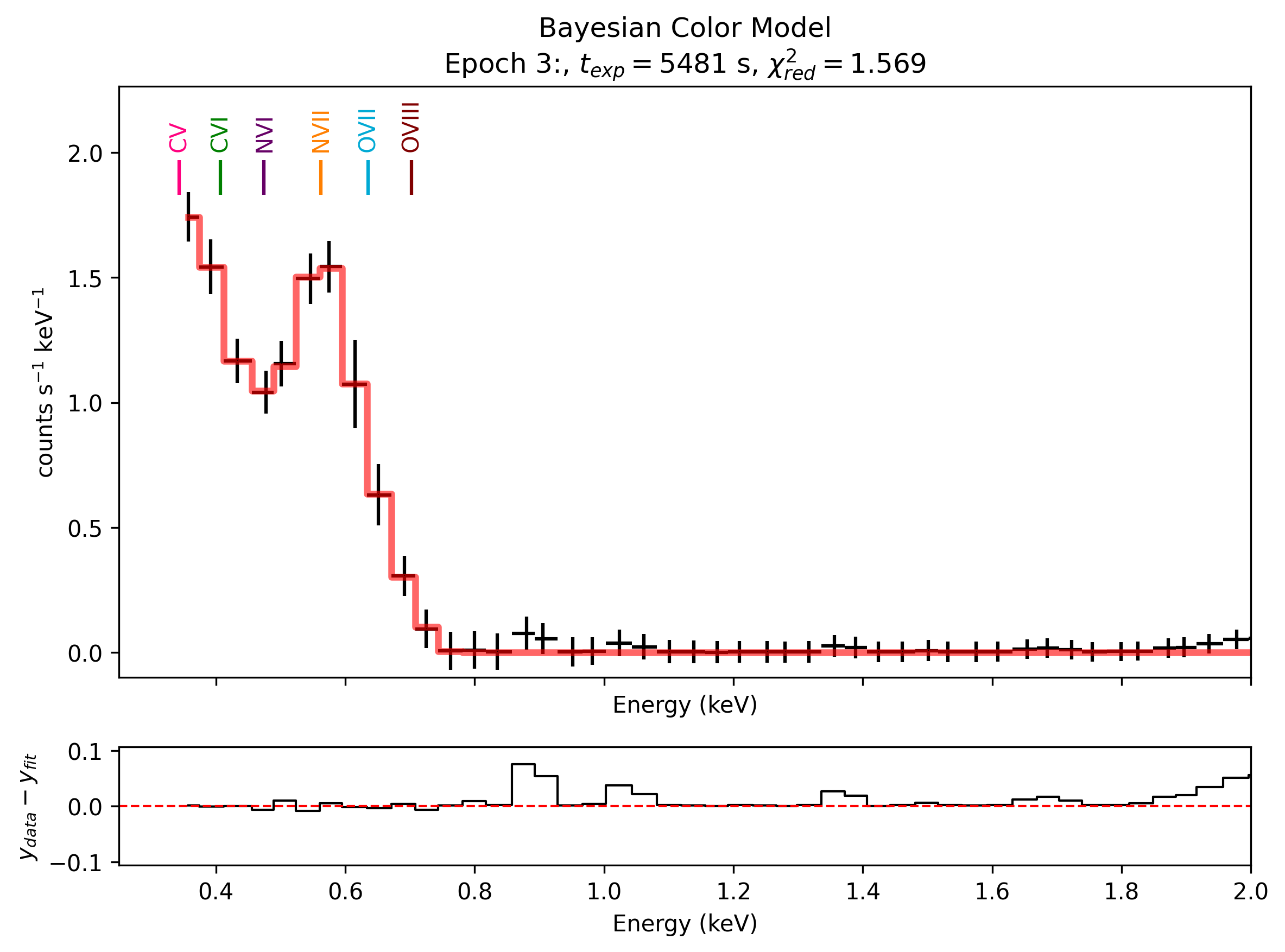}
        \caption{Epoch 3 ($t_{exp}=5481$\,ks).}
        \label{fig:spec_88p_e3-co2}
    \end{subcaptionblock}
    \hfill
    \begin{subcaptionblock}[b]{0.47\textwidth}
        \includegraphics[width=\textwidth,trim={0.35cm 0.25cm 0.2cm 1.35cm},clip]{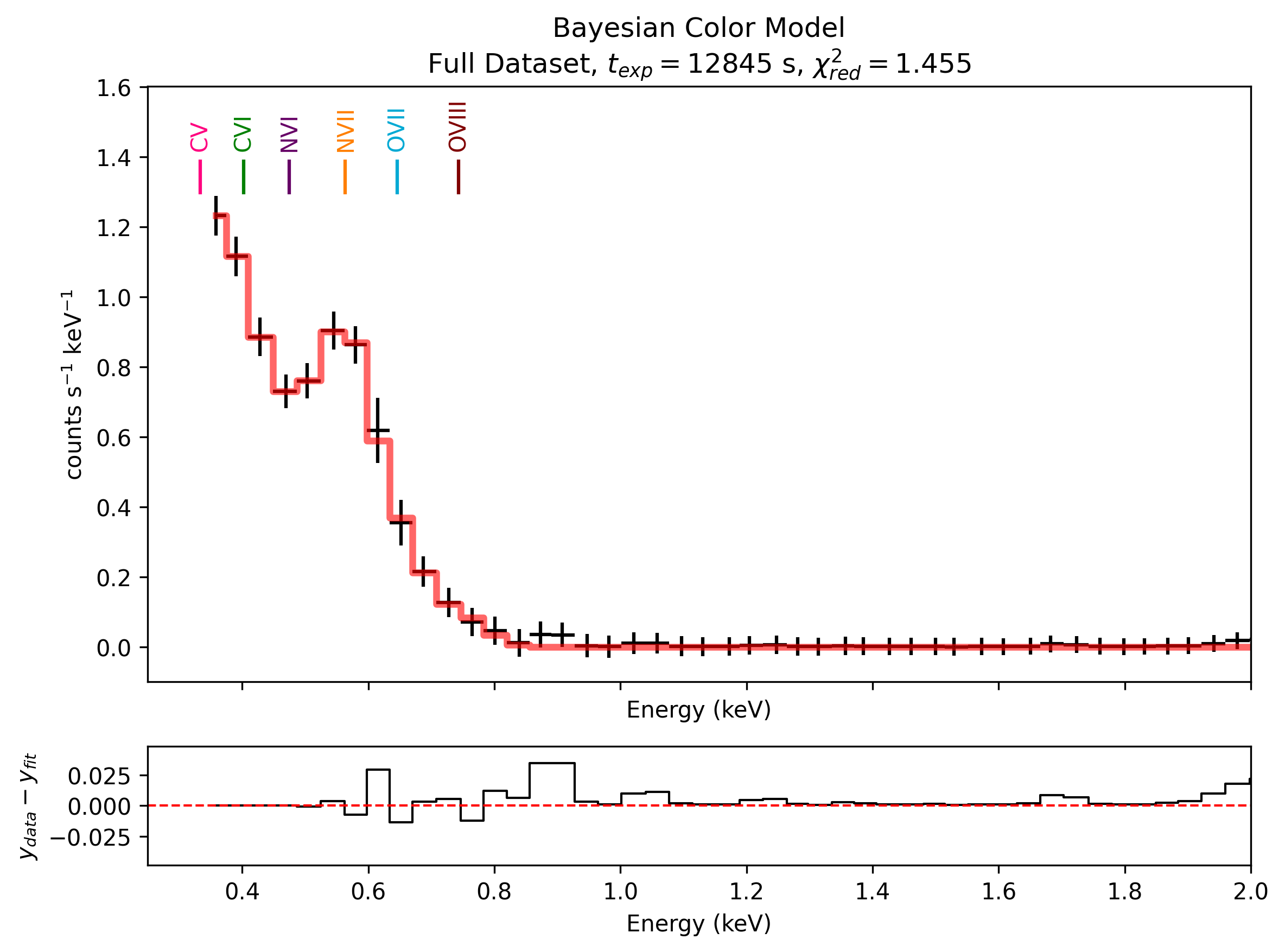}
        \caption{Full dataset ($t_{exp}=12845$\,ks).}
        \label{fig:spec_88p_full-co2}
    \end{subcaptionblock}

    \caption{X-ray spectra from the interaction between the solar wind and the atmosphere of 88P/Howell fitted with the Bayesian Color Model (BCM). (Neutral: CO$_2$).}
    \label{fig:app-spec-88p-co2}
\end{figure*}


\begin{figure*}
    \centering
    \begin{subcaptionblock}[b]{0.47\textwidth}
        \includegraphics[width=\textwidth,trim={0.35cm 0.25cm 0.2cm 1.35cm},clip]{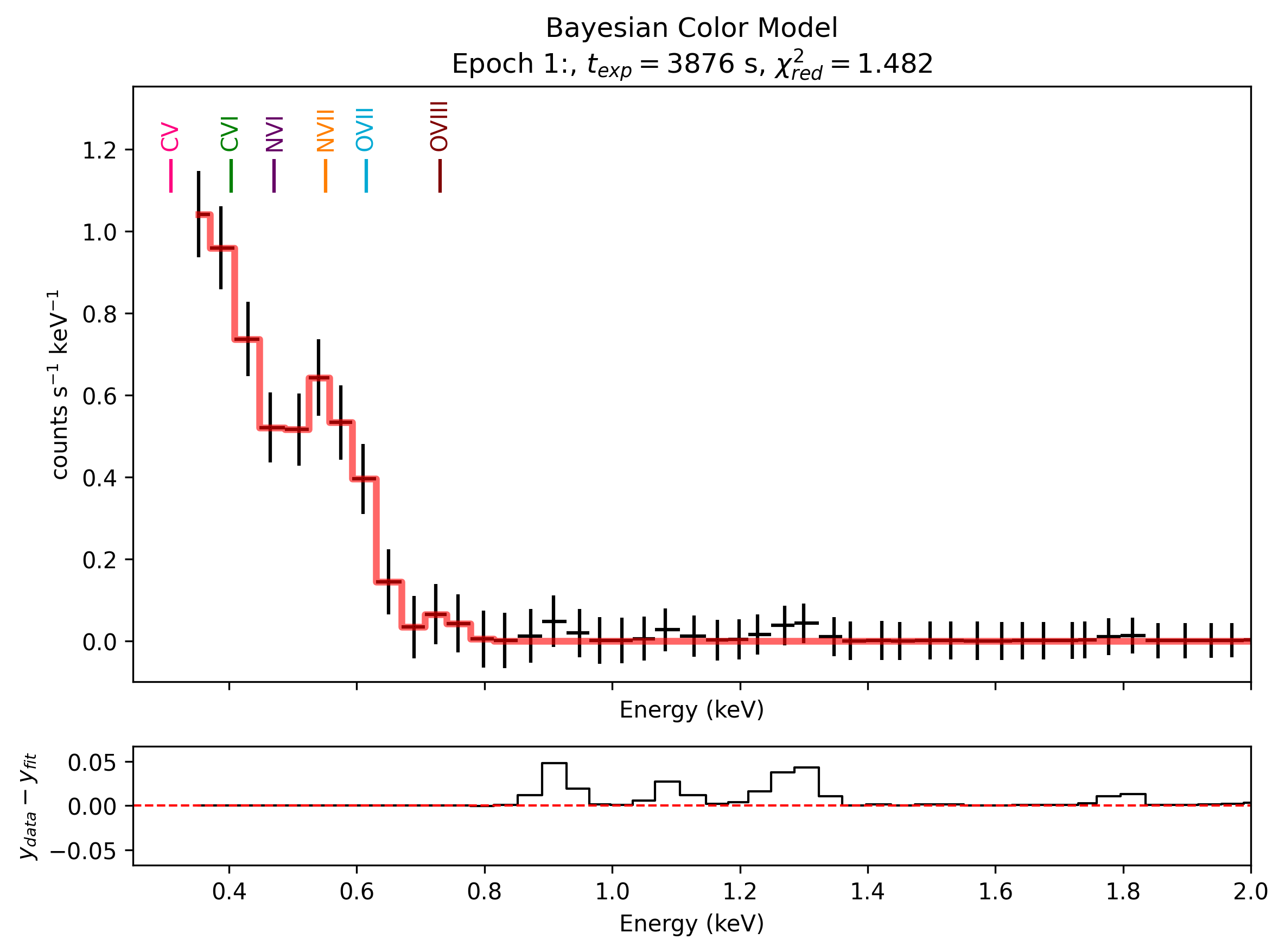}
        \caption{Epoch 1 ($t_{exp}=3876$\,ks).}
        \label{fig:spec_88p_e1-h2o}
    \end{subcaptionblock}
    \hfill
    \begin{subcaptionblock}[b]{0.47\textwidth}
        \includegraphics[width=\textwidth,trim={0.35cm 0.25cm 0.2cm 1.35cm},clip]{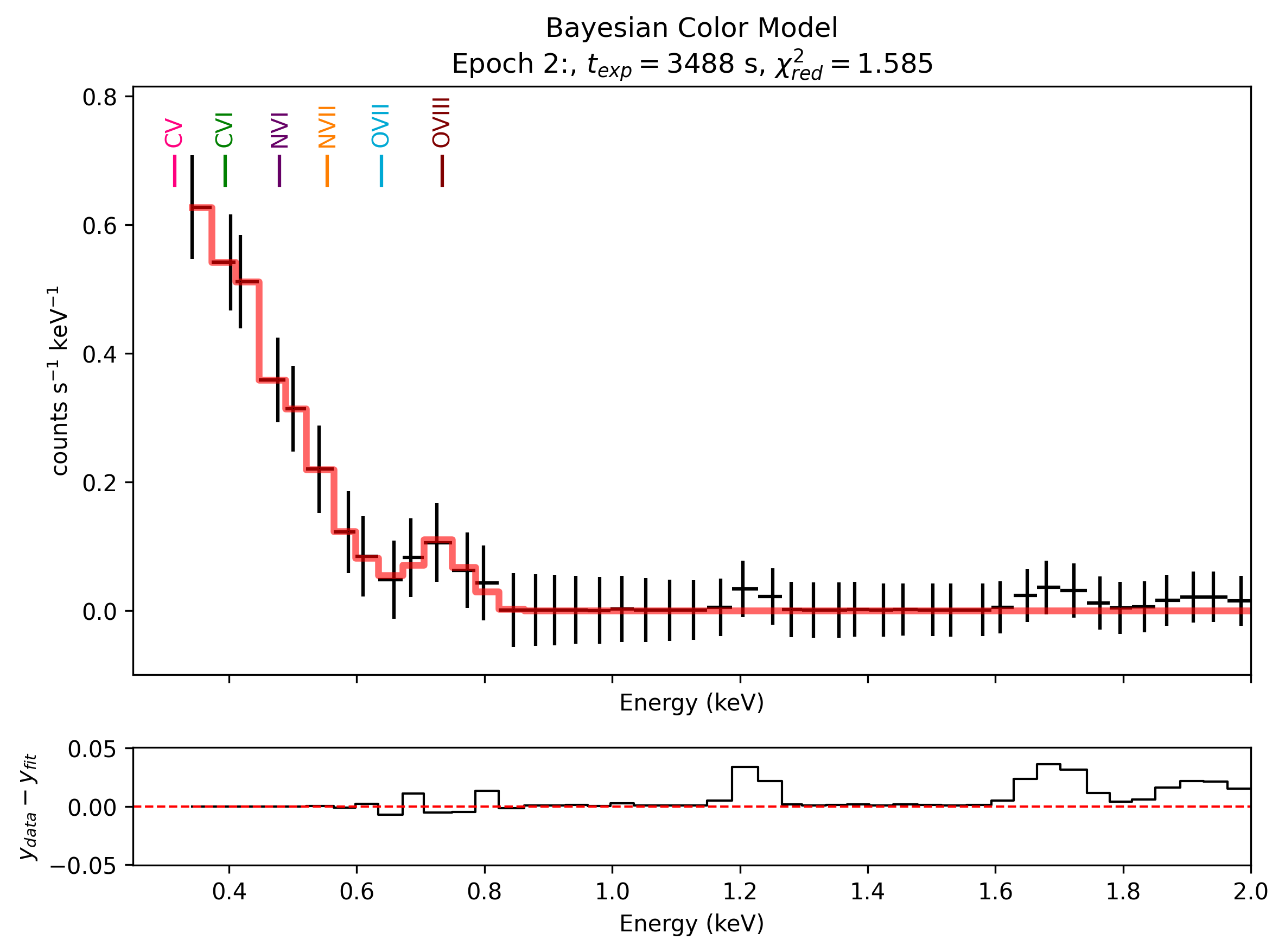}
        \caption{Epoch 2 ($t_{exp}=3488$\,ks).}
        \label{fig:spec_88p_e2-h2o}
    \end{subcaptionblock}
    \hfill
    \begin{subcaptionblock}[b]{0.47\textwidth}
        \includegraphics[width=\textwidth,trim={0.35cm 0.25cm 0.2cm 1.35cm},clip]{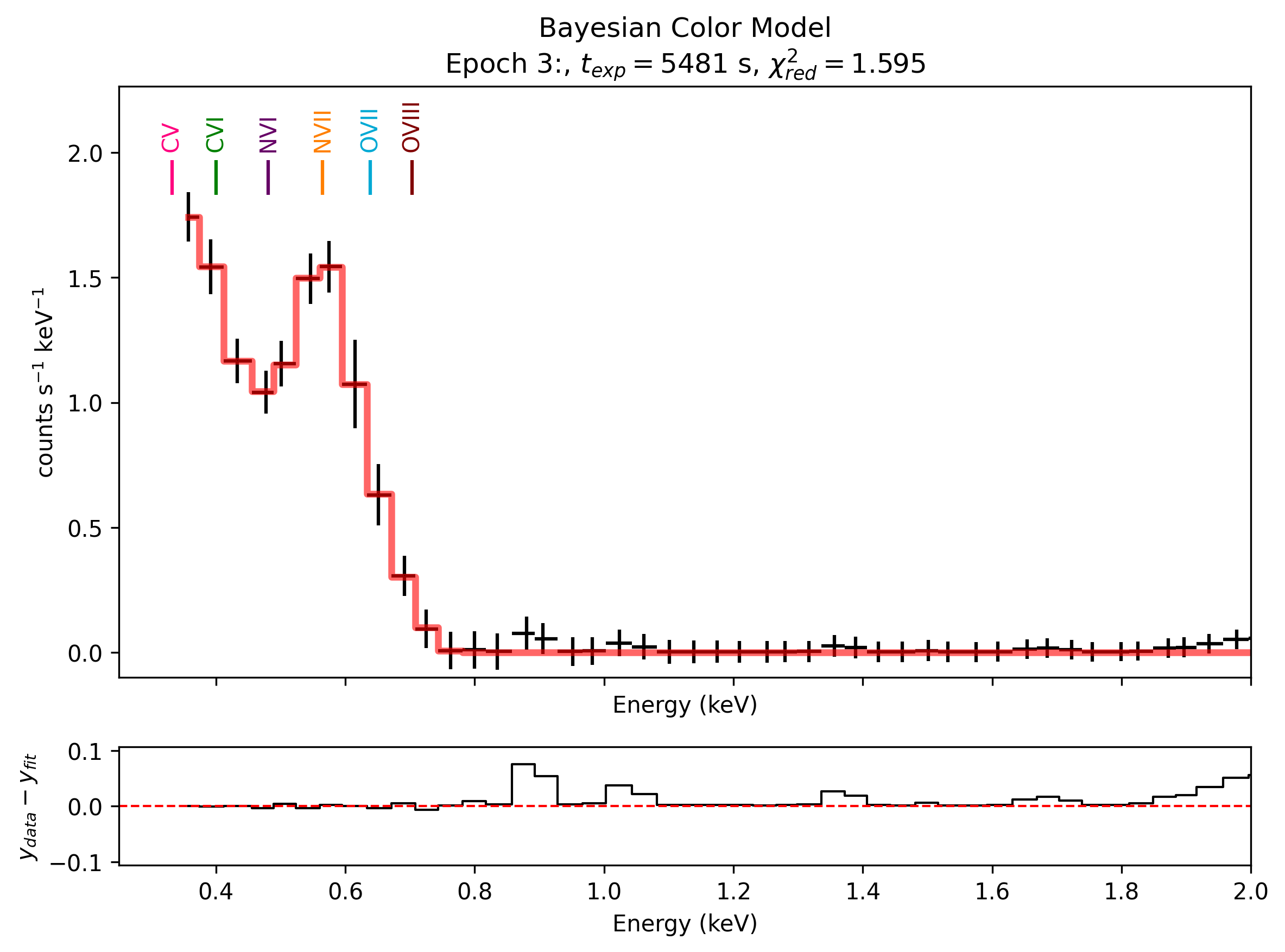}
        \caption{Epoch 3 ($t_{exp}=5481$\,ks).}
        \label{fig:spec_88p_e3-h2o}
    \end{subcaptionblock}
    \hfill
    \begin{subcaptionblock}[b]{0.47\textwidth}
        \includegraphics[width=\textwidth,trim={0.35cm 0.25cm 0.2cm 1.35cm},clip]{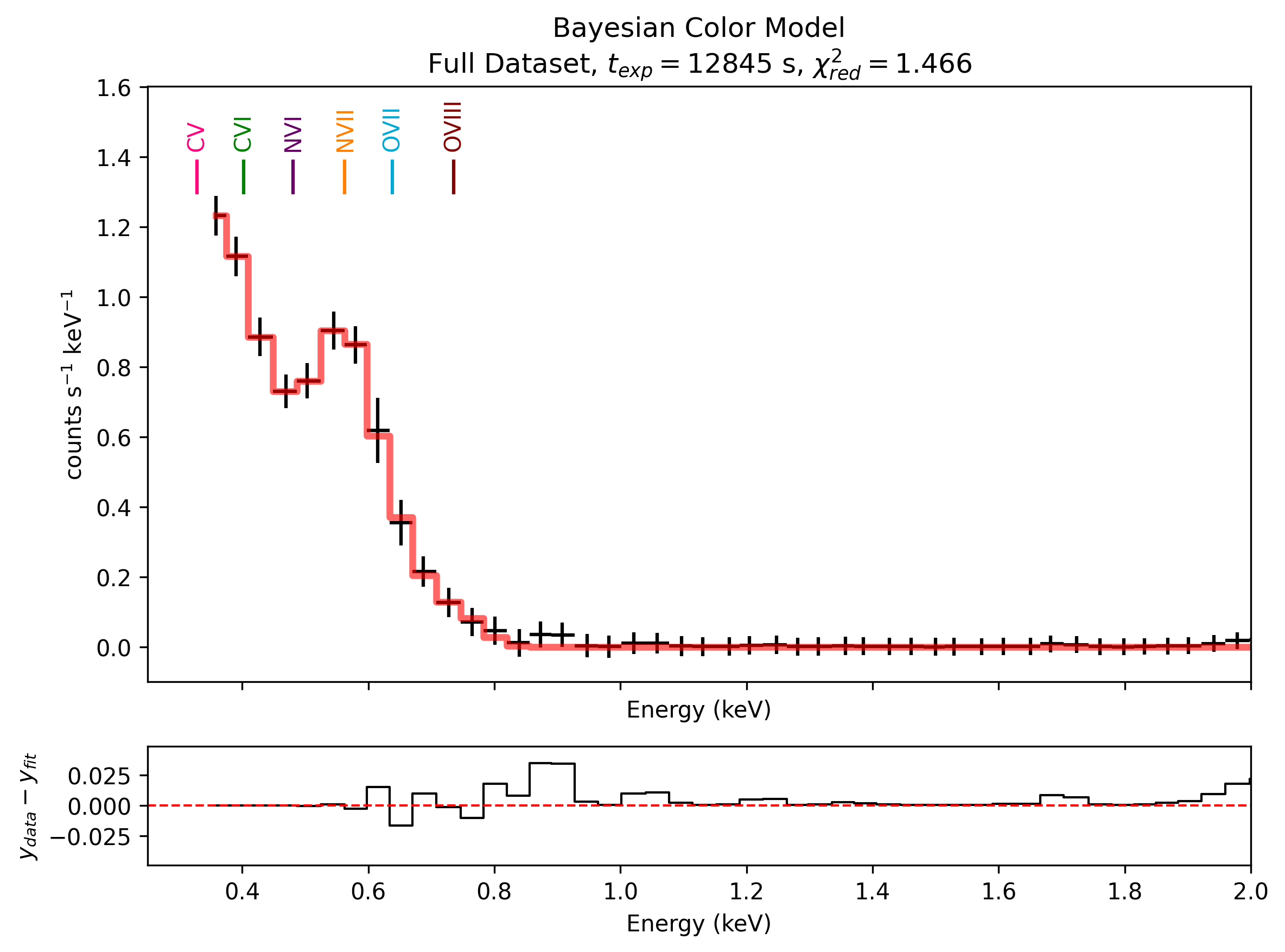}
        \caption{Full dataset ($t_{exp}=12845$\,ks).}
        \label{fig:spec_88p_full-h2o}
    \end{subcaptionblock}

    \caption{X-ray spectra from the interaction between the solar wind and the atmosphere of 88P/Howell fitted with the Bayesian Color Model (BCM). (Neutral: H$_2$O).}
    \label{fig:app-spec-88p-h2o}
\end{figure*}


\begin{figure*}
    \centering
    \begin{subcaptionblock}[b]{0.47\textwidth}
        \includegraphics[width=\textwidth,trim={0.35cm 0.25cm 0.2cm 1.35cm},clip]{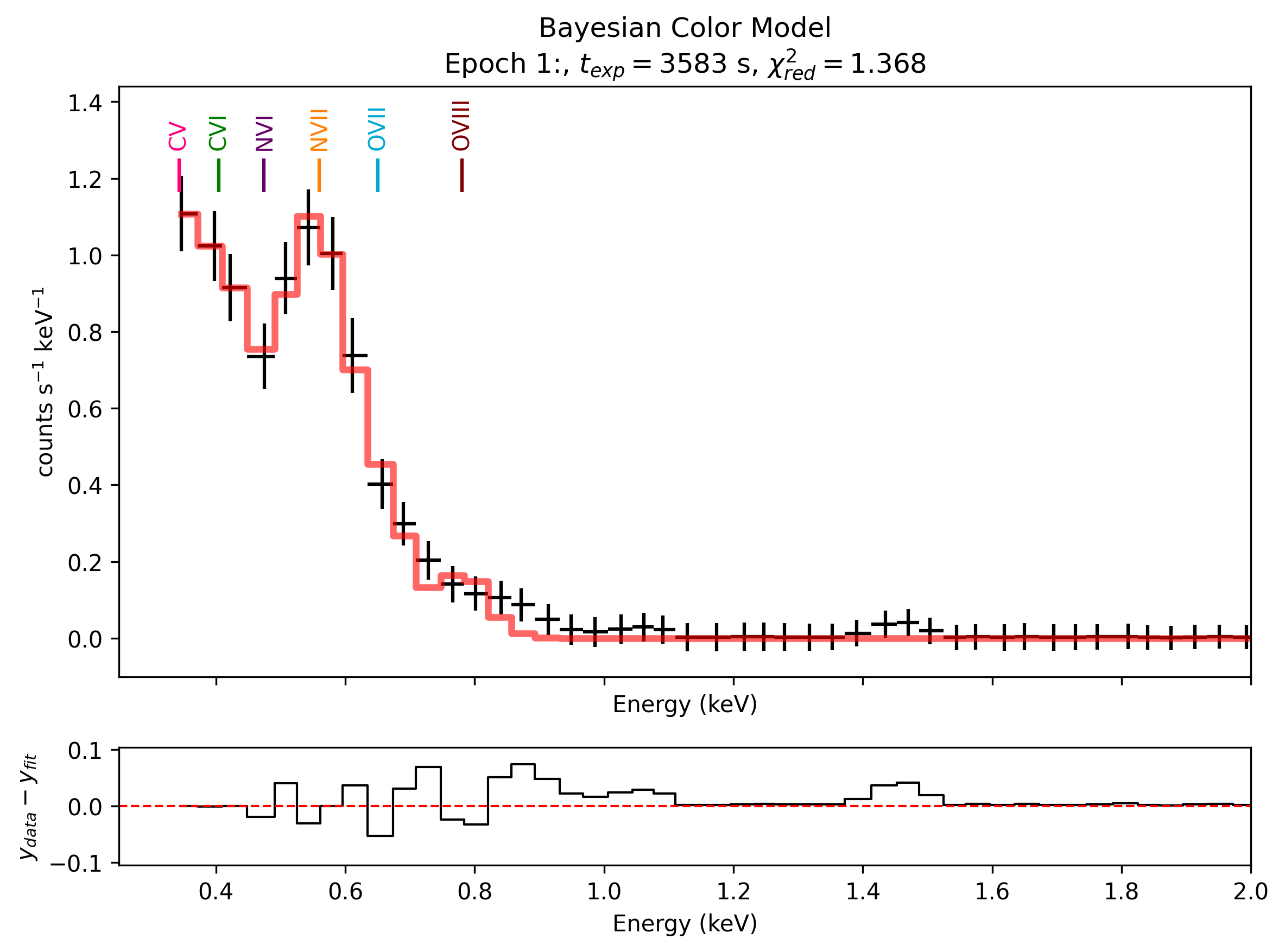}
        \caption{Epoch 1 ($t_{exp}=3583$\,ks).}
        \label{fig:spec_19p_e1-co2}
    \end{subcaptionblock}
    \hfill
    \begin{subcaptionblock}[b]{0.47\textwidth}
        \includegraphics[width=\textwidth,trim={0.35cm 0.25cm 0.2cm 1.35cm},clip]{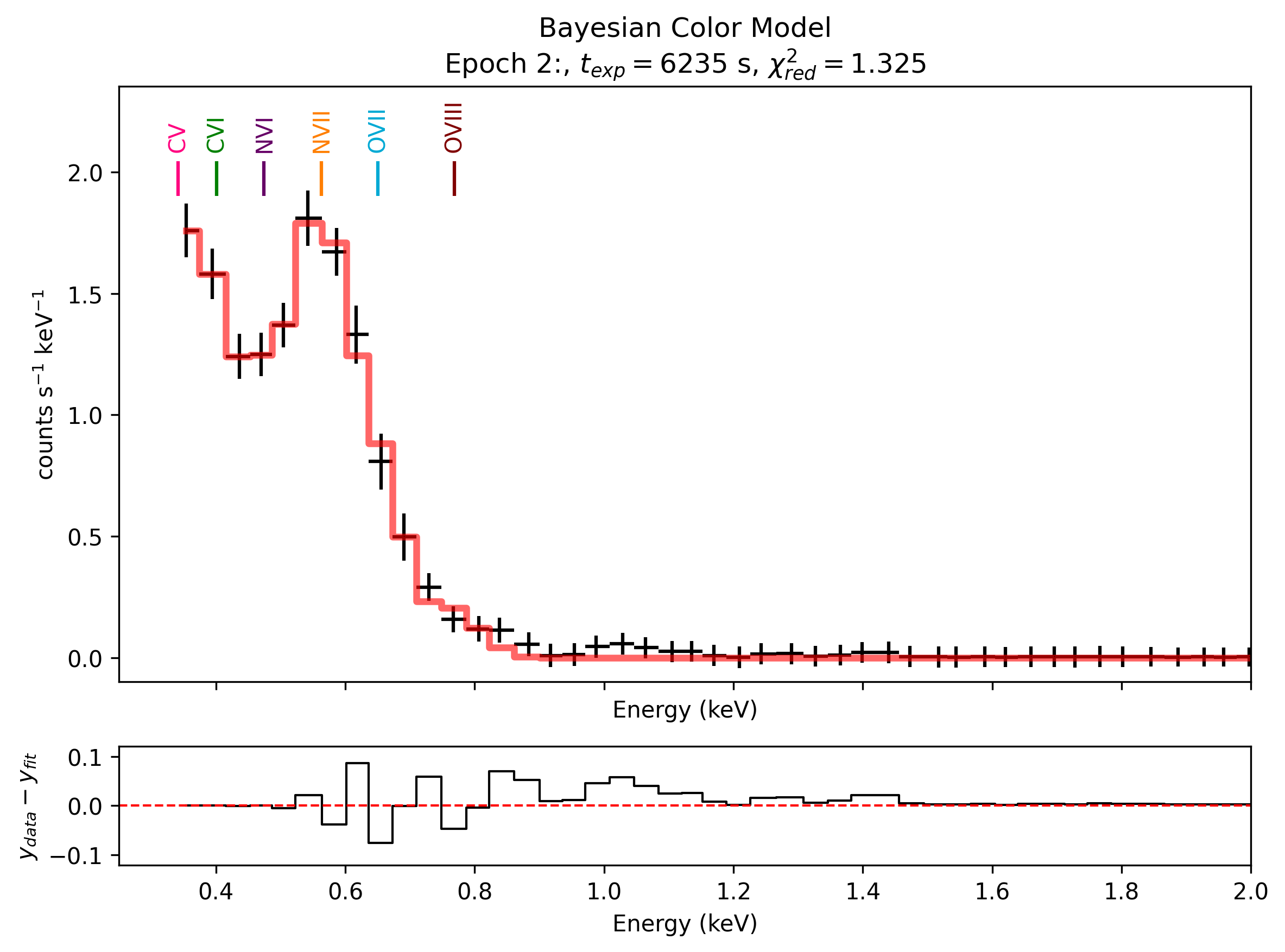}
        \caption{Epoch 2 ($t_{exp}=6235$\,ks).}
        \label{fig:spec_19p_e2-co2}
    \end{subcaptionblock}
    \hfill
    \begin{subcaptionblock}[b]{0.47\textwidth}
        \includegraphics[width=\textwidth,trim={0.35cm 0.25cm 0.2cm 1.35cm},clip]{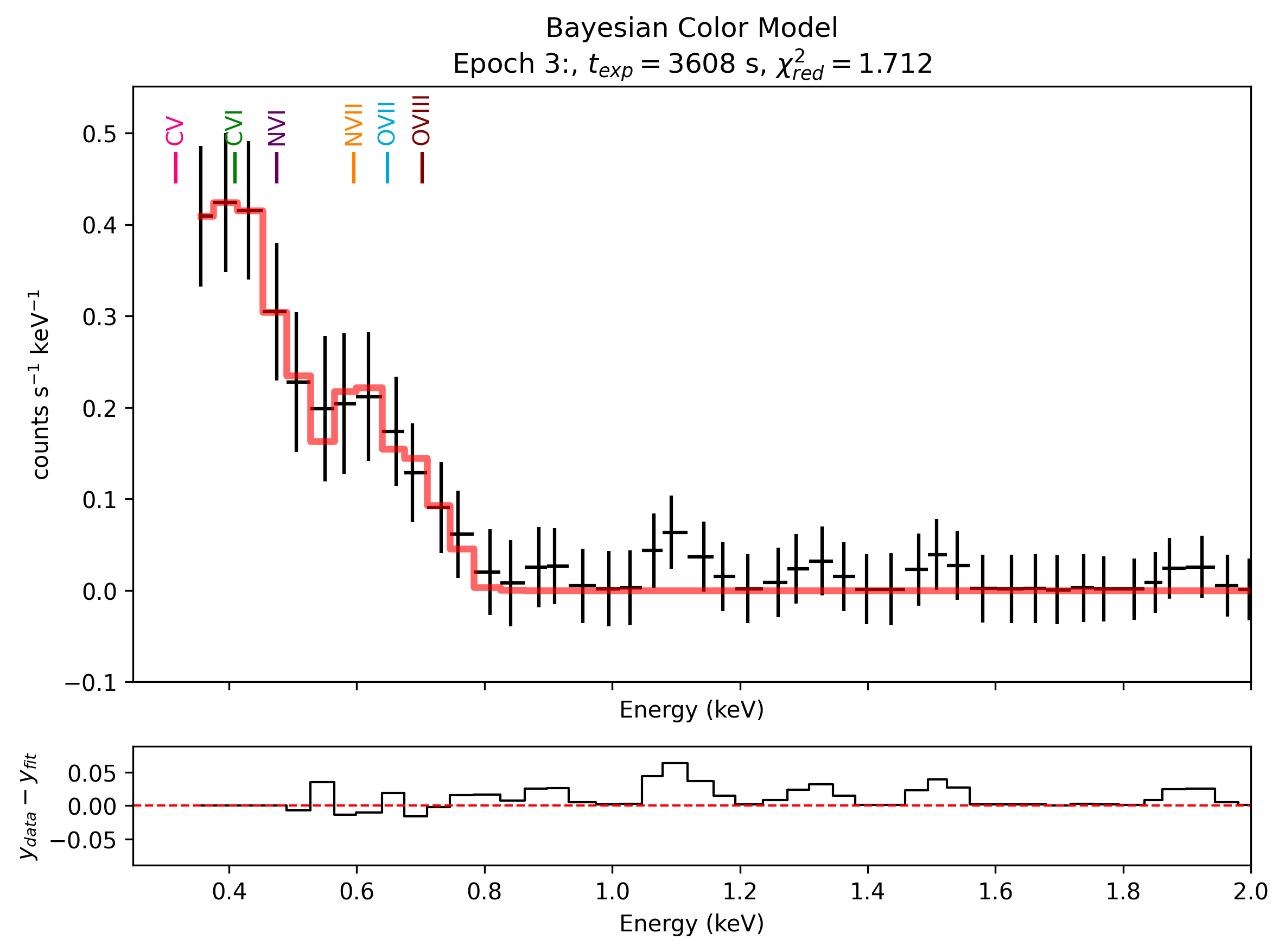}
        \caption{Epoch 3 ($t_{exp}=3608$\,ks).}
        \label{fig:spec_19p_e3-co2}
    \end{subcaptionblock}
    \hfill
    \begin{subcaptionblock}[b]{0.47\textwidth}
        \includegraphics[width=\textwidth,trim={0.35cm 0.25cm 0.2cm 1.35cm},clip]{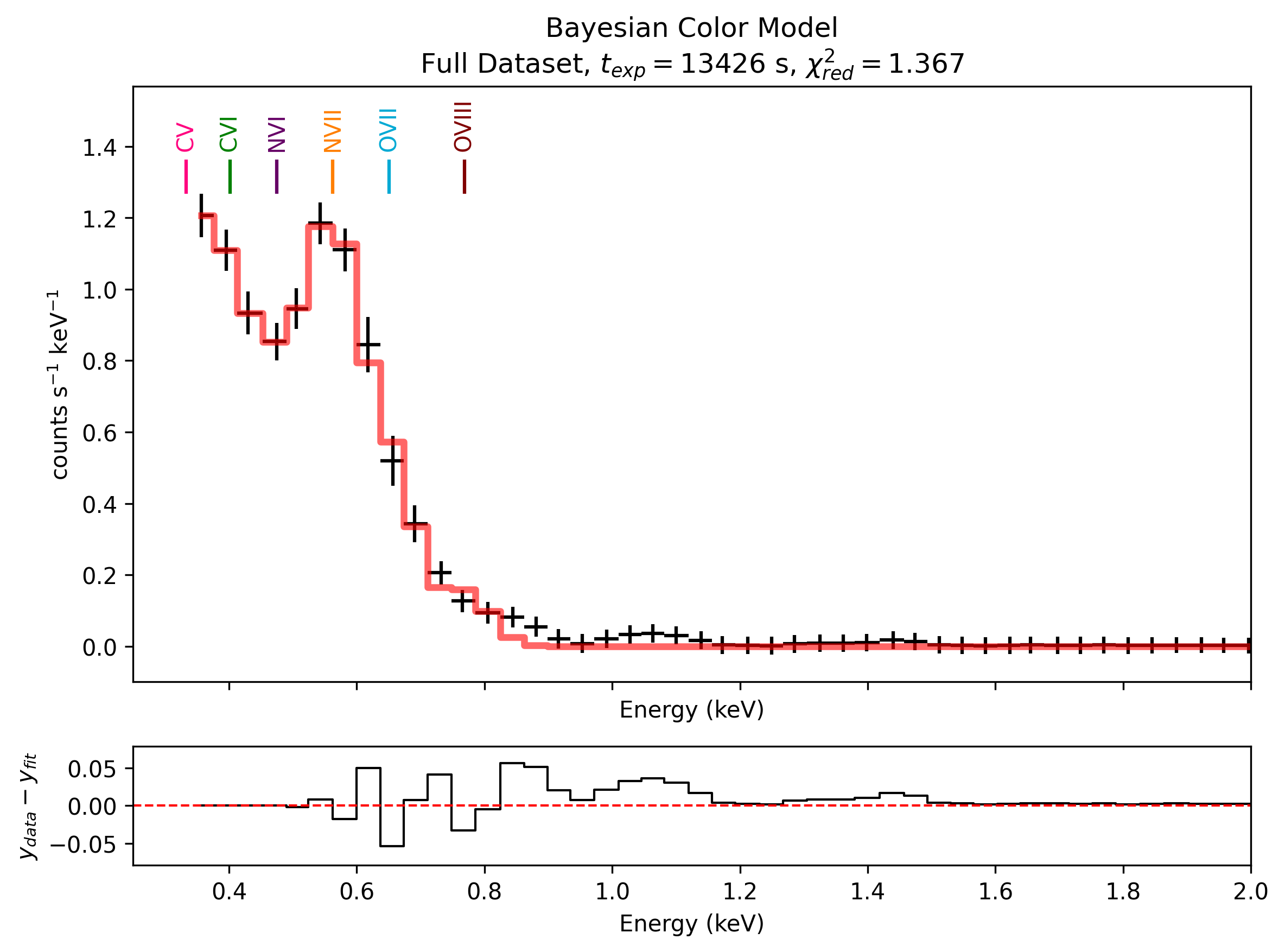}
        \caption{Full dataset ($t_{exp}=13426$\,ks).}
        \label{fig:spec_19p_full-co2}
    \end{subcaptionblock}

    \caption{X-ray spectra from the interaction between the solar wind and the atmosphere of 19P/Borrelly fitted with the Bayesian Color Model (BCM). (Neutral: CO$_2$).}
    \label{fig:app-spec-19p-co2}
\end{figure*}


\begin{figure*}
    \centering
    \begin{subcaptionblock}[b]{0.47\textwidth}
        \includegraphics[width=\textwidth,trim={0.35cm 0.25cm 0.2cm 1.35cm},clip]{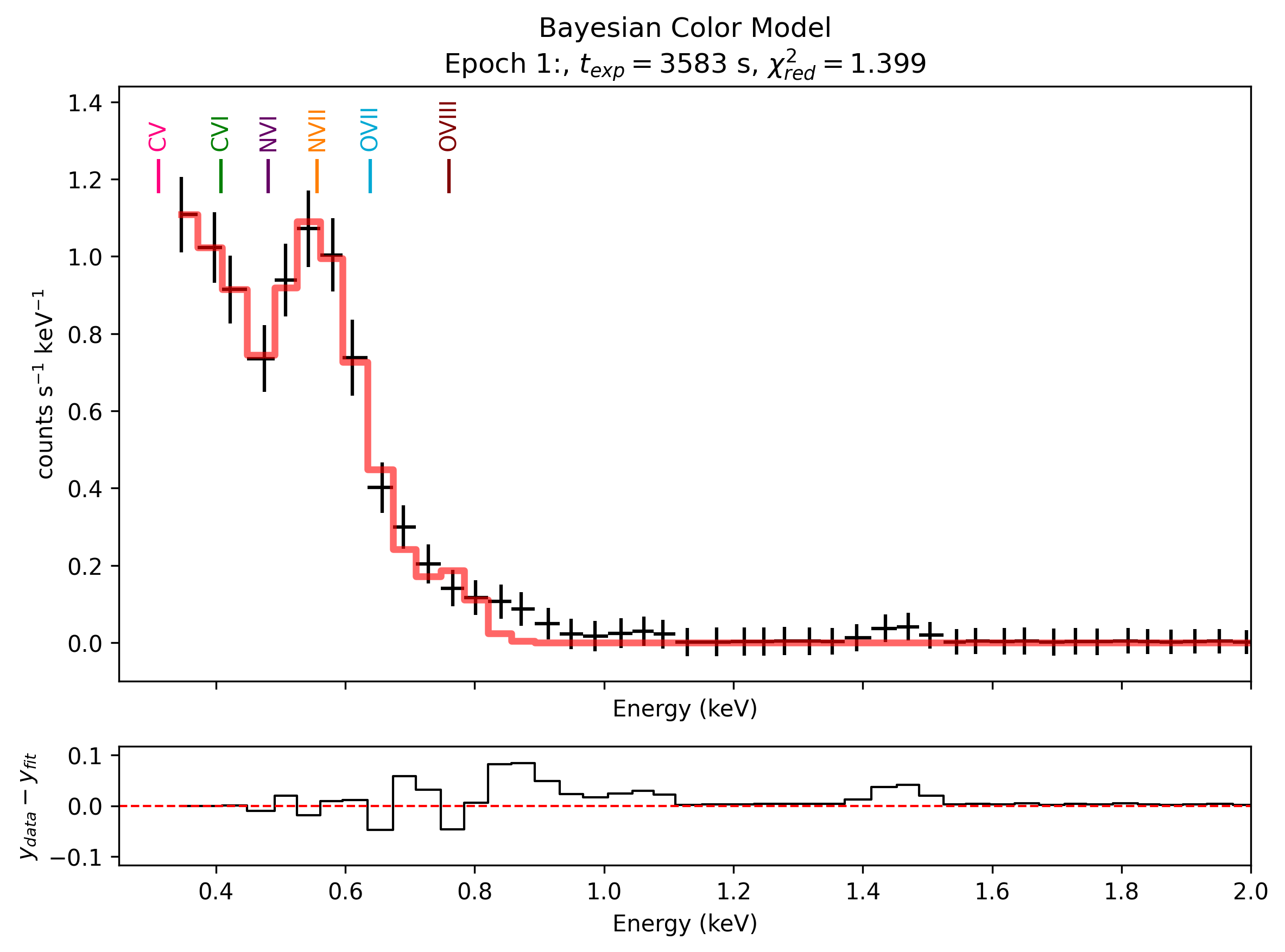}
        \caption{Epoch 1 ($t_{exp}=3583$\,ks).}
        \label{fig:spec_19p_e1-h2o}
    \end{subcaptionblock}
    \hfill
    \begin{subcaptionblock}[b]{0.47\textwidth}
        \includegraphics[width=\textwidth,trim={0.35cm 0.25cm 0.2cm 1.35cm},clip]{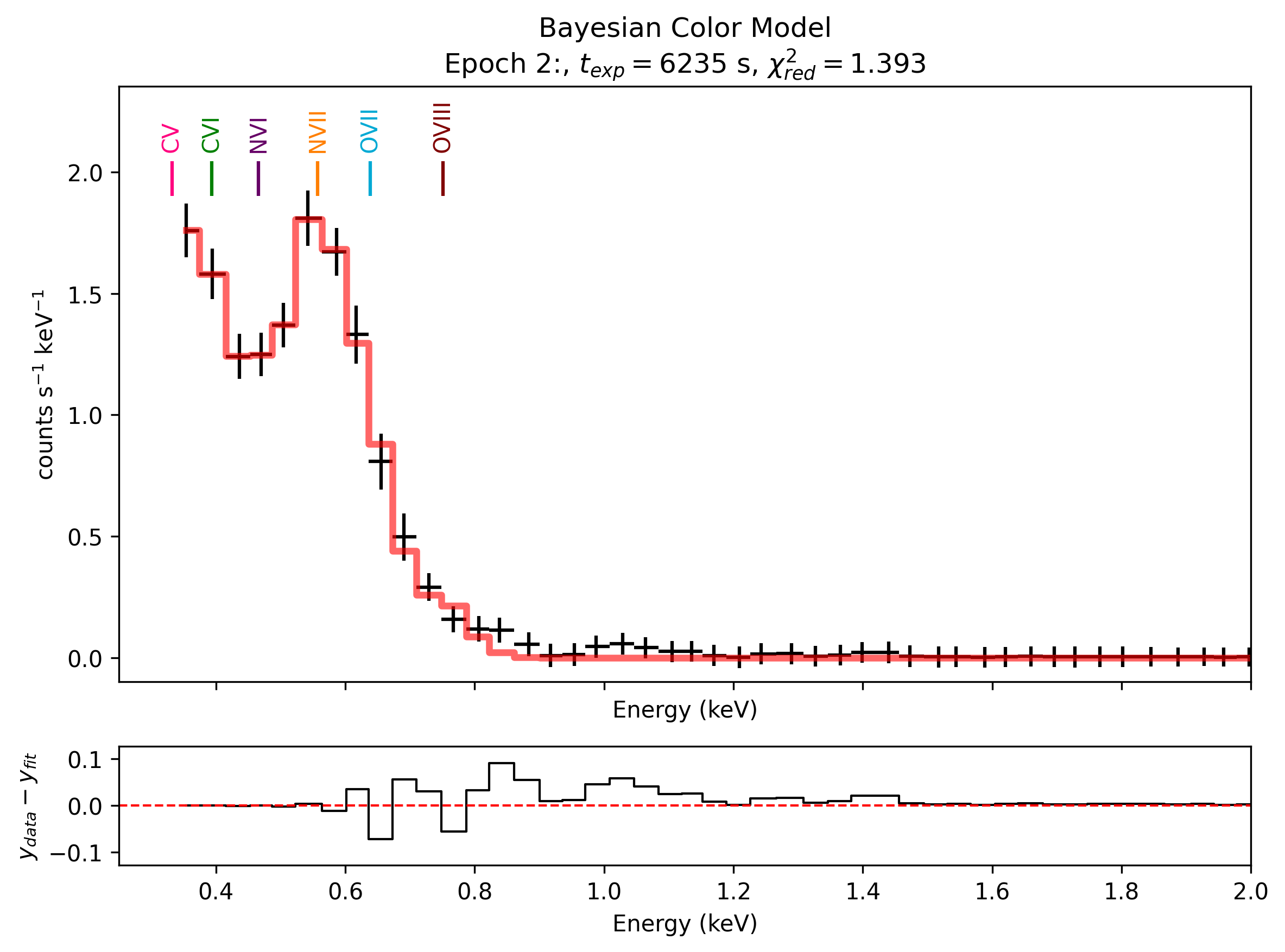}
        \caption{Epoch 2 ($t_{exp}=6235$\,ks).}
        \label{fig:spec_19p_e2-h2o}
    \end{subcaptionblock}
    \hfill
    \begin{subcaptionblock}[b]{0.47\textwidth}
        \includegraphics[width=\textwidth,trim={0.35cm 0.25cm 0.2cm 1.35cm},clip]{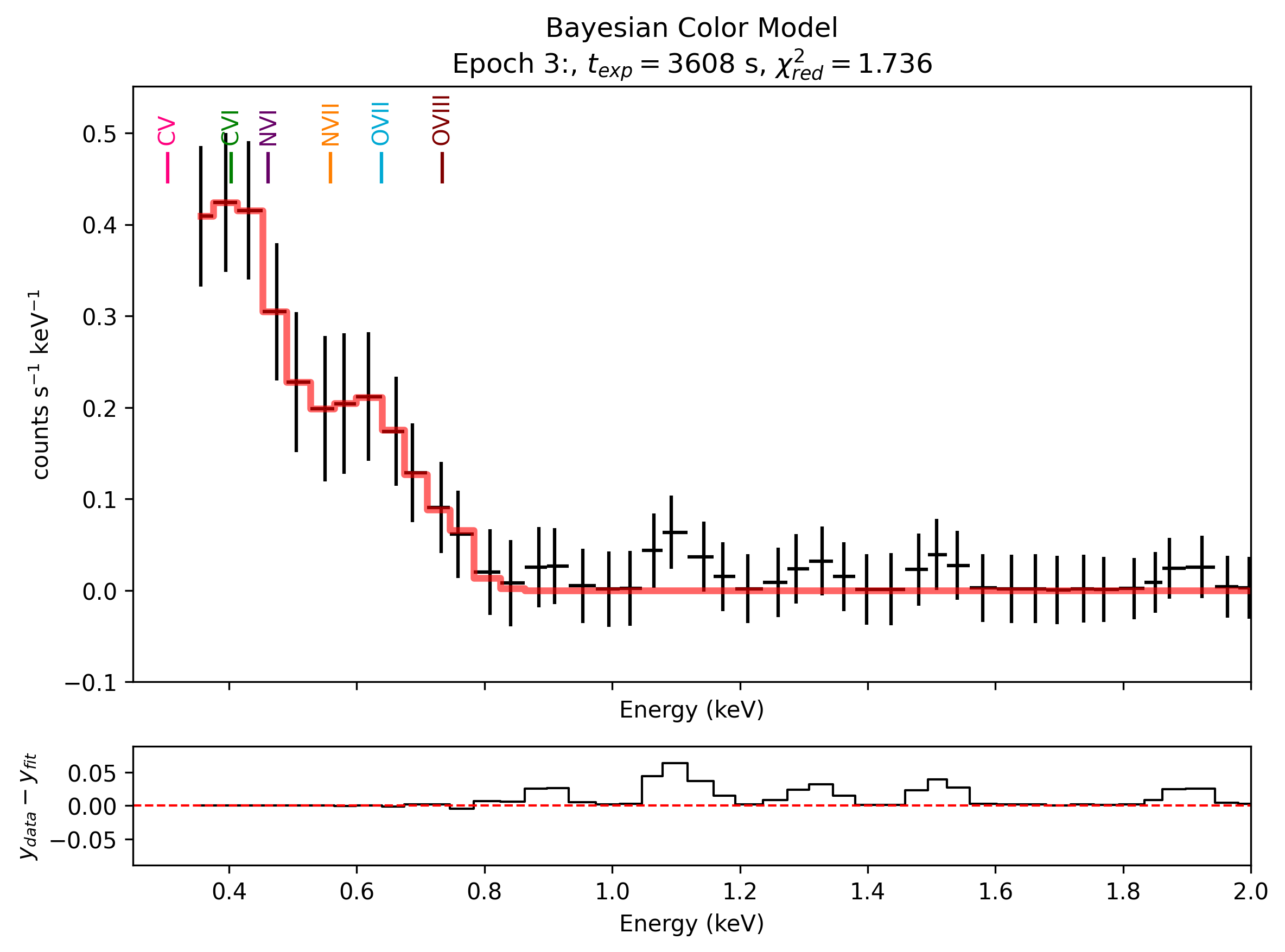}
        \caption{Epoch 3 ($t_{exp}=3608$\,ks).}
        \label{fig:spec_19p_e3-h2o}
    \end{subcaptionblock}
    \hfill
    \begin{subcaptionblock}[b]{0.47\textwidth}
        \includegraphics[width=\textwidth,trim={0.35cm 0.25cm 0.2cm 1.35cm},clip]{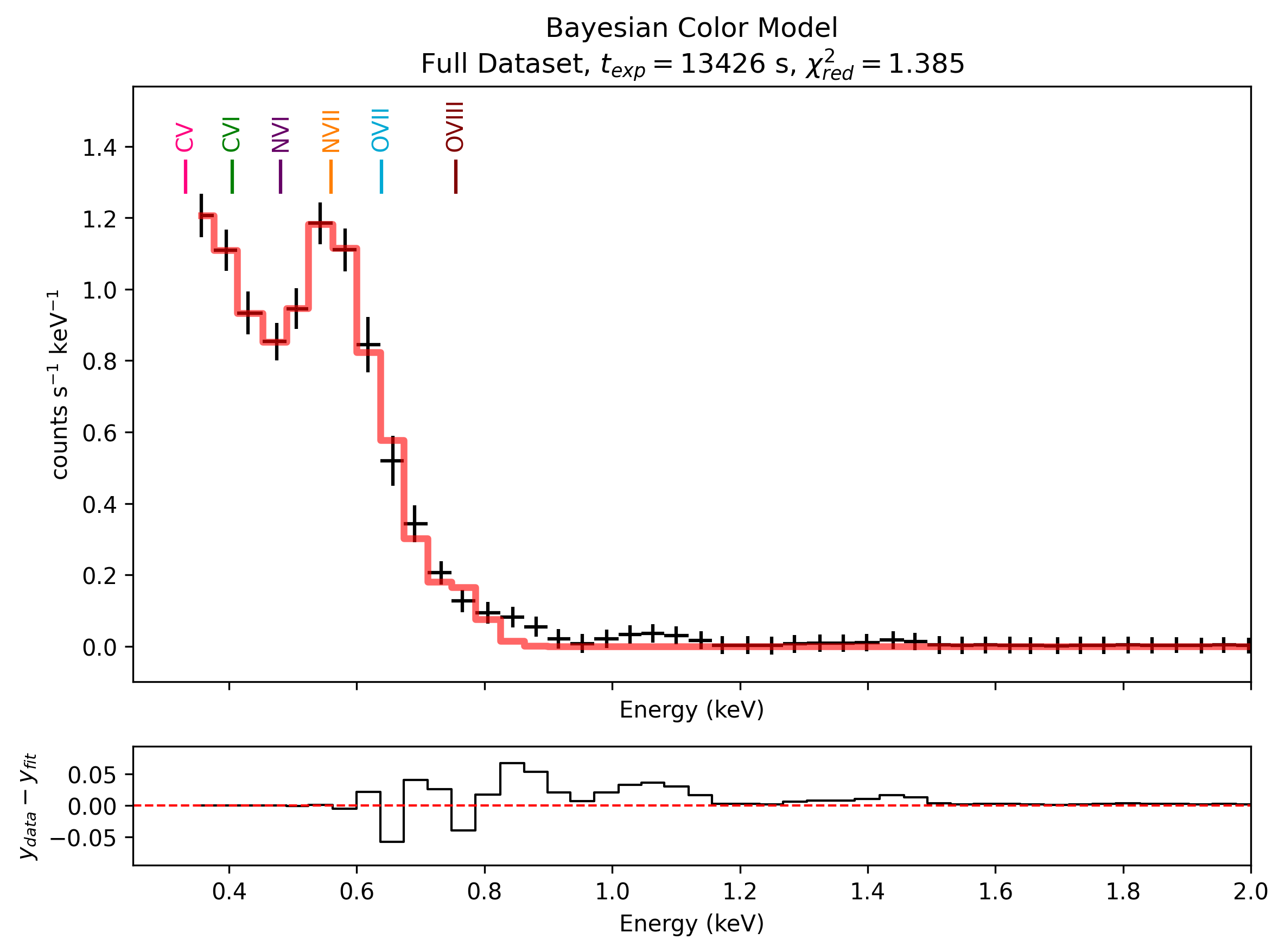}
        \caption{Full dataset ($t_{exp}=13426$\,ks).}
        \label{fig:spec_19p_full-h2o}
    \end{subcaptionblock}

    \caption{X-ray spectra from the interaction between the solar wind and the atmosphere of 19P/Borrelly fitted with the Bayesian Color Model (BCM). (Neutral: H$_2$O).}
    \label{fig:app-spec-19p-h2o}
\end{figure*}


\begin{figure*}
    \centering
    \begin{subcaptionblock}[b]{0.47\textwidth}
        \includegraphics[width=\textwidth,trim={0.35cm 0.25cm 0.2cm 1.35cm},clip]{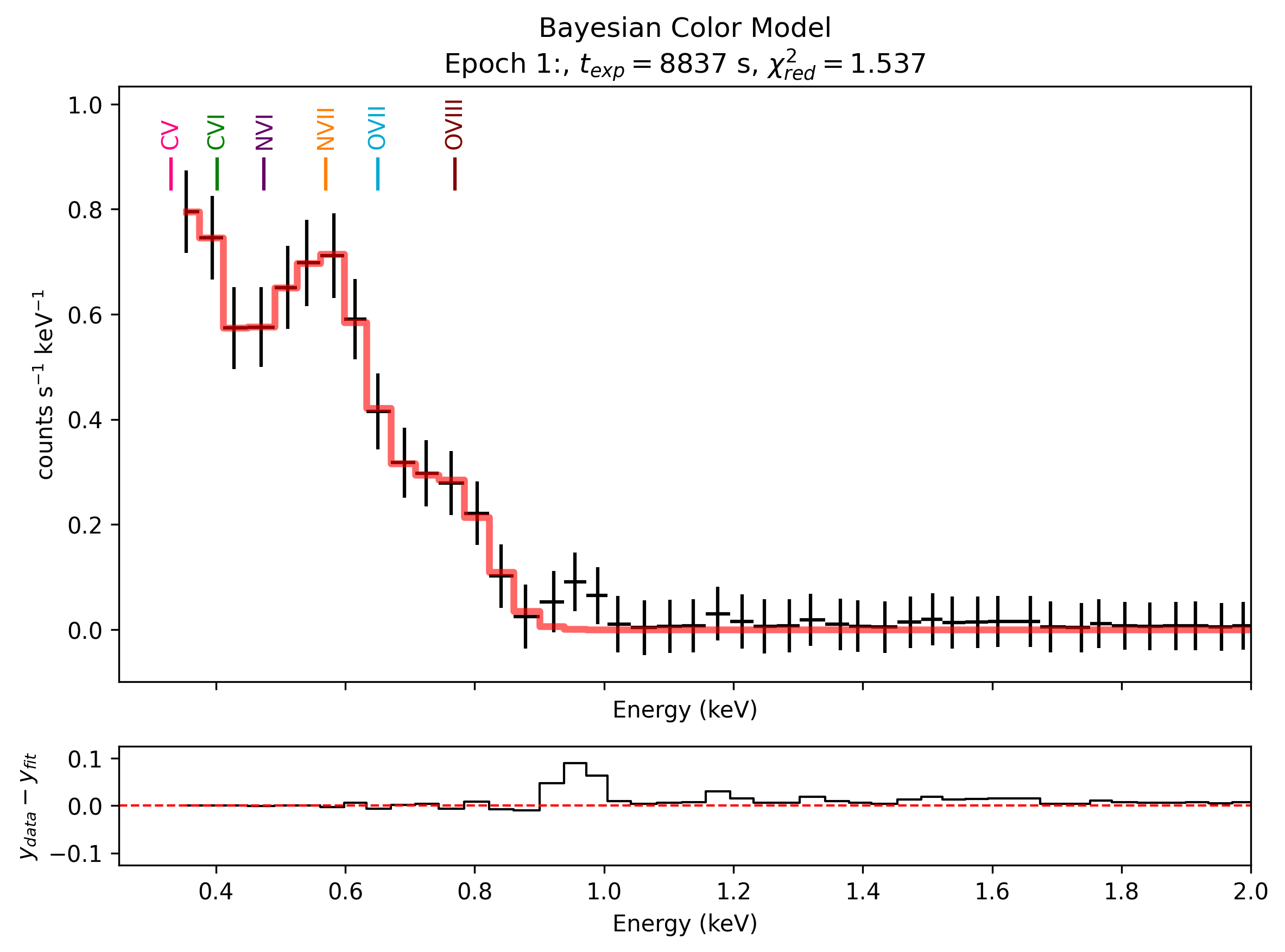}
        \caption{Epoch 1 ($t_{exp}=8837$\,ks).}
        \label{fig:spec_k2_e1-co2}
    \end{subcaptionblock}
    \hfill
    \begin{subcaptionblock}[b]{0.47\textwidth}
        \includegraphics[width=\textwidth,trim={0.35cm 0.25cm 0.2cm 1.35cm},clip]{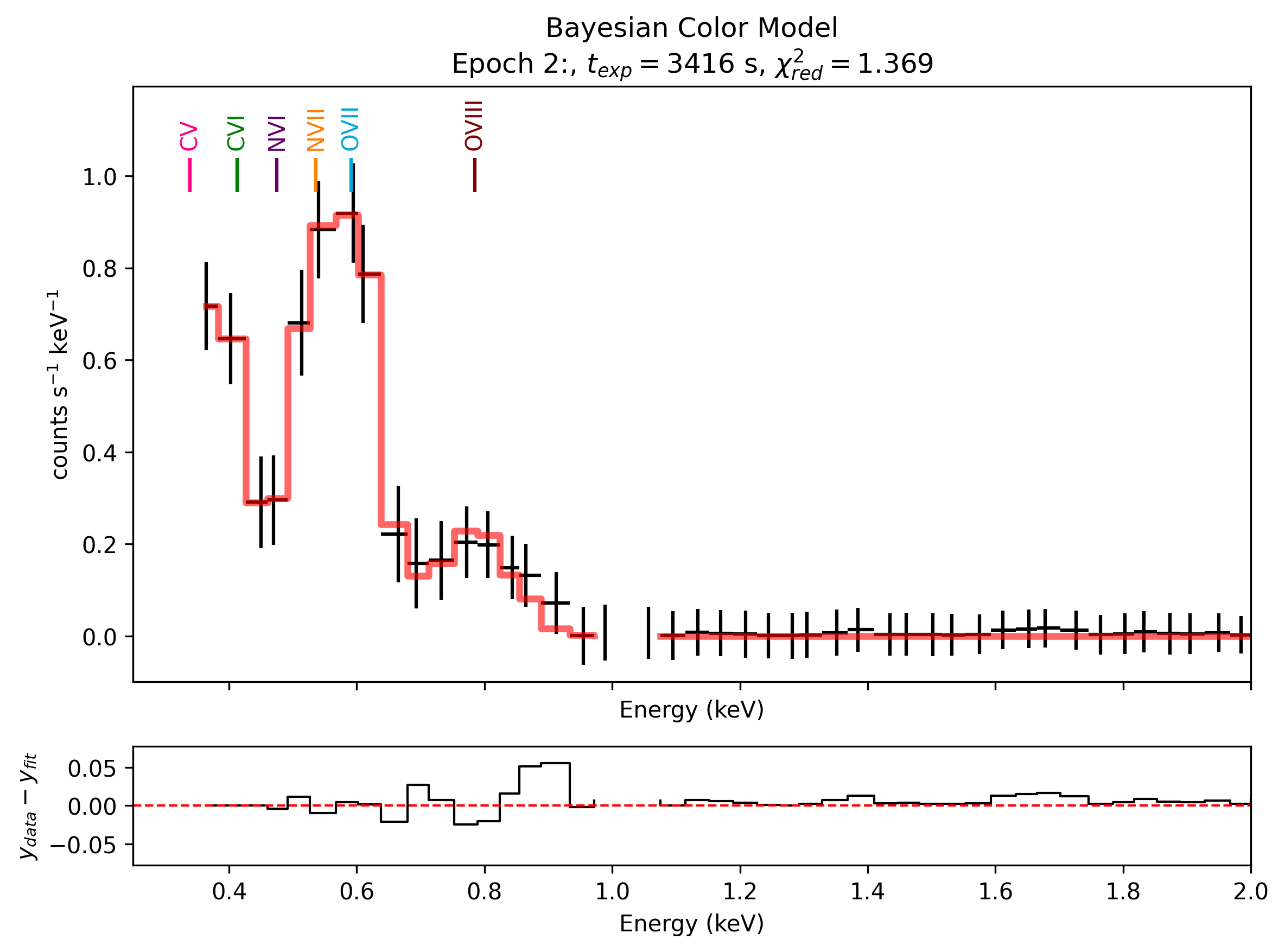}
        \caption{Epoch 2 ($t_{exp}=3416$\,ks).}
        \label{fig:spec_k2_e2-co2}
    \end{subcaptionblock}

    \begin{subcaptionblock}[b]{0.47\textwidth}
        \includegraphics[width=\textwidth,trim={0.35cm 0.25cm 0.2cm 1.35cm},clip]{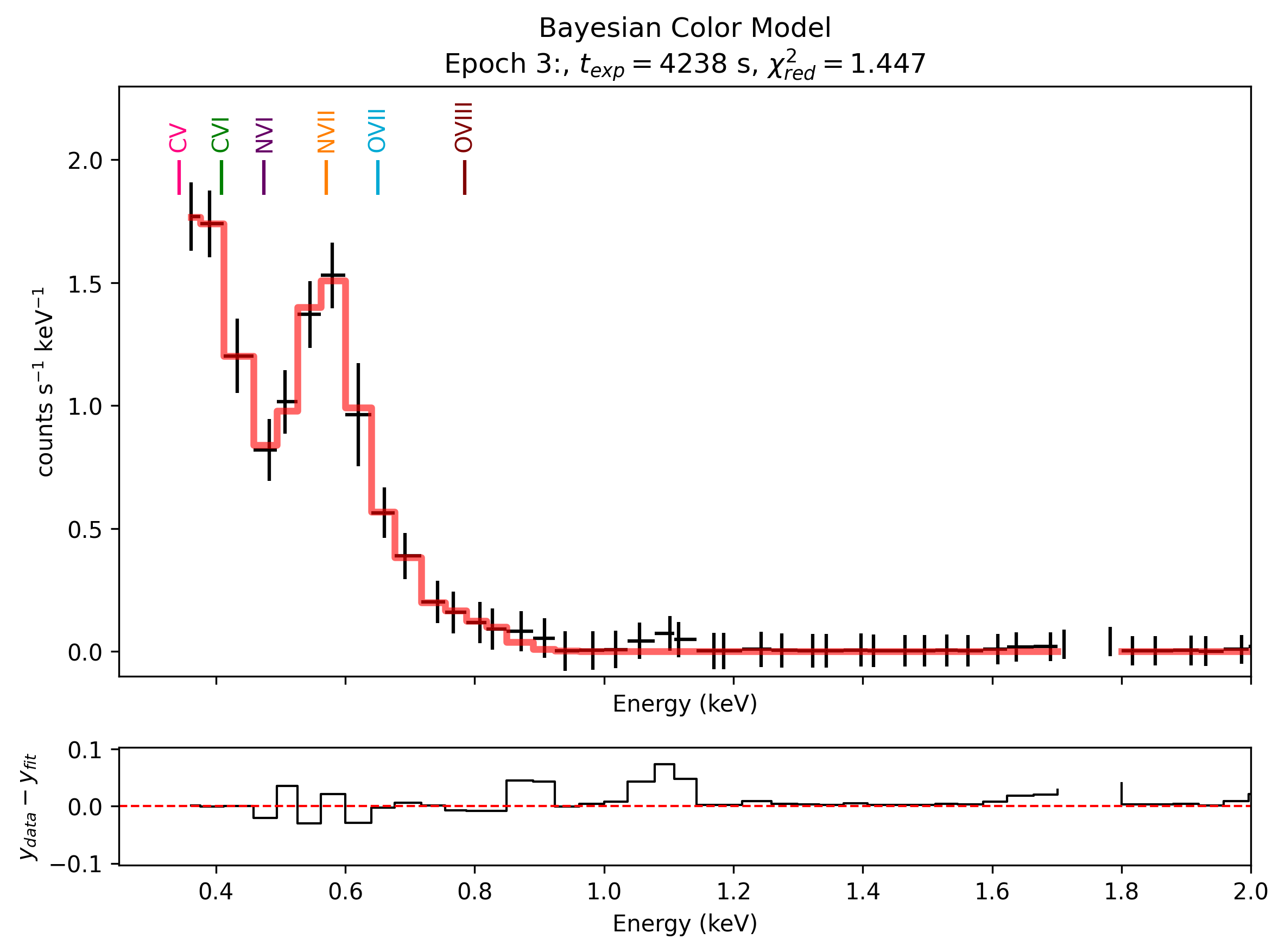}
        \caption{Epoch 3 ($t_{exp}=4238$\,ks).}
        \label{fig:spec_k2_e3-co2}
    \end{subcaptionblock}
    \hfill
    \begin{subcaptionblock}[b]{0.47\textwidth}
        \includegraphics[width=\textwidth,trim={0.35cm 0.25cm 0.2cm 1.35cm},clip]{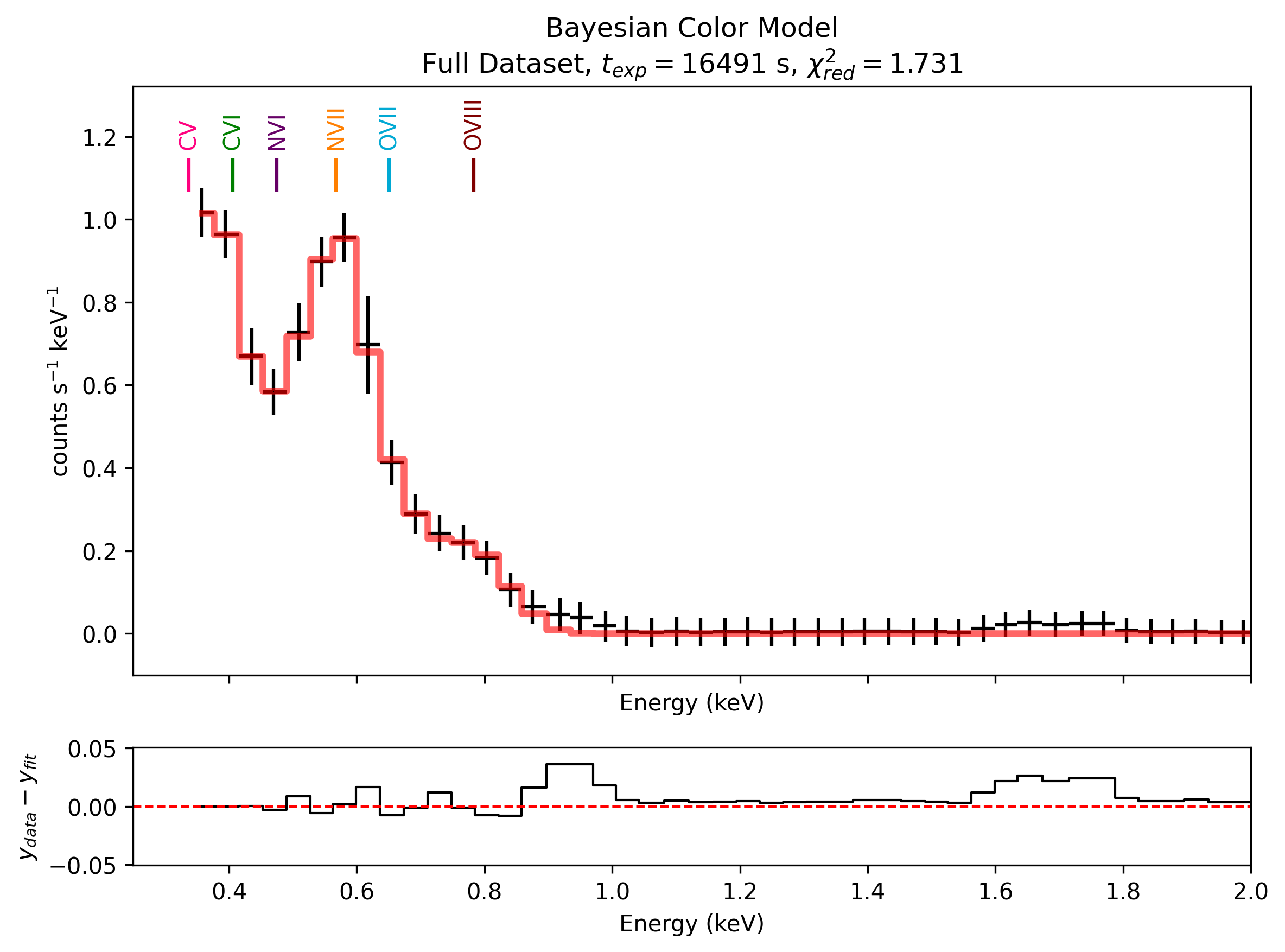}
        \caption{Full dataset ($t_{exp}=16491$\,ks).}
        \label{fig:spec_k2_full-co2}
    \end{subcaptionblock}

    \caption{X-ray spectra from the interaction between the solar wind and the atmosphere of C/2017 K2 (PANSTARRS) fitted with the Bayesian Color Model (BCM). (Neutral: CO$_2$).}
    \label{fig:app-spec-k2-co2}
\end{figure*}


\begin{figure*}
    \centering
    \begin{subcaptionblock}[b]{0.47\textwidth}
        \includegraphics[width=\textwidth,trim={0.35cm 0.25cm 0.2cm 1.35cm},clip]{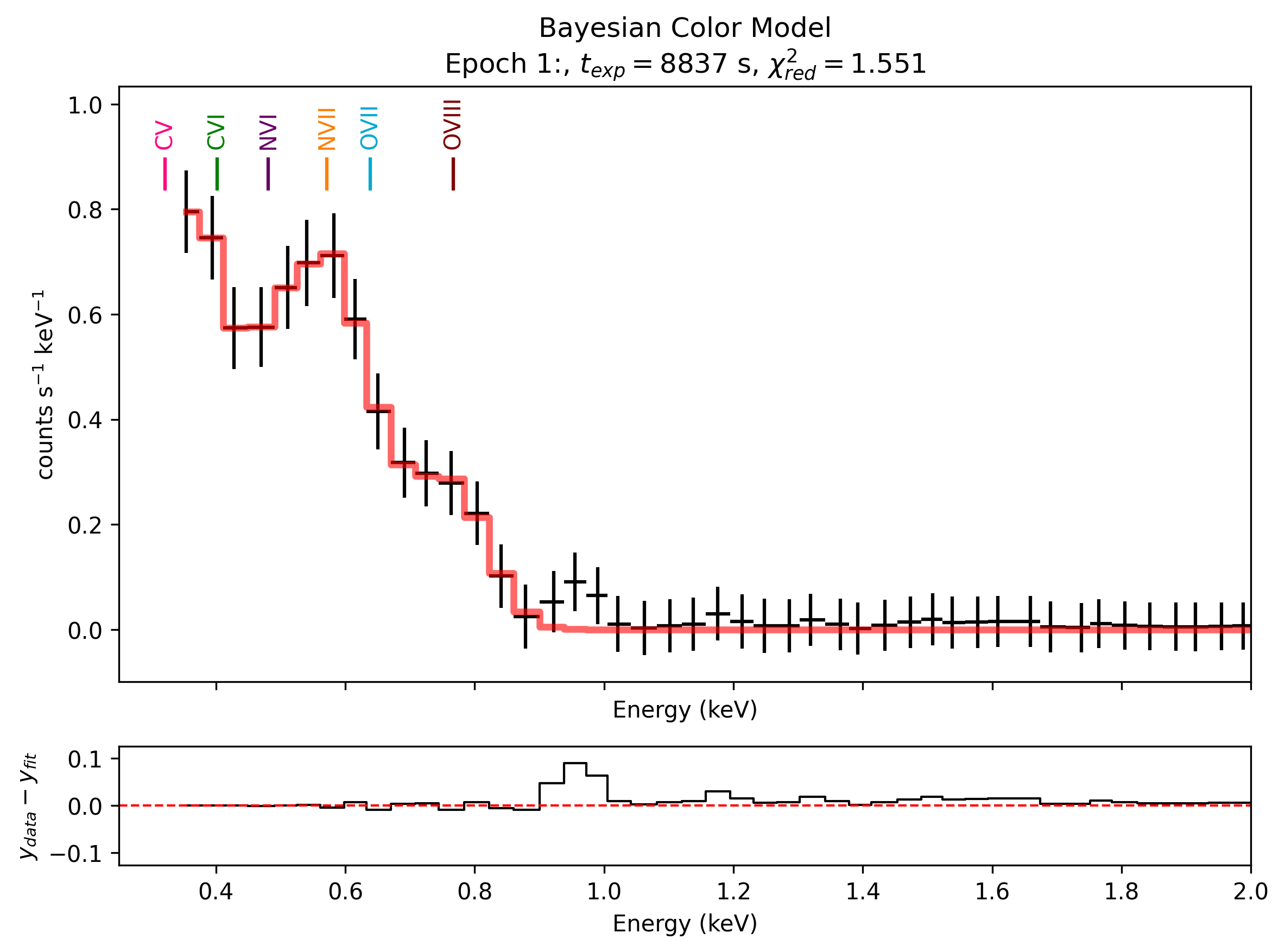}
        \caption{Epoch 1 ($t_{exp}=8837$\,ks).}
        \label{fig:spec_k2_e1-h2o}
    \end{subcaptionblock}
    \hfill
    \begin{subcaptionblock}[b]{0.47\textwidth}
        \includegraphics[width=\textwidth,trim={0.35cm 0.25cm 0.2cm 1.35cm},clip]{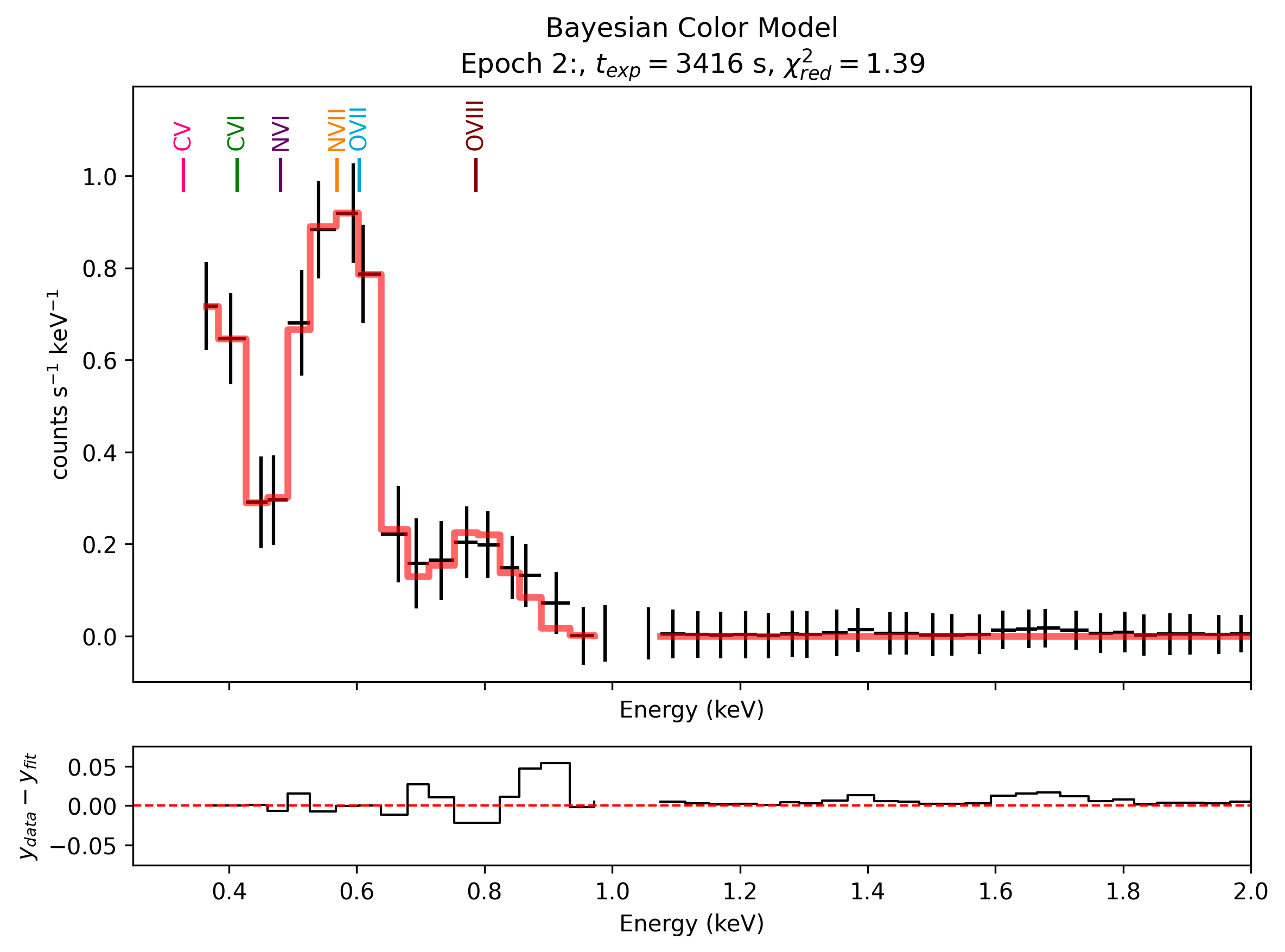}
        \caption{Epoch 2 ($t_{exp}=3416$\,ks).}
        \label{fig:spec_k2_e2-h2o}
    \end{subcaptionblock}

    \begin{subcaptionblock}[b]{0.47\textwidth}
        \includegraphics[width=\textwidth,trim={0.35cm 0.25cm 0.2cm 1.35cm},clip]{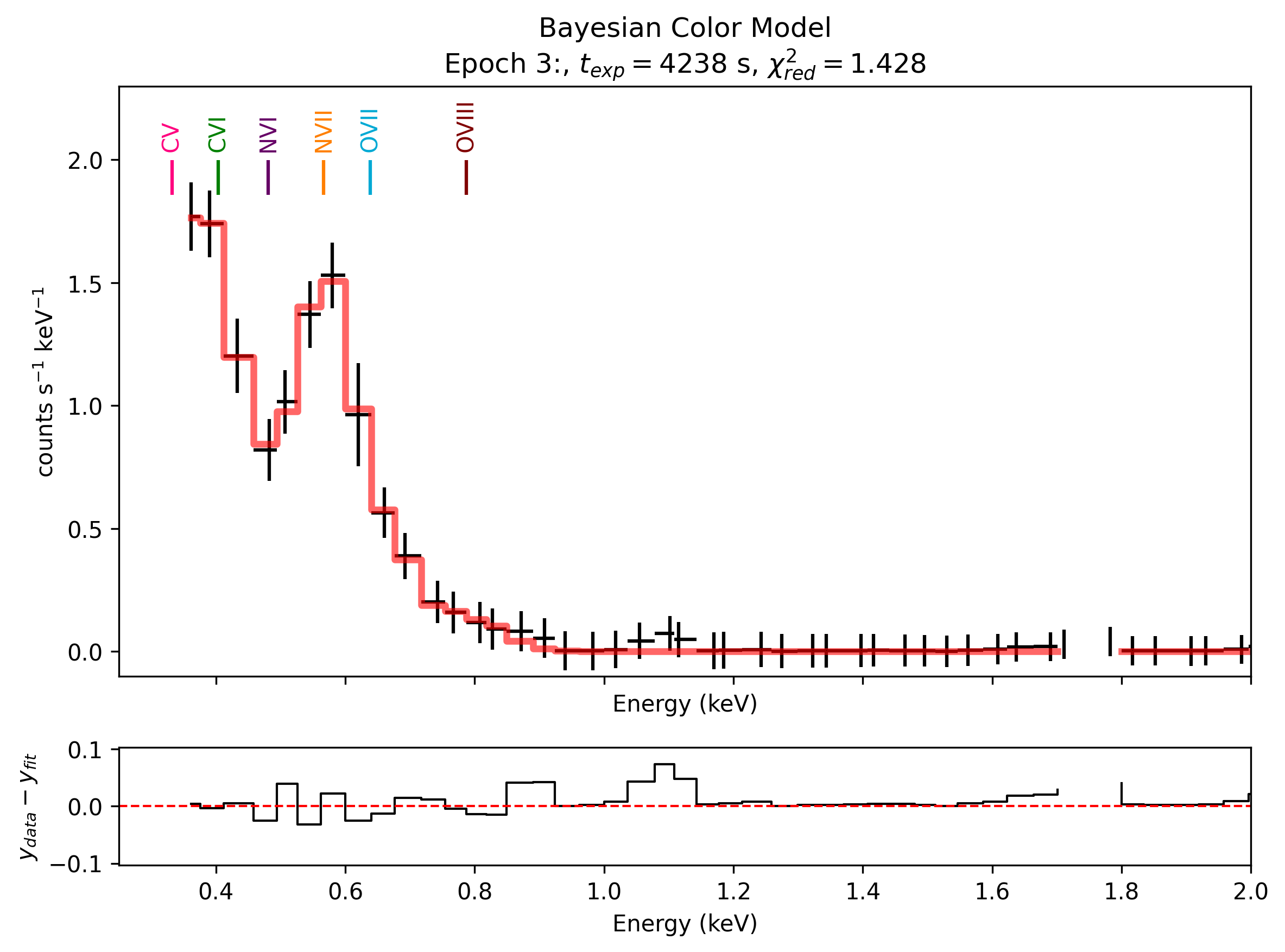}
        \caption{Epoch 3 ($t_{exp}=4238$\,ks).}
        \label{fig:spec_k2_e3-h2o}
    \end{subcaptionblock}
    \hfill
    \begin{subcaptionblock}[b]{0.47\textwidth}
        \includegraphics[width=\textwidth,trim={0.35cm 0.25cm 0.2cm 1.35cm},clip]{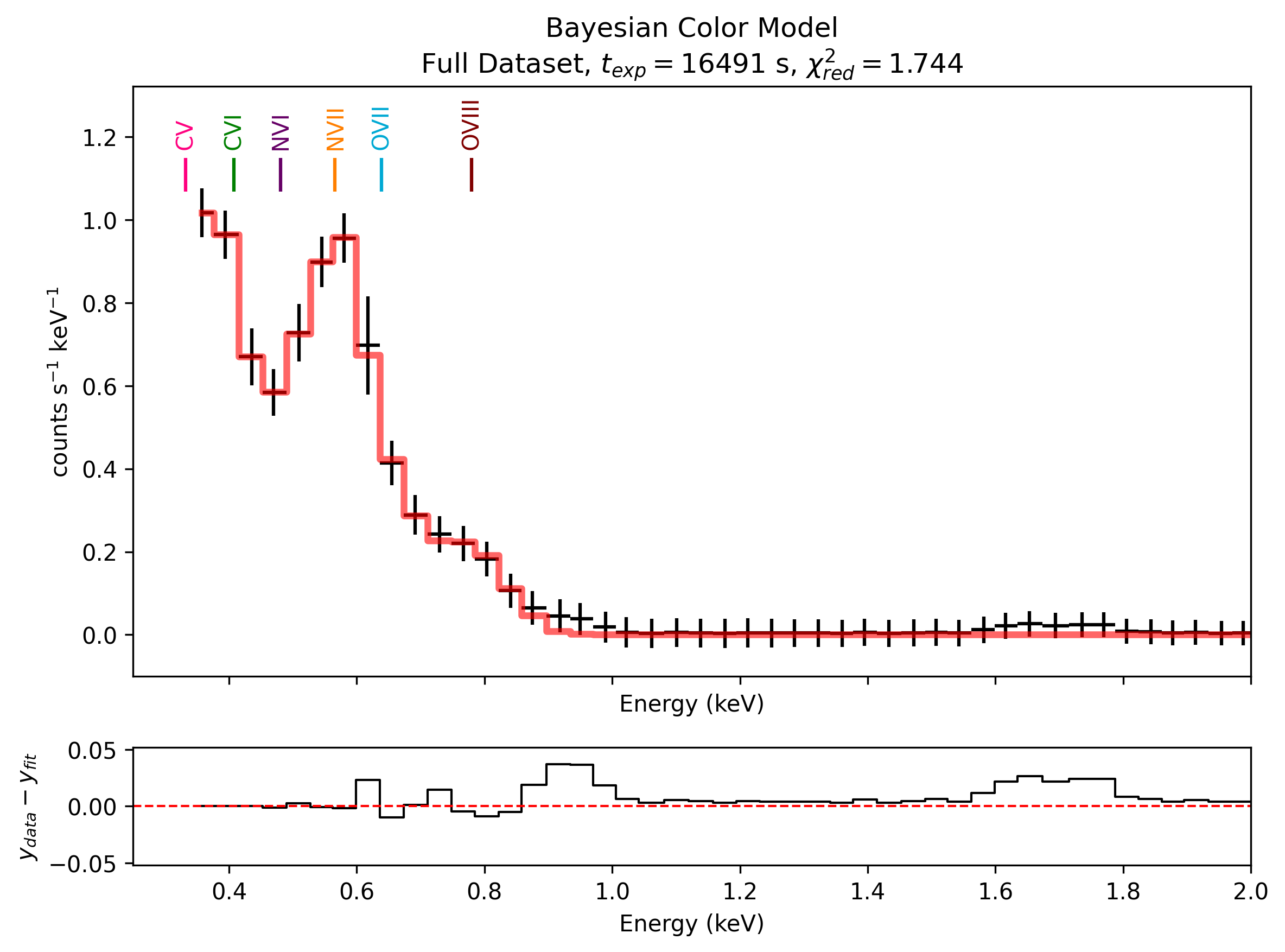}
        \caption{Full dataset ($t_{exp}=16491$\,ks).}
        \label{fig:spec_k2_full-h2o}
    \end{subcaptionblock}

    \caption{X-ray spectra from the interaction between the solar wind and the atmosphere of C/2017 K2 (PANSTARRS) fitted with the Bayesian Color Model (BCM). (Neutral: H$_2$O).}
    \label{fig:app-spec-k2-h2o}
\end{figure*}

\clearpage
\printcredits

\end{document}